\documentclass[aps,prb,twocolumn,groupedaddress,floatfix,showpacs,superscriptaddress]{revtex4}

\usepackage{graphicx}
\usepackage{latexsym}
\usepackage{amsmath,amssymb}
\usepackage{bm}
\DeclareMathOperator{\sgn}{sgn}
\newcommand{\vex}[1]{\bm{\mathrm{#1}}}

\begin{document}

\title{
Metal-insulator transition from combined disorder and interaction effects in
Hubbard-like electronic lattice models with random hopping
}

\author{Matthew S. Foster}
\email{foster@phys.columbia.edu}
\affiliation{Department of Physics, Columbia University, New York, NY 10027}
\author{and Andreas W. W. Ludwig}
\affiliation{Department of Physics, University of California, Santa
Barbara, CA 93106}

\date{\today}

\begin{abstract} 
We uncover a disorder-driven instability in the diffusive Fermi liquid 
phase of a class of many-fermion systems, indicative of a metal-insulator 
transition of first order type, which arises solely from the competition 
between quenched disorder and interparticle interactions. 
Our result is expected to be relevant for sufficiently strong disorder in $d = 3$ 
spatial dimensions. Specifically, we study a class of half-filled, Hubbard-like models 
for spinless fermions with (complex) random hopping and short-ranged interactions 
on bipartite lattices, in $d \geq 2$. In a given realization, the hopping disorder 
breaks time reversal invariance, but preserves the special ``nesting'' symmetry 
responsible for the charge density wave instability of the ballistic Fermi liquid. 
This disorder may arise, e.g., from the application of a random magnetic field to 
the otherwise \emph{clean} model. We derive a low energy effective field theory 
description for this class of disordered, interacting fermion systems, which takes 
the form of a Finkel'stein non-linear sigma model (FNL$\sigma$M) [A. M. Finkel'stein, 
Zh. Eksp. Teor. Fiz. \textbf{84}, 168 (1983), Sov. Phys. JETP \textbf{57}, 97 
(1983)]. We analyze the FNL$\sigma$M using a perturbative, one-loop 
renormalization group analysis controlled via an $\epsilon$-expansion in 
$d = 2 + \epsilon$ dimensions. 
We find that, in $d=2$ dimensions, the interactions destabilize the conducting phase
known to exist in the disordered, non-interacting system.
The metal-insulator transition that we identify in $d>2$ dimensions ($\epsilon >0$) 
occurs for disorder strengths of order $\epsilon$, and is therefore perturbatively 
accessible for $\epsilon \ll 1$. 
We emphasize that the disordered system has no localized phase
in the absence of interactions, so that a localized phase, and the transition into it,
can only appear due to the presence of the interactions. 
\end{abstract}


\maketitle

\section{Introduction}\label{Intro}

In many experimental situations, the effects of \emph{both} static (`quenched') disorder and 
interparticle interactions may play comparatively important roles. 
These issues have, for example, come again to the forefront of debate 
in view of discussions centered around the	 
fascinating, yet still controversial
``metal-insulator transition''  observed in $2\textrm{D}$ semiconductor inversion 
layers.\cite{2DMIT}
Unfortunately, theoretical descriptions of quantum many-particle systems 
incorporating
both disorder and interactions are typically quite 
challenging, and 
difficult to analyze reliably.

With this state of affairs in mind, we apply in this paper a powerful 
analytical technique, known as the Finkel'stein non-linear sigma model 
(FNL$\sigma$M) formalism,\cite{Finkelstein,BK} to a model of spinless 
lattice fermions subject simultaneously to 
both, static
disorder
and short-ranged interactions. The isolated effects of 
disorder or interactions upon the lattice model 
that
we
consider
are already 
well-understood, yet still non-trivial. Our goal is to gain insight into 
the interplay possible between the Fermi liquid, Mott insulating, and 
what we term ``Anderson-Mott''
\cite{footnote-a}
insulating phases of 
interacting many-body Fermi systems
in spatial dimensions $d \geq 2$.
The FNL$\sigma$M formalism 
admits a renormalization group
analysis
controlled via an 
$\epsilon$-expansion in $d = 2 + \epsilon$ dimensions, 
permitting us to address the following general questions 
within the context of our model:
in the simultaneous presence of 
both, disorder and interactions,
({\bf{i}}) does 
a conducting phase
occur in $d=2$, 
and
({\bf{ii}})
is there
a metal-insulator transition (MIT) in dimensions $d>2$? 
If yes, what is its nature?

The primary purpose of this paper is to present a detailed derivation
and a thorough
discussion of 
results previously announced
briefly in Ref.~\onlinecite{AIIIshortpaper}. 
A short summary of these results and an outline of this work 
appears below in Sec.~\ref{Outline}.

\subsection{Interactions and sublattice symmetry}\label{SLSNestingCDW}

\begin{figure}
\includegraphics[width=0.2\textwidth]{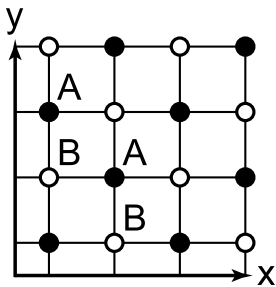}
\caption{The square lattice. Labels `$A$' and `$B$' denote sites belonging to the
black and blank square sublattices, respectively.
\label{SquareLattice}}
\end{figure}

We study a class of ``Hubbard-like'' models\cite{Auerbach} for spinless 
fermions \emph{at half filling} on bipartite lattices, possessing 
short-ranged interparticle interactions and quenched disorder. We work in 
$d\geq 2$ spatial dimensions throughout. For concreteness, we 
consider
in this introduction
the hypercubic lattice.
By definition, any 
bipartite lattice may be subdivided into two interpenetrating sublattices, 
which we will distinguish with the labels `$A$' and `$B$.' The $d=2$ 
dimensional example of the square lattice is depicted in 
Fig.~\ref{SquareLattice}. 

Our starting point is the clean (zero-disorder),
generalized
Hubbard-like 
Hamiltonian
\begin{align}\label{Hclean}
	H_{0} =& - t \sum_{\langle i j \rangle}
	c_{A i}^{\dagger} c_{B j}^{\phantom{\dagger}} + \mathrm{H.c.} 
	+ V \sum_{\langle i j \rangle} 
	\delta\hat{n}_{A i}
	\delta\hat{n}_{B j} \nonumber \\
	& + U \bigl (
	\sum_{\langle\langle i i' \rangle\rangle} 
	\delta\hat{n}_{A i} \delta\hat{n}_{A i'} 
	  + 
	\sum_{\langle\langle j j' \rangle\rangle} 
	\delta\hat{n}_{B j} \delta\hat{n}_{B j'}
	\bigl )
	,
\end{align}
where $c_{A i}^{\dagger}$ and $c_{B j}^{\phantom{\dagger}}$ are creation 
and annihilation operators for spinless fermions on the $A$ and $B$ 
sublattices of the bipartite lattice, respectively. Here, $i$ and $j$ 
respectively index the $A$ and $B$ sublattice sites, and the sums on 
$\langle i j \rangle$ run over all nearest neighbor $A$-$B$ lattice bonds, 
while the sums on $\langle\langle i i' \rangle\rangle$ and 
$\langle\langle j j' \rangle\rangle$ run over all next-nearest neighbor 
(same sublattice) pairs of sites. The homogeneous hopping amplitude $t$ is 
taken to be real. The operators 
$\delta\hat{n}_{A/B} \equiv (c_{A/B}^{\dagger} c_{A/B}^{\phantom{\dagger}} - \frac{1}{2}) $ 
denote 
deviations
of the local sublattice
($A$ or $B$)
fermion densities from their 
value at half filling. Finally, the interaction strengths $V$ and $U$ 
appearing in Eq.~(\ref{Hclean}) couple to nearest neighbor and next-nearest 
neighbor density-density interactions, respectively.

\begin{figure}[b]
\includegraphics[width=0.3\textwidth]{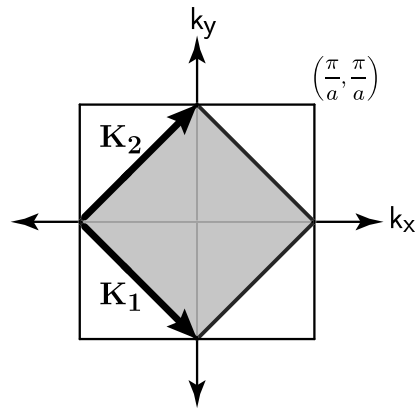}
\caption{Brillouin zone (BZ) associated with the square lattice shown in Fig.~\ref{SquareLattice}. 
The shaded subregion is the sublattice BZ (sBZ) appropriate to the $A$ and $B$ sublattices. This 
subregion also indicates the Fermi sea at half filling, an attribute particular to the square 
lattice model. The nesting wavevectors $\bm{\mathrm{K}_{1}}$ and $\bm{\mathrm{K}_{2}}$ 
are reciprocal lattice vectors for the sBZ.\label{BZandNesting}}
\end{figure}

The 
model
at half-filling,
 given by Eq.~(\ref{Hclean}),
 possesses
the following
special symmetry, 
which we refer to here as ``sublattice'' 
symmetry (SLS)
[this symmetry is 
termed ``chiral'' in the classification scheme of Ref.~\onlinecite{Zirnbauer}
(see also 
Refs.~\onlinecite{1DChiral,Gade,LeeFisher,Furusaki,FC,GLL,RyuHatsugai,MDH2D,BocquetChalker,MRF,BDIpaper})]:
the Hubbard-like 
Hamiltonian in Eq.~(\ref{Hclean}) is invariant under the transformation
\begin{equation}\label{SLS}
	c_{A i}^{\phantom{\dagger}} \rightarrow c_{A i}^{\dagger},
	\qquad
	c_{B j}^{\phantom{\dagger}} \rightarrow - c_{B j}^{\dagger},
\end{equation}
where we simultaneously complex conjugate all
scalar terms in the Hamiltonian. 
[This transformation, like that of time-reversal, is 
antiunitary. In the presence of time-reversal invariance (TRI), SLS is 
equivalent to the usual particle-hole symmetry.] As is well known, the 
Fermi surface\cite{footnote-b}
of the half-filled, non-interacting model, Eq.~(\ref{Hclean}) with $U = V = 0$, 
possesses perfect ``nesting,''
\begin{equation}\label{Nesting}
	\varepsilon(\bm{\mathrm{k}}+\bm{\mathrm{K_{N}}}) = - \varepsilon(\bm{\mathrm{k}}),
\end{equation}
where $\varepsilon(\bm{\mathrm{k}})$ is the non-interacting bandstructure, 
and $\bm{\mathrm{K_{N}}}$ is a nesting wavevector. For the hypercubic lattice 
with lattice spacing $a = 1$, $\bm{\mathrm{K_{N}}}$ takes the form
\begin{equation}\label{NestingK}
	\bm{\mathrm{K_{N}}} \equiv \pi (n_{1}, n_{2}, \ldots, n_{d}),
\end{equation}
where the numbers $n_{s} = \pm 1$, with $s \in \{1, \ldots, d\}$.
Fig.~\ref{BZandNesting} depicts the Brillouin zone (BZ) associated with the 
square lattice shown in Fig.~\ref{SquareLattice}. The set of nesting wavevectors 
$\{\bm{\mathrm{K_{N}}}\}$ defined in Eq.~(\ref{NestingK}) span the 
\emph{sublattice} Brillouin zone (sBZ), appropriate to the $A$ and $B$ 
sublattices of the composite bipartite lattice. For the special case of 
the square lattice, the boundary of the sBZ also serves as the Fermi line 
at half filling, shown in Fig.~\ref{BZandNesting}.

Nesting and SLS are tied together. Under the transformation given by 
Eq.~(\ref{SLS}), the hopping part of the Hamiltonian in Eq.~(\ref{Hclean}) transforms as
\begin{align}\label{SLSandNesting}
	\int_{\textrm{BZ}} \frac{d^{d} \bm{\mathrm{k}}}{(2 \pi)^{d}} \,
	\varepsilon(\bm{\mathrm{k}}) 
	c^{\dagger}_{\bm{\mathrm{k}}} c^{\phantom{\dagger}}_{\bm{\mathrm{k}}}
	& \rightarrow
	\int_{\textrm{BZ}} \frac{d^{d} \bm{\mathrm{k}}}{(2 \pi)^{d}} \,
	\varepsilon^{*}(\bm{\mathrm{k}}) 
	c^{\phantom{\dagger}}_{\bm{\mathrm{k}}- \bm{\mathrm{K_{N}}}} 
	c^{\dagger}_{\bm{\mathrm{k}}- \bm{\mathrm{K_{N}}}} \nonumber\\
	& = \int_{\textrm{BZ}} \frac{d^{d} \bm{\mathrm{k}}}{(2 \pi)^{d}} \,
	\left[ - \varepsilon(\bm{\mathrm{k}}+\bm{\mathrm{K_{N}}}) \right] 
	c^{\dagger}_{\bm{\mathrm{k}}} c^{\phantom{\dagger}}_{\bm{\mathrm{k}}},
\end{align}
where $\bm{\mathrm{K_{N}}}$ is a nesting wavevector as in Eq.~(\ref{NestingK}), and 
\begin{equation}\label{cFT}
	c^{\phantom{\dagger}}_{\bm{\mathrm{k}}} \equiv \sum_{i \in A} 
	e^{-i \bm{\mathrm{k}} \cdot \bm{\mathrm{R_{i}}}} c_{A i}^{\phantom{\dagger}} 
	+ \sum_{j \in B} e^{-i \bm{\mathrm{k}} \cdot \bm{\mathrm{R_{j}}}} c_{B j}^{\phantom{\dagger}}.
\end{equation}

Fermi surface nesting is, in a sense, the \emph{defining} property of Hubbard-like 
models for interacting lattice fermions in $d\geq2$ dimensions. It is the
nesting condition which makes the ballistic Fermi liquid phase at half filling in such models 
unstable to Mott insulating order in the presence of generic, arbitrarily weak interparticle 
interactions.\cite{Hirsch,GSST,Shankar} ``Nesting'' instabilities can occur for 
microscopically repulsive
interparticle interaction strengths, and arise through the exchange 
of the nesting momenta $\{\bm{\mathrm{K_{N}}}\}$ through the particle-hole channel of these 
interactions. Such models may also exhibit the BCS superconducting instability,\cite{Schrieffer}
which exists for any TRI Fermi liquid with (effectively) attractive pairing interactions. 
The 
ground state
of such a half-filled, Hubbard-like model with weak, but 
non-vanishing interactions is typically a Mott insulator, with lattice translational 
symmetry spontaneously broken at the nesting wavelength, a superconductor, or a mixture 
of these, such as a ``supersolid.''\cite{Hirsch,SLGMWSSD,MSSWB,Shankar}       

A simple calculation\cite{GSST} 
using
the random phase approximation (RPA) predicts that the Fermi 
liquid phase of the
(not disordered)
model defined by Eq.~(\ref{Hclean}) is unstable to charge density wave 
(CDW) order for any $V > U \geq 0$. The CDW state is a Mott insulator, in which a greater 
proportion of the fermion density resides on one sublattice than the other. For the 
$2\textrm{D}$ case of the square lattice, the RPA calculation predicts a transition to 
the CDW state at a temperature\cite{footnote-c}
\begin{equation}\label{CDWTemp}
	T_{c} \sim 2 t \exp\left[- \pi \sqrt{\frac{t}{V - U}}\right],
\end{equation}
in the weak coupling limit $0 \leq U < V \ll t$.
Alternatively, a one-loop renormalization group (RG) calculation,\cite{Shankar} performed upon 
a low-energy effective field theory description of the model given by 
Eq.~(\ref{Hclean}), shows that the effective CDW interaction strength, which 
we define here as 
\begin{equation}\label{CDWcoupling}
	W_{c} \sim \frac{1}{2}(U - V), 
\end{equation}
(corresponding to a staggered charge density)
grows to large \emph{negative} values under renormalization if its initial value 
$W_{c}^{(0)} < 0$,
i.e.  was negative
to begin with. This run off to strong 
interaction coupling is taken to signal the onset of CDW formation.\cite{footnote-d}
These and analogous results regarding the N\'eel ground state of the half-filled, 
spin-$1/2$ Hubbard model in $d \geq 2$ are well established, and the latter have 
been further confirmed with extensive numerical work. (See e.g.\ Refs.~\onlinecite{Hirsch,MSSWB,3DHubbard}.) 
The spinless Hubbard-like Hamiltonian in Eq.~(\ref{Hclean}), 
on the other hand,
has received less attention in the numerical literature; early Monte Carlo studies\cite{GSST} 
of the $2\textrm{D}$ version of this Hamiltonian, with $U \equiv 0$, show the existence 
of a CDW ground state for positive $V$, but were unable to access the weak coupling 
limit $V \rightarrow 0^{+}$. 

Sublattice symmetry plays a crucial role in establishing the CDW ground state of 
the model in Eq.~(\ref{Hclean}) (for $V > U \geq 0$). Nesting can
also
occur 
in the absence of SLS, since most aspects of Fermi surface geometry typically depend 
strongly upon microscopic details. Under the repeated application of an appropriate 
renormalization group transformation,\cite{Shankar} however, this geometry is expected to deform 
continuously. SLS protects the nesting condition even as other details (such as the 
presence Fermi surface van Hove singularities) mutate through the RG process. The 
persistence of nesting and unbroken SLS, up to the onset of Mott or supersolid order, 
allows arbitrarily weak repulsive interparticle interactions to destabilize the 
Fermi liquid phase.

\subsection{Quenched disorder}\label{RandomHopping}

Now we turn to the incorporation of quenched (static) disorder into the model given by 
Eq.~(\ref{Hclean}). First, note that the addition of on-site (or: ``diagonal'') 
randomness, characterized by a strength $\lambda_{D}$, breaks sublattice 
symmetry [Eq.~(\ref{SLS})] in every realization of disorder.
Thus, turning on diagonal disorder is expected
to destroy the CDW ground state,\cite{VUV,UV} at
least for sufficiently weak interactions. 
[Keeping the disorder strength $\lambda_D$ fixed, for example,
we expect the absence of zero temperature CDW order 
in a window of (small) interaction strengths, e.g.
for 
$0 \leq V < V_{c}(\lambda_{D})$, with $U = 0$)].
The proximity of a Mott insulating phase to the \emph{non}-interacting (ballistic) 
Fermi liquid is the most essential characteristic of the Hubbard-like lattice model 
defined in Eq.~(\ref{Hclean}). We conclude that this characteristic is \emph{lost} upon 
the incorporation of diagonal disorder. Studies of Hubbard-like models subject to 
diagonal disorder include those listed in Refs.~\onlinecite{VUV,UV,BKAHB} and 
Refs.~\onlinecite{Ma,ScalettarD,HeidarianTrivedi} for spinless and spin-$1/2$ lattice 
fermions, respectively.

Instead, we consider the Hamiltonian in Eq.~(\ref{Hclean}), weakly perturbed by 
purely 
``off-diagonal''
 randomness, which is taken to occur only in the intersublattice 
hopping amplitudes. Such a model is invariant under the sublattice symmetry (SLS) 
transformation [Eq.~(\ref{SLS})] for each and every static realization of the disorder. 
We consider \emph{complex} random (nearest-neighbor) hopping that breaks TRI. [Without 
interactions, the model is in the ``chiral'' symmetry class AIII of 
Ref.~\onlinecite{Zirnbauer}.] The full lattice Hamiltonian is given by $H = H_{0} + \delta H$, 
where $H_{0}$ was defined in Eq.~(\ref{Hclean}), and
\begin{equation}\label{Hdis}
	\delta H = -\sum_{\langle i j \rangle} \delta t^{\phantom{*}}_{i, j} \;
	c_{A i}^{\dagger} c_{B j}^{\phantom{\dagger}} + \mathrm{H.c.}
\end{equation}
Here, the
random part of the hopping matrix element,
$\delta t^{\phantom{*}}_{i, j}$,
is taken to be a Gaussian 
complex random variable with zero mean, independent on different lattice links. 
For our system of spinless fermions, this is consistent with the application of a 
random magnetic field to the otherwise \emph{clean} model. 
For non-interacting spin-$1/2$ fermion systems, with nearest neighbor 
hopping on bipartite lattices at half filling, in random orbital and/or Zeeman magnetic fields, 
a general classification is (briefly) mentioned in Sec.~\ref{Spin1/2},
with more details being provided in Appendix \ref{Spin1/2RHopping}.
We will conclude from this classification that the results of this
paper also apply to a related Hubbard-like model of spin-$1/2$ fermions, subject to
an orbital magnetic field and to (homogeneous or random) spin-orbit
coupling.
Independent work on the effects of interparticle
interactions upon these and several other classes of disordered spin-$1/2$ fermions
has very recently appeared in Ref.~\onlinecite{DellAnna}.

\begin{figure}
\includegraphics[width= 0.25\textwidth]{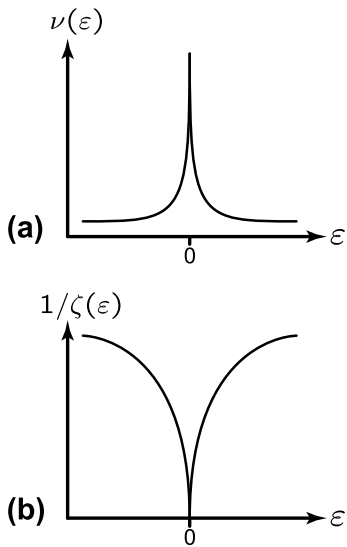}
\caption{
Features of random hopping model (RHM) physics in 2D, in the \emph{absence} of
interparticle
interactions. Subfigures (a) and (b) depict the qualitative energy
($\varepsilon$)-dependence
of the density of states $\nu$ and inverse localization length $1/\zeta$, respectively,
in the chiral\cite{Zirnbauer} orthogonal (BDI) and unitary (AIII) class RHMs in
$\textrm{d} = 2$. Both $\nu$ and $\zeta$ diverge upon approaching the band
center, taken to occur at $\varepsilon = 0$.\cite{Gade,GLL,FC}
\label{FigRandomHopPhys2D}}
\end{figure}

\begin{figure}
\includegraphics[width=0.25\textwidth]{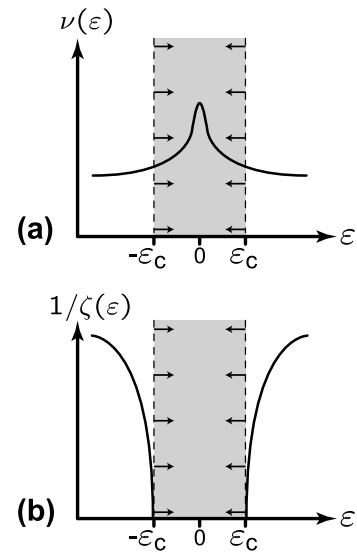}
\caption{
Same as Fig.~\ref{FigRandomHopPhys2D}, but for spatial dimensionalities $d > 2$,
e.g.\ non-interacting random hopping model physics in 3D. Here, the shaded area
represents a band of extended (delocalized) states; $\varepsilon_{\mathsf{c}}$
is the mobility edge. Increasing the strength of the random hopping disorder narrows
the region of extended states in $\textrm{d} > 2$, as indicated by the arrows in
subfigures (a) and (b). The DOS is finite, albeit parametrically enhanced at the
band center.\cite{Gade,FC}
\label{FigRandomHopPhys3D}}
\end{figure}

Our principal motivation for studying the model in Eqs.~(\ref{Hclean}) and (\ref{Hdis})
is 
that,
due to the presence of SLS,
we expect \emph{both} disorder and interparticle interactions to play important 
roles in the description of the low-energy physics. Because random hopping preserves 
the special SLS, our disordered model retains the ``nesting'' CDW instability of the 
associated \emph{clean} system. This instability can therefore compete with the unusual 
localization physics
 arising from
 SLS 
in the
disordered, but non-interacting model (see below).
The further 
assumption of broken TRI guarantees that we do not have to confront an 
additional superconducting 
instability.\cite{BK,JengLudwigSenthilChamon,Hirsch,GSST,Shankar}
We note
that the effects of hopping disorder upon the N\'eel ground state of the 
(slightly more complex) spin-$1/2$ Hubbard model at half filling were studied 
numerically in Refs.~\onlinecite{ScalettarOD1--BDI} and \onlinecite{ScalettarOD2--BDI+C}, 
although these studies were limited to $d = 2$. 

A second motivating factor is that, interestingly,
and as alluded to above,
the presence of SLS radically
changes the localization physics of the disordered, \emph{non-interacting} random
hopping model [Eqs.~(\ref{Hclean}) and (\ref{Hdis}) with $V = U = 0$].
SLS enables the random hopping
model (RHM) to \emph{evade} the phenomenon of Anderson localization. Specifically, the
non-interacting system exhibits a critical, delocalized phase
at the band center (half filling) in one, two, and three dimensions for finite disorder
strength, with a strongly divergent low-energy density of states in
$d
=1,2$.\cite{LeeFisher,1DChiral,Gade,Furusaki,FC,GLL,RyuHatsugai,MDH2D,BocquetChalker,MRF}
In particular, there is no MIT and no Anderson insulating phase in $d = 3$
(in the absence of interactions).
The essential features of random hopping model physics are summarized in
Figs.~\ref{FigRandomHopPhys2D} and \ref{FigRandomHopPhys3D}. These figures
apply, for example, to spinless lattice fermions with sublattice symmetry, 
with or without TRI
(classes BDI and AIII, respectively),\cite{Zirnbauer} in $\textrm{d} \geq 2$.
Upon approaching the band center ($\varepsilon = 0$) in $2\textrm{D}$, both the
density of states $\nu(\varepsilon)$ and the localization length $\zeta(\varepsilon)$
strongly diverge, as indicated in Fig.~\ref{FigRandomHopPhys2D}. 
By contrast, RHMs in $\textrm{d} > 2$ support a diffusive metallic phase, \cite{Gade,FC}
characterized
by a band of delocalized states with energies $|\varepsilon| < \varepsilon_{\mathsf{c}}$,
represented by the shaded regions in Fig.~\ref{FigRandomHopPhys3D}.
($\varepsilon = \pm \varepsilon_{\mathsf{c}}$ are particle and hole mobility edges.)
While $\varepsilon_{\mathsf{c}}$ is expected to decrease monotonically
with increasing disorder, as indicated by the arrows
along the dashed boundaries of the shaded region in Figs.~\ref{FigRandomHopPhys3D}(a) and
\ref{FigRandomHopPhys3D}(b), it is believed that in $\textrm{d} > 2$, there remains always
a region of delocalized states of finite thickness in energy, centered around
$\varepsilon = 0$, for any finite disorder strength in a non-interacting system with SLS.

Indeed, random
hopping models have been of significant theoretical
interest in the recent past, both because of the unusual delocalization physics described
above,
but also because these models have proven amenable to a variety of powerful analytical
techniques in $d \leq 2$, with many exact and/or non-perturbative features
now understood.\cite{1DChiral,GLL,MDH2D,MRF} This situation should be contrasted with
our understanding of the conventional non-interacting (``Wigner-Dyson'') MIT,
which is based largely upon perturbative analytical results
in $d > 2$, using the $\epsilon$-expansion.\cite{LeeRamakrishnan}

\subsection{Summary of results and outline}\label{Outline}

In this work, we analyze the stability of the diffusive Fermi liquid phase of the Hubbard-like 
model defined by Eqs.~(\ref{Hclean}) and (\ref{Hdis}), in the simultaneous presence of both 
disorder and interactions. We derive a low-energy, continuum field theory description of 
this system, which takes the form of a (class $\mathrm{AIII}$)\cite{Zirnbauer} Finkel'stein 
non-linear sigma model (FNL$\sigma$M).\cite{Finkelstein,BK} We employ the 
Schwinger-Keldysh\cite{KELDYSH} method in order to ensemble average over realizations of the 
hopping disorder. The FNL$\sigma$M contains parameters which specify the random hopping
 disorder and the
interparticle interaction strengths. We compute the one-loop renormalization group 
flow equations for these parameters, using a Wilsonian frequency-momentum shell background 
field methodology.\cite{Finkelstein} We then discuss the physics of our 
model in ({\bf i}) 
$d = 2$ and ({\bf ii}) $d > 2$ dimensions. 
We now briefly summarize our results:

For the $2\textrm{D}$ case, 
we find that the conducting phase of the disordered, non-interacting
system is destabilized by the interactions.
By contrast,
such a phase, 
a metallic `diffusive Fermi liquid,'
does exist (trivially) in $d > 2$;
we identify
what we call an ``Anderson-Mott''
disorder-driven instability of
this metallic `diffusive Fermi liquid.'
This instability arises \emph{solely} from the competition
between disorder and short-ranged interactions,
and is perturbatively controlled via an 
$\epsilon$-expansion in $d = 2 + \epsilon$.  
The instability that we find 
is indicative of a first-order metal-insulator transition (MIT); it occurs 
for disorder strengths of order $\epsilon$, and is therefore perturbatively accessible 
for $\epsilon \ll 1$. We expect our result to be relevant for the Hubbard-like model in 
Eqs.~(\ref{Hclean}) and (\ref{Hdis}) in three spatial dimensions
($\epsilon=1$)
for sufficiently strong 
disorder, and we stress that 
this
``Anderson-Mott''
instability is clearly distinct from the pure Mott 
``nesting'' instability,
which is driven solely by interactions (and not by disorder),
although the latter also appears in 
the phase diagram of  our model 
(see Fig.~\ref{PhaseDiag}, below). 
The non-interacting 
random hopping model, 
Eqs.~(\ref{Hclean}) and (\ref{Hdis}) with $V = U = 0$, has no localized phase with disorder 
in the absence of interactions
(see e.g.\ Section \ref{NonIntDiscuss}); 
therefore, a localized phase
can only appear due to 
the presence of the interactions. The discovery of this 
disorder-driven, interaction-mediated
diffusive Fermi liquid instability was previously announced in Ref.~\onlinecite{AIIIshortpaper}. 
In this paper, we present a derivation of this result,
a detailed analysis of the phase diagram of the model, 
and we discuss our findings in view of previously known results for 
other, related disordered and interacting fermion systems.

We note that the
$\epsilon$-expansion
is employed in this work as a technical tool in the 
continuum FNL$\sigma$M description, and should be thought 
of as a
controlled approximation 
scheme to access the physics of the disordered, interacting Hubbard-like Hamiltonian in 
Eqs.~(\ref{Hclean}) and (\ref{Hdis}) in $3\textrm{D}$. 
Clearly, we
cannot
easily define a bipartite 
lattice fermion model in a fractional number of $d = 2 + \epsilon$ dimensions; 
instead, we work with a continuum field theory, the FNL$\sigma$M, argued to capture the 
low energy physics of the lattice model in \emph{both} $2\textrm{D}$ ($\epsilon = 0$) and 
$3\textrm{D}$ ($\epsilon = 1$), whose internal structure is constrained by the crucial SLS 
[Eq.~(\ref{SLS})].
The
field theory
action of the continuum FNL$\sigma$M
[displayed in Eqs.~(\ref{Z})--(\ref{QmatrixUnitary2}), Section \ref{NLSMSummary}]
can be analytically continued 
between integer dimensions in the usual way. The internal structure of the FNL$\sigma$M, 
and thus SLS, is preserved under this continuation; it is SLS, then, that gives meaning 
to such an interpolation between lattice models in disparate integer dimensions.
The 
``Anderson-Mott''
instability of the
diffusive Fermi liquid 
identified in this work occurs for perturbatively 
accessible, weak disorder strengths only
for $\epsilon \ll 1$, i.e.\
in $2 < d \ll 3$. We \emph{conjecture} that 
this 
instability
also exists in the $3\textrm{D}$ FNL$\sigma$M, and should therefore be found 
(i.e.\ via numerics) in the $3\textrm{D}$ Hubbard-like lattice fermion model. This conjecture 
cannot be directly proven here, since the instability
of the $3\textrm{D}$ FNL$\sigma$M,
if it exists, would occur in the strong 
coupling regime. 

The organization of this paper is as follows:

In Sec.~\ref{NLSM}, we formulate a Schwinger-Keldysh\cite{KELDYSH} path integral representation for the 
lattice model given by Eqs.~(\ref{Hclean}) and (\ref{Hdis}); we then derive the low-energy, 
continuum FNL$\sigma$M description. 
The technical content of our work appears in Sections \ref{PandFR} and \ref{oneloop}. We 
set up
our one-loop, frequency-momentum shell renormalization group calculation in 
Sec.~\ref{PandFR}, specifying the parameterization of the FNL$\sigma$M and stating the necessary 
diagrammatic Feynman rules. The actual one-loop calculation is chronicled in Sec.~\ref{oneloop}. 
We analyze and discuss our results in Sec.~\ref{Results}. The reader less interested in 
calculational details may skip Secs. \ref{PandFR} and \ref{oneloop} entirely, and proceed 
from the end of Sec.~\ref{NLSM} directly to Sec.~\ref{Results}. 

A variety of elaborations, extensions, and technical details are relegated to the Appendices.
In Appendix \ref{KeldyshSym}, we show how the structure of the continuum FNL$\sigma$M 
for the model in the \emph{absence} of interparticle interactions\cite{Gade} follows immediately 
from a symmetry analysis of the non-interacting Keldysh action for the random hopping model 
[Eqs.~(\ref{Hclean}) and (\ref{Hdis}), with $U = V = 0$]. Appendix \ref{ExpTrick} details an 
expansion used in Sec.~\ref{NLSM}, while Appendix \ref{Spin1/2RHopping} describes the random 
matrix symmetry classification (along the lines of Ref.~\onlinecite{AltlandZirnbauer}) 
of disordered, bipartite lattice models for spin-$1/2$ electrons. 
[See also 
Sec.~\ref{Spin1/2}]. Appendix \ref{SC} provides a surprising alternative interpretation of 
the class AIII Finkel'stein NL$\sigma$M, studied in this paper, in terms of the spin-$1/2$ 
quasiparticles of a spin-triplet, p-wave superconductor. This quasiparticle system may be 
defined directly in the continuum, without reference to a ``microscopic'' lattice model or 
an additional sublattice symmetry. Finally, Appendix \ref{Integrals} collects the loop 
integrals required in the RG calculation presented in Sec.~\ref{oneloop}.


\section{FNL$\sigma$M formulation}\label{NLSM}

In this section we derive the class AIII Finkel'stein non-linear sigma model 
(FNL$\sigma$M) description of the Hubbard-like lattice model given by 
Eqs.~(\ref{Hclean}) and (\ref{Hdis}), using the Schwinger-Keldysh\cite{KELDYSH} 
method to perform the disorder averaging. The results of the 
derivation are provided below in Eqs.~(\ref{Z}), (\ref{SD}), and (\ref{SI}), 
and interpreted in the discussion following these equations.

\subsection{Class AIII Finkel'stein NL$\sigma$M}

\subsubsection{Schwinger-Keldysh path integral}\label{SKPI}

To begin, we envisage a zero temperature, $d+1$-dimensional real 
time-ordered ($T$-ordered) path integral $Z_{T}$ for the model defined in 
Eqs.~(\ref{Hclean}) and (\ref{Hdis}). As usual, we need to normalize this 
path integral to unity in order to perform the ensemble average over 
realizations of the hopping disorder. We employ the Schwinger-Keldysh 
(Keldysh) method,\cite{KELDYSH} which exploits the identity 
$1/Z_{T} = Z_{\bar{T}}$, where $Z_{\bar{T}}$ is an anti-time ordered 
($\bar{T}$-ordered) path integral for the same model. We write the Keldysh 
generating function\cite{KELDYSH}
\begin{equation}\label{Zmicro}
	Z \equiv Z_{T} Z_{\bar T} = \int \mathcal{D}\bar{c} \mathcal{D}c \, e^{i (S_{1} + S_{2})},
\end{equation}
where the non-interacting action is given by
\begin{align}\label{S1}
	S_{1} = \sum_{a = 1, 2} \xi^{a} \int & d\omega \bigg\lgroup
	\sum_{\langle i j \rangle} t^{\phantom{*}}_{i, j} \, \bar{c}_{A i}^{\,a}(\omega) c_{B j}^{\,a}(\omega) + \textrm{H.c.} \nonumber\\
	& + \sum_{i \in A} \bar{c}_{A i}^{\,a}(\omega)\left[\omega + i \xi^{a} \eta \sgn(\omega)\right]{c}_{A i}^{\,a}(\omega) \nonumber\\
	& + \sum_{j \in B} \bar{c}_{B j}^{\,a}(\omega)\left[\omega + i \xi^{a} \eta \sgn(\omega)\right]{c}_{B j}^{\,a}(\omega) \bigg\rgroup,
\end{align}
with $t^{\phantom{*}}_{i, j} \equiv t + \delta t^{\phantom{*}}_{i, j}$.
The interactions reside in
\begin{align}\label{S2}
	S_{2} = \sum_{a = 1, 2} \xi^{a} \int d t \bigg\lgroup
	& - V \sum_{\langle i j \rangle} n_{A i}^{\,a}(t) n_{B j}^{\,a}(t) \nonumber\\
	& - U \! \! \sum_{\langle\langle i i' \rangle\rangle} n_{A i}^{\,a}(t) n_{A i'}^{\,a}(t)\nonumber\\
	& - U \! \! \sum_{\langle\langle j j' \rangle\rangle} n_{B j}^{\,a}(t) n_{B j'}^{\,a}(t) \bigg\rgroup.
\end{align}
The generating functional defined by Eq.~(\ref{Zmicro}) is an integral 
over the Grassmann fields $\bar{c}_{i A}^{\, a}$, ${c}_{i A}^{\, a}$, and 
$\bar{c}_{j B}^{\, a}$, ${c}_{j B}^{\, a}$, defined on the $A$ and $B$ sublattices 
of the bipartite lattice. Note the locality of the non-interacting sector 
of the theory [$S_{1}$ in Eq.~(\ref{S1})] in frequency $\omega$, versus the 
locality of the interacting sector [$S_{2}$ in Eq.~(\ref{S2})] in time $t$. 
In Eqs.~(\ref{S1}) and (\ref{S2}), indices $\{i, i'\}$ and $\{j, j'\}$ label 
$A$ and $B$ sublattice sites, respectively, while the ``Keldysh'' species index 
$a \in 1,2$ denotes the $T$-ordered ($a = 1$) and $\bar{T}$-ordered ($a = 2$) 
branches of the theory. The number $\xi^{a}$ takes the values
\begin{equation}\label{KeldyshFactor}
	\xi^{a} = \left\{ \begin{array}{lll}
			\phantom{-}1,\; & a = 1	& (T\text{-ordered}), \\
			-1,\; 		& a = 2 & (\bar{T}\text{-ordered}).
			\end{array} \right.
\end{equation}
The factors $i \xi^{a} \eta \sgn(\omega)$ in Eq.~(\ref{S1}) 
are frequency integration pole prescriptions appropriate to $T$- and 
$\bar{T}$-ordered correlation functions, with $\eta \rightarrow 0^{+}$. 
Finally, the density fields in Eq.~(\ref{S2}) are defined via 
$n_{A i / B j}^{\,a}(t) \equiv \bar{c}_{A i / B j}^{\,a}(t) {c}_{A i / B j}^{\,a}(t)$.\cite{footnote-e}

The FNL$\sigma$M that we are after is a matrix field theory; it therefore 
makes good sense to introduce compactifying matrix notation at this early stage.
We think of the field $c_{A i/B j}$ ($\bar{c}_{A i/B j}$) as a column (row) 
vector with a single ``superindex,'' which is a direct product of frequency 
$\{\omega\}$ and Keldysh $\{a\}$ indices: 
e.g.~$c_{A i} \rightarrow c_{A i}^{\, a}(\omega)$. In the Keldysh formalism, 
it is often useful to further divide frequency into a product of (discrete) 
sign and (continuous) modulus spaces: $\omega~=~|\omega|\,\sgn(\omega)$.
Next, we introduce two commuting sets of Pauli matrices:
the matrix $\hat{\Sigma}_{m}$ acts in the $\sgn(\omega)$ space, while the matrix 
$\hat{\xi}_{n}$ acts in the Keldysh species ($T$/$\bar{T}$) space, with $m,n \in \{1,2,3\}$. 
Employing the conventional basis for all Pauli matrices, we identify 
$\hat{\Sigma}_{3} \rightarrow \sgn{\omega} \, \delta_{\omega, \omega'} \, \delta^{\, a, a'}$ 
and $\hat{\xi}_{3} \rightarrow \xi^{a}  \, \delta_{\omega, \omega'} \, \delta^{\, a, a'}$, with 
$\delta_{\omega, \omega'} \equiv 2 \pi \delta(\omega - \omega')$. 
[$\xi^{a}$ was defined in Eq.~(\ref{KeldyshFactor}).] Finally, we define the 
single particle energy matrix $\hat{\omega} \equiv |\hat{\omega}|\, \hat{\Sigma}_{3}$, 
with $|\hat{\omega}| \rightarrow |\omega| \, \delta_{\omega, \omega'} \, \delta^{\, a, a'}.$ 

We make a change of variables 
$\bar{c}_{A i/B j} \rightarrow \bar{c}_{A i/B j} \, \hat{\xi}_{3}$, and rewrite 
the non-interacting sector [Eq.~(\ref{S1})] of the Keldysh action as 
\begin{equation}\label{S1final}
	S_{1} \equiv S_{1}^{\, 0} + \delta S_{1}.
\end{equation}
The clean bipartite hopping model appears in the term
\begin{align}
	S_{1}^{\, 0} =& 
		\int_{\textrm{sBZ}} \frac{d^{d} \bm{\mathrm{k}}}{(2 \pi)^{d}} 	
		\bigg\lgroup
		\begin{bmatrix}
			\bar{c}_{A \, \bm{\mathrm{k}}} & \bar{c}_{B \, \bm{\mathrm{k}}}
		\end{bmatrix}\!\!
		\begin{bmatrix}
			i \eta \hat{\Sigma}_{3} \hat{\xi}_{3} & -\varepsilon(\bm{\mathrm{k}}) \\
			-\varepsilon(\bm{\mathrm{k}}) & i \eta \hat{\Sigma}_{3} \hat{\xi}_{3}
		\end{bmatrix}\!\!
		\begin{bmatrix}
			{c}_{A \, \bm{\mathrm{k}}} \\
			{c}_{B \, \bm{\mathrm{k}}} 
		\end{bmatrix} \nonumber\\
		&\quad\quad\quad\;\,+ 
		\begin{bmatrix}
			\bar{c}_{A \, \bm{\mathrm{k}}} & \bar{c}_{B \, \bm{\mathrm{k}}}
		\end{bmatrix}\!\!
		\begin{bmatrix}
			\hat{\Sigma}_{3} |\hat{\omega}|  & 0 \\
			0 & \hat{\Sigma}_{3} |\hat{\omega}| 
		\end{bmatrix}\!\!
		\begin{bmatrix}
			{c}_{A \, \bm{\mathrm{k}}} \\
			{c}_{B \, \bm{\mathrm{k}}} 
		\end{bmatrix}
		\bigg\rgroup
	\label{S10Exp}\\
	\equiv& 
	\bar{c} \left[
	\hat{\Sigma}_{3} |\hat{\omega}| 
	+ i \eta \hat{\Sigma}_{3} \hat{\xi}_{3} 
	- \hat{\sigma}_{1} \varepsilon(\hat{\bm{\mathrm{k}}})
	\right]
	c. \label{S10}
\end{align}
The momentum integration in Eq.~(\ref{S10Exp}) is taken over the \emph{sublattice} 
Brillouin zone (sBZ); $\varepsilon({\bm{\mathrm{k}}})$ is the clean energy
band structure.\cite{footnote-f}
Equation (\ref{S10}) gives the most compact representation that we will use for the 
clean, non-interacting action. Here we have introduced a third set of Pauli matrices, 
$\hat{\sigma}_{m}$, $m \in \{1,2,3\}$, acting in the sublattice flavor $\{A,B\}$ 
space, and we have also promoted momentum to an operator, 
$\bm{\mathrm{k}} \rightarrow \hat{\bm{\mathrm{k}}}$. The column vector $c$ carries 
indices in the momentum $\bm{\mathrm{k}}$, mod-energy $|\omega|$, 
$\sgn(\omega)$ ($\Sigma$), sublattice flavor ($\sigma$), and Keldysh ($\xi$) 
spaces, i.e.\ $c \rightarrow c_{A/B \, \bm{\mathrm{k}}}^{\, a}(\omega)$ with all 
indices displayed. Here, $a = 1,2$ denotes the Keldysh species.
The disorder is relegated to the perturbation
\begin{equation}\label{DeltaS1}
	\delta S_{1} = \sum_{\langle i j \rangle}
	\left( \bar{c}_{A i} \, \delta t^{\phantom{*}}_{i, j} \, c_{B j} 
	+ \bar{c}_{B j} \, \delta t_{i, j}^{*} \, c_{A i} \right).
\end{equation}

\subsubsection{Disorder averaging and Hubbard Stratonovich decoupling}\label{DisAvgHubStrat}

\begin{figure}
\includegraphics[width=0.2\textwidth]{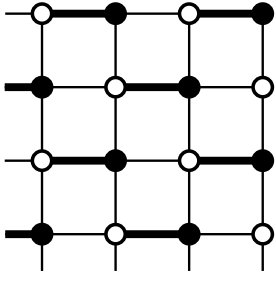}
\caption{Homogeneous pairing (dimer covering) of nearest neighbor sites on the square lattice.  
\label{SquareLatticeDimer}}
\end{figure}

We now ensemble average over realizations of the complex random hopping amplitudes 
$\{\delta t^{\phantom{*}}_{i,j}\}$ appearing in Eq.~(\ref{DeltaS1}). In order to simplify the 
derivation of the FNL$\sigma$M, we employ the following artifice: we assume that 
of the $z_{c}$ nearest neighbors bonds surrounding a given site $i$ belonging 
to the $A$ sublattice of the bipartite lattice under study, only \emph{one} such 
bond is disordered. ($z_{c}$ is the coordination number.) We assume further 
that the same type of bond (specified by its orientation) is made random at 
each and every $A$ site, thus allowing a unique, orientationally homogeneous 
pairing (dimer covering) of the $A$ and $B$ sublattice sites of the bipartite lattice. 
An example of such a pairing is provided for the square lattice in 
Fig.~\ref{SquareLatticeDimer}. This seemingly pathological constraint upon 
the disorder distribution allows for the quickest derivation of the low-energy
effective field theory.\cite{footnote-g}

With such a pairing established between each $A$ sublattice site $i$ and its associated $B$ 
sublattice site $j(i)$, we have 
\begin{equation}\label{DimerizedRandomHopping}
	\delta t^{\phantom{*}}_{i, j} = \left\{
	\begin{array}{rl}
		\delta t^{\phantom{*}}_{i, j(i)},\;& j = j(i) \\
		0,\;& j \neq j(i)
	\end{array}
	\right..
\end{equation}
We take the ``dimerized'' bond amplitudes $\{\delta t^{\phantom{*}}_{i, j(i)}\}$ 
to be Gaussian random variables, independent on different nearest-neighbor lattice links 
$\langle i \, j(i)\rangle$, and identically distributed with the following purely real mean 
and variance:
\begin{equation}\label{DimerBondMean}
	\overline{\delta t^{\phantom{*}}_{i, j(i)}} = \delta t_{\mathsf{dim}}
\end{equation}
and
\begin{equation}\label{DimerBondVar}
	\overline{\delta t_{i, j(i)}^{*} \delta t^{\phantom{*}}_{i, j(i)}} 
	= \lambda^{{\scriptscriptstyle(}\mathsf{m}{\scriptscriptstyle)}}, 
\end{equation}
respectively. The overbars in Eqs.~(\ref{DimerBondMean}) and (\ref{DimerBondVar}) 
denote disorder averaging. Although we will be ultimately interested in the limit 
of zero mean bond dimerization, it will prove convenient in the interim 
to retain $\delta t_{\mathsf{dim}} \neq 0$. The superscript 
``${\scriptstyle(}\mathsf{m}{\scriptstyle)}$'' appearing on the right-hand side of 
Eq.~(\ref{DimerBondVar}) stands for ``microscopic,'' indicating that the quantity 
$\lambda^{{\scriptscriptstyle(}\mathsf{m}{\scriptscriptstyle)}}$ 
is defined at the lattice scale. Using Eqs.~(\ref{DeltaS1})--(\ref{DimerBondVar}), 
we define the disorder-averaged action $\overline{\delta S_{1}}$ via 
\begin{align}\label{DisAvg}
	e^{i \overline{\delta S_{1}}} &\equiv \int |\mathcal{D}\delta t|^{2} \,
	\exp\left[
	{-1\over{
	\lambda^{{\scriptscriptstyle(}\mathsf{m}{\scriptscriptstyle)}}
	}} 
	\sum_{i \in A} 
	|\delta t^{\phantom{*}}_{i,j(i)} - \delta t_{\mathsf{dim}}|^{2}
	+ i \delta S_{1}
	\right] \nonumber\\
	&= \exp\left\lgroup 
	\frac{
	\lambda^{{\scriptscriptstyle(}\mathsf{m}{\scriptscriptstyle)}}
	}{2} 
	\sum_{i \in A} \mathrm{Tr}\left[\hat{\mathcal{A}}_{i}^{2}\right]
	+ i \delta S_{\mathsf{dim}}
	\right\rgroup,
\end{align}
with
\begin{equation}\label{FermMatBilComp}
	\hat{\mathcal{A}}_{i} \equiv
	\begin{bmatrix}
		0 & \hat{A}_{i} \\
		\hat{A}^{\dagger}_{j(i)} & 0
	\end{bmatrix}
	= \frac{\hat{\sigma}_{1} + i \hat{\sigma}_{2}}{2} \hat{A}_{i} + 
	\frac{\hat{\sigma}_{1} - i \hat{\sigma}_{2}}{2} \hat{A}^{\dagger}_{j(i)},
\end{equation} 
where we have defined
\begin{subequations}\label{FermMatBil}
\begin{align}
	\hat{A}_{i} & \rightarrow A_{i \, \omega, \omega'}^{a, a'} \equiv c_{A i}^{\, a}(\omega) \bar{c}_{A i}^{\, a'}(\omega'),\\
	\hat{A}^{\dagger}_{j(i)} & \rightarrow A^{\dagger \, a, a'}_{j \, \omega, \omega'} \equiv c_{B j}^{\, a}(\omega) \bar{c}_{B j}^{\, a'}(\omega').
\end{align}
\end{subequations}
In Eq.~(\ref{DisAvg}), $\mathrm{Tr}$ denotes a matrix trace over indices in the 
sublattice flavor ($\sigma$), frequency ($\omega$), and Keldysh ($\xi$) spaces. 
The field $\hat{\mathcal{A}}_{i}$, introduced in Eq.~(\ref{DisAvg}), is a matrix 
of fermion bilinears; Eq.~(\ref{FermMatBilComp}) details the sublattice flavor space 
decomposition of $\hat{\mathcal{A}}_{i}$ in terms of the purely off-diagonal Pauli 
matrices $\hat{\sigma}_{1}$ and $\hat{\sigma}_{2}$. Although they occupy the 
off-diagonal blocks of $\hat{\mathcal{A}}_{i}$, the fields $\hat{A}_{i}$ and 
$\hat{A}^{\dagger}_{j(i)}$ actually describe fermion degrees of freedom residing 
\emph{entirely} on the $A$ and $B$ sublattices, respectively, as shown in 
Eq.~(\ref{FermMatBil}). $\hat{A}_{i}$ and $\hat{A}^{\dagger}_{j(i)}$ each carry 
frequency $\{\omega, \omega'\}$ and Keldysh species $\{a, a'\}$ indices. 
Finally, a non-zero mean bond dimerization $\delta t_{\mathsf{dim}}$ 
[Eq.~(\ref{DimerBondMean})] gives rise to the following homogeneous term
in Eq.~(\ref{DisAvg}):
\begin{align}\label{DeltaSdim}
	\delta S_{\mathsf{dim}} =& \delta t_{\mathsf{dim}} \sum_{i}
	\left( \bar{c}_{A i} c_{B j(i)} 
	+ \bar{c}_{B j(i)} c_{A i} \right)
	\nonumber\\
	=& - \bar{c}\left[
	\hat{\sigma}_{1} \, \phi_{R}(\hat{\bm{\mathrm{k}}}) 
	+ \hat{\sigma}_{2} \, \phi_{I}(\hat{\bm{\mathrm{k}}}) 
	\right]c.
\end{align}
The second line of Eq.~(\ref{DeltaSdim}) expresses $\delta S_{\mathsf{dim}}$ in
momentum space; here we have adopted the same compact notation employed in Eq.~(\ref{S10}),
wherein $\hat{\sigma}_{1}$ and $\hat{\sigma}_{2}$ act in sublattice flavor space, and
$\hat{\bm{\mathrm{k}}}$ is the (crystal) momentum operator.
The functions $\phi_{R}(\bm{\mathrm{k}})$ and $\phi_{I}(\bm{\mathrm{k}})$ in 
Eq.~(\ref{DeltaSdim}) are real and imaginary components of the ``dimerization'' function
\begin{equation}\label{DimerFuncDef}
	\phi_{R}(\bm{\mathrm{k}}) + i \phi_{I}(\bm{\mathrm{k}}) 
	\equiv 
	- \delta t_{\mathsf{dim}} e^{i \bm{\mathrm{k}} \cdot \bm{\mathrm{n}}}.
\end{equation}
In this equation, $\bm{\mathrm{n}}$ is a unit vector pointing in the direction 
determined by the chosen bond dimerization [see Eqs.~(\ref{DimerizedRandomHopping})  
and (\ref{DimerBondMean}), and Fig.~(\ref{SquareLatticeDimer})].

We have adopted a suggestive notation in Eq.~(\ref{FermMatBil}) to denote 
the pure sublattice fields $\hat{A}_{i}$ and $\hat{A}^{\dagger}_{j(i)}$. 
Consider the following spatially uniform deformation of these fermion bilinears:
\begin{equation}\label{FermMatBilXfm}
	\hat{A}_{i} \rightarrow \hat{U}_{A}^{\phantom{\dagger}} \hat{A}_{i} \hat{U}_{B}^{\dagger},
	\qquad
	\hat{A}^{\dagger}_{j(i)} \rightarrow \hat{U}_{B}^{\phantom{\dagger}} \hat{A}^{\dagger}_{j(i)} \hat{U}_{A}^{\dagger},
\end{equation}
where $\hat{U}_{A}^{\phantom{\dagger}}$ and $\hat{U}_{B}^{\phantom{\dagger}}$ are independent unitary transformations in 
$|\omega|\otimes\Sigma\otimes\xi$ (frequency$\otimes$Keldysh) space:
\begin{equation}\label{UAandUBunitary}
	\hat{U}_{A}^{\dagger}\hat{U}_{A}^{\phantom{\dagger}} 
	= \hat{U}_{B}^{\dagger}\hat{U}_{B}^{\phantom{\dagger}}
	= \hat{1}_{|\omega|\otimes\Sigma\otimes\xi}.
\end{equation}
The transformation in Eq.~(\ref{FermMatBilXfm}) is clearly a symmetry of the 
disorder-averaged action $\overline{\delta S_{1}}$, as can be seen from 
Eqs.~(\ref{DisAvg})--(\ref{DeltaSdim}). In fact, 
there is a direct relationship between this transformation and the symmetry 
structure of the non-interacting Keldysh action [Eqs.~(\ref{S10}) and 
(\ref{DeltaS1})], in every fixed realization of the static disorder; the 
connection is articulated in Appendix \ref{KeldyshSym}. Equation 
(\ref{FermMatBilXfm}) suggests that we may regard $\hat{A}^{\dagger}_{j(i)}$ 
as the ``Hermitian adjoint'' of its associated nearest neighbor $\hat{A}_{i}$, 
justified on length scales much larger than the lattice spacing. The identification 
$\hat{A}^{\dagger}_{j(i)}\sim\hat{A}_{i}^{\dagger}$ implies the Hermiticity 
of the composite matrix field in Eq.~(\ref{FermMatBil}): 
$\hat{\mathcal{A}}_{i}^{\dagger}\sim\hat{\mathcal{A}}_{i}$. 

Next, we decouple all four fermion terms appearing in the disorder-averaged 
and interacting sectors of the theory with bosonic Hubbard-Stratonovich fields. 
In the disorder-averaged sector [Eq.~(\ref{DisAvg})], we write
\begin{multline}\label{HSdis}
	\exp\left\lgroup 
	\frac{\lambda^{{\scriptscriptstyle(}\mathsf{m}{\scriptscriptstyle)}}}
	{2} \sum_{i \in A} \mathrm{Tr}\left[\hat{\mathcal{A}}_{i}^{2}\right] \right\rgroup 
	\\
	= \int \mathcal{D}\hat{\mathcal{Q}}\,
	\exp\left\lgroup
	\sum_{i \in A}
	\mathrm{Tr}\left[ 
	\frac{-1}{2 \lambda^{{\scriptscriptstyle(}\mathsf{m}{\scriptscriptstyle)}}} 
	\hat{\mathcal{Q}}_{i}^{2}
	+ \hat{\mathcal{Q}}_{i} \hat{\mathcal{A}}_{i} 
	\right] 
	\right\rgroup.
\end{multline}
The matrix field $\hat{\mathcal{Q}}_{i}$ is taken to be Hermitian, and purely 
off-diagonal in sublattice flavor space, i.e.\ 
\begin{equation}\label{QComp}
	\hat{\mathcal{Q}}_{i} \equiv
	\begin{bmatrix}
		0 & \hat{Q}_{i} \\
		\hat{Q}^{\dagger}_{j(i)} & 0
	\end{bmatrix}
	= \frac{\hat{\sigma}_{1} + i \hat{\sigma}_{2}}{2} \hat{Q}_{i} + 
	\frac{\hat{\sigma}_{1} - i \hat{\sigma}_{2}}{2} \hat{Q}^{\dagger}_{j(i)},
\end{equation}
with
\begin{equation}\label{Qlattice}
	\hat{Q}_{i} \rightarrow Q_{i \, \omega, \omega'}^{a, a'}, 
	\qquad
	\hat{Q}^{\dagger}_{j(i)} \sim \hat{Q}^{\dagger}_{i} \rightarrow Q^{\dagger \, a, a'}_{i \, \omega, \omega'}.
\end{equation}
[Compare Eqs.~(\ref{FermMatBilComp}) and (\ref{FermMatBil})].

Turning  to the interacting sector, we re-write Eq.~(\ref{S2}) as 
\begin{equation}\label{S2matrix}
	S_{2} = -\sum_{a = 1, 2} \frac{\xi^{a}}{2} \int d t \,
	{N}^{{\mathsf{T}} \,a}(t) \, \hat{\mathcal{X}} \, {N}^{\,a}(t),
\end{equation}
where superscript ``${\mathsf{T}}$'' denotes the matrix transpose operation, and
\begin{equation}\label{CouplingMatrix}
	\hat{\mathcal{X}} \equiv 	
	\begin{bmatrix}
		U \hat{\mathsf{x}}^{\phantom{\mathsf{T}}} & V \hat{\mathsf{y}} \\
		V \hat{\mathsf{y}}^{\mathsf{T}} & U \hat{\mathsf{x}}
	\end{bmatrix}
\end{equation}
is a matrix in sublattice flavor space ($\sigma$), with elements involving the 
position space coupling functions $\hat{\mathsf{x}}$ and $\hat{\mathsf{y}}$ (defined below),
while the fermion sublattice densities are encoded in the vector
\begin{equation}\label{DensityVector}
	{N}^{\,a}(t) \rightarrow
	\begin{bmatrix}
		n_{A}^{\,a}(t) \\
		n_{B}^{\,a}(t)
	\end{bmatrix}.
\end{equation}
We have suppressed all position space indices in 
Eqs.~(\ref{S2matrix})--(\ref{DensityVector}). In Eq.~(\ref{CouplingMatrix}), the function 
$\hat{\mathsf{y}}~\rightarrow~\mathsf{y}_{i, j}$ 
($\hat{\mathsf{x}}~\rightarrow~\mathsf{x}_{i, i'}$) equals unity for pairs of nearest 
neighbor (next-nearest neighbor) lattice sites $\{i,j\}$ ($\{i,i'\}$), and vanishes 
otherwise. Now we decouple
\begin{align}\label{HSint}
	e^{i S_{2}} = \int \mathcal{D}\rho \,
	\exp \bigg\lgroup
	\sum_{a = 1, 2} i \int d t \,
	\bigg[&
	\frac{\xi^{a}}{2}{\rho}^{{\mathsf{T}} \,a}(t) \, \hat{\mathcal{X}}^{-1} \, {\rho}^{\,a}(t) \nonumber\\ 
	& + {\rho}^{{\mathsf{T}} \,a}(t) {N}^{\,a}(t) \bigg] 
	\bigg\rgroup,
\end{align}
where 
\begin{equation}\label{RhoVector}
	{\rho}^{\,a}(t) \rightarrow
	\begin{bmatrix}
		\rho_{A}^{\,a}(t) \\
		\rho_{B}^{\,a}(t)
	\end{bmatrix}.
\end{equation}
Again, we have suppressed all position space indices in Eqs.~(\ref{HSint}) 
and (\ref{RhoVector}).

\subsubsection{Saddle point and gradient expansion}\label{SPGradient}

Gathering together the homogeneous hopping [Eq.~(\ref{S10})], mean bond 
dimerization [Eq.~(\ref{DeltaSdim})], disorder-averaged [Eq.~(\ref{HSdis})], 
and interacting [Eq.~(\ref{HSint})] pieces of our theory, 
we perform the Gaussian integral over the fermion fields, and 
arrive finally at the following effective field theory:
\begin{equation}\label{EffTheory}
	Z = \int \mathcal{D}\hat{\mathcal{Q}} \mathcal{D}\rho \, \exp\left[- S_{Q} - S_{\rho} - S_{\mathsf{DET}}\right],
\end{equation}
where
\begin{equation}\label{SQ}
S_{Q} = 
	\frac{1}{2 \lambda^{{\scriptscriptstyle(}\mathsf{m}{\scriptscriptstyle)}}} \sum_{i \in A}
	\mathrm{Tr}\left[\hat{\mathcal{Q}}_{i}^{2}\right],
\end{equation}
\begin{equation}\label{Srho}
S_{\rho} = 
	\sum_{a = 1, 2} \frac{- i \xi^{a}}{2} \int d t \,
	{\rho}^{{\mathsf{T}} \,a}(t) \, \hat{\mathcal{X}}^{-1} \, {\rho}^{\,a}(t),
\end{equation}
and
\begin{equation}\label{SDET}
S_{\mathsf{DET}} =
	- \mathrm{Tr}
	\left\lgroup
		\ln\left[
		{G}_{0}^{-1} + i \hat{\sigma}_{1} \hat{\mathcal{Q}} + {\rho}
		\right]
	\right\rgroup.
\end{equation}
The trace in Eq.~(\ref{SDET}) is performed over indices in the position, 
sublattice flavor ($\sigma$),  mod-energy $|\omega|$, $\sgn(\omega)$ ($\Sigma$), 
and Keldysh species ($\xi$) spaces. The operator ${G}_{0}^{-1}$ in Eq.~(\ref{SDET}) 
represents the inverse of the (Keldysh) single particle Green's function for 
the clean, non-interacting hopping model described by Eq.~(\ref{S10}), incorporating,
in addition, the mean bond dimerization from Eq.~(\ref{DeltaSdim}):
\begin{equation}\label{CleanGF}
G_{0}^{-1} =
	\hat{\Sigma}_{3} |\hat{\omega}| 
	+ i \eta \hat{\Sigma}_{3} \hat{\xi}_{3} 
	- \hat{\sigma}_{1} 
	\left[\varepsilon(\hat{\bm{\mathrm{k}}}) + \phi_{R}(\hat{\bm{\mathrm{k}}})\right]
	- \hat{\sigma}_{2} \, \phi_{I}(\hat{\bm{\mathrm{k}}}).
\end{equation}
The functions $\phi_{R}$ and $\phi_{I}$ are real and imaginary parts of the
``dimerization'' function defined by Eq.~(\ref{DimerFuncDef}).

We look for a spatially homogeneous saddle point solution to the action given 
by Eqs.~(\ref{SQ}) and (\ref{SDET}), in terms of the matrix field 
$\hat{\mathcal{Q}}$, with $\rho = \phi_{R} = \phi_{I} = 0$ in 
Eqs.~(\ref{SDET}) and (\ref{CleanGF}) (i.e.\ ignoring the interparticle 
interactions, and considering the limit of zero mean bond dimerization).
In the low-energy/long-time limit $|\omega| \rightarrow 0$, 
the structure of the saddle point solution is determined by the pole prescription 
piece (the term proportional to $\eta$) in Eq.~(\ref{CleanGF}). We make the 
standard ansatz
\begin{equation}\label{QCompSP}
	\hat{\mathcal{Q}}_{\mathsf{SP}} \equiv \frac{1}{2 \tau} \hat{\sigma}_{1} \hat{\Sigma}_{3} \hat{\xi}_{3},
\end{equation}
with $\tau$ the elastic scattering lifetime due to the disorder. 
Then the saddle point condition reduces to the usual self-consistent Born 
approximation (SCBA) for the elastic decay rate $1/\tau$:
\begin{equation}\label{SCBA}
	\frac{1}{\lambda^{{\scriptscriptstyle(}\mathsf{m}{\scriptscriptstyle)}}} 
	= \int_{\textrm{sBZ}} \frac{d^{d} \bm{\mathrm{k}}}{(2 \pi)^{d}}
	\left[\varepsilon^{2}(\bm{\mathrm{k}}) + \frac{1}{(2 \tau)^{2}}\right]^{-1}.
\end{equation}
For the $2\textrm{D}$ case of the square lattice (with its concomitant van 
Hove singularities at half filling), Eq.~(\ref{SCBA}) gives 
\begin{equation}
\label{DEFtau}
	\frac{1}{2 \tau} \sim \frac{\lambda^{{\scriptscriptstyle(}\mathsf{m}{\scriptscriptstyle)}}}{2 \pi t} 
	\ln\left(\frac{t^{2}}{\lambda^{{\scriptscriptstyle(}\mathsf{m}{\scriptscriptstyle)}}}\right),
\end{equation}
in the weak disorder limit $\lambda^{{\scriptscriptstyle(}\mathsf{m}{\scriptscriptstyle)}} \ll t^{2}$.

Next we consider fluctuations about the saddle point solution given by 
Eq.~(\ref{QCompSP}). Dominant within the diffusive metallic phase are the 
long-wavelength, low-energy Goldstone (diffusion) modes that preserve the 
saddle point norm $1/2\tau$. These modes are generated by applying a slowly 
spatially varying generalization of the symmetry transformation in 
Eqs.~(\ref{FermMatBilXfm}) and (\ref{UAandUBunitary})
to $\hat{\mathcal{Q}}_{\mathsf{SP}}$. 
(See also Appendix \ref{KeldyshSym}.) In terms of the sublattice flavor 
space decomposition [Eq.~(\ref{QComp})], we define
\begin{align}\label{QCompDef}
	\hat{\mathcal{Q}}(\bm{\mathrm{r}}) \equiv
	\frac{\hat{\sigma}_{1} + i \hat{\sigma}_{2}}{4 \tau} \hat{Q}(\bm{\mathrm{r}}) + 
	\frac{\hat{\sigma}_{1} - i \hat{\sigma}_{2}}{4 \tau} \hat{Q}^{\dagger}(\bm{\mathrm{r}}),
\end{align}
where
\begin{subequations}\label{QDef}
\begin{align}
	\hat{Q}(\bm{\mathrm{r}}) &= 
	\hat{U}_{A}^{\phantom{\dagger}}(\bm{\mathrm{r}})\, 
	\hat{Q}_{\mathsf{SP}}\,
	\hat{U}_{B}^{\dagger}(\bm{\mathrm{r}}), \\ 
	\hat{Q}^{\dagger}(\bm{\mathrm{r}}) &= 
	\hat{U}_{B}^{\phantom{\dagger}}(\bm{\mathrm{r}})\,
	\hat{Q}_{\mathsf{SP}}\,
	\hat{U}_{A}^{\dagger}(\bm{\mathrm{r}}),
\end{align}
\end{subequations}
and
\begin{equation}\label{QmatrixSP1}
	\hat{Q}_{\mathsf{SP}} = \hat{\Sigma}_{3} \hat{\xi}_{3}.
\end{equation}
Alternatively, we note that Eqs.~(\ref{UAandUBunitary}), (\ref{QDef}), and (\ref{QmatrixSP1}) 
imply the unitary constraint
\begin{equation}\label{QmatrixUnitary1}
	\hat{Q}^{\dagger}({\bm{\mathrm{r}}}) \hat{Q}({\bm{\mathrm{r}}}) = \hat{1}_{|\omega|\otimes\Sigma\otimes\xi}.
\end{equation}
Our FNL$\sigma$M will therefore be a field theory of the unitary matrix 
$\hat{Q}({\bm{\mathrm{r}}})\rightarrow \hat{Q}_{\omega, \omega'}^{a, a'}({\bm{\mathrm{r}}})$.

Finally, we assemble the action for the FNL$\sigma$M from the components 
given by Eqs.~(\ref{SQ})--(\ref{SDET}). The unitary constraint 
[Eq.~(\ref{QmatrixUnitary1})] renders $S_{Q}$ in Eq.~(\ref{SQ}) an irrelevant 
constant, so we concentrate upon $S_{\mathsf{DET}}$ in Eq.~(\ref{SDET}). 
Keeping only the most relevant terms in a gradient expansion, one obtains 
\begin{align}\label{SDETexp}
	S_{\mathsf{DET}} \sim&
	i \, 2\tau \int d^{d}{\bm{\mathrm{r}}} \,
	\mathrm{Tr}
	\left\lgroup
	(\hat{\Sigma}_{3} |\hat{\omega}| + i \eta \hat{\Sigma}_{3} \hat{\xi}_{3})
	\,[\hat{Q}^{\dagger}({\bm{\mathrm{r}}})+\hat{Q}({\bm{\mathrm{r}}})]
	\right\rgroup
	\nonumber \\
	&+ (2 \, \overline{v}_{F} \tau)^{2} \int d^{d}{\bm{\mathrm{r}}} \,
	\mathrm{Tr}
	\left\lgroup
	\bm{\nabla}\hat{Q}^{\dagger}({\bm{\mathrm{r}}}) 
	\cdot \bm{\nabla}\hat{Q}({\bm{\mathrm{r}}}) \right\rgroup
	\nonumber \\
	&+ S_{\mathsf{DET}}^{\mathsf{dim}} + S_{I},
\end{align}
where the terms $S_{\mathsf{DET}}^{\mathsf{dim}}$ and $S_{I}$ are defined below 
in Eqs.~(\ref{SDETdim}) and (\ref{SIrough1}), respectively. Eqs.~(\ref{SDETexp}), 
(\ref{SDETdim}), and (\ref{SIrough1}) are derived in Appendix \ref{ExpTrick}.

Versions of the ``energy'' (symmetry breaking) and ``stiffness'' terms 
generic to localization sigma models, but specialized here to the Keldysh 
class AIII FNL$\sigma$M, appear on the first and second lines of 
Eq.~(\ref{SDETexp}), respectively. Here, $\overline{v}_{F}$ is the average 
Fermi velocity (at half filling), while the elastic scattering lifetime $\tau$ 
is determined by the SCBA [Eq.~(\ref{SCBA})]. We emphasize that the structure of these 
two terms, written out explicitly in Eq.~(\ref{SDETexp}), 
is independent\cite{footnote-h}
of the peculiar ``dimerized'' bond distribution
[Eq.~(\ref{DimerizedRandomHopping})] assumed for the random hopping.

The first term on the third line of Eq.~(\ref{SDETexp}) is a perturbation
arising from the presence of a non-zero mean bond dimerization 
$\delta t_{\mathsf{dim}} \neq 0$ [Eq.~(\ref{DimerBondMean})]:
\begin{equation}\label{SDETdim}
	S_{\mathsf{DET}}^{\mathsf{dim}} \equiv  
	\mathfrak{c} \frac{\delta t_{\mathsf{dim}}}{t}
	\int d^{d}{\bm{\mathrm{r}}} \,
	\mathrm{Tr}
	\left[\hat{Q}^{\dagger}({\bm{\mathrm{r}}})
	\bm{\nabla}\hat{Q}({\bm{\mathrm{r}}})\right]
	\cdot {\bm{\mathrm{n}}},
\end{equation}
where the unit vector ${\bm{\mathrm{n}}}$, originally introduced in 
Eq.~(\ref{DimerFuncDef}), specifies the orientation of the mean bond 
dimerization, and $\mathfrak{c}$ is a constant. Eq.~(\ref{SDETdim}) 
is derived in Appendix \ref{ExpTrick}.

Consider the local vector operator
\begin{equation}\label{U(1)TermDef}
	i \bm{\nabla} \Phi(\bm{\mathrm{r}}) \equiv 
	\mathrm{Tr}
	\left[\hat{Q}^{\dagger}({\bm{\mathrm{r}}})
	\bm{\nabla}\hat{Q}({\bm{\mathrm{r}}})\right],
\end{equation}
where $\Phi({\bm{\mathrm{r}}})$ is the U(1) phase of the unitary matrix
field $\hat{Q}({\bm{\mathrm{r}}})$. Eq.~(\ref{SDETdim}) implies that 
$i \bm{\nabla} \Phi(\bm{\mathrm{r}})$ provides a coarse-grained measure 
of the local bond dimerization, i.e.\ of the orientation of the strongest 
nearest neighbor hopping bonds within a neighborhood of size the elastic 
scattering length, in the sublattice symmetric random hopping model. This 
interpretation can be argued on very general symmetry grounds, at least for the  
case of the hypercubic lattice: here, a bond dimerization homogeneous in 
\emph{both} magnitude and orientation preserves both SLS and sublattice 
translational symmetry, while breaking lattice rotation, reflection, 
and composite (intersublattice) translational symmetries: see e.g.\ 
Fig.~\ref{SquareLatticeDimer}. Eq.~(\ref{SDETdim}) is the most relevant
term (in the sense of the renormalization group proximate to the diffusive 
metallic phase) consistent with these conditions.
[A homogeneous background dimerization, as in Eqs.~(\ref{DeltaSdim}) and
(\ref{DimerFuncDef}), breaks lattice reflection invariance through a lattice 
plane perpendicular to the direction of the dimerization vector $\bm{\mathrm{n}}$; 
a composite sublattice translation involves an exchange of sublattice degrees 
of freedom, represented by $\hat{Q} \leftrightarrow \hat{Q}^{\dagger}$ in the 
continuum theory.]

Given this interpretation, we can generalize Eq.~(\ref{SDETdim}) to the 
case of a perturbation involving a static random vector field  
${\bm{\mathrm{n}}} \rightarrow {\bm{\mathrm{n}}}_{\mathsf{dim}}(\bm{\mathrm{r}})$:
\begin{equation}\label{SDETdim2}
	S_{\mathsf{DET}}^{\mathsf{dim}} \rightarrow  
	\int d^{d}{\bm{\mathrm{r}}} \,
	\mathrm{Tr}
	\left[\hat{Q}^{\dagger}({\bm{\mathrm{r}}})
	\bm{\nabla}\hat{Q}({\bm{\mathrm{r}}})\right]
	\cdot {\bm{\mathrm{n}}}_{\mathsf{dim}}(\bm{\mathrm{r}}).
\end{equation}
In Eq.~(\ref{SDETdim2}), 
${\bm{\mathrm{n}}}_{\mathsf{dim}}(\bm{\mathrm{r}})$ is taken to vary in 
both orientation and magnitude, and represents long-wavelength, quenched
orientation fluctuations in the bond strength dimerization of the random hopping.
Since Eq.~(\ref{SDETdim2}) is consistent with the sublattice symmetry of the underlying
lattice model, its effects should be included in the low energy effective theory upon the 
grounds of universality. We see that in order to obtain a final effective field theory 
that possesses rotational, translational, and reflection invariances on average, it is 
necessary to specify a \emph{second} measure of the random hopping strength, in addition to
the parameter
$\lambda^{{\scriptscriptstyle(}\mathsf{m}{\scriptscriptstyle)}}$, the latter of which 
appears in the SCBA, Eq.~(\ref{SCBA}), and was introduced in Eqs.~(\ref{DimerBondVar}) 
and (\ref{DisAvg}). We take ${\bm{\mathrm{n}}}_{\mathsf{dim}}(\bm{\mathrm{r}})$ 
to be a Gaussian random variable of vanishing mean, delta function correlated with the variance
\begin{equation}\label{lambdaAMicroDef}
	\overline{
	n^{i}_{\mathsf{dim}}(\bm{\mathrm{r}}) \,
	n^{j}_{\mathsf{dim}}(\bm{\mathrm{r'}})
	}
	= \lambda_{A}\,
	\delta^{i, j} \delta^{(d)}(\bm{\mathrm{r}}-\bm{\mathrm{r'}}).
\end{equation}
Using Eqs.~(\ref{SDETdim2}) and (\ref{lambdaAMicroDef}), we define the 
disorder-averaged action $\overline{S_{\mathsf{DET}}^{\mathsf{dim}}}$ 
via\cite{footnote-i}
\begin{align}\label{DisAvg2}
	e^{-\overline{S_{\mathsf{DET}}^{\mathsf{dim}}}} 
	&\equiv \int \mathcal{D}{\bm{\mathrm{n}}}_{\mathsf{dim}} \,
	\exp\left[
	{-1\over{2
	\lambda_{A}
	}} 
	\int d^{d}{\bm{\mathrm{r}}} \,
	{\bm{\mathrm{n}}}_{\mathsf{dim}}^{2}({\bm{\mathrm{r}}})
	+ S_{\mathsf{DET}}^{\mathsf{dim}}
	\right] \nonumber\\
	&= \exp\left\lgroup 
	\frac{
	\lambda_{A}
	}{2} 
	\int d^{d}{\bm{\mathrm{r}}} \,
	\left[i \bm{\nabla} \Phi(\bm{\mathrm{r}})\right]^{2}
	\right\rgroup,
\end{align}
where the operator $i \bm{\nabla} \Phi$ was defined in Eq.~(\ref{U(1)TermDef}).
In the low energy effective field theory, then, the random hopping is characterized
by the two parameters
$\lambda^{{\scriptscriptstyle(}\mathsf{m}{\scriptscriptstyle)}}$ [Eq.~(\ref{DimerBondVar})] and
$\lambda_{A}$ [Eq.~(\ref{lambdaAMicroDef})];
the former sets the elastic scattering lifetime and the conductance,\cite{Gade,GLL,FC} and therefore
reflects the ``microscopic'' structure of the random hopping up to small distance scales of order 
the Fermi wavelength (responsible for elastic backscattering events involving large crystal momentum 
transfers), while the latter characterizes the orientational fluctuations of the random hopping at larger 
distance scales (of order the mean free path).

The last term on the third line of Eq.~(\ref{SDETexp}) is due to the interactions, and is given by
(see Appendix \ref{ExpTrick})
\begin{align}\label{SIrough1}
S_{I} \equiv i \, 2\tau \sum_{a = 1,2} 
	\int d t \, d^{d}{\bm{\mathrm{r}}} \,
	\big[
	& \rho_{A}^{a}(t,{\bm{\mathrm{r}}}) \, {Q}^{a, a}_{t, t}({\bm{\mathrm{r}}})
	\nonumber \\
	&+\rho_{B}^{a}(t,{\bm{\mathrm{r}}}) \, {Q}^{\dagger \, a, a}_{t, t}({\bm{\mathrm{r}}})
	\big].
\end{align}
Using Eq.~(\ref{Srho}), we can now perform the Gaussian integral over the auxiliary field $\rho$,
leading to the result
\begin{equation}\label{SIrough2}
	S_{I} \rightarrow i \sum_{a = 1, 2} \frac{\xi^{a}}{2} 
	\int d t \, d^{d}{\bm{\mathrm{r}}} \,
	\begin{bmatrix}
		Q^{a, a}_{t, t} & Q^{\dagger \, a, a}_{t, t}
	\end{bmatrix}\!\!
	\begin{bmatrix}
		\Gamma_{U} & \Gamma_{V} \\
		\Gamma_{V} & \Gamma_{U} 
	\end{bmatrix}\!\!
	\begin{bmatrix}
		Q^{a, a}_{t, t}\\
		Q^{\dagger \, a, a}_{t, t}
	\end{bmatrix}\!\!,
\end{equation}
where 
\begin{equation}\label{DEFGammaUGammaV}
	\Gamma_{U} \sim (2 \tau)^{2} U,
	\qquad
	\Gamma_{V} \sim (2 \tau)^{2} V 
\end{equation}
[c.f.\ Eqs.~(\ref{S2matrix}) and (\ref{CouplingMatrix})],
and
$1/\tau$ is the decay rate, Eqs.~(\ref{SCBA}) and (\ref{DEFtau}).
In moving between Eqs.~(\ref{Srho}),
(\ref{SIrough1}), and (\ref{SIrough2}), we have approximated the short-ranged position 
space functions $\hat{\mathsf{x}}$ and $\hat{\mathsf{y}}$, introduced as elements of 
the sublattice matrix $\hat{\mathcal{X}}$ in Eq.~(\ref{CouplingMatrix}) and defined 
below Eq.~(\ref{DensityVector}), as  
$\hat{\mathsf{x}} \sim \hat{\mathsf{y}} \rightarrow \delta^{(d)}(\bm{\mathrm{r}}-\bm{\mathrm{r'}})$.

\subsubsection{The FNL$\sigma$M and its coupling constants}\label{NLSMSummary}

Combining the results of the previous subsection, Eqs.~(\ref{SDETexp}), (\ref{DisAvg2}),
and (\ref{SIrough2}), we arrive at last to the final form of the Finkel'stein 
NL$\sigma$M (FNL$\sigma$M) description of the Hubbard-like model defined in Eqs.~(\ref{Hclean}) 
and (\ref{Hdis}). The FNL$\sigma$M is given by the functional integral
\begin{equation}\label{Z}
	Z = \int \mathcal{D}\hat{Q} \; e^{-S_{D}-S_{I}},
\end{equation}
where
\begin{align}
	S_{D} = 
	\frac{1}{2 \lambda} & \int d^{d}{\bm{\mathrm{r}}} \,
	\mathrm{Tr}
	\left\lgroup
	\bm{\nabla}\hat{Q}^{\dagger}({\bm{\mathrm{r}}}) 
	\cdot \bm{\nabla}\hat{Q}({\bm{\mathrm{r}}}) \right\rgroup
	\nonumber \\
	+ i \, h & \int d^{d}{\bm{\mathrm{r}}} \,
	\mathrm{Tr}
	\left\lgroup
	(\hat{\Sigma}_{3} |\hat{\omega}| + i \eta \hat{\Sigma}_{3} \hat{\xi}_{3})
	\,[\hat{Q}^{\dagger}({\bm{\mathrm{r}}})+\hat{Q}({\bm{\mathrm{r}}})]
	\right\rgroup
	\nonumber \\
	- & \frac{\lambda_{A}}{2 \lambda^{2}} \int d^{d}{\bm{\mathrm{r}}} 
	\left\lgroup
	\mathrm{Tr}
	\left[\hat{Q}^{\dagger}({\bm{\mathrm{r}}})
	\bm{\nabla}\hat{Q}({\bm{\mathrm{r}}})\right]
	\right\rgroup^{2}
	, \label{SD}
\end{align}
and
\begin{align}
	S_{I} = & i \sum_{a=1,2} \xi^{a} 
	\int dt \, d^{d}{\bm{\mathrm{r}}} 
	\left\lgroup
	2 \Gamma_{V} \, {Q}^{\dagger \, a, a}_{t, t}({\bm{\mathrm{r}}}) 
	{Q}^{a, a}_{t, t}({\bm{\mathrm{r}}})
	\right. 
	\nonumber \\
	& \left.
	+ \Gamma_{U}[{Q}^{a, a}_{t, t}({\bm{\mathrm{r}}}){Q}^{a, a}_{t, t}({\bm{\mathrm{r}}}) 
	+ {Q}^{\dagger \, a, a}_{t, t}({\bm{\mathrm{r}}})
	{Q}^{\dagger \, a, a}_{t, t}({\bm{\mathrm{r}}})]
	\right\rgroup.
	\label{SI}
\end{align}

The field variable $\hat{Q}(\bm{\mathrm{r}})$ in Eqs.~(\ref{Z})--(\ref{SI}) is a 
complex, infinite-dimensional unitary matrix, 
\begin{equation}\label{QmatrixUnitary2}
	\hat{Q}^{\dagger}({\bm{\mathrm{r}}}) \hat{Q}({\bm{\mathrm{r}}}) = \hat{1},
\end{equation}
carrying Keldysh species ($\{a,a'\}$) and time ($\{t,t'\}$) or frequency ($\{\omega,\omega'\}$) indices: 
\begin{equation}\label{QmatrixDef2}
	\hat{Q}({\bm{\mathrm{r}}}) \rightarrow Q^{a, a'}_{\omega, \omega'}({\bm{\mathrm{r}}})
	\;\;\textrm{ or }\;\;
	\hat{Q}({\bm{\mathrm{r}}}) \rightarrow Q^{a, a'}_{t, t'}({\bm{\mathrm{r}}}).
\end{equation}
In Eq.~(\ref{Z}), $\mathcal{D}\hat{Q}$ is the invariant (Haar) functional 
measure for the group manifold associated with $\hat{Q}$.
The matrix $\hat{Q}$ and its adjoint $\hat{Q}^{\dagger}$ may be interpreted
as continuum versions of the \emph{same-sublattice} fermion bilinears
\begin{subequations}\label{QmatrixID}
\begin{align}
	Q^{a, a'}_{t, t'} & \sim c^{a}_{A}(t) \bar{c}^{a'}_{A}(t'), \\
	Q^{\dagger \, a, a'}_{t, t'} & \sim c^{a}_{B}(t) \bar{c}^{a'}_{B}(t'),
\end{align} 
\end{subequations}
which follows from Eqs.~(\ref{FermMatBilComp}), (\ref{FermMatBil}), (\ref{HSdis}), and (\ref{QComp}).  
It is a peculiar feature of the random hopping model that the low energy 
fluctuations on a given sublattice are described by a unitary, rather than 
a Hermitian matrix, familiar from other models.

The $\mathrm{Tr}$ in Eq.~(\ref{SD}) denotes a matrix trace over Keldysh species 
and time or frequency indices; the diagonal Pauli matrices 
$\hat{\Sigma}_{3} \rightarrow \sgn{\omega} \, \delta_{\omega, \omega'} \, \delta^{\, a, a'}$ and 
$\hat{\xi}_{3} \rightarrow \xi^{a}  \, \delta_{\omega, \omega'} \, \delta^{\, a, a'}$ 
act in the $\sgn(\omega)$ and Keldysh species spaces, respectively, while
$|\hat{\omega}| \rightarrow |\omega| \, \delta_{\omega, \omega'} \, \delta^{\, a, a'}$ 
is a matrix of absolute frequencies. 
Here and in Eq.~(\ref{SI}), the number $\xi^{a}$ was defined in 
Eq.~(\ref{KeldyshFactor}).
The saddle point configuration $\hat{Q}(\bm{\mathrm{r}}) = \hat{Q}_{\mathsf{SP}}$, 
determined above in Eq.~(\ref{QmatrixSP1}), is repeated here for convenience:
\begin{equation}\label{QmatrixSP2}
	\hat{Q}_{\mathsf{SP}} = \hat{\Sigma}_{3} \hat{\xi}_{3}.
\end{equation}
The structure of $\hat{Q}_{\mathsf{SP}}$ is set by the symmetry breaking 
(Keldysh pole prescription) term, proportional to $\eta \rightarrow 0^{+}$,
in Eq.~(\ref{SD}).

The action given by Eq.~(\ref{SD}) describes the low-energy diffusive physics 
of the \emph{non}-interacting random hopping model;\cite{Gade,GLL,FC} a replica 
version of the non-interacting sigma model with action $S_{D}$ was originally studied by Gade 
and Wegner.\cite{Gade} This non-interacting sector of the FNL$\sigma$M is 
parameterized by three coupling strengths: $\lambda$, $\lambda_{A}$, and $h$. 
The coupling constant $\lambda$ is a coarse-grained measure of the microscopic 
hopping disorder strength, 
$\lambda^{{\scriptscriptstyle(}\mathsf{m}{\scriptscriptstyle)}}$ [Eq.~(\ref{DimerBondVar})];
we do not provide a precise relationship between $\lambda$ and 
$\lambda^{{\scriptscriptstyle(}\mathsf{m}{\scriptscriptstyle)}}$, since such a connection
depends upon various non-universal lattice-scale details.
It can be shown that $1/{\lambda}$ is proportional to the dimensionless dc conductance 
$g$ of the system.\cite{LeeRamakrishnan,Finkelstein,BK} 
The parameter $\lambda_{A}$ gives a second measure of the disorder strength, 
unique to this ``sublattice symmetry class,'' which strongly influences the behavior of the 
low-energy, disorder-averaged single-particle density of states.\cite{Gade,FC,GLL} 
[In Eq.~(\ref{SD}), we have rescaled $\lambda_{A}$ by a factor of $1/\lambda^{2}$ relative to
Eq.~(\ref{DisAvg2}), so that $\lambda$ and $\lambda_{A}$ now share the same naive
``engineering'' dimension. See also Sec.~\ref{DimAnalysis}.]
The parameter $\lambda_{A}$ may be simply interpreted
as characterizing the strength of long-wavelength, quenched orientational
fluctuations of bond strength dimerization in the microscopic random hopping disorder.
[See the discussion following Eq.~(\ref{SDETdim}) in the previous subsection for details.]
Finally, the parameter $h$ in Eq.~(\ref{SD}) is a dynamic scale factor, introduced here 
in order to track the scaling relationship between length and time as the model is 
renormalized.\cite{Finkelstein,BK}  

The interparticle interactions appear in the second term $S_{I}$, given by 
Eq.~(\ref{SI}). With the advent of Eq.~(\ref{QmatrixID}), we may interpret 
$Q^{a, a}_{t, t}({\bm{\mathrm{r}}})$ and 
$Q^{\dagger \, a, a}_{t, t}({\bm{\mathrm{r}}})$ as continuum local density 
operators on the `$A$' and `$B$' sublattices, respectively. Then the 
coarse-grained interaction strengths $\Gamma_{V} \propto V$ and 
$\Gamma_{U} \propto U$ in Eq.~(\ref{SI}) couple to generic, short-ranged 
intersublattice and same-sublattice density-density interactions, respectively 
[compare to Eq.~(\ref{Hclean}) above]. 

We use the renormalization group (RG) to study the model defined by Eqs.~(\ref{Z})--(\ref{SI}). 
It will prove convenient to define the linear combinations of $\Gamma_{U}$ and $\Gamma_{V}$,
\begin{equation}\label{SingletCDWGamma}
	\Gamma_{s} \equiv \frac{1}{2} (\Gamma_{U} + \Gamma_{V}),\qquad
	\Gamma_{c} \equiv \frac{1}{2} (\Gamma_{U} - \Gamma_{V}).
\end{equation}
In the continuum FNL$\sigma$M [Eq.~(\ref{SI})], the interaction strength 
$\Gamma_{s}$ couples to the squared (\emph{smooth}) local charge density, 
$[Q^{a, a}_{t, t}({\bm{\mathrm{r}}}) + Q^{\dagger \, a, a}_{t, t}({\bm{\mathrm{r}}})]^{2}$,
while $\Gamma_{c}$ couples to the squared sublattice \emph{staggered} charge density,
$[Q^{a, a}_{t, t}({\bm{\mathrm{r}}}) - Q^{\dagger \, a, a}_{t, t}({\bm{\mathrm{r}}})]^{2}$. 
In accordance with the discussion in the paragraph below Eq.~(\ref{CDWcoupling}) 
in Sec.~\ref{Intro}, we expect $\Gamma_{c} < 0$ to promote charge density wave 
formation, while $\Gamma_{c} > 0$ should suppress it.

We compute the one loop flow equations for the coupling constants $\lambda$, 
$\lambda_{A}$, $h$, $\Gamma_{s}$, and $\Gamma_{c}$ in the following two sections. We 
analyze and discuss our results in Sec.~\ref{Results}. The reader less interested 
in calculational details may skip Secs. \ref{PandFR} and \ref{oneloop} entirely, 
and proceed immediately to Sec.~\ref{Results}.

\subsection{Related models for spin-$1/2$ fermions}\label{Spin1/2}

To conclude this section, we briefly discuss some connections between various 
models of disordered (and possibly interacting) spin-$1/2$ fermions, and their 
corresponding random matrix theory classification and sigma model descriptions. 
Details are provided in Appendices \ref{Spin1/2RHopping} and \ref{SC}. 

Consider a clean system of spinless or spinful fermions with homogeneous, 
real nearest neighbor hopping on a bipartite lattice at half filling. In 
both the spinless and spinful cases, such a tight-binding model possesses 
three additional (non-spacetime) discrete symmetries: time-reversal invariance 
(TRI), sublattice symmetry (SLS), and particle-hole symmetry (PH). For spin-$1/2$ 
electrons, we also have spin SU(2) rotational symmetry. Here, we 
\emph{define} the unitary particle-hole transformation as a product of 
antiunitary time-reversal and sublattice symmetry transformations; as a result, 
in the spin-$1/2$ case, the PH transformation involves a spin flip--see 
Appendix \ref{Spin1/2RHopping} for details.  The introduction of quenched 
disorder may break or preserve each of these internal invariances, and the 
resulting disordered (non-interacting) Hamiltonian can be classified using 
random matrix theory.\cite{Zirnbauer,AltlandZirnbauer}

Consider now the effect of random magnetic fields upon the otherwise clean 
spin-$1/2$ hopping model. Here, it is crucial to distinguish between the 
cases of random orbital and random Zeeman magnetic fields. A random orbital 
field preserves spin SU(2) rotational symmetry and SLS, while breaking TRI 
and particle-hole symmetry (PH) in every static disorder realization. A 
random Zeeman field, on the other hand, preserves PH, but breaks TRI, SLS, 
and spin SU(2) rotational symmetry (completely). These two cases actually 
fall into \emph{different}
symmetry classes
of non-interacting, 
disordered quantum systems. In the classification scheme of 
Ref.~\onlinecite{Zirnbauer}, the random orbital field model, with SLS 
[and spin SU(2)] only, belongs to class AIII, 
while the random Zeeman field model, with PH only, belongs to class C.
These results are derived in Appendix \ref{Spin1/2RHopping}.
A model with both random Zeeman and orbital fields falls into the standard 
unitary Wigner-Dyson class A, since all three of the discrete symmetries TRI, SLS, 
and PH are broken. As discussed in the Introduction, known (non-interacting) 
realizations of the ``chiral'' class 
AIII,\cite{Gade,LeeFisher,GLL,FC,Furusaki,RyuHatsugai,BocquetChalker}
including the non-interacting version of the (spinless) model [Eqs.~(\ref{Hclean}) 
and (\ref{Hdis}) with $U = V = 0$] studied in this paper, possess 
a conducting phase
with extended states in $d = 2$ spatial dimensions. 
A system of non-interacting, spin-$1/2$ superconductor quasiparticles, with
broken TRI, and with spin SU(2) rotational symmetry preserved in every instance 
of the static disorder, furnishes a better known realization of class 
C.\cite{Zirnbauer,AltlandZirnbauer,ClassCsc,VishveshwaraSenthilFisher,JengLudwigSenthilChamon,FabrizioDellAnnaCastellani,SpinQH}
Non-interacting systems belonging to classes C or A do not possess extended states 
in $2\textrm{D}$,\cite{LeeRamakrishnan,ClassCRG,ClassCsc} except 
at their respective quantum Hall\cite{QH,SpinQH} transitions.

From this discussion, we conclude that the FNL$\sigma$M that we study in this paper 
also applies to a related Hubbard-like model for spin-$1/2$ fermions, subject to a random 
\emph{orbital} magnetic field, and possessing strong, homogeneous spin-orbit coupling. 
Spin-orbit coupling is needed to suppress an additional hydrodynamic spin diffusion 
channel, which we do not treat in this work. 
The $2\textrm{D}$ half-filled, spin-$1/2$ Hubbard model subject to a random Zeeman field was
studied numerically in Ref.~\onlinecite{ScalettarOD2--BDI+C}. The effects of interparticle interactions 
in the context of the superconductor quasiparticle interpretation of class C were analyzed 
using the FNL$\sigma$M in Ref.~\onlinecite{JengLudwigSenthilChamon}; see also 
Ref.~\onlinecite{FabrizioDellAnnaCastellani}. Very recently, Dell'Anna\cite{DellAnna} has 
independently studied several universality classes of Finkel'stein NL$\sigma$Ms, including 
realizations of both the particle-hole symmetric class C and the sublattice symmetric class 
AIII for electrically charged spin-$1/2$ fermions; his calculations should capture the low 
energy physics of the above-described spinful Hubbard model in random Zeeman and orbital magnetic 
fields, respectively; in the latter case, Dell'Anna includes the spin diffusion channel 
(i.e.\ assumes no spin-orbit coupling). His results are discussed briefly in the conclusion 
to this paper, Sec.~\ref{Conclusion}. Finally, we note that different classes appear in the 
presence of TRI.\cite{Zirnbauer,AltlandZirnbauer}

Surprisingly, our AIII Finkel'stein NL$\sigma$M equivalently
describes a system of spin-$1/2$ superconductor quasiparticles,
subject to disorder and 
interactions,
with TRI and a U(1) remnant of the spin SU(2) rotational 
symmetry preserved in every static disorder realization. Such a (gapless) quasiparticle 
system could occur, e.g., in the polar phase of a p-wave, spin-triplet 
superconductor.\cite{VollhardtWoelfle,Ho}
This quasiparticle system may be defined directly in the continuum, without reference 
to a lattice model or an additional sublattice/chiral symmetry. The connection is 
derived in Appendix \ref{SC}.


\section{Parameterization and Feynman Rules}\label{PandFR}

We turn now to the setup of our perturbative, one-loop renormalization group calculation 
for the FNL$\sigma$M defined by Eqs.~(\ref{Z})--(\ref{SI}). The actual RG computation follows 
in Sec.~\ref{oneloop}. Beyond developing the apparatus necessary for the RG, the material in 
this section serves also to further elucidate the structure of the sigma model description 
of sublattice symmetric disorder and interparticle interactions.

To begin, we shift the saddle point $\hat{\Sigma}_{3} \hat{\xi}_{3}$ [Eq.~(\ref{QmatrixSP2})] 
to the identity $\hat{1}$ via left group translation of the unitary matrix field 
$\hat{Q}(\bm{\mathrm{r}})$,
\begin{equation}\label{QmatrixSPLeft}
	\hat{Q}(\bm{\mathrm{r}}) \rightarrow \hat{\Sigma}_{3} \hat{\xi}_{3} \hat{Q}(\bm{\mathrm{r}}),
	\qquad
	\hat{Q}^{\dagger}(\bm{\mathrm{r}}) \rightarrow \hat{Q}^{\dagger}(\bm{\mathrm{r}}) \hat{\Sigma}_{3} \hat{\xi}_{3}.
\end{equation}
The FNL$\sigma$M action, Eqs.~(\ref{SD}) and (\ref{SI}), becomes
\begin{align}
	S_{D} = 
	& \frac{1}{2 \lambda} \int d^{d}{\bm{\mathrm{r}}} \,
	\mathrm{Tr}
	\left[
	\bm{\nabla}\hat{Q}^{\dagger} 
	\cdot \bm{\nabla}\hat{Q}
	\right]
	\nonumber \\
	& 
	- \frac{\lambda_{A}}{2 \lambda^{2}} \int d^{d}{\bm{\mathrm{r}}} \,
	\left[
	\mathrm{Tr}
	\left(\hat{Q}^{\dagger}
	\bm{\nabla}\hat{Q}
	\right)
	\right]^{2} 
	\nonumber \\
	& + i \, h \int d^{d}{\bm{\mathrm{r}}} \,
	\mathrm{Tr}
	\left\lgroup
	(|\hat{\omega}| \hat{\xi}_{3}  +  \hat{1} i \eta)
	\,(\hat{Q}^{\dagger}+\hat{Q})
	\right\rgroup, \label{SDQmatrixSPLeft}
\end{align}
and
\begin{align}
	S_{I} = \sum_{a} i \xi^{a} 
	& \int 
	\frac{d\omega_{1}}{2 \pi} \frac{d\omega_{2}}{2 \pi} \frac{d\omega_{3}}{2 \pi} \frac{d\omega_{4}}{2 \pi}
	d^{d}{\bm{\mathrm{r}}} \, \delta_{1+3,2+4}
	\nonumber \\
	&\times\bigg\lgroup
	\Gamma_{U}
	\left[
	s_{1} s_{3} \,
	{Q}^{a, a}_{1, 2}{Q}^{a, a}_{3, 4} 
	+ s_{2} s_{4} \,
	{Q}^{\dagger \, a, a}_{1, 2} {Q}^{\dagger \, a, a}_{3, 4}
	\right]
	\nonumber\\
	&\qquad\,+ 2 \Gamma_{V} \, s_{2} s_{3} \, 
	{Q}^{\dagger \, a, a}_{1, 2} {Q}^{a, a}_{3, 4}
	\bigg\rgroup.
	\label{SIQmatrixSPLeft}
\end{align}
Eq.~(\ref{SIQmatrixSPLeft}) expresses $S_{I}$ in frequency space, where we have 
implemented the compact notations 
\begin{align}
	\delta_{1+3,2+4} & \equiv \delta_{\omega_{1}+\omega_{3},\omega_{2}+\omega_{4}}, \nonumber\\
	{Q}^{a, a}_{1, 2}({\bm{\mathrm{r}}}) & \equiv {Q}^{a, a}_{\omega_{1}, \omega_{2}} ({\bm{\mathrm{r}}}), \nonumber\\
	s_{1} & \equiv (\hat{\Sigma}_{3})_{\omega_{1}, \omega_{1}} = \sgn(\omega_{1}), 
\end{align}
etc., i.e.\ numeric subscripts represent associated frequency labels.

The saddle point shift defined by Eq.~(\ref{QmatrixSPLeft}) modifies only the 
symmetry-breaking term, proportional to $h$, in the non-interacting sector of 
the theory [c.f. Eqs.~(\ref{SD}) and (\ref{SDQmatrixSPLeft})]. By contrast, the 
transformation in Eq.~(\ref{QmatrixSPLeft}) inserts explicit factors of 
$s_{i} s_{j} = \sgn(\omega_{i}) \sgn(\omega_{j})$, $i,j \in \{1,2,3,4\}$, into 
all terms inhabiting the interacting sector $S_{I}$ 
[Eqs.~(\ref{SI}) and (\ref{SIQmatrixSPLeft})]. We show below that these factors 
function as projection matrices, dividing the interaction-dressed diffusion 
modes into smooth and sublattice staggered charge density diffuson channels, 
characterized by interaction strengths $\Gamma_{s}$ and $\Gamma_{c}$, respectively. 
[See Eq.~(\ref{SingletCDWGamma}), as well as Eqs.~(\ref{SFast})--(\ref{YPropComponents}), 
below.]

We now parameterize the FNL$\sigma$M for the renormalization group (RG) computation. 
We employ a Wilsonian frequency-momentum shell, background field methodology.\cite{Finkelstein} 
The first step is to split $\hat{Q}$ into ``fast'' $\hat{Q}^{\phantom{\dagger}}_{\mathsf{F}}$ 
and ``slow'' $\hat{Q}^{\phantom{\dagger}}_{\mathsf{S}}$ mode parts,
\begin{align}\label{FastSlow}
	\hat{Q}(\bm{\mathrm{r}}) & \equiv 
	\hat{Q}^{\phantom{\dagger}}_{\mathsf{F}}(\bm{\mathrm{r}}) 
	\hat{Q}^{\phantom{\dagger}}_{\mathsf{S}}(\bm{\mathrm{r}}) 
	\nonumber \\
	& \equiv 
	\hat{Q}^{\phantom{\dagger}}_{\mathsf{F}}(\bm{\mathrm{r}}) 
	\left[\hat{1} + \delta\hat{Q}^{\phantom{\dagger}}_{\mathsf{S}}(\bm{\mathrm{r}})\right],
\end{align}
where both fast and slow matrix fields satisfy the unitary constraint 
[Eq.~(\ref{QmatrixUnitary2})]:
\begin{equation}\label{FastSlowUnitary}
	\hat{Q}_{\mathsf{F}}^{\dagger}(\bm{\mathrm{r}})\hat{Q}^{\phantom{\dagger}}_{\mathsf{F}}(\bm{\mathrm{r}})
	= \hat{Q}_{\mathsf{S}}^{\dagger}(\bm{\mathrm{r}})\hat{Q}^{\phantom{\dagger}}_{\mathsf{S}}(\bm{\mathrm{r}})
	= \hat{1}.
\end{equation}
On the second line of Eq.~(\ref{FastSlow}), we have further 
decomposed\cite{BaranovPruiskenSkoric} the slow mode field 
$\hat{Q}^{\phantom{\dagger}}_{\mathsf{S}}$ into the homogeneous saddle point 
$\hat{1}$ plus the ``small'' perturbation
$\delta\hat{Q}^{\phantom{\dagger}}_{\mathsf{S}}(\bm{\mathrm{r}})\equiv\hat{Q}^{\phantom{\dagger}}_{\mathsf{S}}(\bm{\mathrm{r}})-\hat{1}$. 

We require a parameterization for the unitary fast mode field $\hat{Q}_{\mathsf{F}}$ 
in terms of some unconstrained coordinates; we choose geodetic coordinates on the 
group manifold,
\begin{equation}\label{YFastParam}
	\hat{Q}^{\phantom{\dagger}}_{\mathsf{F}}(\bm{\mathrm{r}}) 
	\equiv \exp\left[i \hat{Y}(\bm{\mathrm{r}}) \right]
	\approx \hat{1} + i \hat{Y}(\bm{\mathrm{r}}) + \ldots,
\end{equation}
where $\hat{Y}^{\dagger} = \hat{Y}$ is a Hermitian matrix belonging to the Lie 
Algebra that generates $\hat{Q}^{\phantom{\dagger}}_{\mathsf{F}}$.

\begin{figure}
\includegraphics[width=0.4\textwidth]{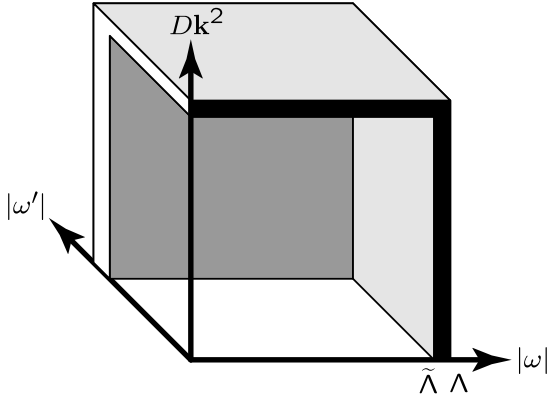}
\caption{Frequency-momentum shell of thickness ($\Lambda - \widetilde{\Lambda}$) 
supporting fast mode field coordinates $Y_{\omega, \omega'}^{a, a'}(\bm{\mathrm{k}})$. 
The cubic volume of linear size $\widetilde{\Lambda}$ enclosed by the shell supports 
the slow mode fields $\delta Q^{a, a'}_{\mathsf{S} \, \omega, \omega'}(\bm{\mathrm{k}})$ 
and $\delta Q^{\dagger \, a, a'}_{\mathsf{S} \, \omega, \omega'}(\bm{\mathrm{k}})$.
\label{FMShellFig}}
\end{figure}

We explain now the meaning of the fast-slow decomposition defined by Eq.~(\ref{FastSlow}).
The spatial Fourier transform of the slow mode fluctuation
\begin{equation}\label{QSlowFourierXfm}
	\delta\hat{Q}^{\phantom{\dagger}}_{\mathsf{S}}(\bm{\mathrm{k}})=
	\int d^{d}{\bm{\mathrm{r}}}\,
	\exp(-i\bm{\mathrm{k}}\cdot\bm{\mathrm{r}})\,
	\delta\hat{Q}^{\phantom{\dagger}}_{\mathsf{S}}(\bm{\mathrm{r}})
	\rightarrow \delta Q^{a, a'}_{\mathsf{S} \, \omega, \omega'}(\bm{\mathrm{k}})
\end{equation}
is taken to possess support within a cube of linear size $\widetilde{\Lambda}$, 
in the (3 dimensional) absolute frequency and squared-momentum space 
$(|\omega|,|\omega'|,D \bm{\mathrm{k}}^{2})$; the cube encompasses the origin 
$|\omega| = |\omega'| = D \bm{\mathrm{k}}^{2} = 0$. Here 
\begin{equation}\label{DDef}
	D \equiv 1/(\lambda h).
\end{equation}
is the effective diffusion constant.
The linear size of the cube satisfies $\widetilde{\Lambda} \ll 1/\tau \sim D/l^{2}$ 
(measured in energy units), where $\tau$ is the elastic scattering lifetime and $l$ 
the mean free path. We take the support of the fast mode coordinate
\begin{equation}\label{YFourierXfm}
	\hat{Y}(\bm{\mathrm{k}}) = 
	\int d^{d}{\bm{\mathrm{r}}}\exp(-i\bm{\mathrm{k}}\cdot\bm{\mathrm{r}})\,\hat{Y}(\bm{\mathrm{r}})
	\rightarrow Y_{\omega, \omega'}^{a, a'}(\bm{\mathrm{k}})
\end{equation}
to lie within a thin frequency-momentum shell enclosing the slow mode cube,
\begin{subequations}\label{FMShell}
\begin{align}	
	&\widetilde{\Lambda} \leq D \bm{\mathrm{k}}^{2} < \Lambda, 	& &0 \leq |\omega| < \Lambda,  	& &0 \leq |\omega'| < \Lambda; \\
	&\widetilde{\Lambda} \leq |\omega| < \Lambda,  	& &0 \leq D \bm{\mathrm{k}}^{2} < \Lambda, 	& &0 \leq |\omega'| < \Lambda; \\
	&\widetilde{\Lambda} \leq |\omega'| < \Lambda, 	& &0 \leq D \bm{\mathrm{k}}^{2} < \Lambda, 	& &0 \leq |\omega| < \Lambda,
\end{align}
\end{subequations}
where $\Lambda/\widetilde{\Lambda} \approx 1 + 2 dl$ is a ratio of energy cutoffs, 
with $0 < dl \ll 1$. The regions of slow and fast mode support are illustrated 
in Fig.~\ref{FMShellFig}. In the next section, we will integrate out the degrees 
of freedom in this shell and determine the resulting effect upon the slow modes 
$\delta\hat{Q}^{\phantom{\dagger}}_{\mathsf{S}}$.

Substituting the fast-slow decomposition [Eq.~(\ref{FastSlow})] into the action 
in Eqs.~(\ref{SDQmatrixSPLeft}) and (\ref{SIQmatrixSPLeft}), and expanding to 
second order\cite{footnote-j}
in the fast mode coordinates $\hat{Y}(\bm{\mathrm{r}})$ using Eq.~(\ref{YFastParam}), 
the Keldysh generating functional $Z$ [Eq.~(\ref{Z})] may be re-written as 
\begin{equation}\label{ZFastSlow}
	Z \sim \int \mathcal{D}\hat{Q}_{\mathsf{S}} \, \mathcal{D} \hat{Y}
	\exp\left\lgroup{-S[\hat{Y},\delta\hat{Q}^{\phantom{\dagger}}_{\mathsf{S}}]}\right\rgroup.
\end{equation}
We divide the action $S$ in Eq.~(\ref{ZFastSlow}) into terms containing (i) only 
slow modes, (ii) only fast modes, and (iii) both fast and slow modes:
\begin{equation}\label{SFastSlow}
	S = 
	S_{\mathsf{S}}[\delta\hat{Q}^{\phantom{\dagger}}_{\mathsf{S}}]
	+S_{\mathsf{F}}[\hat{Y}] 
	+S_{\mathsf{F}/\mathsf{S}}[\hat{Y},\delta\hat{Q}^{\phantom{\dagger}}_{\mathsf{S}}].
\end{equation}
The pure slow mode sector of the theory is
\begin{equation}\label{SSlow}
	S_{\mathsf{S}}[\delta\hat{Q}^{\phantom{\dagger}}_{\mathsf{S}}] = 
	S_{D}[\hat{Q} \rightarrow \hat{1}+\delta\hat{Q}^{\phantom{\dagger}}_{\mathsf{S}}] 
	+ S_{I}[\hat{Q} \rightarrow \hat{1}+\delta\hat{Q}^{\phantom{\dagger}}_{\mathsf{S}}],
\end{equation}
with $S_{D}$ and $S_{I}$ as given by Eqs.~(\ref{SDQmatrixSPLeft}) and 
(\ref{SIQmatrixSPLeft}). Of the remaining terms in Eq.~(\ref{SFastSlow}), 
$S_{\mathsf{F}}[\hat{Y}]$ contains only fast mode degrees of freedom, while 
$S_{\mathsf{F}/\mathsf{S}}[\hat{Y},\delta\hat{Q}^{\phantom{\dagger}}_{\mathsf{S}}]$ 
describes the coupling between the fast and slow mode fields. 

To one loop order, we need only retain terms in $S_{\mathsf{F}}[\hat{Y}]$ 
[Eq.~(\ref{SFastSlow})] to second order in $\hat{Y}(\bm{\mathrm{r}})$, thereby 
obtaining the fast mode Gaussian diffusion propagator. We write
\begin{equation}
	S_{\mathsf{F}} \equiv S_{\mathsf{F}}^{(D)}+ S_{\mathsf{F}}^{(I)},
\end{equation}
where the sector of the fast mode theory independent of the interparticle interactions is
\begin{widetext}
\begin{subequations}\label{SFast}
\begin{align}
	S_{\mathsf{F}}^{(D)} = 
	& \frac{1}{2} \int d^{d}{\bm{\mathrm{r}}}
	\left\lgroup
	\mathrm{Tr}
	\left[
	\frac{1}{\lambda}
	\left(\bm{\nabla}\hat{Y}\right)^{2}
	-i h 
	\left(|\hat{\omega}| \hat{\xi}_{3}  +  \hat{1} i \eta\right)
	\hat{Y}^{2}	
	-i h 
	\hat{Y}
	\left(|\hat{\omega}| \hat{\xi}_{3}  +  \hat{1} i \eta\right)
	\hat{Y}
	\right]
	+ \frac{\lambda_{A}}{\lambda^{2}}
	\left[
	\mathrm{Tr}
	\left(
	\bm{\nabla}\hat{Y}
	\right)
	\right]^{2}
	\right\rgroup, \label{SFastD}\\
\intertext{while the interactions give rise to the term}
	S_{\mathsf{F}}^{(I)} = 
	& -\sum_{a} i \xi^{a} 
	\int 
	\frac{d\omega_{1}}{2 \pi} \frac{d\omega_{2}}{2 \pi} \frac{d\omega_{3}}{2 \pi} \frac{d\omega_{4}}{2 \pi}
	d^{d}{\bm{\mathrm{r}}}	
	\left\lgroup
	\delta_{1+3,2+4}
	\left[
	\Gamma_{s} (s_{1}-s_{2})(s_{3}-s_{4})
	+\Gamma_{c} (s_{1}+s_{2})(s_{3}+s_{4})
	\right]
	Y_{1,2}^{a,a} Y_{3,4}^{a,a}
	\right\rgroup \label{SFastI}.
\end{align}
\end{subequations}
\end{widetext}
We have expressed $S_{\mathsf{F}}^{(I)}$, Eq.~(\ref{SFastI}) containing the 
interparticle interactions, in terms of
the smooth and of the
sublattice staggered 
(CDW) charge density interaction parameters $\Gamma_{s}$ and $\Gamma_{c}$, defined 
by Eq.~(\ref{SingletCDWGamma}). In Eq.~(\ref{SFastI}), $\Gamma_{s}$ and $\Gamma_{c}$ 
couple to the frequency index tensors $(s_{1} - s_{2})(s_{3} - s_{4})\delta_{1+3,2+4}$ 
and $(s_{1} + s_{2})(s_{3} + s_{4})\delta_{1+3,2+4}$, respectively,
which project out the channels of the propagator $\langle Y_{1,2}^{a,a} Y_{3,4}^{a,a}\rangle$ 
off-diagonal and diagonal in $\sgn(\omega)$ space, respectively. These channels 
are orthogonal, so that Eqs.~(\ref{SFastD}) and (\ref{SFastI}) may be simply inverted 
to obtain the fast mode propagator for the theory. Note that all Keldysh indices
must be identical to obtain a non-zero contribution from Eq.~(\ref{SFastI}).

The fast mode propagator cleanly decomposes into four disparate components; in 
terms of the spatial Fourier transform
$\hat{Y}(\bm{\mathrm{k}})$ defined via Eq.~(\ref{YFourierXfm}), 
\begin{equation}\label{YProp}
	\left\langle Y^{a, b}_{1,2}(-\bm{\mathrm{k}}) Y^{c, d}_{3,4}(\bm{\mathrm{k}})\right\rangle
	\equiv \mathsf{P_{\lambda}}+\mathsf{P_{A}}+\mathsf{P_{S}}+\mathsf{P_{C}},
\end{equation}
where the four symbols $\mathsf{P_{\lambda}}$, $\mathsf{P_{A}}$, $\mathsf{P_{S}}$,  
and $\mathsf{P_{C}}$ represent the components
\begin{widetext}
\begin{subequations}\label{YPropComponents}
\begin{align}
	\mathsf{P_{\lambda}} &=
	\Delta_{\mathcal{O}}^{a,b}(|\omega_{1}|,|\omega_{2}|,\bm{\mathrm{k}}) \,
	\delta^{a,d} \delta^{c,b} \delta_{1,4} \delta_{3,2} 
	\label{HProp}\\
	\mathsf{P_{A}} &=
	-\left[\frac{\lambda_{A}}{\lambda^{2}} \bm{\mathrm{k}}^{2} 
	\Delta_{\mathcal{O}}^{a,a}(|\omega_{1}|,|\omega_{1}|,\bm{\mathrm{k}}) 
	\Delta_{\mathcal{O}}^{c,c}(|\omega_{3}|,|\omega_{3}|,\bm{\mathrm{k}})
	\right]\,
	\delta^{a,b} \delta^{c,d} \delta_{1,2} \delta_{3,4} 
	\label{U1Prop}\\
	\mathsf{P_{S}} &=
	\bigg[2 i \xi^{a} \Gamma_{s} (s_{1}-s_{2})(s_{3}-s_{4})\,  
	\Delta_{\mathcal{S}}^{a}(|\omega_{1}-\omega_{2}|,\bm{\mathrm{k}}) 
	\Delta_{\mathcal{O}}^{a,a}(|\omega_{1}-\omega_{2}|,0,\bm{\mathrm{k}})
	\bigg] \delta^{a,b} \delta^{a,d} \delta^{c,d} \delta_{1+3,2+4} 
	\label{SProp}\\
	\mathsf{P_{C}} &= 
	\bigg[ 2 i \xi^{a} \Gamma_{c} (s_{1}+s_{2})(s_{3}+s_{4})\,
	\frac{
	\Delta_{\mathcal{O}}^{a,a}(|\omega_{1}|,|\omega_{2}|,\bm{\mathrm{k}}) 
	\Delta_{\mathcal{O}}^{a,a}(|\omega_{3}|,|\omega_{4}|,\bm{\mathrm{k}})
	}
	{1+ \gamma_{c} f^{a}(|\omega_{1}-\omega_{2}|,\bm{\mathrm{k}})}
	\bigg] \delta^{a,b} \delta^{a,d} \delta^{c,d} \delta_{1+3,2+4}.
	\label{CProp}
\end{align} 
\end{subequations}
\end{widetext}
Keldysh $\{a, b, \ldots\}$ and frequency $\{1, 2, \ldots\}$ indices in 
Eq.~(\ref{YPropComponents}) should be matched to those in Eq.~(\ref{YProp}).
Components $\mathsf{P_{\lambda}}$ and $\mathsf{P_{A}}$ [Eqs.~(\ref{HProp}) and (\ref{U1Prop})] 
follow from the inversion of the disorder-only action in Eq.~(\ref{SFastD}), 
while $\mathsf{P_{S}}$ and $\mathsf{P_{C}}$ [Eqs.~(\ref{SProp}) and (\ref{CProp})] 
incorporate the interactions from Eq.~(\ref{SFastI}). 

\begin{figure}
\includegraphics[width=0.4\textwidth]{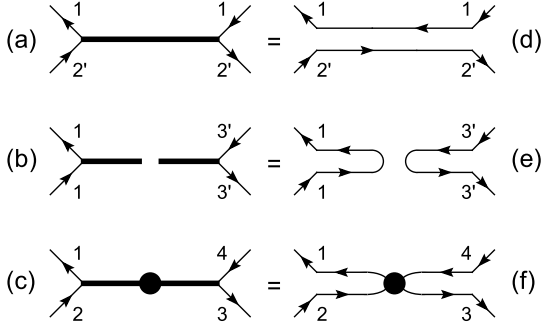}
\caption{Feynman rules I: fast mode propagator, from Eqs.~(\ref{YProp}) and 
(\ref{YPropComponents}). Components of the disorder-only sector
$\mathsf{P_{\lambda}}$ and $\mathsf{P_{A}}$ are depicted in \textsf{(a)} 
and \textsf{(b)}, respectively; \textsf{(c)} represents the sum of interaction-dressed 
components $\mathsf{P_{S}} + \mathsf{P_{C}}$, which vanishes in the non-interacting 
limit. Subfigures \textsf{(d)}, \textsf{(e)}, and \textsf{(f)} show that
each thick line in \textsf{(a)}, \textsf{(b)}, and \textsf{(c)}, respectively, can 
be understood as a \emph{pair} of particle and hole lines, which carry 
counterpropagating arrows to indicate the flow of the conserved electric 
U(1) current. Numeric labels represent the frequency indices of single fermion 
lines, while primes denote the associated Keldysh species indices. 
Numeric labels with the same number of primes share the same Keldysh 
index, while labels with different numbers of primes possess independent 
Keldysh indices. 
\label{FeynmanRulesProps}}
\end{figure}

\begin{figure}
\includegraphics[width=0.4\textwidth]{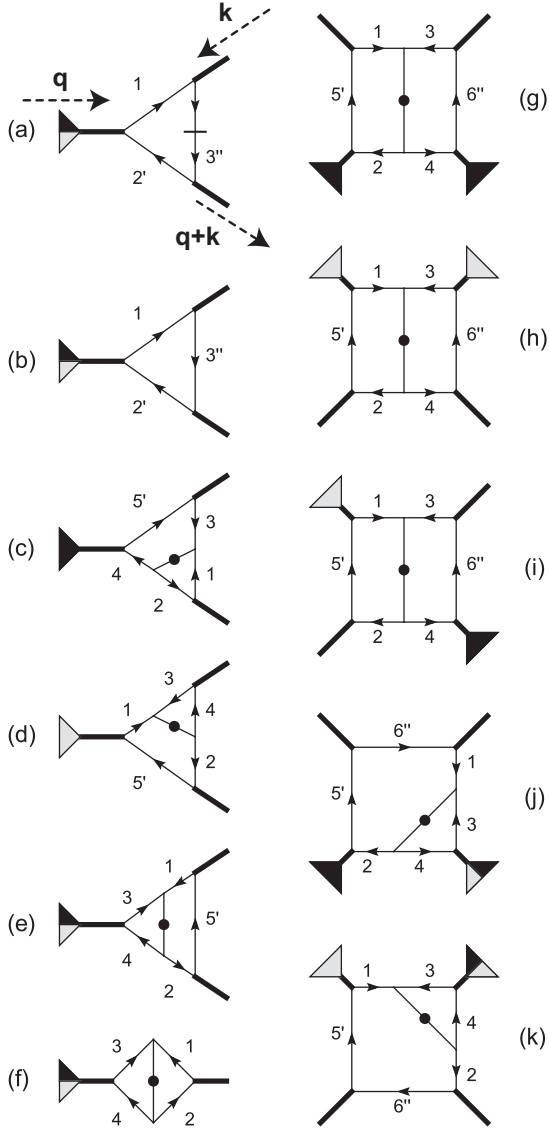}
\caption{Feynman rules II: vertices coupling together fast and slow modes 
arising from the term 
$S_{\mathsf{F}/\mathsf{S}}[\hat{Y},\delta\hat{Q}^{\protect\phantom{\dagger}}_{\mathsf{S}}]$ 
in Eq.~(\ref{SFastSlow}). Vertex $\mathsf{V}(\theta)$ is pictured 
in subfigure ($\theta$), with $\theta \in \{\mathsf{a,b,c},\ldots\}$. 
In this figure, black and blank triangular terminals indicate 
$\delta\hat{Q}^{\protect\phantom{\dagger}}_{s}$ and $\delta\hat{Q}_{s}^{\dagger}$ 
slow mode fields, respectively, while half-black, half-blank terminals 
indicate linear combinations of these. Unterminated thick lines represent 
fast mode fields $\hat{Y}$. Vertices $\mathsf{V(a)}$ and $\mathsf{V(b)}$ 
derive from the stiffness and energy terms, respectively, of the 
non-interacting sigma model action [Eq.~(\ref{SDQmatrixSPLeft})], 
while vertices $\mathsf{V(c)}$--$\mathsf{V(k)}$ arise from the interparticle 
interactions [Eq.~(\ref{SIQmatrixSPLeft})]. Momentum labels 
$\bm{\mathrm{q}}$, $\bm{\mathrm{k}}$, and $\bm{\mathrm{q}}+\bm{\mathrm{k}}$ 
have been furnished for the stiffness vertex $\mathsf{V(a)}$, as this vertex 
depends explicitly upon these momenta. 
\label{FeynmanRulesVertices}}
\end{figure}

The basic (heat) diffuson kernel in Eq.~(\ref{HProp}) is
\begin{equation}\label{HeatDiffuson}
	\Delta_{\mathcal{O}}^{a,b}(|\omega_{1}|,|\omega_{2}|,\bm{\mathrm{k}})
	\equiv \frac{1}{h}\left[D \bm{\mathrm{k}}^{2} -i (\xi^{a}|\omega_{1}| + \xi^{b}|\omega_{2}|)\right]^{-1},
\end{equation}
with the diffusion constant $D$ defined via Eq.~(\ref{DDef}). 
The diffuson matrix propagator $\mathsf{P_{\lambda}}$ in Eqs.~(\ref{HProp}) 
is pictured as a thick line segment in Fig.~\ref{FeynmanRulesProps}\textsf{(a)}, 
with ends that split into pairs of directed thin lines. As shown in 
Figs.~\ref{FeynmanRulesProps}\textsf{(d)}--\ref{FeynmanRulesProps}\textsf{(f)},
each such thick line segment in 
Figs.~\ref{FeynmanRulesProps}\textsf{(a)}--\ref{FeynmanRulesProps}\textsf{(c)} 
can be taken to represent a \emph{pair} of single fermion particle and hole lines, 
carrying counterpropagating arrows indicating the flow of the conserved electric 
U(1) current [See, e.g., Eq.~(\ref{QmatrixID})]. The numeric labels appearing 
along the terminating thin lines in Fig.~\ref{FeynmanRulesProps}\textsf{(a)} 
encode the frequency and Keldysh indices carried by the fast mode matrix fields 
$\hat{Y} \rightarrow Y_{\omega, \omega'}^{a, a'}$. These indices are propagated 
along (unbroken) thick line segments \emph{without} mixing, as in 
Fig.~\ref{FeynmanRulesProps}\textsf{(d)}. In both Figs.~\ref{FeynmanRulesProps} 
and \ref{FeynmanRulesVertices}, we employ the following convention: different 
numerical labels encode independent frequency indices, and numerical labels 
carrying different numbers of primes indicate independent Keldysh species indices; 
numerical labels with the same number of primes represent indices that share 
the same Keldysh species. Each thick line splitting in Fig.~\ref{FeynmanRulesProps} 
corresponds to the insertion of a fast mode field $Y_{\omega, \omega'}^{a, a'}$, 
where the ``left'' indices $\{\omega,a\}$ accompany the arrow flowing \emph{out of} 
the thick line, while the ``right'' indices $\{\omega',a'\}$ accompany the 
arrow flowing \emph{into} the thick line.

Inversion of Eq.~(\ref{SFastD}) also gives the component $\mathsf{P_{A}}$ 
proportional to $\lambda_{A}$, Eq.~(\ref{U1Prop}), which has a different frequency 
and Keldysh index structure than the basic diffuson $\mathsf{P_{\lambda}}$ 
[Eq.~(\ref{HProp})]. $\mathsf{P_{A}}$ is depicted in Fig.~\ref{FeynmanRulesProps}\textsf{(b)}. 
The inversion of Eq.~(\ref{SFastD}) gives only the sum of $\mathsf{P_{\lambda}}$ 
and $\mathsf{P_{A}}$, i.e.\ terms up to first order in $\lambda_{A}$. We can try 
to build terms higher order in $\lambda_{A}$ by cascading together multiple 
such sections, but it is clear from Figs.~\ref{FeynmanRulesProps}\textsf{(b)} 
and \ref{FeynmanRulesProps}\textsf{(e)} that such a construction necessarily 
contains at least one closed ``Keldysh'' loop, defined here as a simultaneous 
Keldysh species index summation and frequency integration along a closed 
single fermion line. As with the replica trick, such closed loops vanish 
in the Keldysh formalism.

The smooth and sublattice staggered interparticle interactions in Eq.~(\ref{SFastI}), 
characterized by $\Gamma_{s}$ and $\Gamma_{c}$, respectively, dress the bare 
diffuson propagator $\mathsf{P_{\lambda}}$ [Eq.~(\ref{HProp})], giving rise 
to the propagator components $\mathsf{P_{S}}$ and $\mathsf{P_{C}}$, Eqs.~(\ref{SProp}) 
and (\ref{CProp}), respectively. These equations show that the component 
$\mathsf{P_{S}}$ ($\mathsf{P_{C}}$) projects upon the propagator channel 
off-diagonal (diagonal) in $\sgn(\omega)$ space. The kernel
\begin{equation}\label{SDiffuson}
	\Delta_{\mathcal{S}}^{a}(|\omega|,\bm{\mathrm{k}})
	\equiv \frac{1}{h}\left[D \bm{\mathrm{k}}^{2} -i (1-\gamma_{s})\xi^{a}|\omega| \right]^{-1}
\end{equation}
appears in $\mathsf{P_{S}}$, the channel of the interaction-dressed propagator 
off-diagonal in $\sgn(\omega)$ space [Eq.(\ref{SProp})], and is related\cite{Finkelstein,BK} 
to the diffusion of the physical, conserved electric U(1) charge. 
In Eqs.~(\ref{SDiffuson}) and (\ref{CProp}), we have introduced the relative 
interaction constants
\begin{subequations}\label{SingletCDW}
\begin{align}
	\gamma_{s} \equiv & \frac{4}{\pi h} \Gamma_{s} = \frac{2}{\pi h}(\Gamma_{U} + \Gamma_{V}), \\
	\gamma_{c} \equiv & \frac{4}{\pi h} \Gamma_{c} = \frac{2}{\pi h}(\Gamma_{U} - \Gamma_{V}),
\end{align}
\end{subequations}
where we have used Eq.~(\ref{SingletCDWGamma}).
Finally, the logarithmic function
\begin{equation}\label{DiffLog}
	f^{a}(|\omega|,\bm{\mathrm{k}}) \equiv 
	\ln\left[\frac{2 \Lambda}{|\omega|+i \xi^{a} D {\bm{\mathrm{k}}}^{2}}\right],
\end{equation}
with $\Lambda$ a hard frequency cutoff [c.f.\ Eq.~(\ref{FMShell})], appears in 
$\mathsf{P_{C}}$, the channel of the interaction-dressed propagator diagonal 
in $\sgn(\omega)$ space [Eq.~(\ref{CProp})]. The structure\cite{BK,KotliarSorella,KBCooperon}
 of $\mathsf{P_{C}}$ does \emph{not} follow from any conservation law, as the 
sublattice \emph{staggered} electric charge density (the `CDW' order parameter 
associated with the interaction strength $\Gamma_{c}$) does not represent 
a conserved quantity. In this paper, we will work only to the lowest non-trivial 
order in $\gamma_{c}$; practically, this means ignoring\cite{Finkelstein,KotliarSorella} 
the logarithmic denominator [Eq.~(\ref{DiffLog})] in $\mathsf{P_{C}}$ [Eq.~(\ref{CProp})]. 
This approximation is adequate for all of the results presented in Sec.~\ref{Results}. 
The sum of the interaction-dressed propagator components $\mathsf{P_{S}}+\mathsf{P_{C}}$ 
will be depicted as a thick line with a black dot, as shown in 
Fig.~\ref{FeynmanRulesProps}\textsf{(c)}.

The channel of the fast mode propagator diagonal in $\sgn(\omega)$ space, 
including the components $\mathsf{P_{A}}$, proportional to the disorder 
strength $\lambda_{A}$ [Eq.~(\ref{U1Prop})], and $\mathsf{P_{C}}$, proportional 
to the CDW interaction strength $\Gamma_{c}$ [Eq.~(\ref{CProp})], are 
special to a system with sublattice symmetry [Eq.~(\ref{SLS})]. (Recall
from Secs.~\ref{SPGradient} and \ref{NLSMSummary} that $\lambda_{A}$ measures 
the strength of quenched orientational fluctuations of bond dimerization 
in the intersublattice hopping disorder.) The addition of SLS breaking, e.g.\ in the form of 
on-site disorder, changes the random matrix class\cite{Zirnbauer} of the (non-interacting 
version of the) random hopping model studied in this paper, Eqs.~(\ref{Hclean}) 
and (\ref{Hdis}) with $U = V = 0$, from the ``sublattice/chiral'' class AIII 
to the ordinary unitary metal class $\mathrm{A}$. Equivalently, breaking SLS 
reduces the size of the sigma model target manifold, as discussed in Appendix 
\ref{KeldyshSym}. The crossover is marked by the appearance of a ``mass'' in 
the FNL$\sigma$M propagator channel diagonal in $\sgn(\omega)$, which gaps out 
the propagator components $\mathsf{P_{A}}$ and $\mathsf{P_{C}}$ [as well as 
``half'' of the basic diffuson modes $\mathsf{P_{\lambda}}$ in Eq.~(\ref{HProp})]. 
This is the reason that the coupling strengths $\lambda_{A}$ and $\Gamma_{c}$ 
do not appear in the spinless unitary class Finkel'stein NL$\sigma$M,\cite{BK} 
which retains the parameters $\lambda$, $h$, and $\Gamma_{s}$.

\begin{table}[b]
\caption{
	\label{VertexTable} Feynman rules III: factors associated with vertices 
	$\{\mathsf{V(a)},\mathsf{V(b)},\ldots\}$, pictured in 
	Fig.~\ref{FeynmanRulesVertices}, coupling together fast and slow modes. 
	Frequency $\{1,2,3,\dots\}$ and Keldysh species $\{a, a', a''\}$ indices 
	in the factor $\mathsf{V}(\theta)$ should be matched to those in the 
	corresponding Figure \ref{FeynmanRulesVertices}($\theta$).
	Fast mode fields $\hat{Y}$ have been amputated from the vertex factors 
	in this table for brevity; the fast-slow mode coupling structure may be 	
	understood from Fig.~\ref{FeynmanRulesVertices}. The (spatial Fourier 
	transform of the) slow mode operator 
	$\bm{\mathrm{L}}_{\mathsf{S}\, 1,2}^{a, a'}(\bm{\mathrm{q}})$ is defined 
	by Eq.~(\ref{LSDef}).
	}
\begin{ruledtabular}
\begin{tabular}{l @{} l @{} l}
$\mathsf{V(a)}$ &$=$ 
&
$
	-({i}/{2 \lambda}) \int \left[{d^{d} \bm{\mathrm{q}}}/{(2 \pi)^{d}}\right]
	(2\bm{\mathrm{k}} + \bm{\mathrm{q}}) \cdot \bm{\mathrm{L}}_{\mathsf{S}\, 1,2}^{a, a'}(\bm{\mathrm{q}})
$
\medskip\\
$\mathsf{V(b)}$ &$=$
&
$	
	({i h}/{2}) \int {d^{d} \bm{\mathrm{r}}}
	\left(
	|{\omega_{2}}| {\xi}^{a'} \delta Q_{\mathsf{S}\,1,2}^{a, a'} +
	|{\omega_{1}}| {\xi}^{a}  \delta Q_{\mathsf{S}\,1,2}^{\dagger \, a, a'}
	\right)
$
\medskip\\
$\mathsf{V(c)}$ &$=$ 
&	
$
	2 i \xi^{a} \delta_{1+3,2+4} 
	\left(\Gamma_{U} s_{1} s_{3} - \Gamma_{V} s_{2} s_{3}\right)
	\int {d^{d} \bm{\mathrm{r}}} \,
	\delta Q_{\mathsf{S}\,5,4}^{a', a}
$
\medskip\\
$\mathsf{V(d)}$ &$=$
&
$	2 i \xi^{a} \delta_{1+3,2+4} 
	\left(\Gamma_{U} s_{2} s_{4} - \Gamma_{V} s_{2} s_{3}\right)
	\int {d^{d} \bm{\mathrm{r}}} \,
	\delta Q_{\mathsf{S}\,1,5}^{\dagger \, a, a'}
$
\medskip\\
$\mathsf{V(e)}$ &$=$
&
$
	i \xi^{a} \delta_{1+3,2+4}
	\int {d^{d} \bm{\mathrm{r}}} \,
	\left[ 
	\Gamma_{s} (s_{1} + s_{2})(s_{3}\delta Q_{\mathsf{S}\,3,4}^{a, a}+s_{4}\delta Q_{\mathsf{S}\,3,4}^{\dagger \, a, a}) \right.
$
\smallskip\\
& & 	
$
	\qquad\qquad\qquad\quad\left.+
	\Gamma_{c} (s_{1} - s_{2})(s_{3}\delta Q_{\mathsf{S}\,3,4}^{a, a}-s_{4}\delta Q_{\mathsf{S}\,3,4}^{\dagger \, a, a})
	\right]
$
\medskip\\
$\mathsf{V(f)}$ &$=$
&
$
	2 \xi^{a} \delta_{1+3,2+4}
	\int {d^{d} \bm{\mathrm{r}}} \,
	\left[ 
	\Gamma_{s} (s_{1} - s_{2})(s_{3}\delta Q_{\mathsf{S}\,3,4}^{a, a}+s_{4}\delta Q_{\mathsf{S}\,3,4}^{\dagger \, a, a}) \right.
$
\smallskip\\
& & 	
$
	\qquad\qquad\qquad\quad\left.+
	\Gamma_{c} (s_{1} + s_{2})(s_{3}\delta Q_{\mathsf{S}\,3,4}^{a, a}-s_{4}\delta Q_{\mathsf{S}\,3,4}^{\dagger \, a, a})
	\right]
$
\medskip\\
$\mathsf{V(g)}$ &$=$
&
$
	2 i \xi^{a} \delta_{1+3,2+4} 
	\Gamma_{U} s_{1} s_{3}
	\int {d^{d} \bm{\mathrm{r}}} \,
	\delta Q_{\mathsf{S}\,5,2}^{a', a}
	\delta Q_{\mathsf{S}\,6,4}^{a'', a}
$
\medskip\\
$\mathsf{V(h)}$ &$=$
&
$
	2 i \xi^{a} \delta_{1+3,2+4} 
	\Gamma_{U} s_{2} s_{4}
	\int {d^{d} \bm{\mathrm{r}}} \,
	\delta Q_{\mathsf{S}\,1,5}^{\dagger \, a, a'}
	\delta Q_{\mathsf{S}\,3,6}^{\dagger \, a, a''}
$
\medskip\\
$\mathsf{V(i)}$ &$=$
&
$
	-2 i \xi^{a} \delta_{1+3,2+4} 
	\Gamma_{V} s_{2} s_{3}
	\int {d^{d} \bm{\mathrm{r}}} \,
	\delta Q_{\mathsf{S}\,1,5}^{\dagger \, a, a'}
	\delta Q_{\mathsf{S}\,6,4}^{a'', a}
$
\medskip\\
$\mathsf{V(j)}$ &$=$
&
$	
	i \xi^{a} \delta_{1+3,2+4}
	\int {d^{d} \bm{\mathrm{r}}} \left[
	\Gamma_{U} s_{1} s_{3} \delta Q_{\mathsf{S}\,5,2}^{a', a} \delta Q_{\mathsf{S}\,3,4}^{a, a} \right.
$\smallskip\\
& &
$ 
	\qquad\qquad\qquad\qquad{\left. +
	\Gamma_{V} s_{1} s_{4} \delta Q_{\mathsf{S}\,5,2}^{a', a} \delta Q_{\mathsf{S}\,3,4}^{\dagger \, a, a} \right]}
$
\medskip\\
$\mathsf{V(k)}$ &$=$
&
$	
	i \xi^{a} \delta_{1+3,2+4}
	\int {d^{d} \bm{\mathrm{r}}} \left[
	\Gamma_{U} s_{2} s_{4} \delta Q_{\mathsf{S}\,1,5}^{\dagger \, a, a'} \delta Q_{\mathsf{S}\,3,4}^{\dagger \, a, a} \right.
$\smallskip\\
& &
$ 
	\qquad\qquad\qquad\qquad{\left. +
	\Gamma_{V} s_{2} s_{3} \delta Q_{\mathsf{S}\,1,5}^{\dagger \, a, a'} \delta Q_{\mathsf{S}\,3,4}^{a, a} \right]}
$
\medskip\\
\end{tabular}
\end{ruledtabular}
\end{table}

The fast and slow mode fields are coupled together by the term 
$S_{\mathsf{F}/\mathsf{S}}[\hat{Y},\delta\hat{Q}^{\phantom{\dagger}}_{\mathsf{S}}]$ 
in Eq.~(\ref{SFastSlow}). 
$S_{\mathsf{F}/\mathsf{S}}[\hat{Y},\delta\hat{Q}^{\phantom{\dagger}}_{\mathsf{S}}]$ 
gives rise to the Feynman vertices pictured in Figs.~\ref{FeynmanRulesVertices}. 
We will refer to the vertex pictured in Fig.~\ref{FeynmanRulesVertices}($\theta$) 
as ``$\mathsf{V}(\theta)$,'' with $\theta \in \{\mathsf{a,b,c},\ldots\}$ In these 
figures, the vertex corners adorned by thick line ``spokes'' that split into 
directed thin lines represent insertions of the slow ($\delta\hat{Q}^{\phantom{\dagger}}_{\mathsf{S}}$ 
and/or $\delta\hat{Q}^{\dagger}_{\mathsf{S}}$) or fast ($\hat{Y}$) mode matrix 
fields. Specifically, the triangular black and blank ``spoke'' terminals denote 
the slow mode fields $\delta\hat{Q}^{\phantom{\dagger}}_{\mathsf{S}}$ and 
$\delta\hat{Q}^{\dagger}_{\mathsf{S}}$, respectively, while the half-black, 
half-blank terminals denote linear combinations of these. Each unterminated 
thick line ``spoke'' in Fig.~\ref{FeynmanRulesVertices} represents a fast 
mode field $\hat{Y}$, which may be interconnected in pairs using the propagators 
shown in Fig.~\ref{FeynmanRulesProps}, with amplitudes given by the associated 
expressions in Eq.~(\ref{YPropComponents}). As in Fig.~\ref{FeynmanRulesProps}, 
we use numerical labels to indicate frequency indices, with primes to distinguish 
independent Keldysh species indices. At a given fast or slow mode field (thick 
line ``spoke'') insertion, e.g.\ the 
$\delta\hat{Q}^{\phantom{\dagger}}_{\mathsf{S}} \rightarrow \delta{Q}^{a',a}_{\mathsf{S} \, 5,4}$ 
slow mode associated with the black triangular terminal in 
Fig.~\ref{FeynmanRulesVertices}\textsf{(c)}, the ``left'' indices $\{5,a'\}$ accompany 
the arrow flowing \emph{out of} the thick line ``spoke,'' into the vertex, while the 
``right'' indices $\{4,a\}$ accompany the arrow flowing \emph{into} the thick 
line, out of the vertex.

The factors associated with the vertices $\{\mathsf{V(a)},\mathsf{V(b)},\ldots\}$ 
are listed in Table \ref{VertexTable}. In this table, the fast mode fields 
$\hat{Y}$ have been amputated from the vertex expressions; the structure of the 
fast-slow mode coupling should be understood from Fig.~(\ref{FeynmanRulesVertices}). 
Vertices $\mathsf{V(a)}$ and $\mathsf{V(b)}$ obtain from the non-interacting 
sector of the FNL$\sigma$M action, Eq.~(\ref{SDQmatrixSPLeft}), while vertices 
$\mathsf{V(c)}$--$\mathsf{V(k)}$ derive from the interparticle interactions, 
Eq.~(\ref{SIQmatrixSPLeft}), indicated by the black dots in 
Figs.~\ref{FeynmanRulesVertices}\textsf{(c)}--\ref{FeynmanRulesVertices}\textsf{(k)}. 
The first entry of Table \ref{VertexTable} gives the factor associated with 
the ``stiffness vertex'' $\mathsf{V(a)}$, shown in 
Fig.~\ref{FeynmanRulesVertices}\textsf{(a)}. Here we have introduced the slow 
mode vector operator 
\begin{equation}\label{LSDef}
	\hat{\bm{\mathrm{L}}}_{\mathsf{S}}(\bm{\mathrm{r}}) 
	\equiv 
	\hat{Q}^{\phantom{\dagger}}_{\mathsf{S}}(\bm{\mathrm{r}})
	\bm{\nabla}
	\hat{Q}^{{\dagger}}_{\mathsf{S}}(\bm{\mathrm{r}})
	= \left[\hat{1} + \delta\hat{Q}^{\phantom{\dagger}}_{\mathsf{S}}(\bm{\mathrm{r}})\right]
	\bm{\nabla}
	\delta\hat{Q}^{{\dagger}}_{\mathsf{S}}(\bm{\mathrm{r}}).
\end{equation}

Note that all fast-slow vertices pictured in Fig.~\ref{FeynmanRulesVertices}, 
except $\mathsf{V(f)}$, are bilinear in the fast mode fields 
$\hat{Y}(\bm{\mathrm{r}})$ (represented by unterminated thick lines); the 
latter vertex involves only a single fast mode field. Such a vertex would 
vanish in a pure momentum shell treatment, in which one integrates undetermined 
fast mode loop momenta over a thin shell, while simultaneously integrating 
undetermined fast mode loop frequencies over the entire real line.  
We show in Sec.~\ref{oneloop} that a diagram involving two copies of 
$\mathsf{V(f)}$ produces the term in the RG flow equations that gives rise to 
the CDW instability in the \emph{clean} Hubbard-like model [Eq.~(\ref{Hclean})].  
As the CDW term must be present in the advent of sublattice symmetry [e.g.\ in 
the ballistic limit $\lambda, \lambda_{A} \rightarrow 0$], we are forced to work 
with the frequency-momentum shell method in this paper.
A similar term responsible for the BCS superconducting instability of the diffusive
Fermi liquid also arises in the FNL$\sigma$M description of normal, TRI metals
only in the frequency-momentum shell scheme.\cite{Finkelstein,KotliarSorella,BK,KBCooperon}


\section{One-loop Calculation}\label{oneloop}

With the setup outlined in Sec.~\ref{PandFR} complete, we commence here the 
renormalization group (RG) calculation proper for the FNL$\sigma$M originally 
defined by Eqs.~(\ref{Z})--(\ref{SI}). In two dimensions, the disorder parameters 
$\lambda$ and $\lambda_{A}$, as well as the ratios $\gamma_{s}$ and $\gamma_{c}$, 
defined by Eq.~(\ref{SingletCDW}), carry zero ``engineering'' dimension (demonstrated 
explicitly in Sec.~\ref{DimAnalysis}, below). In the language of the RG, we say 
that these parameters are marginal (in $d=2$) at tree level. In order 
to understand the infrared physics of the theory, we must therefore go beyond 
the reach of dimensional analysis. We compute here the one-loop RG flow equations 
for the coupling strengths $\lambda$, $\lambda_{A}$, and $h$, as well as 
$\Gamma_{U}$ and $\Gamma_{V}$, or equivalently, $\Gamma_{s}$ and $\Gamma_{c}$
[Eq.~(\ref{SingletCDWGamma})]. The loop expansion is performed in 
$d = 2 + \epsilon$ dimensions, with $0 \leq \epsilon \ll 1$, and is 
formally organized as an expansion in powers of the disorder strength $\lambda$, 
which is inversely proportional to the dimensionless DC conductance. As 
discussed below Eq.~(\ref{DiffLog}), we choose to work only to the lowest non-trivial 
order in the CDW interaction parameter $\Gamma_{c}$ (or $\gamma_{c}$); by contrast, 
the loop expansion incorporates contributions from $\lambda_{A}$, $h$, and $\Gamma_{s}$ 
to \emph{all} orders. We need make no assumptions about the smallness of these 
latter three parameters. The diagrammatics appear in Sec.~\ref{Diagrammatica}; 
the results of this section are combined with dimensional analysis in 
Sec.~\ref{DimAnalysis}, yielding the desired one-loop flow equations. 
The flow equations obtained here are summarized and interpreted in Sec.~\ref{Results}.


\subsection{Renormalization}\label{Diagrammatica}

The calculation is performed in the frequency-momentum shell background 
field formalism established in the previous section. We integrate out 
the fast modes $Y^{a,a'}_{\omega, \omega'}(\bm{\mathrm{k}})$ 
[Eq.~(\ref{YFastParam})] lying within the shell defined by Eq.~(\ref{FMShell}), 
where 
\begin{equation}\label{ShellRatio}
	\frac{\Lambda}{\widetilde\Lambda} \approx 1 + 2 dl,
\end{equation}
and $0 < dl \ll 1$, as below Eq.~(\ref{FMShell}).
[The quantity $dl$ may be understood as an infinitesimal change in 
the log of the \emph{length} scale $L$, with $L$ the linear system 
size. In the diffusive metallic regime, energy $\omega$ scales like 
the square of an inverse length: $\omega \sim D/L^{2}$; hence, the 
factor of ``2'' in the definition (\ref{ShellRatio}).] 
The fast mode integration produces corrections to the action 
$S_{\mathsf{S}}$ [Eq.~(\ref{SSlow})] for the slow mode fields
$\delta {Q}^{a,a'}_{\mathsf{S} \, \omega, \omega'}(\bm{\mathrm{k}})$ 
and 
$\delta {Q}^{\dagger \, a,a'}_{\mathsf{S} \, \omega, \omega'}(\bm{\mathrm{k}})$ 
[Eq~(\ref{FastSlow})]. The regions of fast and slow mode support 
were shown in Fig.~\ref{FMShellFig}. Note that in the case of the 
fast mode field $Y^{a,a'}_{\omega, \omega'}(\bm{\mathrm{k}})$, 
\emph{at least} one of the three variables 
$\alpha \in \{D \bm{\mathrm{k}}^{2}, |\omega|, |\omega'|\}$ is always 
``fast,'' lying within the range $\widetilde\Lambda < \alpha < \Lambda$.

In the pure slow mode sector of the theory, with action $S_{\mathsf{S}}$ 
defined via Eqs.~(\ref{SSlow}), (\ref{SDQmatrixSPLeft}), and 
(\ref{SIQmatrixSPLeft}), all five parameters $\lambda$, $\lambda_{A}$, 
$h$, $\Gamma_{U}$, and $\Gamma_{V}$ couple to local operators which 
contain non-vanishing terms quadratic in 
$\delta \hat{Q}^{\phantom{\dagger}}_{\mathsf{S}}$ and/or
$\delta \hat{Q}_{\mathsf{S}}^{\dagger}$.\cite{footnote-k}
In order to renormalize the parameters of the theory, then, we need 
only consider corrections to terms up to second order in 
$\delta \hat{Q}_{\mathsf{S}}^{\protect\phantom{\dagger}}$ (and 
its adjoint) in the pure slow mode FNL$\sigma$M action.

The vertices necessary for the one loop calculation are provided by 
Fig.~\ref{FeynmanRulesVertices} and Table \ref{VertexTable}. To 
compute all one loop corrections up to second homogeneous order in 
$\delta \hat{Q}^{\phantom{\dagger}}_{\mathsf{S}}(\bm{\mathrm{r}})$ and 
$\delta \hat{Q}^{\dagger}_{\mathsf{S}}(\bm{\mathrm{r}})$,
we must calculate diagrams involving one copy of each of the vertices 
pictured in Figs.~\ref{FeynmanRulesVertices}(\textsf{a})--\ref{FeynmanRulesVertices}(\textsf{k}), 
diagrams involving pairs of the vertices (in two graph topologies) in 
Figs.~\ref{FeynmanRulesVertices}(\textsf{a})--\ref{FeynmanRulesVertices}(\textsf{e}), 
and finally a diagram involving two copies of the vertex depicted in 
Fig.~\ref{FeynmanRulesVertices}(\textsf{f}). Taking into account the 
three propagator sections shown in Fig.~\ref{FeynmanRulesProps}, a 
naive estimate gives $3\times11 + 3\times3\times15\times2 + 3 = 306$ 
possible diagrams! However, many of these turn out to individually 
vanish, give corrections to higher order in $\Gamma_{c}$, or correct 
only irrelevant operators in the FNL$\sigma$M action. Among the diagrams 
that individually vanish are those that contain one or more closed 
``Keldysh'' loops. A Keldysh loop is defined as a simultaneous Keldysh 
species index summation and frequency integration along a closed single 
fermion (thin) line. The contributions from the two Keldysh species exactly 
cancel in such a loop, as required by the normalization condition $Z = 1$. 
[See also Eq.~(\ref{Zmicro}) and the discussion preceding it.] 

In this subsection we furnish the subset of $83$ non-vanishing diagrams 
that correct the marginal parameters in the FNL$\sigma$M and/or correspond 
to the generation of marginal operators not originally present in the 
FNL$\sigma$M action. All contributions of the latter type must cancel if 
the theory is to be renormalizable to one loop; we will show that this 
is indeed the case. The required $83$ diagrams are organized into 14 
categories, shown in Figs.~\ref{D1Fig}--\ref{D13Fig}, renormalizing each 
of the coupling strengths appearing in the disorder-only and interacting 
sectors of the FNL$\sigma$M. To each category of diagrams, we provide the label 
$\mathfrak{D}\mathsf{m}$, with $\mathsf{m} \in \mathsf{\{1,\ldots,14\}}$. 
Category labels $\{\mathfrak{D}\mathsf{1}, \mathfrak{D}\mathsf{2}, \ldots\}$ 
appear in the captions of the associated figures \{\ref{D1Fig}, 
\ref{D2Fig}, \ldots\}. Individual diagrams \emph{and} their associated 
amplitudes will be referred to by a category label, a letter, and 
if necessary, a subscript Roman numeral. We give two examples: 
$\mathfrak{D}\mathsf{1(a)}$ refers to the diagram labeled \textsf{(a)} 
in the left-hand column of Fig.~\ref{D1Fig} (category 
$\mathfrak{D}\mathsf{1}$), while $\mathfrak{D}\mathsf{2(b)_{iii}}$ 
refers to the diagram labeled \textsf{(iii)} in the right-hand 
column of Fig.~\ref{D2Fig} (category $\mathfrak{D}\mathsf{2}$).
Below we examine each category of corrections in turn. We also 
calculate the one loop renormalization of the single particle 
density of states (DOS) $\nu(\omega)$. Given the rather large 
number of non-vanishing diagrams, we explain in detail the computation 
of only a handful of the associated one loop corrections. The goal 
here is to illustrate the process; calculation of the remaining diagrams 
is straight-forward, if time consuming. Explicit frequency and Keldysh 
indices (using the same conventions employed in Figs.~\ref{FeynmanRulesProps} 
and \ref{FeynmanRulesVertices}) distinguish the diagrams in 
Figs.~\ref{D1Fig}--\ref{D13Fig} whose detailed evaluation is provided in this section. 

The key ingredients for the renormalization process are the fast mode 
propagators  $\mathsf{P_{\lambda}}$, $\mathsf{P_{A}}$, $\mathsf{P_{S}}$,  
and $\mathsf{P_{C}}$, depicted in Figs.~\ref{FeynmanRulesProps}\textsf{(a)}--\textsf{(c)}, 
with the associated amplitudes provided by Eq.~(\ref{YPropComponents}), 
and the vertices $\mathsf{V}(\theta)$ coupling together fast and slow 
modes, with $\theta \in \{\mathsf{a,b,c},\ldots\}$, cataloged in Table 
\ref{VertexTable} and pictured in Fig.~\ref{FeynmanRulesVertices}. Explicit 
formulae for the necessary frequency-momentum shell loop integrations are 
relegated to Appendix \ref{Integrals}.

\subsubsection{Propagator with a twist}

\begin{figure}[b]
\includegraphics[width=0.35\textwidth]{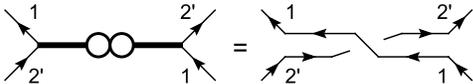}
\caption{Basic diffuson propagator $\mathsf{P_{\lambda}}$ 
[Fig.~\ref{FeynmanRulesProps}\textsf{(a)} and Eq.~(\ref{HProp})] with a twist.\label{TwistedProp}}
\end{figure}

Before we begin, we need to introduce one additional piece of diagrammatic 
notation. The basic diffuson propagator $\mathsf{P_{\lambda}}$, given by 
Eq.~(\ref{HProp}), is represented as the thick line segment shown in 
Fig.~\ref{FeynmanRulesProps}\textsf{(a)}. As shown in 
Fig.~\ref{FeynmanRulesProps}\textsf{(d)} and discussed in the paragraph 
below Eq.~(\ref{HeatDiffuson}), each such thick line segment can be thought 
of as a pair of counterdirected thin lines, corresponding to the 
propagation of a particle-hole pair. The constituent particle and hole 
lines carry Keldysh and frequency indices that traverse $\mathsf{P_{\lambda}}$ 
without mixing. In Fig.~\ref{TwistedProp}, we picture the same basic 
diffuson propagator shown in Fig.~\ref{FeynmanRulesProps}\textsf{(a)}, 
but with a twist of the right end relative to the left. The twist is 
represented by the ``$\bm{\infty}$'' symbol. We will use this twisted 
representation of $\mathsf{P_{\lambda}}$ whenever convenient to simplify 
the 2D diagrammatic representation of the one loop corrections.

\subsubsection{Renormalization of $\lambda_{A}$}\label{lambdaARenorm}

\begin{figure}
\includegraphics[width=0.45\textwidth]{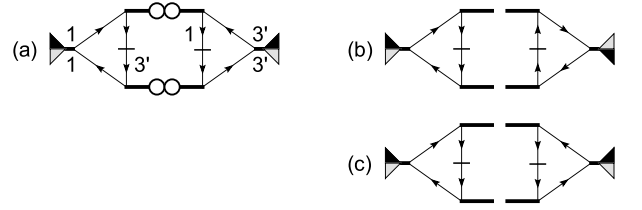}
\caption{Category $\mathfrak{D}\mathsf{1}$: Diagrams renormalizing $\lambda_{A}$.\label{D1Fig}}
\end{figure}

Diagrams $\mathfrak{D}\mathsf{1(a)}$--$\mathfrak{D}\mathsf{1(c)}$ appearing 
in Fig.~\ref{D1Fig} renormalize the disorder parameter $\lambda_{A}$. All 
three diagrams shown in this figure pair together two copies of the stiffness 
vertex $\mathsf{V(a)}$, using pairs of the disorder-only fast mode propagators 
$\mathsf{P_{\lambda}}$ and $\mathsf{P_{A}}$. There are no undetermined 
loop frequencies in these three diagrams, thanks, e.g., to the propagator 
twists in $\mathfrak{D}\mathsf{1(a)}$; the associated amplitudes therefore 
involve pure momentum shell integrations, since all propagator frequency 
indices are slow. [See Eq.~(\ref{FMShell}) and Fig.~\ref{FMShellFig}.] We 
will compute $\mathfrak{D}\mathsf{1(a)}$ explicitly; the labels 
$1 \rightarrow \{\omega_{1},a\}$ and $3' \rightarrow \{\omega_{3},a'\}$ 
in Fig.~\ref{D1Fig} represent external, slow frequencies ($\omega_{1},\omega_{3}$) 
and Keldysh indices ($a,a'$). 

Using the Feynman rules, we find 
\begin{equation}\label{D1(a)}
	\mathfrak{D}\mathsf{1(a)} = \frac{1}{2!} \left(\frac{-i}{2 \lambda}\right)^{2} 
	\int \frac{d^{d} \bm{\mathrm{q}}}{(2 \pi)^{d}}
	\mathrm{L}_{\mathsf{S}\, 1,1}^{i \, a, a}(\bm{\mathrm{q}})
	\mathrm{L}_{\mathsf{S}\, 3,3}^{j \, a', a'}(-\bm{\mathrm{q}})
	\, I_{1}^{i,j}
\end{equation}
where $i$ and $j$ denote vector components, with the pure momentum shell integration
\begin{align}\label{I1}
	I_{1}^{i,j} =& 
	\begin{aligned}[t]
	\int & \frac{d^{\textrm{2}} \bm{\mathrm{k}}}{(2 \pi)^{\textrm{2}}}
	(2\bm{\mathrm{k}} + \bm{\mathrm{q}})^{i}(-2\bm{\mathrm{k}} - \bm{\mathrm{q}})^{j} \\
	& \times\left[
	\Delta_{\mathcal{O}}^{a,a'}(|\omega_{1}|,|\omega_{3}|,\bm{\mathrm{k}})
	\Delta_{\mathcal{O}}^{a,a'}(|\omega_{1}|,|\omega_{3}|,\bm{\mathrm{k}}+\bm{\mathrm{q}})
	\right] 
	\end{aligned}
	\nonumber\\
	\sim& 
	\frac{-4 \delta^{i,j}}{2 h^{2}}\frac{1}{2 D^{2} (2 \pi)} 
	\int_{\widetilde\Lambda}^{\Lambda} \frac{dx}{x} = \frac{-\delta^{i,j} \lambda^{2}}{2\pi}2 dl,
\end{align}
up to irrelevant slow mode frequency ($\omega_{1},\omega_{3}$)- and momentum 
($\bm{\mathrm{q}}$)-dependent terms. On the last line of Eq.~(\ref{I1}), we have 
performed the change of integration variables $x \equiv D \bm{\mathrm{k}}^{2}$, 
with $D = 1/\lambda h$ [Eq.~(\ref{DDef})], and we have used Eq.~(\ref{ShellRatio}). 
Combining Eqs.~(\ref{D1(a)}) and (\ref{I1}), using Eq.~(\ref{LSDef}), and summing 
over slow mode frequency and Keldysh indices, we obtain
\begin{equation}\label{D1(a)--F}
	\mathfrak{D}\mathsf{1(a)} = \left(\frac{\lambda^{2} \, 2 dl}{8 \pi \lambda_{A}}\right)
	\frac{\lambda_{A}}{2\lambda^{2}}
	\int d^{d} \bm{\mathrm{r}}
	\left[
	\mathrm{Tr}
	\left(\hat{Q}^{\dagger}_{\mathsf{S}}
	\bm{\nabla}\hat{Q}^{\phantom{\dagger}}_{\mathsf{S}}
	\right)
	\right]^{2}. 
\end{equation}
$\mathfrak{D}\mathsf{1(b)}$ and $\mathfrak{D}\mathsf{1(c)}$ may be similarly 
evaluated; in fact, these diagrams exactly cancel because of the directional 
dependence of the stiffness vertex $\mathsf{V(a)}$ upon the loop momenta---see 
Table \ref{VertexTable}. Thus the complete $\lambda_{A}$ renormalization is 
given by Eq.~(\ref{D1(a)--F}).

\subsubsection{Renormalization of $\lambda$ and $h$}\label{lambdaandhRenorm}

\begin{figure}[b]
\includegraphics[width=0.45\textwidth]{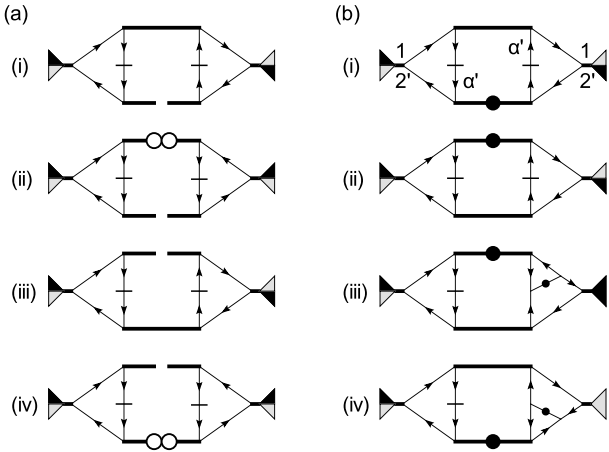}
\caption{Category $\mathfrak{D}\mathsf{2}$: Diagrams renormalizing $\lambda$.\label{D2Fig}}
\end{figure}

The diagrams in category $\mathfrak{D}\mathsf{2}$, Fig.~\ref{D2Fig}, renormalize 
the disorder parameter $\lambda$, proportional to the inverse dimensionless DC conductance. 
$\mathfrak{D}\mathsf{2(a)_{i}}$--$\mathfrak{D}\mathsf{2(a)_{iv}}$, shown in the 
left-hand column of Fig.~\ref{D2Fig}, possess no undetermined loop frequencies, 
and therefore involve pure momentum shell integrations, similar to that in 
Eq.~(\ref{I1}). In fact, $\mathfrak{D}\mathsf{2(a)_{i}}$ and 
$\mathfrak{D}\mathsf{2(a)_{iii}}$ exactly cancel $\mathfrak{D}\mathsf{2(a)_{ii}}$ 
and $\mathfrak{D}\mathsf{2(a)_{iv}}$, due to the presence of twists in the latter 
diagrams and the momentum-dependence of the stiffness vertex $\mathsf{V(a)}$ [Table \ref{VertexTable}].  

On the other hand, the diagrams in the right-hand column of Fig.~\ref{D2Fig} involve 
simultaneous frequency and momentum loop integrations, and their sum indeed yields 
a non-vanishing renormalization of $\lambda$.

Diagrams $\mathfrak{D}\mathsf{2(b)_{i}}$ and $\mathfrak{D}\mathsf{2(b)_{ii}}$ 
pictured in Fig.~\ref{D2Fig} give identical contributions, and each involve two 
copies of the stiffness vertex $\mathsf{V(a)}$, with one basic diffuson 
$\mathsf{P_{\lambda}}$ and one interacting sector propagator 
$\mathsf{P_{S}}+\mathsf{P_{C}}$. We evaluate
\begin{equation}\label{D2(b)i}
	\mathfrak{D}\mathsf{2(b)_{i}} = \frac{2 i \xi^{a'}(-i)^{2}}{2! (2 \lambda)^{2}} 
	\int \frac{d^{d} \bm{\mathrm{q}}}{(2 \pi)^{d}}
	\mathrm{L}_{\mathsf{S}\, 1,2}^{i \, a, a'}(\bm{\mathrm{q}})
	\mathrm{L}_{\mathsf{S}\, 2,1}^{j \, a', a}(-\bm{\mathrm{q}})
	\, I_{2}^{i,j},
\end{equation}
where the frequency-momentum integral divides into two pieces
\begin{equation}\label{I2}
	I_{2}^{i,j} \equiv I_{2\,\mathsf{S}}^{i,j} + I_{2\,\mathsf{C}}^{i,j},
\end{equation}
corresponding to the $\mathsf{P_{S}}$ and $\mathsf{P_{C}}$ components of 
the interaction sector propagator, respectively.
In the former case, up to irrelevant terms we have
\begin{align}\label{I2S}
	I_{2\,\mathsf{S}}^{i,j} &=
	\begin{aligned}[t]
	\int & \frac{d\omega_{\alpha} \, d^{\textrm{2}}\bm{\mathrm{k}}}{(2 \pi)^{\textrm{3}}}
	(2\bm{\mathrm{k}})^{i}(2\bm{\mathrm{k}})^{j}
	\Gamma_{s} (s_{\alpha} - s_{2})(s_{2} - s_{\alpha}) \\
	&\!\!\!\times\Big[
	\Delta_{\mathcal{O}}^{a,a'}(0,|\omega_{\alpha}|,\bm{\mathrm{k}})
	\Delta_{\mathcal{S}}^{a'}(|\omega_{\alpha}|,\bm{\mathrm{k}})
	\Delta_{\mathcal{O}}^{a',a'}(|\omega_{\alpha}|,0,\bm{\mathrm{k}})
	\Big] 
	\end{aligned}
	\nonumber\\
	&
	\begin{aligned}
	=
	\frac{-4^{2} \Gamma_{s} \delta^{i,j}}{2 h^{3}}
	\int\limits_{\omega_{\alpha} > 0} & \frac{d\omega_{\alpha} \, d^{\textrm{2}}\bm{\mathrm{k}}}{(2 \pi)^{\textrm{3}}}
	\left\{
	\bm{\mathrm{k}}^{2}
	[D\bm{\mathrm{k}}^{2} - i \xi^{a'} \omega_{\alpha}]^{-2} \right.
	\\
	&\left.\times
	[D\bm{\mathrm{k}}^{2} - i (1-\gamma_{s})\xi^{a'} \omega_{\alpha}]^{-1}
	\right\}.
	\end{aligned}
\end{align}
The factor $(s_{\alpha} - s_{2})^{2}$ appearing in the first line of Eq.~(\ref{I2S}) 
is inherited from $\mathsf{P_{S}}$, which projects onto the propagator channel 
purely off-diagonal in $\sgn(\omega)$ space [see Eq.~(\ref{SProp}) and the 
discussion below Eq.~(\ref{SFastI})]. Eq.~(\ref{I2S}) yields a result 
\emph{independent} of the slow frequency $\omega_{2}$, so we may take 
$\sgn(\omega_{2}) < 0$ without loss of generality. As a consequence, the 
only effect of the aforementioned ``projection factor'' in Eq.~(\ref{I2S}) 
is the restriction of the loop frequency integration to the half space 
$\omega_{\alpha} > 0$, indicated on the third line of this equation. Following 
a change of variables, Eq.~(\ref{I2S}) gives
\begin{align}\label{I2S--2}
	I_{2\,\mathsf{S}}^{i,j} & = \frac{-4^{2} \Gamma_{s} \delta^{i,j}}{2 h^{3}} 
	J_{3}\biglb(\xi^{a'};(1-\gamma_{s})\xi^{a'}\bigrb) 
	\nonumber\\ 
	& = - \frac{-i \xi^{a'} \delta^{i,j} \lambda^{2}}{4\pi}2 dl 
	\left[
	1 + \frac{1-\gamma_{s}}{\gamma_{s}} \ln(1-\gamma_{s})
	\right].
\end{align}
The frequency-momentum shell integral $J_{3}(z;z')$ in Eq.~(\ref{I2S--2}) is 
defined and evaluated in Appendix \ref{Integrals} [Eq.~(\ref{J3})]. The CDW 
channel contribution $I_{2\,\mathsf{C}}^{i,j}$ to Eq.~(\ref{I2}) may be similarly 
computed, using the propagator $\mathsf{P_{C}}$, and working only to lowest 
order in $\Gamma_{c}$ [i.e.\ ignoring the logarithmic denominator, 
Eq.~(\ref{DiffLog}), in Eq.~(\ref{CProp})].
Using Eq.~(\ref{LSDef}), and summing the identical contributions from 
$\mathfrak{D}\mathsf{2(b)_{i}}$ and $\mathfrak{D}\mathsf{2(b)_{ii}}$, one finds
\begin{multline}\label{D2(b)i+D2(b)ii--F}
	\mathfrak{D}\mathsf{2(b)_{i}} + \mathfrak{D}\mathsf{2(b)_{ii}}
	\\
	\begin{aligned}[b]
	=&\left(\frac{\lambda \, 2 dl}{4 \pi}\right)
	\left[
	1 + \frac{1-\gamma_{s}}{\gamma_{s}} \ln(1-\gamma_{s}) 
	- \frac{\gamma_{c}}{2}
	\right]
	\\
	&\times
	\frac{1}{2 \lambda} \int d^{d}{\bm{\mathrm{r}}}\,
	\mathrm{Tr}
	\left[
	\bm{\nabla}\hat{Q}^{\dagger}_{\mathsf{S}}
	\cdot \bm{\nabla}\hat{Q}^{\phantom{\dagger}}_{\mathsf{S}}
	\right].
	\end{aligned}
\end{multline}
If we set $\gamma_{c} = 0$ in Eq.~(\ref{D2(b)i+D2(b)ii--F}), the remainder 
is recognized as the usual correction to the ``dimensionless DC resistance'' 
$\lambda$ in the presence of \emph{short-ranged} density-density interparticle 
interactions.\cite{Finkelstein,AltshulerAronov,NPCS,Aleiner} 

The computation of the final two diagrams $\mathfrak{D}\mathsf{2(b)_{iii}}$ 
and $\mathfrak{D}\mathsf{2(b)_{iv}}$ in Fig.~\ref{D2Fig} is more complicated, 
as each graph involves \emph{two} loop frequency integrations. We will 
demonstrate the evaluation of diagrams involving two [$\mathfrak{D}\mathsf{6(a)_{iii}}$] 
and three [$\mathfrak{D}\mathsf{6(a)_{v}}$] frequency loops appearing in 
Fig.~\ref{D6Fig}, below. Here, we simply give the result for the sum
\begin{align}\label{D2(b)iii+D2(b)iv--F}
	\mathfrak{D}\mathsf{2(b)_{iii}} + \mathfrak{D}\mathsf{2(b)_{iv}}
	=& \left(\frac{-\lambda \, 2 dl}{4 \pi}\right)
	\left[
	2 + \frac{2-\gamma_{s}}{\gamma_{s}} \ln(1-\gamma_{s}) 
	\right]
	\nonumber\\
	&\times
	\frac{1}{2 \lambda} \int d^{d}{\bm{\mathrm{r}}}\,
	\mathrm{Tr}
	\left[
	\bm{\nabla}\hat{Q}^{\dagger}_{\mathsf{S}}
	\cdot \bm{\nabla}\hat{Q}^{\phantom{\dagger}}_{\mathsf{S}}
	\right].
\end{align}
Obtaining this result requires the use of the integral formulae given by 
Eqs.~(\ref{J2}) and (\ref{J10}) of Appendix \ref{Integrals}. 
We demonstrate shortly that the ``anomalous'' amplitude in 
Eq.~(\ref{D2(b)iii+D2(b)iv--F}) is precisely canceled by other diagrams.

\begin{figure}[t]
\includegraphics[width=0.4\textwidth]{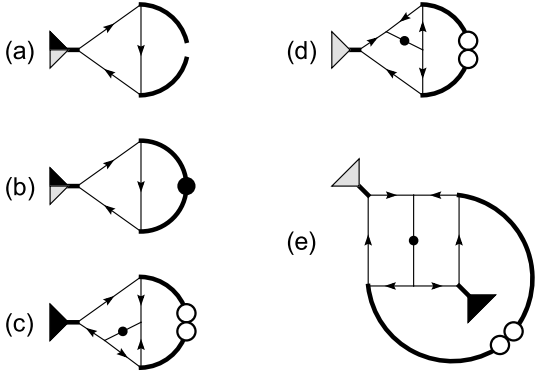}
\caption{Category $\mathfrak{D}\mathsf{3}$: Diagrams renormalizing $h$.\label{D3Fig}}
\end{figure}

We turn now to the renormalization of $h$ by the diagrams in Fig.~\ref{D3Fig}. 
$\mathfrak{D}\mathsf{3(a)}$ and $\mathfrak{D}\mathsf{3(b)}$ prove easy to evaluate, 
involving only the associated symmetry-breaking vertex $\mathsf{V(b)}$, 
fused with the propagators $\mathsf{P_{A}}$ and $\mathsf{P_{S}}+\mathsf{P_{C}}$, 
respectively. $\mathfrak{D}\mathsf{3(a)}$ involves a pure momentum shell 
integration, while $\mathfrak{D}\mathsf{3(b)}$ requires the integral $J_{1}(z,z')$, 
Eq.~(\ref{J1}). 
The result is
\begin{align}\label{D3(a)+D3(b)--F}
	\mathfrak{D}\mathsf{3(a)} + \mathfrak{D}\mathsf{3(b)}
	=& \left(\frac{-2 dl}{8 \pi}\right)
	\left\{
	\lambda_{A} 
	+\lambda\left[\ln(1-\gamma_{s}) + \gamma_{c}\right] 
	\right\} 
	\nonumber\\
	&\times
	i h \int d^{d}{\bm{\mathrm{r}}} \,
	\mathrm{Tr}
	\left[
	|\hat{\omega}| \hat{\xi}_{3}
	\,(\hat{Q}^{\dagger}_{\mathsf{S}}+\hat{Q}^{\phantom{\dagger}}_{\mathsf{S}})
	\right].
\end{align}
Diagrams $\mathfrak{D}\mathsf{3(c)}$--$\mathfrak{D}\mathsf{3(e)}$ are as simple 
to evaluate, but the sum of the associated amplitudes gives zero.

\begin{figure}
\includegraphics[width=0.45\textwidth]{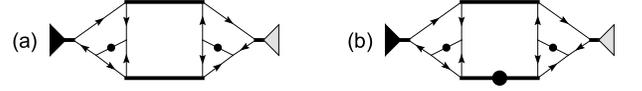}
\caption{Category $\mathfrak{D}\mathsf{4}$: Diagrams renormalizing $\lambda$ and $h$.\label{D4Fig}}
\end{figure}

To complete the renormalization of $\lambda$ and $h$, we must compute the graphs 
pictured in Fig.~\ref{D4Fig}. These diagrams represent frequency-momentum 
integrations quadratically divergent in momentum; evaluation of the required 
integrals produces terms zeroth and first order in the energy cutoff $\Lambda$; 
the latter are presumably canceled by the measure, although we have not checked 
this in detail. 
$\mathfrak{D}\mathsf{4(a)}$ and $\mathfrak{D}\mathsf{4(b)}$ in Fig.~\ref{D4Fig} 
involve the crosspairing of the \emph{interaction} vertices $\mathsf{V(c)}$ and 
$\mathsf{V(d)}$, with both non-interacting $\mathsf{P_{\lambda}}$ and interacting 
$\mathsf{P_{S}}+\mathsf{P_{C}}$ propagator components.
These diagrams are quite lengthy to evaluate, because they require two 
[$\mathfrak{D}\mathsf{4(a)}$] and three [$\mathfrak{D}\mathsf{4(b)}$] frequency 
loop integrations, and necessitate Taylor expansions of the fast mode propagators 
in powers of the external frequencies and momenta. Other diagrams with multifrequency 
loop integrals will be tackled in detail in Sec.~\ref{IntRenorm1}, below. Here we 
merely quote the result
\begin{align}\label{D4(a)+D4(b)--F}
	\mathfrak{D}\mathsf{4(a)} + \mathfrak{D}\mathsf{4(b)}
	=& -\mathfrak{D}\mathsf{2(b)_{iii}} - \mathfrak{D}\mathsf{2(b)_{iv}}\qquad\qquad\qquad\qquad\nonumber\\
	&\begin{aligned}
	+ &\left(\frac{\lambda \, 2 dl}{8 \pi}\right)
	\left[
	\gamma_{s} + \ln(1-\gamma_{s}) 
	\right]i h 
	\\
	&\times
	\int d^{d}{\bm{\mathrm{r}}} \,
	\mathrm{Tr}
	\left[
	|\hat{\omega}| \hat{\xi}_{3}
	\,(\hat{Q}^{\dagger}_{\mathsf{S}}+\hat{Q}^{\phantom{\dagger}}_{\mathsf{S}})
	\right].
	\end{aligned}
\end{align}
In order to obtain Eq.~(\ref{D4(a)+D4(b)--F}), one employs the frequency-momentum 
shell integral formulae given by Eqs.~(\ref{J7}), (\ref{J8}), (\ref{J11}), and 
(\ref{J12}). As promised, the ``anomalous'' correction to $\lambda$ obtained in 
Eq.~(\ref{D2(b)iii+D2(b)iv--F}) is completely canceled by the amplitude 
$\mathfrak{D}\mathsf{4(a)} + \mathfrak{D}\mathsf{4(b)}$. In addition, we pick up 
a crucial renormalization of $h$ from 
Eq.~(\ref{D4(a)+D4(b)--F}).

\subsubsection{Renormalization of the density of states $\nu(\omega)$}

Before we treat the interparticle interaction parameters $\Gamma_{U}$ and $\Gamma_{V}$, 
we pause to consider the local scaling operator
\begin{equation}\label{DOSDef}
	{\nu}^{a}(\omega,\bm{\mathrm{r}}) \equiv
	\int \frac{d \omega'}{2 \pi} \,
	\left[
	Q^{a, a}_{\omega, \omega'}(\bm{\mathrm{r}}) 
	+ Q^{\dagger \, a, a}_{\omega, \omega'}(\bm{\mathrm{r}})
	\right].
\end{equation}
With the aid of Eq.~(\ref{QmatrixID}), it can be seen that the expectation value of 
${\nu}^{a}(\omega,\bm{\mathrm{r}})$ in Eq.~(\ref{DOSDef}) represents a measure of 
the disorder-averaged, coarse-grained, single particle density of states (DOS)
$\nu(\omega)$ in the diffusive Fermi liquid, with $\omega$ measured relative to the 
Fermi energy. [The integral over the auxiliary frequency $\omega'$ in Eq.~(\ref{DOSDef})
is necessary to eliminate an energy-conserving delta function in this expectation.]
The scaling behavior of $\nu(\omega)$ as a function of energy scale $\omega$
or system size $L$ may be determined through the renormalization of 
${\nu}^{a}(\omega,\bm{\mathrm{r}})$. Applying the Wilsonian background field 
decomposition [Eqs.~(\ref{FastSlow}) and (\ref{YFastParam})] and the RG program 
[Eqs.~(\ref{YProp}), (\ref{YPropComponents}) and Table~\ref{VertexTable}] of the 
previous section to Eq.~(\ref{DOSDef}), one encounters the same diagrams responsible 
for \emph{part} of the renormalization of the frequency rescaling factor $h$, graphs 
$\mathfrak{D}\mathsf{3(a)}$ and $\mathfrak{D}\mathsf{3(b)}$, pictured in 
Fig.~\ref{D3Fig}. The category $\mathfrak{D}\mathsf{4}$ ``interaction-interaction'' 
diagrams shown in Fig.~\ref{D4Fig}, which provide a further renormalization of 
$h$ [Eq.~(\ref{D4(a)+D4(b)--F})], do \emph{not} appear in the computation of the 
scaling dimension of 
${\nu}^{a}(\omega,\bm{\mathrm{r}})$.

We obtain the renormalization
\begin{align}\label{DOSCutoff}
	\left\langle {\nu}^{a}(\omega,\bm{\mathrm{r}})  \right\rangle_{\Lambda}
	\sim& \left\langle {\nu}^{a}(\omega,\bm{\mathrm{r}})  \right\rangle_{\widetilde{\Lambda}}
	\nonumber\\
	&\times
	\bigglb(
	1 + \frac{2 dl}{8 \pi}
		\left\{
			\lambda_{A} + \lambda \left[\ln(1-\gamma_{s}) +\gamma_{c}\right]
		\right\}
	\bigglb)
\end{align}
similar to the diagrammatic amplitude expressed in Eq.~(\ref{D3(a)+D3(b)--F}). 
In Eq.~(\ref{DOSCutoff}), the symbols $\langle\ldots\rangle_{\Lambda}$ and
$\langle \ldots \rangle_{\widetilde{\Lambda}}$ denote the expectation value 
taken with respect to the generating functional $Z$ in Eq.~(\ref{ZFastSlow}), 
before and after the elimination of the fast modes 
$Y_{\omega, \omega'}^{a, a'}(\bm{\mathrm{k}})$, respectively. In Sec.~\ref{DimAnalysis}, 
we will use Eq.~(\ref{DOSCutoff}) to derive a flow equation governing the scaling 
behavior of the DOS.


\subsubsection{Renormalization of $\Gamma_{U}$ and $\Gamma_{V}$}\label{IntRenorm1}

\begin{table}[b]
\caption{
Slow mode operators that appear in the renormalization of the interaction parameters $\Gamma_{U}$ and $\Gamma_{V}$ 
(or equivalently,  $\Gamma_{s}$ and $\Gamma_{c}$).
The operators in this table are summed over Keldysh species index ($a$), 
integrated over position space ($\bm{\mathrm{r}}$), and integrated over 
frequency indices $\{1,2,3,4\}\rightarrow\{\omega_{1},\omega_{2},\omega_{3},\omega_{4}\}$; 
we have used the compact notation introduced in Eq.~(\ref{MultiFreqIntegral}) 
for this integration. In the interacting sector of the pure slow mode 
FNL$\sigma$M action, 
Eq.~(\ref{SSlow}), the sum $\mathcal{O}_{U}+\overline{\mathcal{O}}_{U}$ 
couples to the same-sublattice interaction strength $\Gamma_{U}$, while 
$\mathcal{O}_{V}$ couples to the intersublattice interaction constant 
$\Gamma_{V}$. The operators $\mathcal{O}_{X}$, 
$\overline{\mathcal{O}}_{X}$, and $\mathcal{O}_{Y}$ do \emph{not} correspond 
to terms occuring in the original FNL$\sigma$M action [Eq.~(\ref{SIQmatrixSPLeft}), 
\emph{after} the saddle point shift in Eq.~(\ref{QmatrixSPLeft})], but are 
generated at intermediate steps in the renormalization process.
\label{IntOpsTable}}
\begin{ruledtabular}
\begin{tabular}{l @{} l}
$\mathcal{O}_{U}$  
&
$	
	\sum\limits_{a} i \xi^{a}\!\int\limits_{1,2,3,4}\!
	d^{d}{\bm{\mathrm{r}}} \, 
	s_{1} s_{3}
	{\delta Q}^{a, a}_{\mathsf{S} \, 1, 2}{\delta Q}^{a, a}_{\mathsf{S} \, 3, 4}
$
\medskip\\
$\overline{\mathcal{O}}_{U}$  
&
$	
	\sum\limits_{a} i \xi^{a}\!\int\limits_{1,2,3,4}\!
	d^{d}{\bm{\mathrm{r}}} \, 
	s_{2} s_{4}
	{\delta Q}^{\dagger \, a, a}_{\mathsf{S} \, 1, 2}{\delta Q}^{\dagger \, a, a}_{\mathsf{S} \, 3, 4}
$
\medskip\\
$\mathcal{O}_{V}$  
&
$	
	\sum\limits_{a} i \xi^{a}\!\int\limits_{1,2,3,4}\! 
	d^{d}{\bm{\mathrm{r}}} \, 
	2 s_{2} s_{3}
	{\delta Q}^{\dagger \, a, a}_{\mathsf{S} \, 1, 2}{\delta Q}^{a, a}_{\mathsf{S} \, 3, 4}
$
\medskip\\
$\mathcal{O}_{X}$  
&
$	
	\sum\limits_{a} i \xi^{a}\!\int\limits_{1,2,3,4}\!
	d^{d}{\bm{\mathrm{r}}} \, 
	{\delta Q}^{a, a}_{\mathsf{S} \, 1, 2}{\delta Q}^{a, a}_{\mathsf{S} \, 3, 4}
$
\medskip\\
$\overline{\mathcal{O}}_{X}$  
&
$	
	\sum\limits_{a} i \xi^{a}\!\int\limits_{1,2,3,4}\! 
	d^{d}{\bm{\mathrm{r}}} \, 
	{\delta Q}^{\dagger \, a, a}_{\mathsf{S} \, 1, 2}{\delta Q}^{\dagger \, a, a}_{\mathsf{S} \, 3, 4}
$
\medskip\\
$\mathcal{O}_{Y}$  
&
$	
	\sum\limits_{a} i \xi^{a}\!\int\limits_{1,2,3,4}\!
	d^{d}{\bm{\mathrm{r}}} \, 
	2 {\delta Q}^{\dagger \, a, a}_{\mathsf{S} \, 1, 2}{\delta Q}^{a, a}_{\mathsf{S} \, 3, 4}
$
\medskip\\
\end{tabular}
\end{ruledtabular}
\end{table}

With the renormalization of the disorder-only sector parameters 
$\lambda_{A}$, $\lambda$, and $h$ complete, we now turn to the (much 
more involved) renormalization of the interparticle interaction parameters 
$\Gamma_{U}$ and $\Gamma_{V}$. It will prove convenient to introduce a set 
of six slow mode operators 
$\{\mathcal{O}_{U},\overline{\mathcal{O}}_{U},\mathcal{O}_{V},\mathcal{O}_{X},\overline{\mathcal{O}}_{X},\mathcal{O}_{Y}\}$, 
defined by the expressions given in Table~\ref{IntOpsTable}. In this table, 
we have introduced the following compact notation for the position and frequency 
space integral 
\begin{equation}\label{MultiFreqIntegral}
	\int \frac{d\omega_{1}}{2 \pi} \frac{d\omega_{2}}{2 \pi} \frac{d\omega_{3}}{2 \pi} \frac{d\omega_{4}}{2 \pi}
	d^{d}{\bm{\mathrm{r}}} \, \delta_{1+3,2+4}
	\equiv
	\int\limits_{1,2,3,4}\!\!\!\! 
	d^{d}{\bm{\mathrm{r}}}.
\end{equation}
Operators $\mathcal{O}_{U} + \overline{\mathcal{O}}_{U}$ and $\mathcal{O}_{V}$ 
correspond to terms in the original FNL$\sigma$M action, Eq.~(\ref{SIQmatrixSPLeft}) 
[\emph{after} the saddle point shift in Eq.~(\ref{QmatrixSPLeft}), with the 
replacement 
$\hat{Q} \rightarrow \hat{1} + \delta\hat{Q}_{S}^{\phantom{\dagger}}$, as in 
Eq.~(\ref{SSlow})], coupling to the same-sublattice and intersublattice 
interaction strengths 
$\Gamma_{U}$ and $\Gamma_{V}$, respectively. The last three operators listed 
in Table~\ref{IntOpsTable} are new; they do not correspond to terms appearing 
in Eq.~(\ref{SIQmatrixSPLeft}). The structures possessed by the operators 
$\mathcal{O}_{X}$, $\overline{\mathcal{O}}_{X}$, and $\mathcal{O}_{Y}$ closely 
parallel those of $\mathcal{O}_{U}$, $\overline{\mathcal{O}}_{U}$, and 
$\mathcal{O}_{V}$, with a crucial difference: the former lack the frequency-dependent 
vertex factors $s_{1} s_{3} \rightarrow \sgn(\omega_{1})\sgn(\omega_{3})$, etc., 
appearing in the latter. We will see that 
$\mathcal{O}_{X}$, $\overline{\mathcal{O}}_{X}$, and $\mathcal{O}_{Y}$ are 
generated at intermediate steps in the RG process;
in order for the theory to be renormalizable, these terms must completely 
cancel upon summing all one-loop diagrammatic amplitudes.
We will show that this indeed occurs.\cite{footnote-l}

\begin{figure}
\includegraphics[width=0.4\textwidth]{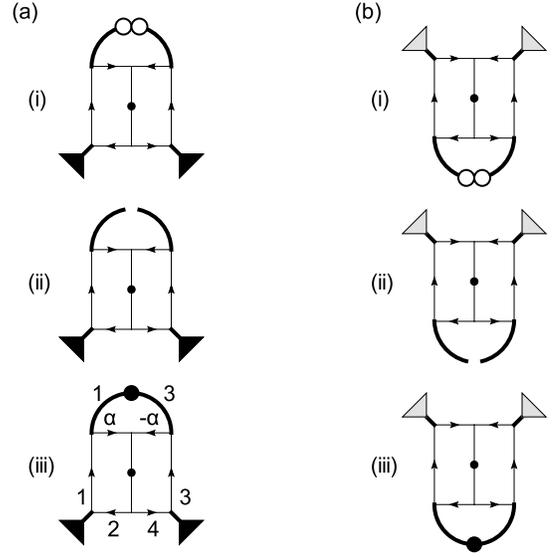}
\caption{Category $\mathfrak{D}\mathsf{5}$: Diagrams renormalizing $\Gamma_{U}$.\label{D5Fig}}
\end{figure}

We consider first diagrams that (nominally) renormalize the same-sublattice 
interaction strength $\Gamma_{U}$, corresponding to operators $\mathcal{O}_{U}$ 
and $\overline{\mathcal{O}}_{U}$ in Table~\ref{IntOpsTable}.  
We begin with the category $\mathfrak{D}\mathsf{5}$ diagrams pictured in 
Fig.~\ref{D5Fig}. The elements of these diagrams include single copies of the 
$\mathsf{V(g)}$ and $\mathsf{V(h)}$  vertices, and one of each ``flavor'' of 
the propagator components $\mathsf{P_{\lambda}}$, $\mathsf{P_{A}}$, and 
$\mathsf{P_{S}}+\mathsf{P_{C}}$. Diagrams $\mathfrak{D}\mathsf{5(a)_{i}}$ and 
$\mathfrak{D}\mathsf{5(a)_{ii}}$ yield simple pure momentum shell integrations, 
with the result
\begin{equation}\label{D5(a)i+D5(a)ii--F}
	\mathfrak{D}\mathsf{5(a)_{i}} + \mathfrak{D}\mathsf{5(a)_{ii}} 
	= \left[\frac{(\lambda - \lambda_{A})2 dl}{4 \pi}\right] \Gamma_{U} \mathcal{O}_{U}.
\end{equation}

$\mathfrak{D}\mathsf{5(a)_{iii}}$ involves an undetermined loop frequency; we 
evaluate this diagram explicitly. Note that in providing the frequency labels 
for the single fermion lines in $\mathfrak{D}\mathsf{5(a)_{iii}}$, Fig.~\ref{D5Fig}, 
we have neglected irrelevant small shifts of the loop frequency
$\mathsf{\alpha} \rightarrow \omega_{\alpha}$ by the slow mode frequencies 
$\{1,3\} \rightarrow \{\omega_{1},\omega_{3}\}$, necessary for strict energy conservation. 
We have
\begin{equation}\label{D5(a)iii}
	\mathfrak{D}\mathsf{5(a)_{iii}} 
	= \frac{(2 i \xi^{a})^{2}}{2!} 
	\delta_{1+3,2+4}\, 
	\Gamma_{U}
	\int {d^{d} \bm{\mathrm{r}}} \,
	{\delta Q}^{a, a}_{\mathsf{S} \, 1, 2}{\delta Q}^{a, a}_{\mathsf{S} \, 3, 4} \, I_{3},
\end{equation}
where the frequency-momentum integral divides into the pieces
\begin{equation}\label{I3}
	I_{3} \equiv I_{3\,\mathsf{S}} + I_{3\,\mathsf{C}}.  
\end{equation}
We calculate $I_{3\,\mathsf{C}}$; up to irrelevant terms, 
\begin{align}\label{I3C}
	I_{3\,\mathsf{C}} =& 
	\int  \frac{d\omega_{\alpha} \, d^{\textrm{2}}\bm{\mathrm{k}}}{(2 \pi)^{\textrm{3}}}
	(-s_{\alpha}^{2})
	\Gamma_{c} (s_{1} + s_{\alpha})(s_{3} - s_{\alpha}) \nonumber\\
	&\times
	\left[\Delta_{\mathcal{O}}^{a,a}(0,|\omega_{\alpha}|,\bm{\mathrm{k}})\right]^{2}.
\end{align}
$\mathsf{P_{C}}$ yields the projection factor $(s_{1} + s_{\alpha})(s_{3} - s_{\alpha})$ 
in Eq.~(\ref{I3C}); this factor gives 
a non-zero contribution only for $s_{1} = - s_{3} = s_{\alpha}$. We therefore obtain
\begin{align}\label{I3C--2}
	I_{3\,\mathsf{C}} =& 
	\frac{\Gamma_{c}(s_{1}-s_{3})^{2}}{h^{2}}\int\limits_{\omega_{\alpha}>0}  
	\frac{d\omega_{\alpha} \, d^{\textrm{2}}\bm{\mathrm{k}}}{(2 \pi)^{\textrm{3}}}
	[D\bm{\mathrm{k}}^{2} - i \xi^{a} \omega_{\alpha}]^{-2} \nonumber\\
	=& \frac{\Gamma_{c}(s_{1}-s_{3})^{2}}{h^{2}} J_{0}(\xi^{a}) 
	=\frac{\lambda i \xi^{a} \, 2 dl}{32 \pi} \gamma_{c} (s_{1}-s_{3})^{2},
\end{align}
where we have used Eqs.~(\ref{SingletCDW}) and (\ref{J0}). Evaluating $I_{3\,\mathsf{S}}$ 
[Eq.~(\ref{I3})], in a similar fashion using Eq.~(\ref{J1}), and combining this with 
Eqs.~(\ref{D5(a)iii}) and (\ref{I3C--2}), we find the result
\begin{equation}\label{D5(a)iii--F}
	\mathfrak{D}\mathsf{5(a)_{iii}} 
	= \left(\frac{\lambda \, 2 dl}{8 \pi}\right)[\gamma_{c} - \ln(1-\gamma_{s})] 
	\Gamma_{U} \left(\mathcal{O}_{U} - \mathcal{O}_{X}\right)
\end{equation}
The amplitude in Eq.~(\ref{D5(a)iii--F}) renormalizes the same sublattice interaction 
operator $\mathcal{O}_{U} \leftrightarrow \Gamma_{U}$, but also generates its ``evil 
twin,'' $\mathcal{O}_{X}$, which does not appear in the original theory 
[Eqs.~(\ref{SDQmatrixSPLeft}) and (\ref{SIQmatrixSPLeft})]. 

The amplitudes for $\mathfrak{D}\mathsf{5(b)_{i}}$--$\mathfrak{D}\mathsf{5(b)_{iii}}$ mirror 
those found in Eqs.~(\ref{D5(a)i+D5(a)ii--F}) and 
(\ref{D5(a)iii--F}):
\begin{equation}\label{D5(b)i+D5(b)ii--F}
	\mathfrak{D}\mathsf{5(b)_{i}} + \mathfrak{D}\mathsf{5(b)_{ii}} 
	= \frac{(\lambda - \lambda_{A})2 dl}{4 \pi} \Gamma_{U} \overline{\mathcal{O}}_{U},
\end{equation}
and
\begin{equation}\label{D5(b)iii--F}
	\mathfrak{D}\mathsf{5(b)_{iii}} 
	= \left(\frac{\lambda \, 2 dl}{8 \pi}\right)[\gamma_{c} - \ln(1-\gamma_{s})] 
	\Gamma_{U} \left(\overline{\mathcal{O}}_{U} - \overline{\mathcal{O}}_{X}\right).
\end{equation}

\begin{figure}
\includegraphics[width=0.45\textwidth]{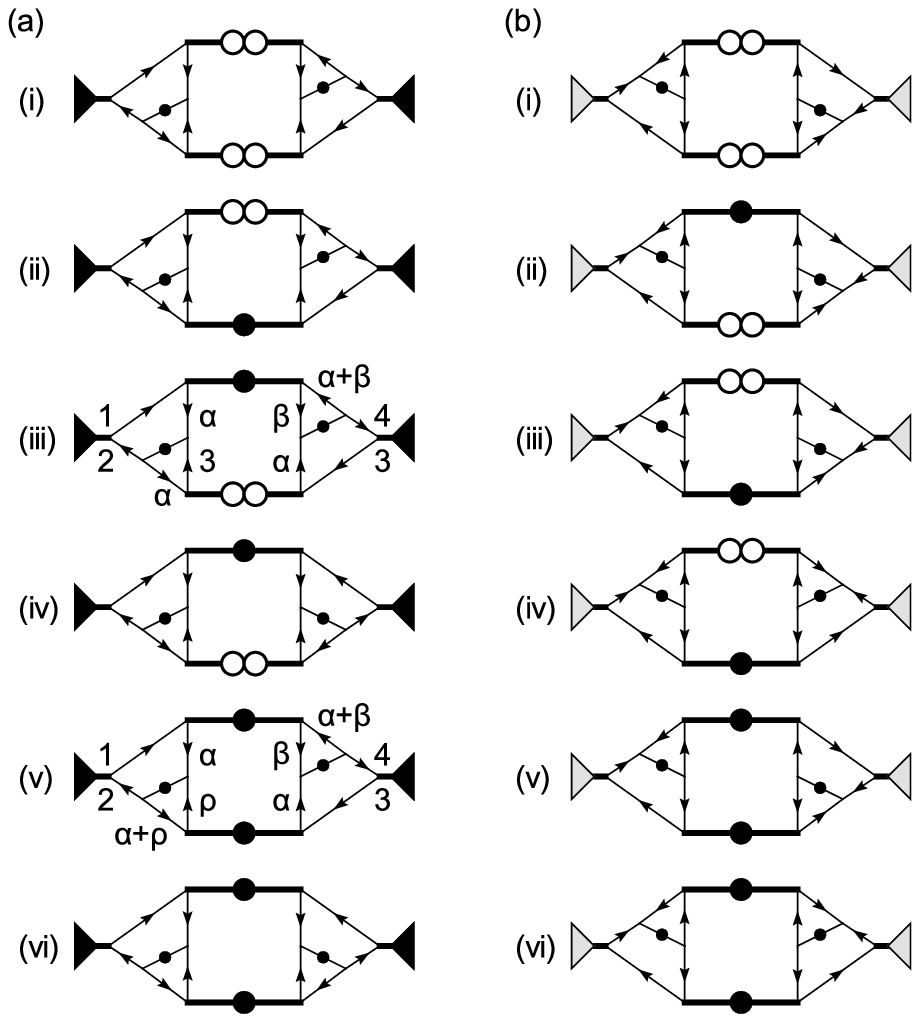}
\caption{Category $\mathfrak{D}\mathsf{6}$: Diagrams renormalizing $\Gamma_{U}$.\label{D6Fig}}
\end{figure}

We next direct our attention to the large array of diagrams shown in Fig.~(\ref{D6Fig}), 
denoted as category $\mathfrak{D}\mathsf{6}$. 
Diagrams in subcategory $\mathfrak{D}\mathsf{6(a)}$ [$\mathfrak{D}\mathsf{6(b)}$] pair 
together two copies of the vertex 
$\mathsf{V(c)}$ [$\mathsf{V(d)}$]. Moreover, as suggested by the figure, diagrams 
$\mathfrak{D}\mathsf{6(a)_{m}}$ and $\mathfrak{D}\mathsf{6(b)_{m}}$, with 
$\mathsf{m \in \{i,\ldots,vi\}}$, give structurally similar results that renormalize 
same-sublattice [Eq.~(\ref{QmatrixID})] local operators  involving 
${\delta Q}^{a, a}_{\mathsf{S} \, 1, 2}{\delta Q}^{a, a}_{\mathsf{S} \, 3, 4}$ 
($\mathcal{O}_{U}$ or $\mathcal{O}_{X}$) and	
${\delta Q}^{\dagger \, a, a}_{\mathsf{S} \, 1, 2}{\delta Q}^{\dagger \, a, a}_{\mathsf{S} \, 3, 4}$ 
($\overline{\mathcal{O}}_{U}$ or $\overline{\mathcal{O}}_{X}$), respectively. We therefore 
evaluate the subcategories $\mathfrak{D}\mathsf{6(a)}$ and $\mathfrak{D}\mathsf{6(b)}$ simultaneously.

$\mathfrak{D}\mathsf{6(a)_{i}}$ and $\mathfrak{D}\mathsf{6(b)_{i}}$ each possess a single 
frequency loop, which may be evaluated using Eq.~(\ref{J0}), giving the results
\begin{align}\label{D6(a)i--F}
	\mathfrak{D}\mathsf{6(a)_{i}}
	=& \left(\frac{-\lambda \, 2 dl}{8 \pi}\right)
	\nonumber\\
	&\times
	\left[\Gamma_{U}(\gamma_{s}+\gamma_{c})\mathcal{O}_{U}
	+\Gamma_{V}(\gamma_{s}-\gamma_{c})\mathcal{O}_{X}\right],
\end{align}
and
\begin{align}\label{D6(b)i--F}
	\mathfrak{D}\mathsf{6(b)_{i}}
	=&\left(\frac{-\lambda \, 2 dl}{8 \pi}\right)
	\nonumber\\
	&\times
	\left[\Gamma_{U}(\gamma_{s}+\gamma_{c})\overline{\mathcal{O}}_{U}
	+\Gamma_{V}(\gamma_{s}-\gamma_{c})\overline{\mathcal{O}}_{X}\right].
\end{align}

$\mathfrak{D}\mathsf{6(a)_{ii}}$ and $\mathfrak{D}\mathsf{6(a)_{iii}}$ give identical contributions. 
We evaluate $\mathfrak{D}\mathsf{6(a)_{iii}}$ as an example of a diagram with two frequency loop 
integrations over $\mathsf{\alpha} \rightarrow \omega_{\alpha}$ and 
$\mathsf{\beta} \rightarrow \omega_{\beta}$. Using the Feynman rules, we have
\begin{equation}\label{D6(a)iii}
	\mathfrak{D}\mathsf{6(a)_{iii}} 
	= \frac{(2 i \xi^{a})^{3}}{2!} 
	\delta_{1+3,2+4} 
	\int {d^{d} \bm{\mathrm{r}}} \,
	{\delta Q}^{a, a}_{\mathsf{S} \, 1, 2}{\delta Q}^{a, a}_{\mathsf{S} \, 3, 4} \, I_{4},
\end{equation}
where the frequency-momentum integral again separates into
\begin{equation}\label{I4}
	I_{4} \equiv I_{4\,\mathsf{S}} + I_{4\,\mathsf{C}},  
\end{equation}
corresponding to the $\mathsf{P_{S}}$ and $\mathsf{P_{C}}$ components of the interaction 
sector propagator in $\mathfrak{D}\mathsf{6(a)_{iii}}$, 
Fig.~\ref{D6Fig}.
We first compute $I_{4\,\mathsf{S}}$; up to irrelevant terms one finds 
\begin{align}\label{I4S}
	I_{4\,\mathsf{S}} =& 
	\int  \frac{d\omega_{\alpha} \, d\omega_{\beta} \, d^{\textrm{2}}\bm{\mathrm{k}}}{(2 \pi)^{\textrm{4}}}
	\Gamma_{s} 
	(s_{1} - s_{\alpha}) (s_{\alpha + \beta} - s_{\beta})
	s_{\alpha}^{2}
	\nonumber\\
	&\times
	(\Gamma_{U} s_{3} - \Gamma_{V} s_{\alpha})
	(\Gamma_{U} s_{\beta} - \Gamma_{V} s_{\alpha + \beta}) \nonumber\\
	&\times
	\left[\Delta_{\mathcal{O}}^{a,a}(0,|\omega_{\alpha}|,\bm{\mathrm{k}})\right]^{2}
	\Delta_{\mathcal{S}}^{a}(|\omega_{\alpha}|,\bm{\mathrm{k}}).
\end{align}
Although this expression entails two frequency loop integrations, the propagator kernels 
$\Delta_{\mathcal{O}}$ and $\Delta_{\mathcal{S}}$ in Eq.~(\ref{I4S}) depend exclusively 
upon a single frequency variable, $\omega_{\alpha}$. The integral over $\omega_{\beta}$ 
yields only a kinematical factor of $|\omega_{\alpha}|$: the projector 
$(s_{\alpha + \beta} - s_{\beta})$ in Eq.~(\ref{I4S}) gives a non-zero contribution as a 
function of $\omega_{\beta}$ only over a frequency segment of length $|\omega_{\alpha}|$; 
over this range, $(\Gamma_{U} s_{\beta} - \Gamma_{V} s_{\alpha + \beta}) = s_\beta 2 \Gamma_{s}$. 
The factor $(s_{1} - s_{\alpha})$ limits the range of the $\omega_{\alpha}$ integration to the 
half line (as usual); one obtains
\begin{align}\label{I4S--2}
	I_{4\,\mathsf{S}}
	=&
	\frac{-4 \Gamma_{s}^{2}}{\pi h^{3}}(\Gamma_{U} s_{1} s_{3} + \Gamma_{V})
	\nonumber\\
	&\times
	\int\limits_{\omega_{\alpha} > 0}  \frac{d\omega_{\alpha} \, d^{\textrm{2}}\bm{\mathrm{k}}}{(2 \pi)^{\textrm{3}}}\omega_{\alpha}
	[D\bm{\mathrm{k}}^{2} - i \xi^{a} \omega_{\alpha}]^{-2}
	\nonumber\\
	&\quad\;\;\times
	[D\bm{\mathrm{k}}^{2} - i (1-\gamma_{s})\xi^{a} \omega_{\alpha}]^{-1}. 
\end{align}
We use Eq.~(\ref{J2}) to evaluate the integral, and we find the result
\begin{align}\label{I4S--3}
	I_{4\,\mathsf{S}}
	=&
	\frac{-4 \Gamma_{s}^{2}}{\pi h^{3}}(\Gamma_{U} s_{1} s_{3} + \Gamma_{V})
	J_{2}\biglb(\xi^{a};(1-\gamma_{s})\xi^{a}\bigrb) 
	\nonumber\\
	=& \frac{-\lambda \, 2 dl }{32 \pi}(\Gamma_{U} s_{1} s_{3} + \Gamma_{V})
	[\gamma_{s} + \ln(1-\gamma_{s})],
\end{align}
What we have found here is well-known (see e.g.\ Ref.~\onlinecite{KELDYSH}) 
from other Finkel'stein NL$\sigma$Ms: in a one-loop calculation, a given diagram may involve 
$n = \{0,1,2,\ldots\}$ undetermined frequency loops; however, for $n \geq 2$, 
$n - 1$ of the associated frequency integrations typically give simple kinematical 
factors, and the final diagrammatic amplitude can almost always be expressed in 
terms of an integration over a single loop frequency and a single loop momentum, 
as in Eq.~(\ref{I4S--2}) 
[but see the evaluation of $\mathfrak{D}\mathsf{9(b)_{i}}$, Eq.~(\ref{I6C}), and 
$\mathfrak{D}\mathsf{13}$, Eq.~(\ref{I7}) below].

We can try to evaluate $I_{4\,\mathsf{C}}$ in a similar fashion, but the associated 
amplitude goes like 
$\lambda \Gamma_{c}^{2}[1 + \bm{\mathit{O}}(\Gamma_{c})]$. In our one-loop results 
[Eqs.~(\ref{FlowEqs1}) and (\ref{FlowEqs2}), below], we keep only terms to second 
homogeneous order in $\lambda$ and $\Gamma_{c}$ in the flow equations for the 
interaction constants $\gamma_{s}$ and $\gamma_{c}$; therefore, we neglect the 
contribution of $I_{4\,\mathsf{C}}$.

Combining Eqs.~(\ref{D6(a)iii}) and (\ref{I4S--2}), and accounting for the identical 
contribution from $\mathfrak{D}\mathsf{6(a)_{ii}}$, we have
\begin{align}\label{D6(a)ii+D6(a)iii--F}
	\mathfrak{D}\mathsf{6(a)_{ii}}+\mathfrak{D}\mathsf{6(a)_{iii}}
	=& \left(\frac{\lambda \, 2 dl}{4 \pi}\right)[\gamma_{s} + \ln(1-\gamma_{s})]
	\nonumber\\
	&\times\left(\Gamma_{U} \mathcal{O}_{U} + \Gamma_{V} \mathcal{O}_{X}\right),	
\end{align}
and similarly 
\begin{align}\label{D6(b)ii+D6(b)iii--F}
	\mathfrak{D}\mathsf{6(b)_{ii}}+\mathfrak{D}\mathsf{6(b)_{iii}}
	=& \left(\frac{\lambda \, 2 dl}{4 \pi}\right)[\gamma_{s} + \ln(1-\gamma_{s})]
	\nonumber\\
	&\times\left(\Gamma_{U} \overline{\mathcal{O}}_{U} + \Gamma_{V} \overline{\mathcal{O}}_{X}\right).	
\end{align}

Diagrams $\mathfrak{D}\mathsf{6(a)_{iv}}$ and $\mathfrak{D}\mathsf{6(b)_{iv}}$ also involve 
two frequency loops, and may be similarly evaluated; the results obtained are
\begin{align}\label{D6(a)iv--F}
	\mathfrak{D}\mathsf{6(a)_{iv}}
	=& \left(\frac{-\lambda \, 2 dl}{8 \pi}\right)
	\left\{
	\Gamma_{s}[\gamma_{s} + \ln(1-\gamma_{s})]-
	\Gamma_{c}\frac{\gamma_{s}^{2}}{2}
	\right\}
	\nonumber\\
	&\times\left(\mathcal{O}_{U} - \mathcal{O}_{X}\right),
\end{align}
and 
\begin{align}\label{D6(b)iv--F}
	\mathfrak{D}\mathsf{6(b)_{iv}}
	=& \left(\frac{-\lambda \, 2 dl}{8 \pi}\right)
	\left\{
	\Gamma_{s}[\gamma_{s} + \ln(1-\gamma_{s})]-
	\Gamma_{c}\frac{\gamma_{s}^{2}}{2}
	\right\}
	\nonumber\\
	&\times\left(\overline{\mathcal{O}}_{U} - \overline{\mathcal{O}}_{X}\right),
\end{align}

$\mathfrak{D}\mathsf{6(a)_{v}}$ and $\mathfrak{D}\mathsf{6(a)_{vi}}$ provide examples 
of the most complicated type of diagram that appears to one loop, incorporating two 
copies of the interaction propagator component sum $\mathsf{P_{S}}+\mathsf{P_{C}}$, 
sandwiched between a pair of vertices [here 
$\mathsf{V(c)}$ and $\mathsf{V(c)}$] that originate from the interacting sector 
[Eq.~(\ref{SIQmatrixSPLeft})] of the FNL$\sigma$M. These graphs involve \emph{three} 
loop frequencies $\mathsf{\alpha} \rightarrow \omega_{\alpha}$, $\mathsf{\beta} 
\rightarrow \omega_{\beta}$, and $\mathsf{\rho} \rightarrow \omega_{\rho}$. We 
evaluate $\mathfrak{D}\mathsf{6(a)_{v}}$ explicitly, as it requires less work than 
$\mathfrak{D}\mathsf{6(a)_{vi}}$. 
Using the Feynman rules, we write
\begin{equation}\label{D6(a)v}
	\mathfrak{D}\mathsf{6(a)_{v}} 
	= \frac{(2 i \xi^{a})^{4}}{2!} 
	\delta_{1+3,2+4} 
	\int {d^{d} \bm{\mathrm{r}}} \,
	{\delta Q}^{a, a}_{\mathsf{S} \, 1, 2}{\delta Q}^{a, a}_{\mathsf{S} \, 3, 4} \, I_{5},
\end{equation}
where now the loop integral $I_{5}$ breaks into \emph{four} pieces, 
\begin{equation}\label{I5}
	I_{5} \equiv 
	I_{5\,\mathsf{S}}^{\,\mathsf{S}} + 
	I_{5\,\mathsf{C}}^{\,\mathsf{S}} + 
	I_{5\,\mathsf{S}}^{\,\mathsf{C}} + 
	I_{5\,\mathsf{C}}^{\,\mathsf{C}}.  
\end{equation}
The two propagators occuring in the fast mode loop of $\mathfrak{D}\mathsf{6(a)_{v}}$  
appear at the ``top'' and ``bottom'' of the associated diagram in Fig.~\ref{D6Fig}. Each 
propagator may represent either $\mathsf{P_{S}}$ or $\mathsf{P_{C}}$. In Eq.~(\ref{I5}), 
the raised index describes the character of the ``top'' propagator component, while the 
lower index describes the character of the ``bottom'' component, i.e.\
$I_{5\,\mathsf{S}}^{\,\mathsf{C}}$ is the amplitude for ``top'' component to be $\mathsf{P_{C}}$ 
and the ``bottom'' component to be $\mathsf{P_{S}}$. Fortunately in this example, 
$I_{5\,\mathsf{C}}^{\,\mathsf{S}}$,  $I_{5\,\mathsf{S}}^{\,\mathsf{C}}$, and 
$I_{5\,\mathsf{C}}^{\,\mathsf{C}}$ all give corrections of order $\lambda \Gamma_{c}^{2}$, 
which we ignore.

We compute $I_{5\,\mathsf{S}}^{\,\mathsf{S}}$ as follows:
\begin{align}\label{I5SS}
	I_{5\,\mathsf{S}}^{\,\mathsf{S}} =& 
	\int  \frac{d\omega_{\alpha} \, d\omega_{\beta} \, d\omega_{\rho} \, d^{\textrm{2}}\bm{\mathrm{k}}}{(2 \pi)^{\textrm{5}}}
	\Gamma_{s}^{2} s_{\alpha}^{2}
	(s_{1} - s_{\alpha}) (s_{\alpha + \beta} - s_{\beta})
	\nonumber\\
	&\times
	(s_{3} - s_{\alpha}) (s_{\alpha + \rho} - s_{\rho})
	(\Gamma_{U} s_{\rho} - \Gamma_{V} s_{\alpha + \rho})
	\nonumber\\
	&\times
	(\Gamma_{U} s_{\beta} - \Gamma_{V} s_{\alpha + \beta}) 
	\Big[\Delta_{\mathcal{O}}^{a,a}(0,|\omega_{\alpha}|,\bm{\mathrm{k}})
	\Delta_{\mathcal{S}}^{a}(|\omega_{\alpha}|,\bm{\mathrm{k}})\Big]^{2}\!.
\end{align}
Similar to the two frequency loop case, Eq.~(\ref{I4S}), the integrals over 
$\omega_{\beta}$ and $\omega_{\rho}$ give only an overall kinematic factor of 
$\omega_{\alpha}^{2}$, thanks to the projectors $(s_{\alpha + \beta} - s_{\beta})$ 
and $(s_{\alpha + \rho} - s_{\rho})$ in the integrand of Eq.~(\ref{I5SS}). The 
factors $(s_{1} - s_{\alpha})$ and $(s_{3} - s_{\alpha})$ in this integrand allow 
for a non-zero amplitude only when 
$s_{1}=s_{3}=-s_{\alpha}$. As a result, we obtain
\begin{align}\label{I5SS--2}
	I_{5\,\mathsf{S}}^{\,\mathsf{S}} 
	=& 
	\frac{4 \Gamma_{s}^{4}}{\pi^{2} h^{4}}(s_{1} + s_{3})^{2}
	\nonumber\\
	&\times
	\int\limits_{\omega_{\alpha} > 0}  \frac{d\omega_{\alpha} \, d^{\textrm{2}}\bm{\mathrm{k}}}{(2 \pi)^{\textrm{3}}}\omega_{\alpha}^{2}
	[D\bm{\mathrm{k}}^{2} - i \xi^{a} \omega_{\alpha}]^{-2}
	\nonumber\\
	&\quad\;\;\times
	[D\bm{\mathrm{k}}^{2} - i (1-\gamma_{s})\xi^{a} \omega_{\alpha}]^{-2},
\end{align}
which may be evaluated using Eq.~(\ref{J5}):
\begin{align}\label{I5SS--3}
	I_{5\,\mathsf{S}}^{\,\mathsf{S}} 
	=& 
	\frac{4 \Gamma_{s}^{4}}{\pi^{2} h^{4}}(s_{1} + s_{3})^{2}
	J_{5}\biglb(\xi^{a};(1-\gamma_{s})\xi^{a}\bigrb)
	\nonumber\\
	=&\frac{- \lambda i \xi^{a} 2 dl}{128\pi}(s_{1} + s_{3})^{2}
	\nonumber\\
	&\times
	\Gamma_{s}\left[
	\gamma_{s}\left(\frac{2 - \gamma_{s}}{1-\gamma_{s}}\right)
	+2\ln(1-\gamma_{s})
	\right].
\end{align}

Combining Eqs.~(\ref{D6(a)v}) and (\ref{I5SS--3}), the resulting amplitude 
for $\mathfrak{D}\mathsf{6(a)_{v}}$ is
\begin{align}\label{D6(a)v--F}
	\mathfrak{D}\mathsf{6(a)_{v}}
	=& \left(\frac{-\lambda \, 2 dl}{8\pi}\right)
	\Gamma_{s}
	\left[
	\gamma_{s}\left(\frac{2 - \gamma_{s}}{1-\gamma_{s}}\right)
	+2\ln(1-\gamma_{s})
	\right]
	\nonumber\\
	&\times
	\left(\mathcal{O}_{U} + \mathcal{O}_{X}\right).	
\end{align}
Similarly, we find
\begin{align}\label{D6(b)v--F}
	\mathfrak{D}\mathsf{6(b)_{v}}
	=& \left(\frac{-\lambda \, 2 dl}{8\pi}\right)
	\Gamma_{s}
	\left[
	\gamma_{s}\left(\frac{2 - \gamma_{s}}{1-\gamma_{s}}\right)
	+2\ln(1-\gamma_{s})
	\right]
	\nonumber\\
	&\times
	\left(\overline{\mathcal{O}}_{U} + \overline{\mathcal{O}}_{X}\right).	
\end{align}

Finally we give the results for the last two diagrams in Fig.~\ref{D6Fig},
\begin{align}\label{D6(a)vi--F}
	\mathfrak{D}\mathsf{6(a)_{vi}}
	=& \left(\frac{\lambda \, 2 dl}{8\pi}\right)
	\left\{
	\Gamma_{s}
	\left[
	\gamma_{s}\left(\frac{2 - \gamma_{s}}{1-\gamma_{s}}\right)
	+2\ln(1-\gamma_{s})
	\right]
	\right.
	\nonumber\\
	&\left.\qquad\qquad-\Gamma_{c}
	\left[
	\ln(1-\gamma_{s}) +\gamma_{s}+\frac{\gamma_{s}^{2}}{2}
	\right]
	\right\}
	\nonumber\\
	&\times
	\left(\mathcal{O}_{U} - \mathcal{O}_{X}\right),
\end{align}
and
\begin{align}\label{D6(b)vi--F}
	\mathfrak{D}\mathsf{6(b)_{vi}}
	=& \left(\frac{\lambda \, 2 dl}{8\pi}\right)
	\left\{
	\Gamma_{s}
	\left[
	\gamma_{s}\left(\frac{2 - \gamma_{s}}{1-\gamma_{s}}\right)
	+2\ln(1-\gamma_{s})
	\right]
	\right.
	\nonumber\\
	&\left.\qquad\qquad-\Gamma_{c}
	\left[
	\ln(1-\gamma_{s}) +\gamma_{s}+\frac{\gamma_{s}^{2}}{2}
	\right]
	\right\}
	\nonumber\\
	&\times
	\left(\overline{\mathcal{O}}_{U} - \overline{\mathcal{O}}_{X}\right).
\end{align}
Obtaining these results requires the use of Eqs.~(\ref{J5}) and (\ref{J7}).

\begin{figure}
\includegraphics[width=0.45\textwidth]{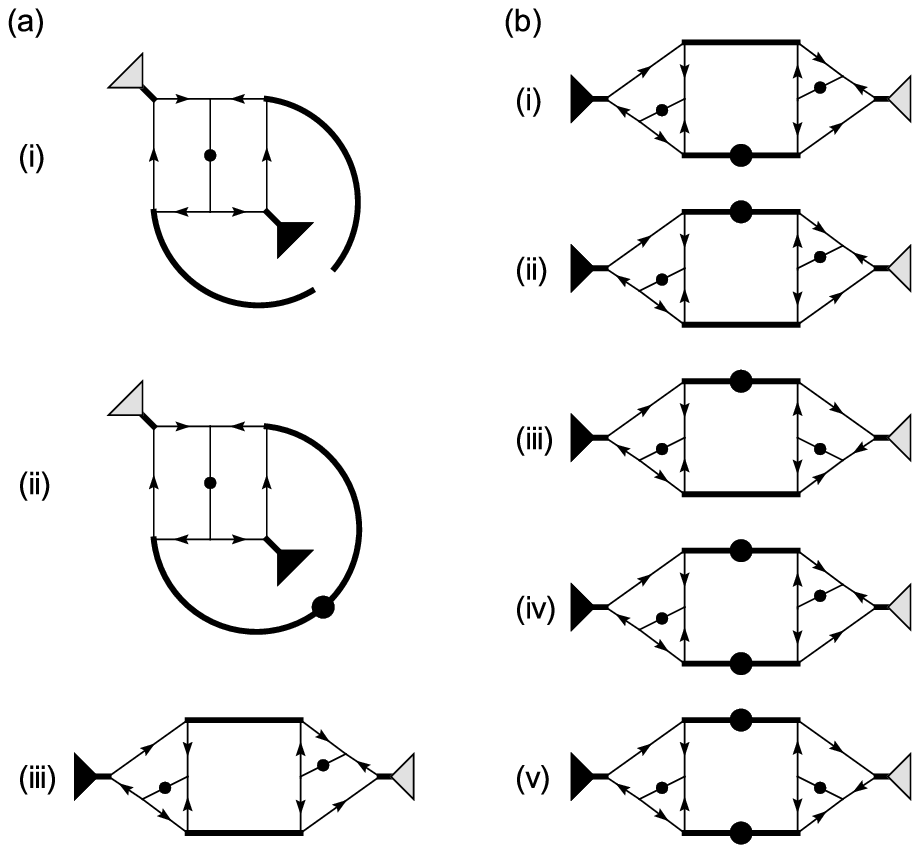}
\caption{Category $\mathfrak{D}\mathsf{7}$: Diagrams renormalizing $\Gamma_{V}$.\label{D7Fig}}
\end{figure}

The diagrams in category $\mathfrak{D}\mathsf{7}$, Fig.~\ref{D7Fig}, nominally 
renormalize the intersublattice interaction strength $\Gamma_{V}$, corresponding 
to the operator $\mathcal{O}_{V}$ in Table \ref{IntOpsTable}. The computation of 
these graphs closely parallels the results we have obtained above in 
Eqs.~(\ref{D5(a)i+D5(a)ii--F}), (\ref{D5(a)iii--F})--(\ref{D6(b)i--F}), 
(\ref{D6(a)ii+D6(a)iii--F})--(\ref{D6(b)iv--F}), and (\ref{D6(a)v--F})--(\ref{D6(b)vi--F}).  
We therefore summarize only the results:
\begin{equation}\label{D7(a)i--F}
	\mathfrak{D}\mathsf{7(a)_{i}} 
	= \frac{\lambda_{A} \, 2 dl}{4 \pi} \Gamma_{V} \mathcal{O}_{V},
\end{equation}
\begin{equation}\label{D7(a)ii--F}
	\mathfrak{D}\mathsf{7(a)_{ii}} 
	= \left(\frac{\lambda \, 2 dl}{8 \pi}\right)[\gamma_{c} + \ln(1-\gamma_{s})] 
	\Gamma_{V} \left(\mathcal{O}_{V} + \mathcal{O}_{Y}\right),
\end{equation}
\begin{equation}\label{D7(a)iii--F}
	\mathfrak{D}\mathsf{7(a)_{iii}}
	=\left(\frac{\lambda \, 2 dl}{8 \pi}\right) 
	\left[\Gamma_{V}(\gamma_{s}-\gamma_{c})\mathcal{O}_{V}
	-\Gamma_{U}(\gamma_{s}+\gamma_{c})\mathcal{O}_{Y}\right],
\end{equation}
\begin{align}\label{D7(b)i+D7(b)ii--F}
	\mathfrak{D}\mathsf{7(b)_{i}}+\mathfrak{D}\mathsf{7(b)_{ii}}
	=& \left(\frac{-\lambda \, 2 dl}{4 \pi}\right)[\gamma_{s} + \ln(1-\gamma_{s})]
	\nonumber\\
	&\times\left(\Gamma_{V} \mathcal{O}_{V} - \Gamma_{U} \mathcal{O}_{Y}\right),	
\end{align}
\begin{align}\label{D7(b)iii--F}
	\mathfrak{D}\mathsf{7(b)_{iii}}
	=& \left(\frac{\lambda \, 2 dl}{8 \pi}\right)
	\left\{
	\Gamma_{s}[\gamma_{s} + \ln(1-\gamma_{s})]+
	\Gamma_{c}\frac{\gamma_{s}^{2}}{2}
	\right\}
	\nonumber\\
	&\times\left(\mathcal{O}_{V} + \mathcal{O}_{Y}\right),
\end{align}
\begin{align}\label{D7(b)iv--F}
	\mathfrak{D}\mathsf{7(b)_{iv}}
	=& \left(\frac{\lambda \, 2 dl}{8\pi}\right)
	\Gamma_{s}
	\left[
	\gamma_{s}\left(\frac{2 - \gamma_{s}}{1-\gamma_{s}}\right)
	+2\ln(1-\gamma_{s})
	\right]
	\nonumber\\
	&\times
	\left(\mathcal{O}_{V} - \mathcal{O}_{Y}\right),
\end{align}
and finally
\begin{align}\label{D7(b)v--F}
	\mathfrak{D}\mathsf{7(b)_{v}}
	=& \left(\frac{-\lambda \, 2 dl}{8\pi}\right)
	\left\{
	\Gamma_{s}
	\left[
	\gamma_{s}\left(\frac{2 - \gamma_{s}}{1-\gamma_{s}}\right)
	+2\ln(1-\gamma_{s})
	\right]
	\right.
	\nonumber\\
	&\left.\qquad\qquad+\Gamma_{c}
	\left[
	\ln(1-\gamma_{s}) +\gamma_{s}+\frac{\gamma_{s}^{2}}{2}
	\right]
	\right\}
	\nonumber\\
	&\times
	\left(\mathcal{O}_{V} + \mathcal{O}_{Y}\right).
\end{align}
Note that in Eqs.~(\ref{D7(a)ii--F})--(\ref{D7(b)v--F}), we obtain both the 
renormalization of the operator $\mathcal{O}_{V} \leftrightarrow \Gamma_{V}$, 
as well as the generation of the unwanted term $\mathcal{O}_{Y}$, just 
as we had the generation of $\mathcal{O}_{X}$ and $\overline{\mathcal{O}}_{X}$ 
in the discussion of $\mathcal{O}_{U}$ and $\overline{\mathcal{O}}_{U}$ renormalization, above.

\begin{figure}[t]
\includegraphics[width=0.45\textwidth]{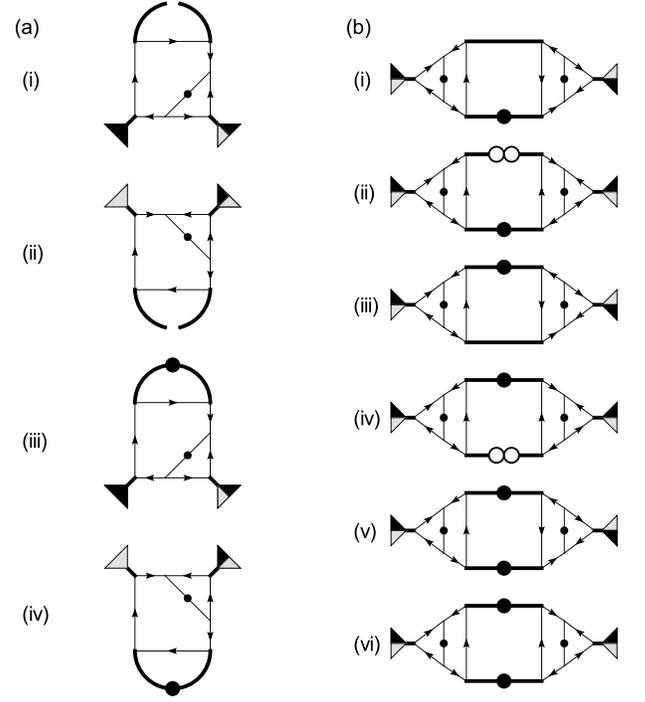}
\caption{Category $\mathfrak{D}\mathsf{8}$: Diagrams renormalizing
$\Gamma_{s}=\frac{\Gamma_{U} + \Gamma_{V}}{2}$.\label{D8Fig}}
\end{figure}

Additional diagrams renormalizing the interparticle interaction strengths appear 
in Figs.~\ref{D8Fig}--\ref{D13Fig}; these will be discussed in the next subsection. 
Let us first pause to take stock of the results so far obtained. We will sum 
the amplitudes of all three diagram categories evaluated in this subsection, 
using the category symbols $\mathfrak{D}\mathsf{5}$, $\mathfrak{D}\mathsf{6}$, 
and $\mathfrak{D}\mathsf{7}$ to denote the associated sums. We express our 
results in terms of the interaction parameters $\Gamma_{s}$ and $\Gamma_{c}$, 
defined via Eq.~(\ref{SingletCDWGamma}), as well as the relative versions 
$\gamma_{s}$ and $\gamma_{c}$, introduced in Eq.~(\ref{SingletCDW}). We drop 
terms higher than first order in $\Gamma_{c}$ (or $\gamma_{c}$). 
A large number of non-trivial cancelations occur, and the results for 
$\mathfrak{D}\mathsf{5} + \mathfrak{D}\mathsf{6}$ and $\mathfrak{D}\mathsf{7}$ prove quite simple.

Summing categories $\mathfrak{D}\mathsf{5}$ and $\mathfrak{D}\mathsf{6}$, Eqs.~(\ref{D5(a)i+D5(a)ii--F}), 
(\ref{D5(a)iii--F})--(\ref{D6(b)i--F}), (\ref{D6(a)ii+D6(a)iii--F})--(\ref{D6(b)iv--F}), 
and (\ref{D6(a)v--F})--(\ref{D6(b)vi--F}), we find
\begin{align}\label{D5+D6--F}
	\mathfrak{D}\mathsf{5} & + \mathfrak{D}\mathsf{6} \nonumber\\
	&\;\;=
	\left(\frac{-2 dl}{8\pi}\right)
	\Big[
	2 \left(\lambda_{A} - \lambda\right) 
	(\Gamma_{s}+\Gamma_{c})
	\Big]
	\left(\mathcal{O}_{U} + \overline{\mathcal{O}}_{U}\right)
	\nonumber\\
	&\;\;\;\,\phantom{=}
	+ \left(\frac{-2 dl}{8\pi}\right) 
	\left[
	\frac{2 \lambda \Gamma_{s} \gamma_{s}}{1-\gamma_{s}}
	\right]
	\left(\mathcal{O}_{X} + \overline{\mathcal{O}}_{X}\right).
\end{align}
Summing the diagrams in category $\mathfrak{D}\mathsf{7}$, Eqs.~(\ref{D7(a)i--F})--(\ref{D7(b)v--F}), we obtain
\begin{align}\label{D7--F}
	\mathfrak{D}\mathsf{7}
	=& 
	\left(\frac{-2 dl}{8\pi}\right)
	\Big[
	-2 \lambda_{A} (\Gamma_{s}-\Gamma_{c})
	\Big]
	\mathcal{O}_{V}
	\nonumber\\
	&+
	\left(\frac{-2 dl}{8\pi}\right) 
	\left[
	\frac{2 \lambda \Gamma_{s} \gamma_{s}}{1-\gamma_{s}}
	\right]
	\mathcal{O}_{Y}.
\end{align}

In the partial results given by these equations, observe the almost complete cancelation 
of all ``junk'' terms involving the operators 
$\mathcal{O}_{X}$, $\overline{\mathcal{O}}_{X}$, and $\mathcal{O}_{Y}$, not present in 
the original FNL$\sigma$M action [Eqs.~(\ref{SDQmatrixSPLeft}) and (\ref{SIQmatrixSPLeft})]. 
In fact, the remaining terms proportional to $\mathcal{O}_{X} + \overline{\mathcal{O}}_{X}$ 
and $\mathcal{O}_{Y}$ in Eqs.~(\ref{D5+D6--F}) and (\ref{D7--F}), respectively, exactly 
cancel, up to terms fourth order in the slow mode fluctuations 
$\delta \hat{Q}_{\mathsf{S}}^{\phantom{\dagger}}$ and $\delta \hat{Q}_{\mathsf{S}}^{\dagger}$. 
To see this, we use Eqs.~(\ref{FastSlow}) and (\ref{FastSlowUnitary}) to write
\begin{equation}\label{DeltaQsAlgebra}
	\delta \hat{Q}_{\mathsf{S}}^{\phantom{\dagger}} + \delta \hat{Q}_{\mathsf{S}}^{\dagger} = 
	- \delta \hat{Q}_{\mathsf{S}}^{\phantom{\dagger}} \delta \hat{Q}_{\mathsf{S}}^{\dagger}.
\end{equation}
Repeated applications of Eq.~(\ref{DeltaQsAlgebra}) prove the identification
\begin{equation}
	\mathcal{O}_{Y} = -\left(\mathcal{O}_{X} + \overline{\mathcal{O}}_{X}\right) + 
	\left(\delta \hat{Q}_{\mathsf{S}}^{\phantom{\dagger}} \delta \hat{Q}_{\mathsf{S}}^{\dagger}\right)^{2}.
\end{equation}
Thus, to our working order in the slow mode fields, the FNL$\sigma$M appears (so far) to be 
renormalizable.\cite{footnote-m}
We must complete the one-loop calculation to verify this.


\subsubsection{(Further) renormalization of $\Gamma_{s}$ and $\Gamma_{c}$}\label{IntRenorm2}

\begin{figure}[b]
\includegraphics[width=0.45\textwidth]{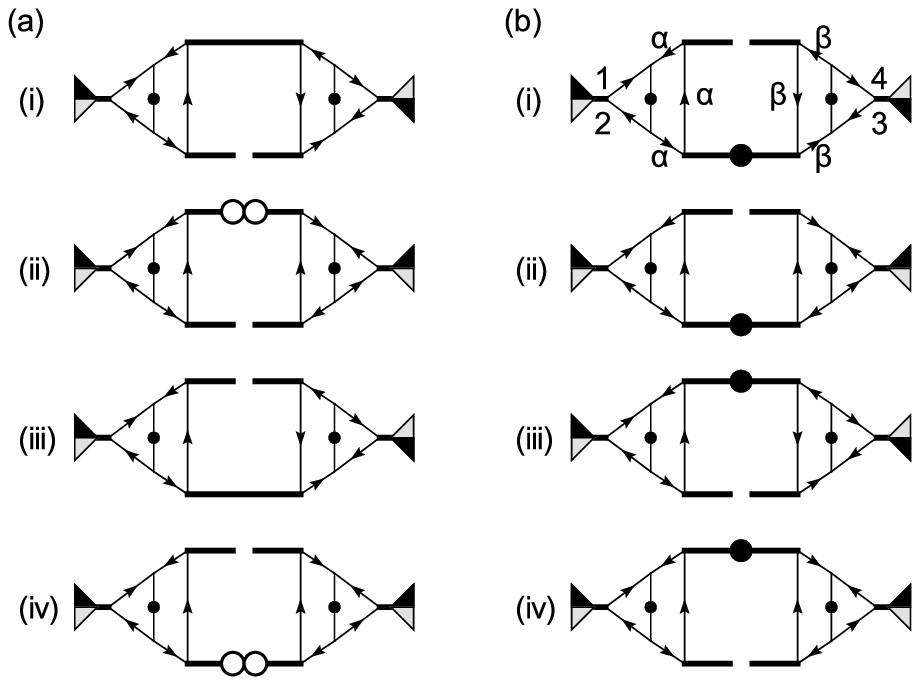}
\caption{Category $\mathfrak{D}\mathsf{9}$: Diagrams renormalizing 
$\Gamma_{s}=\frac{\Gamma_{U} + \Gamma_{V}}{2}$.\label{D9Fig}}
\end{figure}

\begin{figure}[b]
\includegraphics[width=0.45\textwidth]{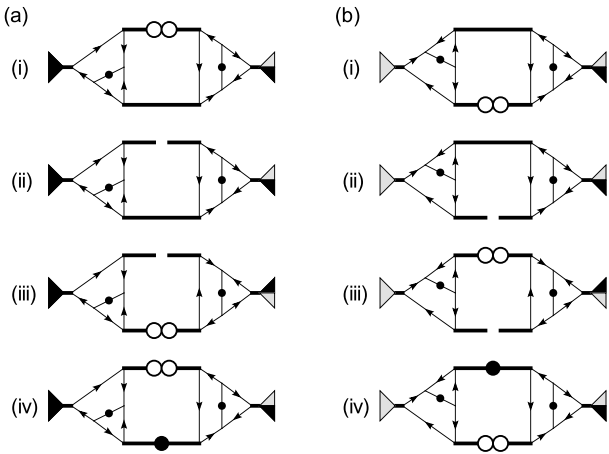}
\caption{Category $\mathfrak{D}\mathsf{10}$: Diagrams renormalizing 
$\Gamma_{s}~=~\frac{\Gamma_{U} + \Gamma_{V}}{2}$.\label{D10Fig}}
\end{figure}

\begin{figure}
\includegraphics[width=0.45\textwidth]{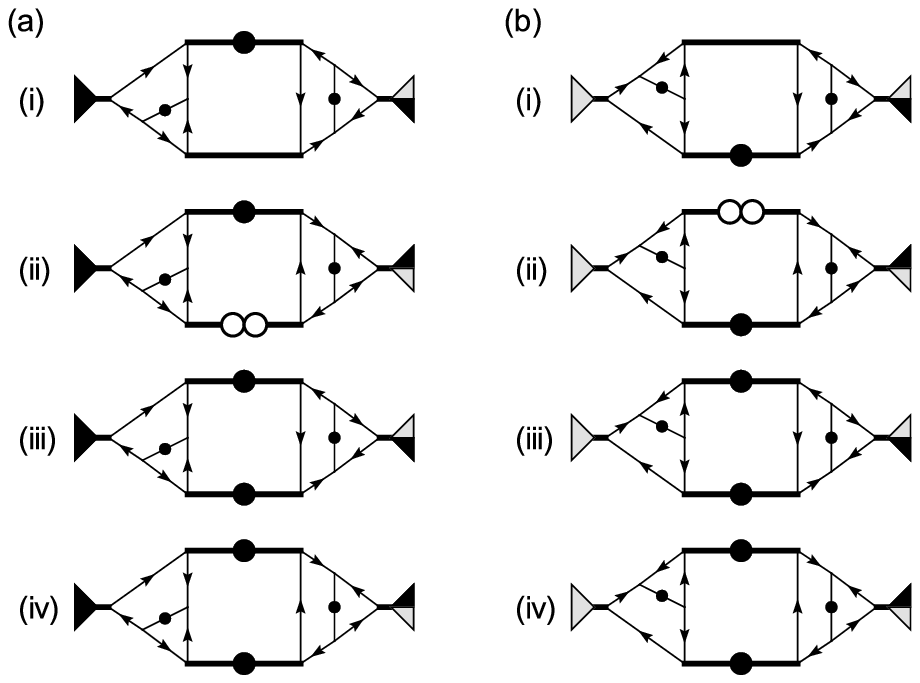}
\caption{Category $\mathfrak{D}\mathsf{11}$: Diagrams renormalizing 
$\Gamma_{s}~=~\frac{\Gamma_{U} + \Gamma_{V}}{2}$.\label{D11Fig}}
\end{figure}

\begin{figure}[b]
\includegraphics[width=0.45\textwidth]{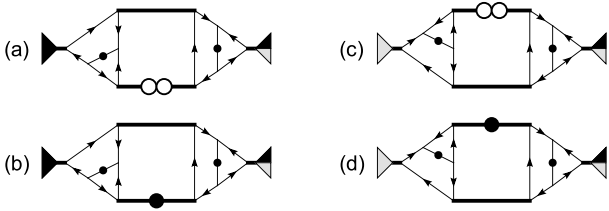}
\caption{Category $\mathfrak{D}\mathsf{12}$: Additional non-zero diagrams, which cancel in pairs. 
Contributions from individual diagrams correspond to the generation of local 
operators not present in the original model.\label{D12Fig}}
\end{figure}

\begin{figure}
\includegraphics[width=0.3\textwidth]{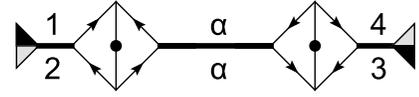}
\caption{Category $\mathfrak{D}\mathsf{13}$: Diagram renormalizing 
$\Gamma_{c}~=~\frac{\Gamma_{U} - \Gamma_{V}}{2}$.\label{D13Fig}}
\end{figure}

The interaction sector renormalizations described by the remaining diagrams in 
Figs.~\ref{D8Fig}--\ref{D13Fig} are most compactly stated in terms of corrections 
to the linear combinations $\Gamma_{s}$ and $\Gamma_{c}$ [Eq.~(\ref{SingletCDWGamma})],
which couple to the slow mode operators 
\begin{align}
	\mathcal{O}_{s} & \equiv \mathcal{O}_{U}+\overline{\mathcal{O}}_{U} + \mathcal{O}_{V}, \nonumber\\
\intertext{and}
	\mathcal{O}_{c} & \equiv \mathcal{O}_{U}+\overline{\mathcal{O}}_{U} - \mathcal{O}_{V},
\end{align}
respectively [Eq.~(\ref{SSlow})]. 
We choose now to quicken our pace, providing calculational details only when 
substantially different from those presented in the previous subsection.

Category $\mathfrak{D}\mathsf{8}$ diagrams are shown in Fig.~\ref{D8Fig}. Diagrams in 
subcategory $\mathfrak{D}\mathsf{8(a)}$ involve vertices 
$\mathsf{V(j)}$ and $\mathsf{V(k)}$, with propagator components $\mathsf{P_{A}}$ and 
$\mathsf{P_{S}}+\mathsf{P_{C}}$. The sum of these four graphs gives
\begin{align}\label{D8(a)--F}
	\mathfrak{D}\mathsf{8(a)} =&
	\left(\frac{-2 dl}{8\pi}\right)
	\left\{
	2 \lambda_{A} + 2 \lambda\left[\gamma_{c} + \ln(1-\gamma_{s}) \right] 
	\right\}
	\nonumber\\
	&\times
	\left(\Gamma_{s} \mathcal{O}_{s} + \Gamma_{c} \mathcal{O}_{c}\right).
\end{align}
The diagrams in subcategory $\mathfrak{D}\mathsf{8(b)}$ each pair together two 
copies of the vertex $\mathsf{V(e)}$, with various combinations of the propagator 
components $\mathsf{P_{\lambda}}$ and $\mathsf{P_{S}}+\mathsf{P_{C}}$. Graphs 
$\mathfrak{D}\mathsf{8(b)_{i}}$--$\mathfrak{D}\mathsf{8(b)_{vi}}$ 
each encompass two frequency loops, and their 
evaluation proceeds along the lines of $\mathfrak{D}\mathsf{6(a)_{iii}}$ 
[Eqs.~(\ref{D6(a)iii})--(\ref{D6(a)ii+D6(a)iii--F})]. Diagrams $\mathfrak{D}\mathsf{8(b)v}$ 
and $\mathfrak{D}\mathsf{8(b)vi}$ involve three frequency loops a la Eq.~(\ref{I5SS}), 
but their amplitudes precisely cancel. The sum of all six graphs in $\mathfrak{D}\mathsf{8(b)}$ 
gives 
\begin{align}\label{D8(b)--F}
	\mathfrak{D}\mathsf{8(b)} =&
	\left(\frac{\lambda \, 2 dl}{8\pi}\right)
	\left(2 \Gamma_{c} \gamma_{s}^{2}\right)
	\mathcal{O}_{s}.
\end{align}

The eight diagrams in category $\mathfrak{D}\mathsf{9}$, pictured in Fig.~\ref{D9Fig}, 
represent only two discernable amplitudes.
These diagrams also involve two copies of the vertex $\mathsf{V(e)}$, with the 
$\lambda_{A}$ propagator component $\mathsf{P_{A}}$ in combination with 
$\mathsf{P_{\lambda}}$ or $\mathsf{P_{S}}+\mathsf{P_{C}}$.
The four graphs in subcategory $\mathfrak{D}\mathsf{9(a)}$ each give identical 
contributions, with the sum
\begin{align}\label{D9(a)--F}
	\mathfrak{D}\mathsf{9(a)} =
	\left(\frac{\lambda_{A} \, 2 dl}{8\pi}\right)
	\Gamma_{s} \gamma_{s}
	\mathcal{O}_{s}.
\end{align}

Likewise, each of the four diagrams in subcategory $\mathfrak{D}\mathsf{9(b)}$ 
gives the same amplitude.
Each of these graphs involves two undetermined frequency loops.
We evaluate $\mathfrak{D}\mathsf{9(b)_{i}}$ explicitly, because the structure 
of the two frequency loop integration is quite different from that of Eq.~(\ref{I4S}), 
studied in the previous subsection. Using the Feynman rules, we write
\begin{multline}\label{D9(b)i}
	\mathfrak{D}\mathsf{9(b)_{i}} 
	=\frac{2(i \xi^{a})^{3}}{2!}
	\left(
	\frac{-\lambda_{A} \Gamma_{s}^{2}}{\lambda^{2}}
	\right)
	\delta_{1+3,2+4} 
	\\
	\begin{aligned}[b]
	\times
	\int {d^{d} \bm{\mathrm{r}}}
	&\left[ 
	(s_{1}\delta Q_{\mathsf{S}\,1,2}^{a, a}+s_{2}\delta Q_{\mathsf{S}\,1,2}^{\dagger \, a, a})
	\right.\\
	&\left.
	\times(s_{3}\delta Q_{\mathsf{S}\,3,4}^{a, a}+s_{4}\delta Q_{\mathsf{S}\,3,4}^{\dagger \, a, a})
	\right]
	I_{6}.
	\end{aligned}
\end{multline}
In this equation, note that both of the two $\mathsf{V(e)}$ vertices contribute only 
the portions of the associated vertex amplitudes [Table \ref{VertexTable}] proportional 
to $\Gamma_{s}$; all other terms vanish. The integral $I_{6}$ separates into parts
\begin{equation}\label{I6}
	I_{6} \equiv I_{6\,\mathsf{S}} + I_{6\,\mathsf{C}},  
\end{equation}
corresponding to interacting propagator components $\mathsf{P_{S}}$ and $\mathsf{P_{C}}$, 
respectively. From the figure, and given the structure of 
$\mathsf{P_{S}}$, Eq.~(\ref{SProp}), it can be seen that $I_{6\,\mathsf{S}}=0$. On the 
other hand, $I_{6\,\mathsf{C}}$ may be written
\begin{align}\label{I6C}
	I_{6\,\mathsf{C}} =& 
	\int  \frac{d\omega_{\alpha} \, d\omega_{\beta} \, d^{\textrm{2}}\bm{\mathrm{k}}}{(2 \pi)^{\textrm{4}}}
	\Gamma_{c}
	(2 s_{\alpha})^{2} (2 s_{\beta})^{2}
	\bm{\mathrm{k}}^{2}
	\nonumber\\
	&\times
	\Big[\Delta_{\mathcal{O}}^{a,a}(|\omega_{\alpha}|,|\omega_{\alpha}|,\bm{\mathrm{k}})
	\Delta_{\mathcal{O}}^{a,a}(|\omega_{\beta}|,|\omega_{\beta}|,\bm{\mathrm{k}})\Big]^{2}.
\end{align}
Distinct from the two frequency loop integration previously evaluated, Eq.~(\ref{I4S}), 
the propagator kernels $\Delta_{\mathcal{O}}$ in 
Eq.~(\ref{I6C}) depend upon \emph{either} of the two loop frequencies $\omega_{\alpha}$ 
or $\omega_{\beta}$. Eq.~(\ref{I6C}) should be understood as an integration over the full 
frequency-momentum shell, Eq.~(\ref{FMShell}), pictured in Fig.~\ref{FMShellFig}, in the 
space 
$(|\omega_{\alpha}|,|\omega_{\beta}|,D \bm{\mathrm{k}}^{2})$. It may be evaluated using 
Eq.~(\ref{J9}):
\begin{equation}\label{I6C--2}
	I_{6\,\mathsf{C}}=
	\frac{2^{6} \, \Gamma_{c}}{h^{4}}
	J_{9}(2\xi^{a}) 
	=\frac{-\lambda^{2} 2 dl}{\pi (\pi h)^{2}}\Gamma_{c}.
\end{equation}
Summing identical contributions from all four diagrams in $\mathfrak{D}\mathsf{9(b)}$, we obtain
\begin{align}\label{D9(b)--F}
	\mathfrak{D}\mathsf{9(b)} =&
	\left(\frac{-\lambda_{A} \, 2 dl}{8\pi}\right)
	\left(2 \Gamma_{c} \gamma_{s}^{2}\right)
	\mathcal{O}_{s}.
\end{align}

The eight category $\mathfrak{D}\mathsf{10}$ diagrams depicted in Fig.~\ref{D10Fig} pair 
$\mathsf{V(e)}$ with the other ``triangular'' interaction vertices, 
$\mathsf{V(c)}$ and $\mathsf{V(d)}$. The evaluation of these graphs proceeds as in 
Sec.~\ref{IntRenorm1}; we give only the result for the entire category:
\begin{align}\label{D10--F}
	\mathfrak{D}\mathsf{10} =&
	\left(\frac{-2 dl}{8\pi}\right)
	\Gamma_{s}
	\Big\{
	\lambda
	\left[
	2 \gamma_{c} - 2 \ln(1 - \gamma_{s})
	\right]
	-4 \lambda_{A}
	\gamma_{c}
	\Big\}
	\mathcal{O}_{s}.
\end{align}

Categories $\mathfrak{D}\mathsf{11}$ and $\mathfrak{D}\mathsf{12}$ appear in 
Figs.~\ref{D11Fig} and \ref{D12Fig}, respectively. The diagrams in these categories 
share the same structural elements as those in $\mathfrak{D}\mathsf{10}$, but there 
is no net contribution to the RG from either 
$\mathfrak{D}\mathsf{11}$ or $\mathfrak{D}\mathsf{12}$. The graphs in Fig.~\ref{D11Fig} 
cancel pairwise: 
\begin{align}
	\mathfrak{D}\mathsf{11(a)_{i}} + \mathfrak{D}\mathsf{11(a)_{ii}}
	&=\mathfrak{D}\mathsf{11(a)_{iii}} + \mathfrak{D}\mathsf{11(a)_{iv}}
	\nonumber\\
	&=\mathfrak{D}\mathsf{11(b)_{i}} + \mathfrak{D}\mathsf{11(b)_{ii}}
	\nonumber\\
	&=\mathfrak{D}\mathsf{11(b)_{iii}} + \mathfrak{D}\mathsf{11(b)_{iv}}
	\nonumber\\
	&=0.
\end{align}
The individual diagrams in category $\mathfrak{D}\mathsf{12}$ correspond to 
the generation of new operators, not present in the original FNL$\sigma$M action 
[Eqs.~(\ref{SDQmatrixSPLeft}) and (\ref{SIQmatrixSPLeft})]. Fortunately, their 
sum gives zero 
[up to terms $\bm{\mathit{O}}(\delta\hat{Q}^{3}_{\mathsf{S}})$, whose cancelation 
we have not checked in detail].

Finally, we evaluate the single diagram in $\mathfrak{D}\mathsf{13}$, Fig.~\ref{D13Fig}. 
In this figure, two copies of the vertex $\mathsf{V(f)}$ are connected by a single, 
basic diffuson propagator $\mathsf{P_{\lambda}}$. $\mathfrak{D}\mathsf{13}$ is unique 
among the graphs presented in this paper, in that it involves a pure frequency loop 
integration, since the momentum carried by $\mathsf{P_{\lambda}}$ between the slow mode 
fields at the vertices is necessarily slow. We will see that $\mathfrak{D}\mathsf{13}$ 
drives the CDW instability in the \emph{clean} limit, $\lambda,\lambda_{A} \rightarrow 0$.
Such a contribution appears naturally in the frequency-momentum shell RG, but not in a 
pure momentum shell scheme.\cite{Finkelstein,KotliarSorella,BK,KBCooperon}
Using the Feynman rules, we write
\begin{multline}\label{D13}
	\mathfrak{D}\mathsf{13} 
	=\frac{(2 \xi^{a})^{2}}{2!}
	\delta_{1+3,2+4} \Gamma_{c}^{2} 
	\\
	\begin{aligned}[b]
	\times
	\int {d^{d} \bm{\mathrm{r}}}
	&\left[ 
	(s_{1}\delta Q_{\mathsf{S}\,1,2}^{a, a}-s_{2}\delta Q_{\mathsf{S}\,1,2}^{\dagger \, a, a})
	\right.\\
	&\left.
	\times(s_{3}\delta Q_{\mathsf{S}\,3,4}^{a, a}-s_{4}\delta Q_{\mathsf{S}\,3,4}^{\dagger \, a, a})
	\right]
	I_{7}.
	\end{aligned}
\end{multline}
In writing the amplitude for $\mathfrak{D}\mathsf{13}$, Eq.~(\ref{D13}), we have exactly 
the \emph{opposite} of the situation we had with 
$\mathfrak{D}\mathsf{9(b)i}$, Eq.~(\ref{D9(b)i}): 
both of the two $\mathsf{V(f)}$ vertices in Eq.~(\ref{D13}) contribute only the portions 
of the associated vertex amplitudes [Table \ref{VertexTable}] proportional to $\Gamma_{c}$; 
all other terms vanish. The pure frequency integral $I_{7}$ may be evaluated as
\begin{align}\label{I7}
	I_{7} =&
	\int  \frac{d\omega_{\alpha}}{2 \pi}
	(2 s_{\alpha})^{2}
	\Delta_{\mathcal{O}}^{a,a}(|\omega_{\alpha}|,|\omega_{\alpha}|,\bm{\mathrm{q}})
	\nonumber\\
	=& \frac{4}{\pi h}\int_{\widetilde{\Lambda}}^{\Lambda} d\omega_{\alpha} 
	\left[D \bm{\mathrm{q}}^{2} - 2 i \xi^{a} \omega_{\alpha}\right]^{-1}
	\nonumber\\
	\approx& \frac{4 i \xi^{a}}{2 \pi h}2 dl. 
\end{align}
Combining Eqs.~(\ref{D13}) and (\ref{I7}), we obtain the result
\begin{equation}\label{D13--F}
	\mathfrak{D}\mathsf{13} =
	\Gamma_{c} \gamma_{c} \,2 dl
	\,\mathcal{O}_{c}.
\end{equation}
This expression is second order in $\Gamma_{c}$, but \emph{zeroth} order in the disorder strength $\lambda$.


\subsection{Dimensional analysis}\label{DimAnalysis}

We obtain the flow equations for the FNL$\sigma$M parameters in the usual way, by 
re-exponentiating the diagrammatic corrections derived in the previous subsection 
and subtracting them from the pure slow mode action $S_{\mathsf{S}}$, 
Eq.~(\ref{SSlow}), and by rescaling position 
$\bm{\mathrm{r}}$ and time $t$ via 
\begin{subequations}\label{RGRescale}
\begin{align}
	\bm{\mathrm{r}} &\rightarrow b \bm{\mathrm{r}}, \\
	t &\rightarrow b^{z} t,
\end{align}
\end{subequations}
where $b \approx 1 + dl$ is the change in length scale, and $z$ is the so far 
undetermined dynamic critical exponent. 
Equivalently, Eq.~(\ref{RGRescale}) implies that momentum $\bm{\mathrm{k}}$ 
carries the engineering dimension
\begin{equation}\label{MomentumDim}
	\left[\bm{\mathrm{k}}\right] = 1,
\end{equation}
while frequency $\omega$ carries the (possibly scale dependent) dimension
\begin{equation}\label{EnergyDim}
	\left[\omega\right] = z,
\end{equation}
both stated in inverse length units. 
Similar analysis of the FNL$\sigma$M action, Eqs.~(\ref{SD}) and (\ref{SI}) 
[or (\ref{SDQmatrixSPLeft}) and (\ref{SIQmatrixSPLeft})], gives the 
engineering dimensions of the field matrix elements
\begin{equation}\label{QmatrixFreqTimeDim}
	[Q^{a, a'}_{\omega, \omega'}({\bm{\mathrm{r}}})] = -z,
	\qquad
	[Q^{a, a'}_{t, t'}({\bm{\mathrm{r}}})] = z,
\end{equation}
and of the model coupling constants,
\begin{subequations}\label{ParameterDim}
\begin{equation}\label{ParameterDimDisorder}
	[\lambda] = [\lambda_{A}] = 2 - d,
\end{equation}
\begin{equation}
	[h] = [\Gamma_{s}] = [\Gamma_{c}] = d - z,
\end{equation}
and
\begin{equation}
	[\gamma_{s}] = [\gamma_{c}] = 0,
\end{equation}
\end{subequations}
where $\Gamma_{s}$ and $\Gamma_{c}$ ($\gamma_{s}$ and $\gamma_{c}$) were 
defined in Eq.~(\ref{SingletCDWGamma}) [Eq.~(\ref{SingletCDW})].

Combining the above dimensional analysis with the diagrammatic results,  
Eqs.~(\ref{D1(a)--F}), (\ref{D2(b)i+D2(b)ii--F})--(\ref{D4(a)+D4(b)--F}), (\ref{D5+D6--F}), (\ref{D7--F}), 
(\ref{D8(a)--F})--(\ref{D9(a)--F}), (\ref{D9(b)--F}), (\ref{D10--F}), and (\ref{D13--F}), 
we obtain the one loop RG equations
\begin{subequations}\label{FlowEqs1}
\begin{align}
	\frac{d \lambda}{d l} = & -\epsilon \, \lambda - \frac{\lambda^{2}}{4\pi}\gamma_{c}
	+ \frac{2 \lambda^{2} }{4\pi}
	\left[
	1 + \frac{1-\gamma_{s}}{\gamma_{s}} \ln(1-\gamma_{s})
	\right] \qquad	
	\nonumber\\
	& + \bm{\mathit{O}}(\lambda^{2} \gamma_{c}^{2}, \lambda^{3}), 
		\label{lambdaFlow1}\\
	\frac{d \lambda_{A}}{d l} = & \epsilon \, \lambda_{A} + \frac{\lambda^{2}}{4\pi}
	+ 2 \frac{\lambda_{A}}{\lambda} \frac{d \lambda}{d l}
	+\bm{\mathit{O}}(\lambda^{3}), 
		\label{lambdaAFlow1}\\
	\frac{d \Gamma_{s}}{d l} = & (d - z)\Gamma_{s}
	+ \frac{1}{4\pi}
	\left(2 \lambda_{A} - \lambda\right)\left(\Gamma_{s} + \Gamma_{c}\right)
	\nonumber\\
	&
	+ \frac{1}{4\pi}
	\Big[
	2 \left(\lambda - \lambda_{A}\right)\left(2 - \gamma_{s}\right)\Gamma_{s} \gamma_{c} 
	- \lambda_{A} \Gamma_{s} \gamma_{s}
	\Big] 
	\nonumber\\
	& + \bm{\mathit{O}}(\lambda \Gamma_{c} \gamma_{c},\lambda_{A} \Gamma_{c} \gamma_{c},
		\lambda^{2} \Gamma_{s/c}, \lambda \lambda_{A} \Gamma_{s/c}), 
		\label{GammasFlow1}
\end{align}
\begin{align}
	\frac{d \Gamma_{c}}{d l} = & (d - z)\Gamma_{c}
	+ \frac{1}{4\pi}
	\left(2 \lambda_{A} - \lambda\right)\left(\Gamma_{s} + \Gamma_{c}\right)
	\nonumber\\
	&+ \frac{1}{4\pi} 
	\Big[2 \lambda \Gamma_{c} \ln(1 - \gamma_{s})\Big]
	- 2 \Gamma_{c} \gamma_{c}
	\nonumber\\
	& + \bm{\mathit{O}}(\lambda \Gamma_{c} \gamma_{c},\lambda_{A} \Gamma_{c} \gamma_{c},
		\lambda^{2} \Gamma_{s/c}, \lambda \lambda_{A} \Gamma_{s/c}), 
		\label{GammacFlow1}\\
	\frac{d \ln h}{d l} = & (d - z) +  \frac{1}{4\pi}\Big[\lambda_{A} + \lambda(\gamma_{c} - \gamma_{s})\Big]
	\nonumber\\
	& + \bm{\mathit{O}}(\lambda \gamma_{c}^{2}, \lambda^{2}, \lambda \lambda_{A}). \label{lnhFlow1}
\end{align}
\end{subequations}
In Eqs.~(\ref{lambdaFlow1}) and (\ref{lambdaAFlow1}), we have set $d = 2 + \epsilon$ 
explictly. The reader may be puzzled by the right-hand side of Eq.~(\ref{lambdaAFlow1}), 
which seems to imply the wrong engineering dimension ($\epsilon$) for $\lambda_{A}$. 
However, the full engineering dimension (-$\epsilon$) arises through the sum of the 
first term in this equation with the third, which is itself proportional to 
Eq.~(\ref{lambdaFlow1}). This third term in Eq.~(\ref{lambdaAFlow1}) originates from 
the definition of $\lambda_{A}$: in the non-interacting sector of the Keldysh
FNL$\sigma$M action, Eq.~(\ref{SD}), the local ``dimerization'' disorder\cite{footnote-n}
operator 
\begin{equation}\label{DimerDisorder}
	\left\lgroup
	\mathrm{Tr}
	\left[\hat{Q}^{\dagger}({\bm{\mathrm{r}}})
	\bm{\nabla}\hat{Q}({\bm{\mathrm{r}}})\right]
	\right\rgroup^{2}
\end{equation}
couples not to $\lambda_{A}$, but to the \emph{ratio} $\lambda_{A}/\lambda^{2}$. Application
of the chain rule to this ratio leads to the third term on the right-hand side of 
Eq.~(\ref{lambdaAFlow1}).

The flow equations (\ref{lambdaFlow1})--(\ref{lnhFlow1}) are given to the lowest non-trivial 
orders in both the ``dimensionless DC resistance'' parameter $\lambda$ and the CDW interaction 
strength $\Gamma_{c}$ (or $\gamma_{c}$). Specifically, we retain terms in 
Eq.~(\ref{lambdaFlow1})--(\ref{GammasFlow1}), and 
(\ref{lnhFlow1}) only to first order in $\Gamma_{c}$, i.e.\ terms 
$\bm{\mathit{O}}(\lambda^{2} \gamma_{c})$ [Eq.~(\ref{lambdaFlow1})] and terms  
$\bm{\mathit{O}}(\lambda \Gamma_{c}, \lambda_{A} \Gamma_{c})$ 
[Eqs.~(\ref{GammasFlow1}) and (\ref{lnhFlow1})]. In principle, since we 
make no assumption about the smallness of $\lambda_{A}$, we should retain 
terms $\bm{\mathit{O}}(\lambda_{A}^{n} \Gamma_{c})$, with 
$n \in \{1, 2, 3\ldots\}$, but only $n = 1$ appears in the RG equations.
The second order pure CDW self-interaction, $(- 2 \Gamma_{c} \gamma_{c})$ in 
Eq.~(\ref{GammacFlow1}), is the sole exception to the rule; this term is tied 
to the ballistic Fermi liquid physics in the presence of sublattice symmetry (and 
therefore nesting), and must be retained in the weakly disordered limit. We have 
checked by numerical integration that all of the results presented in Sec.~\ref{Results}, 
including the identification of a disorder-driven, interaction-mediated instability of the diffusive Fermi 
liquid phase in $d = 2 + \epsilon$ dimensions, are un-modified (within the 
perturbatively accessible coupling strength regime) by the inclusion of higher 
order terms in $\Gamma_{c}$. 

Using Eq.~(\ref{DOSCutoff}), one may also obtain a flow equation governing the scaling 
behavior of the disorder-averaged DOS $\nu(\omega)$. In the (Keldysh) sigma model 
formalism, Eqs.~(\ref{DOSDef}) and (\ref{QmatrixFreqTimeDim}) imply that the DOS carries 
zero engineering dimension, so that Eq.~(\ref{DOSCutoff}) implies the flow equation 
\begin{equation}\label{DOSScaling1}
	\frac{d\ln\nu}{d l} = - \frac{1}{4\pi}
	\left[\lambda_{A} + \lambda \ln(1-\gamma_{s}) + \lambda \gamma_{c}\right].
\end{equation}


\section{Results and Discussion}\label{Results}

In this section we summarize and interpret the results of the RG calculation, 
set up in Sec.~\ref{PandFR} and performed in Sec.~\ref{oneloop}, for the 
Finkel'stein NL$\sigma$M (FNL$\sigma$M) originally defined by Eqs.~(\ref{Z})--(\ref{SI}) 
in Sec.~\ref{NLSM}. The FNL$\sigma$M action given by Eqs.~(\ref{SD}) 
and (\ref{SI}) is parameterized by five coupling constants: $\lambda$, proportional 
to the dimensionless DC resistance of the system; $\lambda_{A}$, which gives a second measure
of the hopping disorder (associated with quenched orientational fluctuations of bond 
strength dimerization, as discussed in Secs.~\ref{SPGradient} and \ref{NLSMSummary});
$h$, which tracks the relative scaling of length and time in the 
theory; $\Gamma_{U}$ 
and $\Gamma_{V}$, which 
characterize the strengths of generic short-ranged same-sublattice and 
intersublattice interparticle interactions, respectively,
in the coarse-grained FNL$\sigma$M description.
Alternatively, we have introduced 
the interaction parameters $\Gamma_{s}$ and $\Gamma_{c}$
[the sum and difference of $\Gamma_{U}$ and $\Gamma_{V}$, see Eq.~(\ref{SingletCDWGamma})],
which couple to
the square of
the smooth and of the sublattice staggered local charge densities,
respectively.
As discussed below Eq.~(\ref{SingletCDWGamma}), a 
staggered
interaction $\Gamma_{c} < 0$ 
is expected to
promote the CDW instability in the clean 
limit [i.e.\ we identify $\Gamma_{c} \sim W_{c}$, with $W_{c}$ defined by 
Eq.~(\ref{CDWcoupling}) for the Hubbard-like model given by Eq.~(\ref{Hclean})]. 
The RG calculation has been performed with the aid of an 
epsilon expansion in
$d = 2 + \epsilon$ spatial dimensions, with 
$0 \leq \epsilon \ll 1$.  

The one-loop
RG flow equations for the coupling strengths $\lambda$, $\lambda_{A}$, $\Gamma_{s}$, 
$\Gamma_{c}$, and $h$ were obtained at the end of the previous section, 
in Eqs.~(\ref{lambdaAFlow1})--(\ref{GammacFlow1}).
We restate them below in a slightly more streamlined form. We will need 
the ``relative interaction parameters'' 
$\gamma_{s}$ and $\gamma_{c}$, defined in Eq.~(\ref{SingletCDW}), 
repeated here for convenience:
\begin{subequations}\label{SingletCDWR}
\begin{align}
	\gamma_{s} \equiv & \frac{4}{\pi h} \Gamma_{s} = \frac{2}{\pi h}(\Gamma_{U} + \Gamma_{V}), \\
	\gamma_{c} \equiv & \frac{4}{\pi h} \Gamma_{c} = \frac{2}{\pi h}(\Gamma_{U} - \Gamma_{V}).
\end{align}
\end{subequations}
Next, we perform a trivial rescaling of the 
coupling constants
appearing in Eqs.~(\ref{SD}) 
and (\ref{FlowEqs1}),
\begin{equation}\label{Disorder4pi}
	\lambda \rightarrow 4 \pi \lambda,
	\qquad
	\lambda_{A} \rightarrow 4 \pi \lambda_{A},
\end{equation}
after which
the
one-loop RG flow equations for the couplings 
$\lambda$, $\lambda_{A}$, $\gamma_{s}$, $\gamma_{c}$, and $h$
take the following form
in $d = 2 + \epsilon$ dimensions
\begin{subequations}\label{FlowEqs2}
\begin{eqnarray}
	\frac{d \lambda}{d l} & = & -\epsilon \, \lambda - \lambda^{2}\gamma_{c} 
	\nonumber\\
	& & + 2 \lambda^{2} 
	\left[
	1 + \frac{1-\gamma_{s}}{\gamma_{s}} \ln(1-\gamma_{s})
	\right], \qquad\quad
		\label{lambdaFlow2}\\
	\frac{d \lambda_{A}}{d l} & = & \epsilon \, \lambda_{A} + \lambda^{2} 
	+ 2 \frac{\lambda_{A}}{\lambda} \frac{d \lambda}{d l}, 
		\label{lambdaAFlow2}\\
	\frac{d \gamma_{s}}{d l} & = & \lambda_{A} (1-\gamma_{s})
	\left(\gamma_{s} + 2 \gamma_{c} - 2 \gamma_{s} \gamma_{c} \right) 
	\nonumber \\
	& & - \lambda (1-\gamma_{s}) \left( \gamma_{s} + \gamma_{c} 
	- 2 \gamma_{s} \gamma_{c} \right),
		\label{gammasFlow2}\\
	\frac{d \gamma_{c}}{d l} & = & \lambda_{A} \left(\gamma_{c} 
	+ 2 \gamma_{s}\right) - \lambda \left(\gamma_{s} 
	+ \gamma_{c}\right) \nonumber\\
	& & + \lambda \left[ 2 \gamma_{c} \ln(1 - \gamma_{s}) + \gamma_{s} 
	\gamma_{c}\right] - 2 \gamma_{c}^{2}, 
		\label{gammacFlow2}\\
	\frac{d \ln h}{d l} & = & (d - z) +  \lambda_{A} + \lambda(\gamma_{c} 
	- \gamma_{s}), \label{lnhFlow2}
\end{eqnarray}
\end{subequations}
where $l$ is the logarithm of the spatial length scale. The parameter 
$z$ in Eq.~(\ref{lnhFlow2}) is the (as yet undetermined, possibly scale-dependent) 
``dynamic critical exponent.''\cite{footnote-o}
We have also calculated the 
scale dependence
of the disorder-averaged, single particle
density of states (DOS) $\nu(\omega)$. Implementing the rescaling of 
Eq.~(\ref{Disorder4pi}) in Eq.~(\ref{DOSScaling1}), the 
one-loop
flow equation for $\nu$ is
\begin{equation}\label{DOSScaling2}
	\frac{d\ln\nu}{d l} =
	-\left[\lambda_{A} + \lambda \ln(1-\gamma_{s}) + \lambda \gamma_{c}\right].
\end{equation}

Flow equations (\ref{lambdaFlow2})--(\ref{lnhFlow2}) and (\ref{DOSScaling2}) are given 
to the lowest non-trivial order in the parameters $\lambda$ and $\gamma_{c}$, but contain 
contributions from $\lambda_{A}$ and $\gamma_{s}$ to \emph{all} orders.
This is an advantage of the Finkel'stein NL$\sigma$M formulation, which provides\cite{BK} a 
loop expansion that is controlled perturbatively by the (small) dimensionless 
resistance $\lambda$, but which does not require the coupling strengths $\lambda_{A}$ or 
$\gamma_{s}$ to be small. 
We will
discuss
the physics that emerges from an analysis 
of our results [Eq.~(\ref{FlowEqs2})],
in turn 
for  $d = 2$ and $d > 2$
dimensions.

\subsection{Structure of the one-loop flow equations}\label{StructureDiscuss}

Before turning to such an analysis, we pause to consider the structure of 
Eqs.~(\ref{lambdaFlow2})--(\ref{lnhFlow2}) and (\ref{DOSScaling2}).
To gain an understanding of the competing mechanisms driving the RG flow, we must attempt 
to isolate the effects of the various terms. Fortunately, the most important 
structures in these equations either occur generically in the perturbative 
description of interacting, 
diffusive Fermi liquids,\cite{Finkelstein,BK,AltshulerAronov,NPCS,Aleiner} 
or can
be tied specifically to the special sublattice symmetry (SLS) [Eq.~(\ref{SLS})] 
and its effects upon the well-understood limits of (i) disorder with vanishing interparticle 
interactions,\cite{Gade,FC,GLL} and (ii) interactions with vanishing disorder.\cite{GSST,Shankar} 
Interestingly, as we
will see below,
the \emph{competition} between the disorder and 
interaction effects gives 
nevertheless
rise to completely new physics.

\subsubsection{Non-interacting limit}\label{NonIntDiscuss}

We consider first the non-interacting limit.
In $d = 2$ spatial dimensions, the FNL$\sigma$M [Eqs.~(\ref{Z})--(\ref{SI})] 
with $\Gamma_{U} = \Gamma_{V} = 0$ ($\gamma_{s} = \gamma_{c} = 0$) 
and non-zero disorder 
couplings
$\lambda$ and $\lambda_{A}$ describes an 
unusual
low-dimensional delocalized 
state,\cite{Gade,FC,GLL}
analogous in many ways to the type of 
critical state\cite{LeeRamakrishnan,Wegner,Mirlin} that occurs at a continuous, 
\emph{three-dimensional} (``Anderson'') metal insulator transition (MIT) of 
non-interacting diffusive electrons, in the \emph{absence} of SLS. At such an 
ordinary Anderson transition, the single particle wave functions at the Fermi 
energy are extended, and thus capable of transporting 
charge, heat,
etc.\ across 
the system, but these states are also very far from the plane waves of a clean 
Fermi gas, 
and exhibit
so-called ``multifractal'' scaling\cite{Wegner,Mirlin}
(due to the presence of the disorder). 
In $d = 2$, the extended single particle wave functions reside only 
at the band center (energy $\omega = 0$)
for arbitrarily weak disorder 
[i.e.\ 
wave functions 
\emph{are} exponentially
localized, with a localization length that diverges 
upon approaching the band center ($\omega \rightarrow 0$)],\cite{Gade,FC} 
so that the critical state appears only at 
half filling [a necessary but not sufficient condition for SLS, Eq.~(\ref{SLS})].  
The critical state at the band center ($\omega=0$) described above
characterizes the $2\textrm{D}$ non-interacting random hopping model 
defined by the Hamiltonian in Eqs.~(\ref{Hclean}) and (\ref{Hdis}), with $U=V=0$.\cite{Gade,FC,GLL} 
This delocalized state turns into a somewhat more conventional metallic one
in $d > 2$, but it is still distinguished by SLS (and broken TRI). 
[See also the discussion in the Introduction, Sec.~\ref{RandomHopping},
especially
Figs.~\ref{FigRandomHopPhys2D} and \ref{FigRandomHopPhys3D}.]

These conclusions follow from the fact that a variety of aspects of the class AIII
non-interacting sigma model, argued in Sec.~\ref{NLSM} and in Refs.~\onlinecite{Gade,FC,GLL} 
to capture the low-energy physics of the sublattice symmetric random hopping model 
lacking TRI, can be solved\cite{GLL} exactly in $d = 2$ using conformal field theory techniques.
In particular, in the absence of interactions, it is possible\cite{GLL}
to obtain the \emph{exact} renormalization group equations (to all orders in $\lambda$ and $\lambda_{A}$)
in $2\textrm{D}$ for the disorder-only sector sigma model parameters $\lambda$ and $\lambda_{A}$, and for the 
average density of states $\nu$, in a particular RG scheme. These are of the following form\cite{footnoteRGEquationsIn2D}
	\begin{subequations}\label{FlowEqsNI}
	\begin{eqnarray}
		\frac{d \lambda}{d l} & = & 0,
		\label{lambdaFlowNI}\\
		\frac{d \lambda_{A}}{d l} & = & f_{1}(\lambda)+ 2 \frac{\lambda_{A}}{\lambda} \frac{d \lambda}{d l},
		\nonumber\\
		& = & f_{1}(\lambda)
		\label{lambdaAFlowNI}\\
		\frac{d \ln \nu}{d l} & = & - \lambda_{A} f_{2}(\lambda) - f_{3}(\lambda),	
		\label{DOSScalingNI}
	\end{eqnarray}
	\end{subequations}
with $f_{i}(\lambda)$, $i \in \{1,2,3\}$ 
real analytic
(RG scheme dependent)
functions of $\lambda$. 
The second term on the first line of the right-hand side of Eq.~(\ref{lambdaAFlowNI}), which 
vanishes here via Eq.~(\ref{lambdaFlowNI}), arises through the chain rule applied
to the definition of $\lambda_{A}$ in Eq.~(\ref{SD}), as discussed below
Eq.~(\ref{FlowEqs1}) in Sec.~\ref{DimAnalysis}. Using the results of our perturbative analysis 
[Eqs.~(\ref{lambdaFlow2}), (\ref{lambdaAFlow2}), and (\ref{DOSScaling2})] and setting 
$\epsilon = 0$, we find agreement with Eq.~(\ref{FlowEqsNI}), obtaining the lowest 
order terms in the expansions
	\begin{subequations}\label{fiExpansions}
	\begin{eqnarray}
		f_{1}(\lambda) & = & \lambda^{2} + {\bm{\mathit{O}}}(\lambda^{3}),  
		\label{f1}\\
		f_{2}(\lambda) & = & 1 + {\bm{\mathit{O}}}(\lambda),  
		\label{f2}\\
		f_{3}(\lambda) & = & 0 + {\bm{\mathit{O}}}(\lambda^{2}).  
		\label{f3}
	\end{eqnarray}
	\end{subequations}
[The term in square brackets on the second line of Eq.~(\ref{lambdaFlow2}) 
vanishes when $\gamma_{s} \rightarrow 0$].
The
lowest order (one-loop)
results in Eqs.~(\ref{FlowEqsNI}) and (\ref{fiExpansions}) 
are universal and
were originally obtained via perturbative methods by Gade and Wegner in 
Ref.~\onlinecite{Gade}. 
Eq.~(\ref{lambdaFlowNI}) states 
that the ``dimensionless DC resistance'' $\lambda$ is exactly marginal in 
$d = 2$; equivalently, the dimensionless conductance $g$ scales 
classically, since all quantum corrections to it vanish.
The
exact marginality 
of $\lambda$ demonstrates that Anderson localization effects 
are `disabled' by 
SLS at the band center in the $2\textrm{D}$ random hopping model. 
We can extend these results for the one-loop RG equations  of
the NL$\sigma$M to $d = 2 + \epsilon$ dimensions, by simply adding the appropriate 
``engineering'' dimension terms, i.e.\ $-\epsilon \lambda$, to Eq.~(\ref{lambdaFlowNI}), 
and $- \epsilon \lambda_{A}$ to the second line of Eq.~(\ref{lambdaAFlowNI}).\cite{footnote-p}
[See Eq.~(\ref{ParameterDimDisorder}).] 
Moreover, this procedure yields\cite{footnoteDimensionalRegularization,ZinnJustin} 
the exact result for the RG equation for $\lambda$, which [from Eq.(\ref{lambdaFlowNI})]
reads in $d = 2 + \epsilon$ dimensions:
$ {d \lambda}/{d l}  =  -\epsilon \lambda$.
We conclude from this field theory treatment of the system that,
in the absence of interactions and in dimensions $d \geq 2$, 
the 
delocalized phase,
characterized by a finite conductivity,
persists even for `strong' disorder; in particular,
even the strongly disordered system exhibits no transition into
a localized phase (as already mentioned in the
Introduction).

The behavior of the system in $d = 2$ 
is special in that
the ``dimerization'' disorder\cite{footnote-q}
parameter $\lambda_{A}$ is (logarithmically) driven to strong coupling at a rate determined by 
$f_{1}(\lambda)$ [Eq.~(\ref{lambdaAFlowNI})]. The running $\lambda_{A}$ feeds into the average 
density of states $\nu(\omega)$, as shown in Eq.~(\ref{DOSScalingNI}). This equation in 
turn implies the divergence of the low-energy DOS upon approaching the band center, taken to 
reside at $\omega = 0$. Specifically, one finds that in $2\textrm{D}$,\cite{Gade,FC,GLL}
\begin{equation}\label{DOSNI}
	\nu(\omega) \sim \frac{1}{\omega} \exp\left(-c \, {\left| \ln \omega \right|}^{\alpha} \right),
\end{equation} 
in the limit as $\omega \rightarrow 0$, where $c = c(\lambda)$ is a scale 
independent constant and the exponent $\alpha$ is $1/2$.\cite{footnote-r}
Note that the DOS divergence in Eq.~(\ref{DOSNI}) that occurs in the $2\textrm{D}$ 
random hopping model has nothing to do with Fermi surface van Hove singularities,
which may appear only in the clean limit,\cite{footnote-s}
e.g.\ in the case of pure nearest neighbor hopping on the square lattice, 
where such a singularity at half filling gives a weaker, logarithmic divergence. 
Returning to the random hopping model, in $d>2$ the low energy DOS 
is finite, but parametrically enhanced at the band center. By contrast, the 
density of states in a non-interacting ordinary diffusive metal 
(lacking SLS, and being in one of the three Wigner-Dyson classes)
is typically not renormalized by the disorder.\cite{LeeRamakrishnan} 

Summarizing, we have the following picture of the non-interacting random 
hopping model, which we have argued
to be described
by the NL$\sigma$M 
in Eqs.~(\ref{Z})--(\ref{SI}), with $\Gamma_{U} = \Gamma_{V} = 0$: a delocalized 
phase exists (at the band center, i.e.\ for half filling)
in $d = 2$ and
(as always, and trivially)
in $d > 2$; in
neither case 
does a 
transition
into an Anderson insulating 
state occur as the disorder strength is increased from zero. The single 
particle DOS diverges strongly upon approaching the band center in $d = 2$, 
while it is parametrically enhanced in $d > 2$.
[See Figs.~\ref{FigRandomHopPhys2D} and \ref{FigRandomHopPhys3D}
in Sec.~\ref{RandomHopping}.]

\subsubsection{Interparticle interactions: diffusive Fermi liquid and clean Hubbard-like model physics}\label{IntDiscuss}

We turn now to the interpretation of various pieces of Eq.~(\ref{FlowEqs2}) 
involving the interparticle interactions.
We consider first the term in square brackets on the second line of 
Eq.~(\ref{lambdaFlow2}). This term can be recognized as the usual perturbative 
correction to the inverse conductance in a diffusive Fermi liquid, in the 
presence of \emph{short-ranged} interparticle interactions,\cite{Finkelstein,AltshulerAronov,NPCS} 
and may be interpreted\cite{Aleiner} as coherent backscattering of carriers off of 
disorder-induced Friedel oscillations in the background electronic charge 
density. Background density fluctuations become a source of 
\emph{on-site} disorder in the presence of electron-electron interactions, 
so that we may attribute this non-trivial correction to ``dynamic SLS breaking.'' 

Next we note the non-trivial zero of the 
one-loop
RG Eq.~(\ref{gammasFlow2}) at $\gamma_{s} = 1$.
[A factor of $(1-\gamma_{s})$ is expected in all orders.]
This zero follows from the established representation of the thermodynamic 
compressibility $\partial n/\partial \mu \equiv \kappa$ in terms of the 
Finkel'stein model parameters,\cite{BK}
\begin{equation}\label{kappaDef}
	\kappa \sim h(1 - \gamma_{s}) = h - \frac{4}{\pi} \Gamma_{s},
\end{equation}
valid in the diffusive Fermi liquid regime,
where we have used Eq.~(\ref{SingletCDWR}) and where we set $z = d$ 
[see Eq.~(\ref{lnhFlow2})]. Eq.~(\ref{kappaDef}) shows that the incompressible limit 
$\kappa \rightarrow 0$ is attained by sending $\gamma_{s} \rightarrow 1$ 
(for finite $h$). From the definition Eq.~(\ref{SingletCDWR}), we have
\begin{align}\label{gammasDotZero}
	\frac{d\gamma_{s}}{dl} =& -\frac{1}{h}\frac{d}{dl}\left(h - \frac{4}{\pi} \Gamma_{s}\right)
				+ \left(1 - \gamma_{s}\right)\frac{d \ln h}{dl}
	\nonumber\\
	& \sim -\frac{1}{h}\frac{d \kappa}{dl} + \left(1 - \gamma_{s}\right)\frac{d \ln h}{dl}.
\end{align}
In 
an interacting, disordered normal metal, 
$\kappa$ receives no divergent 
corrections,\cite{BK} so that the condition 
$({d\gamma_{s}}/{dl})(\gamma_{s}=1)=0$ is satisfied automatically. In the 
advent of sublattice symmetry, however, $\kappa$ does renormalize, so that 
the first term on the right-hand side of Eq.~(\ref{gammasDotZero}) is 
typically non-zero. This can be seen from the non-interacting limit, where 
$\kappa$ is equivalent to the single particle density of states $\nu$; as 
shown in Eq.~(\ref{DOSNI}) for the $2\textrm{D}$ case, $\nu(\omega)$ is 
strongly renormalized upon approaching the band center ($\omega \rightarrow 0$). 
Regardless, it is plausible to expect that in the limit $\gamma_{s} =1$ and 
$\kappa = 0$, the incompressibility of the diffusive Fermi liquid is preserved 
under the RG flow, so that ${d \kappa}/{dl}(\gamma_{s} = 1) = 0$, and therefore 
$({d\gamma_{s}}/{dl})(\gamma_{s}=1)=0$, as we have found in Eq.~(\ref{gammasFlow2}).

Finally, we note that the last term on the second line of Eq.~(\ref{gammacFlow2}) 
drives the CDW instability, which is a remnant of the \emph{clean} Hubbard-like 
model (recall that in our conventions $\gamma_{c} < 0$ signals this instability).

\subsection{Results (i): No metallic phase in $d = 2$}\label{Results2D}

At last we
analyze our results, considering first the $2\textrm{D}$ case. As 
discussed above in Sec.~\ref{NonIntDiscuss}, the non-interacting NL$\sigma$M 
describes in $d = 2$ an unusual critical, delocalized 
state of the random hopping model in Eqs.~(\ref{Hclean}) 
and (\ref{Hdis}) with $U = V = 0$. 
Other non-interacting models of disordered 
electrons are also known to
possess delocalized states in $d=2$, 
such as the spin-orbit (``symplectic'') normal metal class, 
where
e.g.,  a delocalized (metallic) phase occurs for sufficiently weak disorder.\cite{LeeRamakrishnan}
The latter delocalized phase
is however destabilized\cite{BK} by the introduction of 
(in this case: {\it long-ranged})
3D Coulomb interactions, yielding an insulator in the presence of both disorder and
interactions in $d = 2$. What is the analogous outcome 
in the presence of sublattice symmetry, i.e.\ for the Hubbard-like models
which are the focus of this work?
Are the short-ranged interparticle interactions characterized by the 
(relative) parameters $\gamma_{s}$ and $\gamma_{c}$ relevant or irrelevant 
perturbations to the critical, non-interacting 
system?
The answer may be obtained by specializing Eqs.~(\ref{lambdaFlow2})--(\ref{gammacFlow2}) 
to the vicinity of  $\gamma_{s} = \gamma_c = 0.$ The linearized flow equations 
are given by (\ref{lambdaFlowNI}), (\ref{lambdaAFlowNI}) with (\ref{f1}), and 
\begin{subequations}\label{FlowEqsNearOrigin}
	\begin{eqnarray}
	\frac{d \gamma_{U}}{d l} & \sim & \left(3 \lambda_{A} - 2 \lambda\right) \gamma_{U},  
	\label{gammaUFlow}\\
	\frac{d \lambda_{V}}{d l} & \sim & -\lambda_{A} \gamma_{V},
	\label{gammaVFlow}
	\end{eqnarray}
\end{subequations}
where we have introduced the relative same-sublattice (intersublattice) interaction strength 
$\gamma_{U}\equiv\gamma_{s}+\gamma_{c}$ ($\gamma_{V}\equiv\gamma_{s}-\gamma_{c}$) 
[c.f.\ Eqs.~(\ref{SingletCDWGamma}) and (\ref{SingletCDWR})].
(Note that while  this result is valid
only to first order in $\lambda$, no such restriction
is placed on $\lambda_A$.)
Eq.~(\ref{gammaUFlow}) 
shows that the \emph{same}-sublattice interaction strength $\gamma_{U}$ feeds back 
upon itself positively via the special sublattice class disorder
coupling
$\lambda_{A} \geq 0$. Since the latter is always driven to strong coupling 
[see Eqs.~(\ref{lambdaAFlowNI}) and (\ref{f1})] in $d = 2$ 
for any $\lambda >  0$, we see 
that same-sublattice interactions are rapidly enhanced as we renormalize the 
FNL$\sigma$M. At the very least, our result implies that the non-interacting 
description is unsuitable for describing the ground state of the $2\textrm{D}$ 
version of the full interacting, disordered Hubbard-like lattice model. 
This should be compared to an analogous result\cite{BDIpaper} previously obtained for a TRI, 
interacting random hopping model on the honeycomb lattice. In both models, the non-interacting 
phase is initially destabilized by the growth of short-ranged same-sublattice interparticle interactions, 
as in Eq.~(\ref{gammaUFlow}), and this growth is mediated by a special disorder 
coupling,
here $\lambda_{A}$, which occurs 
in general
in the description of
random
lattice models with an underlying SLS.
We note that Eqs.~(\ref{gammaUFlow}) and (\ref{gammaVFlow}) may alternatively be 
obtained from an analysis performed in the (much simpler) \emph{non}-interacting $2\textrm{D}$ 
(but disordered) theory
 (using 
e.g.\
replicas or supersymmetry), since these equations describe only 
the scaling of the interaction operators near the non-interacting 
fixed point
(i.e.\ are of linear order in the interaction strengths).

Turning to an analysis of the full flow equations (\ref{lambdaFlow2})--(\ref{lnhFlow2}),
we observe that Eq.~(\ref{lambdaAFlow2}) precludes the existence of \emph{any} perturbatively accessible, 
non-trivial fixed points occuring at non-zero resistance $\lambda >0$, 
for small $\epsilon \geq 0$
($d \geq 2$). Numerically integrating Eq.~(\ref{FlowEqs2}) for generic initial 
conditions in $d = 2$, we observe that either $\gamma_{c} \rightarrow - \infty$, 
signaling CDW formation, or that $\lambda,\, \lambda_{A} \rightarrow \infty$ and 
$\gamma_{c} \rightarrow + \infty$, indicating a 
flow toward 
\emph{simultaneously}
strong disorder
and strong interactions.
We demonstrate below that in $d>2$ dimensions, 
these two flow directions away from the non-interacting state evolve into two distinguishable 
instabilities of the diffusive Fermi liquid, which exists as a stable phase throughout
a region of finite volume in the four-dimensional $(\lambda,\lambda_{A},\gamma_{s},\gamma_{c})$
coupling constant space. By contrast, since we find no stable metallic region for $\epsilon = 0$,
we expect that the $2\textrm{D}$ disordered and interacting Hubbard-like model in 
Eqs.~(\ref{Hclean}) and (\ref{Hdis}) is always an insulator at zero temperature.  

We see that in $2\textrm{D}$, sublattice symmetry is both the genesis of
delocalized
(critical) 
behavior
in the absence of interactions, as well as the doom of such behavior in the presence 
of interactions. Moreover, in the Hubbard-like model studied in this work, SLS is also 
responsible for the Mott insulating charge density wave ground state in the \emph{clean} limit.
The physics that we have found is consistent with numerical studies 
\cite{ScalettarOD1--BDI,ScalettarOD2--BDI+C} of the half-filled spin-$1/2$ Hubbard model in 
$d=2$, which have shown that TRI random hopping disorder \emph{preserves} the charge 
compressibility gap of the clean Mott insulator, and that the disordered and interacting system 
shows no signs of metallic behavior.


\subsection{Results (ii):
		Fermi liquid instability in $d = 2 + \epsilon$}\label{Results>2D}

\begin{figure}[b]
\includegraphics[width=0.4\textwidth]{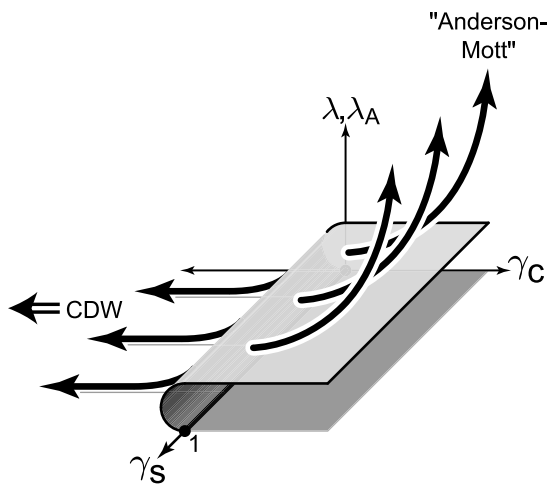}
\caption{Schematic phase diagram in $d = 2 + \epsilon$ dimensions, with $0 < \epsilon \ll 1$. 
$\lambda$ and $\lambda_A$ are both measures of the disorder, whereas $\gamma_s$ and $\gamma_c$ 
characterize the interaction strengths. The full phase diagram resides in the $4\textrm{D}$ 
$(\lambda,\lambda_{A},\gamma_{s},\gamma_{c})$ coupling constant space; 
this figure can be thought of as a 3D projection of this space, where the vertical
axis (perpendicular to the interaction axes) measures the total disorder strength
for some fixed ratio of $\lambda/\lambda_{A}$. The stable metallic phase resides 
between the ballistic plane ($\lambda=\lambda_{A}=0$) and the shaded sheath; 
the thick arrows indicate the two instabilities of the (metallic) diffusive Fermi liquid 
discussed in the text.
\label{PhaseDiag}}
\end{figure}

\begin{figure*}
\includegraphics[width=0.8\textwidth]{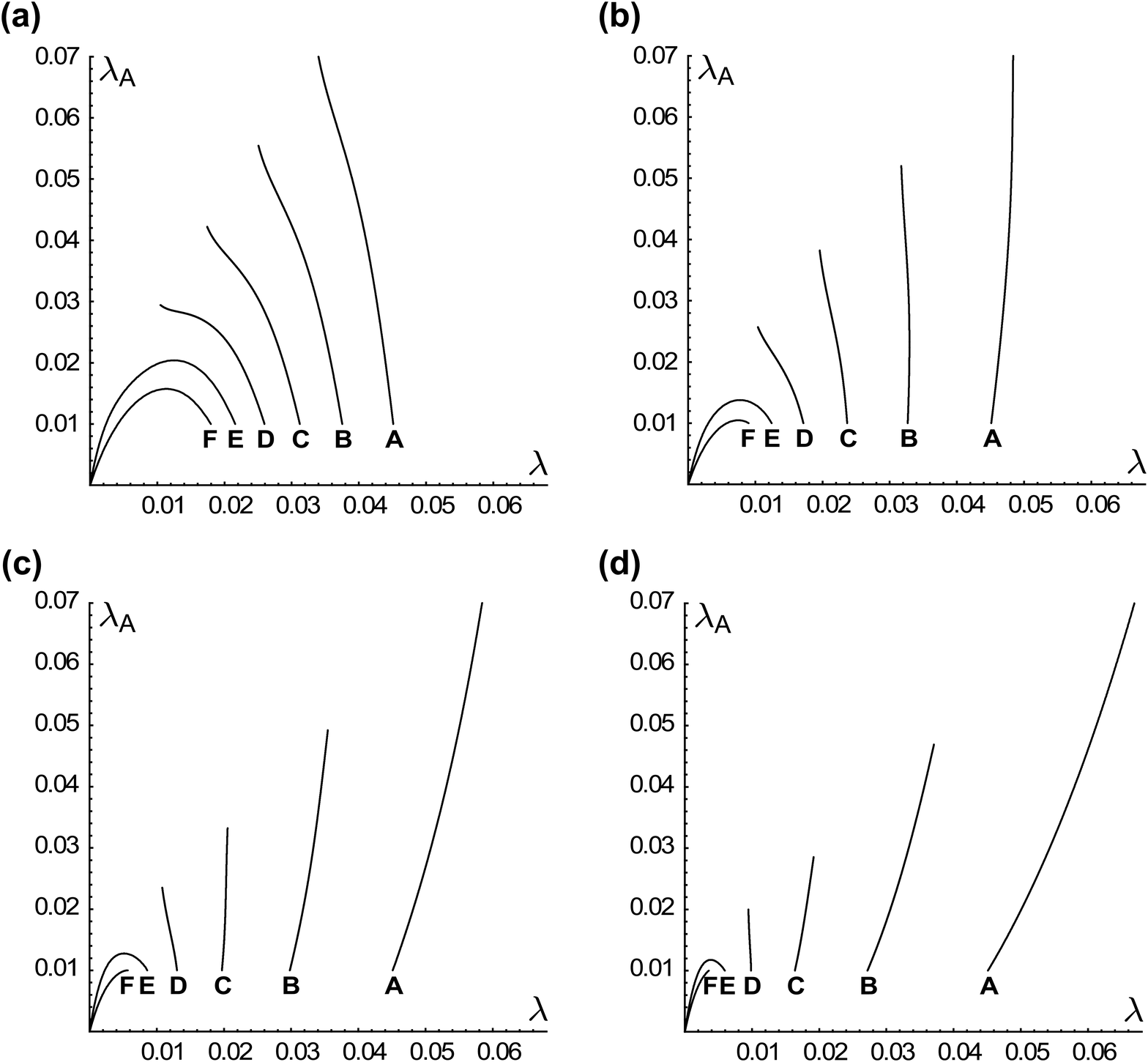}
\caption{
RG flow trajectories in $d = 2 + \epsilon$ dimensions, projected into
the disorder $(\lambda,\lambda_{A})$ plane. In this and Fig.~\ref{DISINSTIntF}, 
$\epsilon = 0.01$. 
Each subfigure {\bf(a)}--{\bf(d)} depicts six different initial $\lambda$ values, 
distinguished by the labels `{\bf\textsf{A}},'\ldots,`{\bf\textsf{F}}',
located at the {\it initial conditions} of the trajectories. 
The chosen values tune through the phase boundary enclosing the stable diffusive 
Fermi liquid state, demonstrating the disorder-driven instability
discussed in the text. In each of these subfigures, the trajectories
labeled `{\bf\textsf{E}}' and `{\bf\textsf{F}}' flow into the diffusive metal.
As for the other coupling strengths, all trajectories in this figure share 
the initial conditions (ICs) $\lambda_{A} = - \gamma_{c} = \epsilon = 0.01$; 
trajectories in subfigures {\bf(a)}, {\bf(b)}, {\bf(c)}, and {\bf(d)} 
share the IC $\gamma_{s}$ = $0.1$, $0.3$, $0.5$, and $0.7$, respectively.
The sequence {\bf(a)}--{\bf(d)} thus exhibits the $\gamma_{s}$-evolution
of the phase boundary.
\label{DISINSTDisF}}
\end{figure*}

\begin{figure*}
\includegraphics[width=0.8\textwidth]{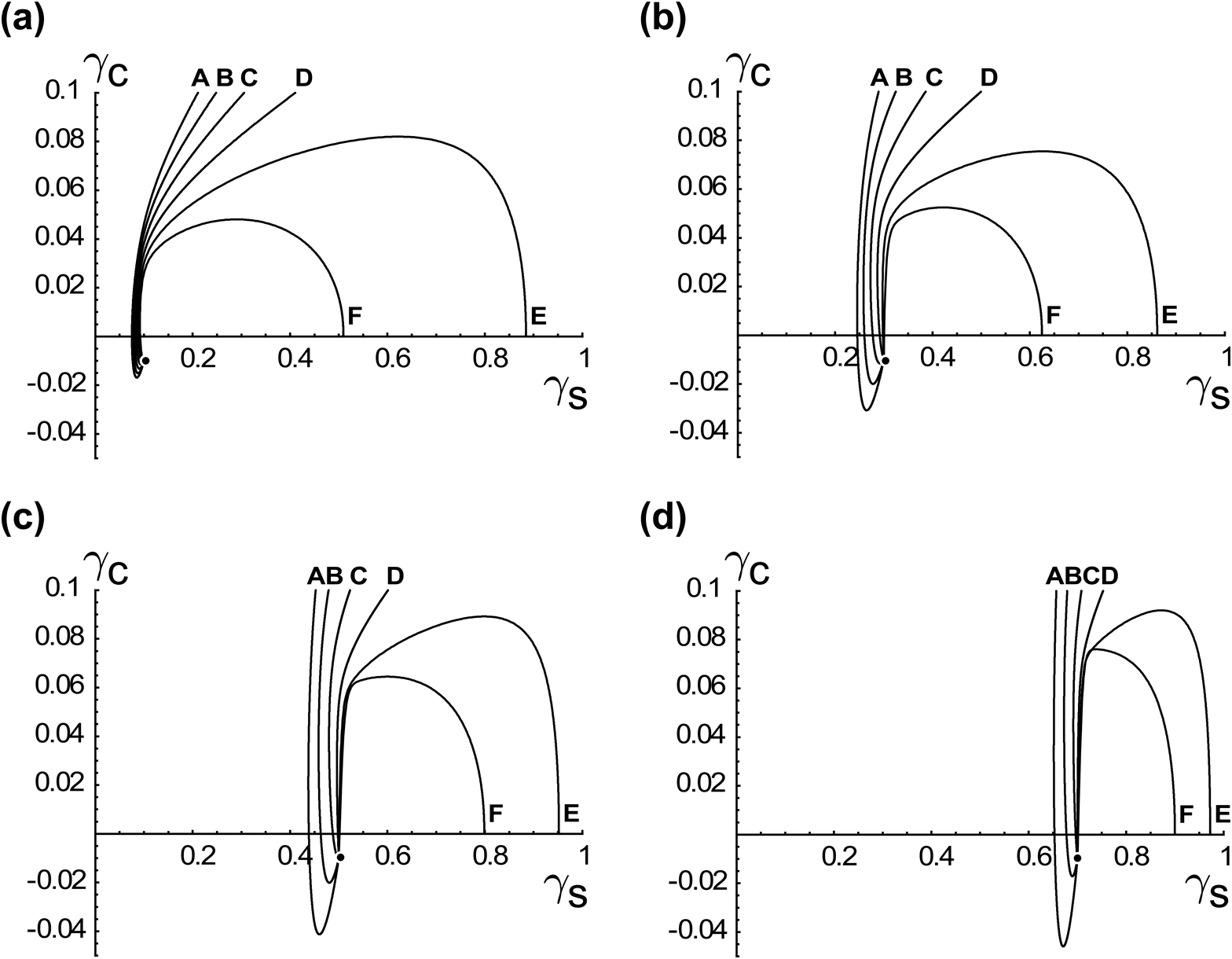}
\caption{
Same RG trajectories shown in Fig.~\ref{DISINSTDisF}, here projected
into the interaction $(\gamma_{s},\gamma_{c})$ plane. In each subfigure, the labels
`{\bf\textsf{A}},'\ldots,`{\bf\textsf{F}}' denote
[in contrast to Fig.~\ref{DISINSTDisF}]
the \emph{endpoints} of the trajectories
bearing the same labels in the corresponding subfigure of Fig.~\ref{DISINSTDisF}.
\emph{Initial} conditions (ICs) in the $(\gamma_{s},\gamma_{c})$ plane are shared 
by all trajectories in a given subfigure, indicated by the black dots.
ICs in the disorder $(\lambda,\lambda_{A})$ plane may be read off from the plots in
Fig.~\ref{DISINSTDisF}. In each of the subfigures {\bf(a)}--{\bf(d)}, the trajectories
labeled `{\bf\textsf{E}}' and `{\bf\textsf{F}}' terminate in the diffusive metal.
\label{DISINSTIntF}}
\end{figure*}

The situation in $d = (2 + \epsilon) > 2$ dimensions is more interesting.
Upon increasing $\epsilon$ from zero, a narrow, irregularly shaped sliver corresponding to a 
\emph{stable} metallic, diffusive Fermi liquid state opens up in the four-dimensional 
$(\lambda,\lambda_{A},\gamma_{s},\gamma_{c})$ coupling constant space. 
[A metallic, diffusive Fermi liquid is a state characterized by
the condition that both disorder parameters $\lambda$ and $\lambda_A$ tend to zero
at large length scales, 
whereas the interaction parameters approach $\gamma_s \to const.$ and
$\gamma_c \to 0$ in the same limit; see Figs.~\ref{DISINSTDisF}, \ref{DISINSTIntF}.
In this limit, the FNL$\sigma$M discussed in Section \ref{PandFR}
becomes a simple, weakly coupled Gaussian theory.\cite{BK}]
The sliver
encloses the line $\lambda = \lambda_{A} = \gamma_{c} = 0$, with $-\infty < \gamma_{s} < 1$, 
the entirety of which is perturbatively accessible because the FNL$\sigma$M does not require 
the interaction strength $\gamma_{s}$ to be small. A highly schematic 3D ``projected'' phase 
diagram is depicted in Fig.~\ref{PhaseDiag}. In this figure, the interaction 
constants
$(\gamma_s, \gamma_c)$
reside 
in the horizontal plane, while the vertical direction schematically represents (both) disorder 
strengths,
$(\lambda, \lambda_A$). For
example, we could take the vertical direction to measure the geometric average
$\sqrt{\lambda \,\lambda_{A}}$, while fixing the ratio $\lambda/\lambda_{A}$ to a constant.
The shaded sheath in Fig.~\ref{PhaseDiag} is a cartoon for the boundary of the stable metallic 
volume, which resides between it and the ballistic ($\lambda = \lambda_{A} = 0$) plane.
Over the range of perturbatively small values of $\gamma_{c}$, the stable
metallic, diffusive
 Fermi liquid phase resides in the region $\gamma_{c} \gtrsim 0$, terminating near $\gamma_{c} = 0$.
For $\gamma_{c} > 0$, the ``height'' of the stable metallic volume in the vertical 
``disorder'' direction is controlled by $\epsilon$, and is approximately independent of 
$\gamma_{c}$ (as indicated in Fig.~\ref{PhaseDiag}), although the precise shape and size of 
the phase boundary does vary with the ratio $\lambda/\lambda_{A}$ and with $\gamma_{s}$, 
making it difficult to characterize analytically. 

The flow equations (\ref{lambdaFlow2})--(\ref{gammacFlow2}) 
possess no nontrivial RG fixed points in dimensions $d > 2$, which would be accessible
in an epsilon expansion about $d = 2$. Thus no continuous metal-insulator transition
can be identified.
However, the two instabilities described 
in the previous subsection
for the $d=2$ case persist
for $d > 2$, 
and become clearly distinct roads out of the metallic state. The conventional CDW 
instability is found to always occur
for initial $\gamma_{c} < 0$ and sufficiently weak disorder, 
i.e. when $\lambda, \lambda_{A} \ll \epsilon$, and is represented by the flow 
$\gamma_{c} \rightarrow - \infty$.
[Recall Eqs.~(\ref{DEFGammaUGammaV}) and (\ref{SingletCDWR}),
and Section \ref{SLSNestingCDW}.]
This flow is accompanied by a decay in both disorder 
strengths $\lambda, \lambda_A$,
indicative of the clean limit. 
In Fig.~\ref{PhaseDiag}, the CDW instability is 
schematically indicated by the thick arrows that emanate from the terminating
surface of the metallic phase near $\gamma_{c} = 0$, running off toward large negative 
$\gamma_{c}$.

The primary result of this paper is the identification of a \emph{second} route
out of the diffusive Fermi liquid phase of the class AIII FNL$\sigma$M in 
$d = (2 + \epsilon) > 2$ dimensions, 
different from
the Mott CDW instability,
arising solely from the competition of disorder and interaction effects. 
As in $d=2$ dimensions,
this second route is characterized by a flow off to
simultaneously
strong disorder ($\lambda, \lambda_{A} \rightarrow \infty$) and strong 
interactions ($\gamma_{c} \rightarrow +\infty$), as indicated by the thick arrows 
emerging from
the $\gamma_{c}>0$ portion of the phase boundary shown in 
Fig.~\ref{PhaseDiag}; we therefore dub it an ``Anderson-Mott'' 
instability. 
\cite{footnote-a}
Even though there is no perturbatively accessible fixed point, this 
second
instability is nonetheless perturbatively controlled in $d = (2 + \epsilon)$ 
over a wide range of initial conditions when $\epsilon << 1$; in particular,
it is 
accessible over the entire range $0 \leq \gamma_{s} < 1$.\cite{footnote-t}
Numerically integrating Eqs.~(\ref{lambdaFlow2})--(\ref{gammacFlow2}) for small 
$\epsilon \ll 1$, we find that the ``Anderson-Mott'' 
instability\cite{footnote-a}
can
apparently always be reached by increasing {\it only} 
the ``dimensionless resistance'' $\lambda$ 
beyond some small threshold value $\lambda_{\textrm{T}}$,
while keeping the other three parameters
$\lambda_{A},\gamma_{s},\gamma_{c}$
fixed.
The threshold value $\lambda_{\textrm{T}}$
is a function 
of $\epsilon$ (and of the other coupling strengths)
which vanishes continuously in the limit $\epsilon \rightarrow 0$.
By contrast, we find that it is difficult 
to access the ``Anderson-Mott'' route out of the Fermi liquid phase by varying the CDW 
interaction strength $\gamma_{c}$ alone, despite the fact the ensuing instability 
is characterized by the rapid flow of $\gamma_{c} \rightarrow + \infty$. We 
therefore interpret the boundary separating the flow toward the stable metallic regime
from that toward the regime of the ``Anderson-Mott'' instability as a disorder-driven, 
first order metal-insulator transition (MIT). We emphasize that a MIT does not exist 
in the non-interacting random hopping model, which possesses only a delocalized phase 
at half-filling for finite disorder in $d\geq1$,\cite{1DChiral,Gade,FC,GLL} 
while the clean spinless Hubbard model possesses only the Mott CDW instability.  

We demonstrate the physical picture described in the previous paragraph with a 
selection of RG flow trajectory plots. We have numerically integrated 
Eqs.~(\ref{lambdaFlow2})--(\ref{gammacFlow2}) for the case of $d = 2 + \epsilon$
dimensions, with $\epsilon \equiv 0.01$, for a variety of initial conditions lying
within the perturbatively accessible volume of the 
four-dimensional coupling constant space
$(\lambda,\lambda_{A},\gamma_{s},\gamma_{c})$.
Projected traces of these flow trajectories in the disorder $(\lambda,\lambda_{A})$ 
and interaction $(\gamma_{s},\gamma_{c})$ planes are shown in Figs.~\ref{DISINSTDisF} 
and \ref{DISINSTIntF}, respectively. 
An individual trajectory is identified by a label {\bf\textsf{A}}--{\bf\textsf{F}}, 
drawn at the initial (final) condition of the corresponding projected trace 
in Fig.~\ref{DISINSTDisF} (Fig.~\ref{DISINSTIntF}),
and by a subfigure label {\bf(a)}--{\bf(d)}. Traces with matching labels in
these two figures describe
different projections of
the same RG trajectory,
which resides in the four-dimensional space
$(\lambda, \lambda_A, \gamma_s, \gamma_c)$.
In a given subfigure of Fig.~\ref{DISINSTDisF}, six different initial values
of the ``dimensionless dc resistance'' $\lambda$ have been chosen that
straddle the phase boundary between the diffusive Fermi liquid state and the
``Anderson-Mott'' instability. In each of these subfigures, the trajectories
labeled `{\bf\textsf{E}}' and `{\bf\textsf{F}}' flow into the
metallic, diffusive Fermi liquid state (`diffusive metal').
As for the other coupling strengths, all trajectories in 
Figs.~\ref{DISINSTDisF} and \ref{DISINSTIntF} share the initial conditions 
(ICs) $\lambda_{A} = - \gamma_{c} = \epsilon = 0.01$; trajectories in 
subfigures {\bf(a)}, {\bf(b)}, {\bf(c)}, and {\bf(d)} share the IC 
$\gamma_{s}$ = $0.1$, $0.3$, $0.5$, and $0.7$, respectively, 
as indicated by the black dots in
Fig.~\ref{DISINSTIntF}.
The subfigure sequence {\bf(a)}--{\bf(d)} 
thus exhibits the $\gamma_{s}$-evolution of the phase boundary.

It is interesting to speculate upon the nature of the insulating state
ultimately obtained upon breaching the boundary of the stable,
metallic diffusive Fermi liquid
along the ``Anderson-Mott'' route. Although we cannot rule out
the possibility that both this and the CDW instability, while clearly
distinct along the boundary of the diffusive Fermi liquid, 
eventually terminate into the \emph{same} insulating phase, it is important
to stress that this 
need not
be the case. As shown in Figs.~\ref{DISINSTDisF}
and \ref{DISINSTIntF}, the ``Anderson-Mott'' instability is characterized
by a rapid flow to large $\lambda_{A}$ and large positive $\gamma_{c}$.
Since $\lambda_{A}$ describes quenched orientational fluctuations in bond
strength dimerization (as discussed in Secs.~\ref{SPGradient} and \ref{NLSMSummary}),
while $\gamma_{c}$ is responsible for the Mott CDW instability 
($\gamma_{c} \rightarrow -\infty$), we might expect the insulator reached 
by the ``Anderson-Mott'' instability to possess local Mott insulating 
order, broken-up on longer length scales into randomly oriented domains 
pinned by the strong bond dimerization disorder. Note that a 
flow to large positive $\gamma_{c}$ is \emph{not} consistent with local 
CDW formation [i.e.\ order at a nesting wavevector $\bm{\mathrm{K_{N}}}$, 
as in Eq.~(\ref{NestingK})], but other types of Mott insulating order may occur
in the limit of strong interparticle interactions ($\gamma_{c} \rightarrow +\infty$). 
For example,
sending the same sublattice interaction strength $U \rightarrow + \infty$, for 
fixed $V$ in the $2\textrm{D}$ square lattice version of the clean Hubbard-like 
model defined via Eq.~(\ref{Hclean}), favors charge ``stripes,'' i.e.\ order 
at $\bm{\mathrm{K}} = (0,\pi)$, rather than at $\bm{\mathrm{K_{N}}} = (\pi,\pi)$.
[Recall from Eqs.~(\ref{DEFGammaUGammaV}) and (\ref{SingletCDW}),
 that this limit corresponds to $\gamma_c \to +\infty$.
In the same model, large negative values of $\gamma_c$ would arise, e.g., from
the different limit $V \rightarrow + \infty$ for fixed $U$.]

Finally, let us put the ``Anderson-Mott'' instability, and the associated
first order metal-insulator transition that we have found into 
some context. As we have stressed, the SLS possessed by the Hubbard-like model 
in Eqs.~(\ref{Hclean}) and (\ref{Hdis})
leads to special properties in both,
the non-interacting,
but disordered limit (delocalized phase at band center in 
$d \geq 1$),\cite{1DChiral,Gade,FC,GLL} as well as in the interacting, 
but non-disordered (ballistic) limit (Mott insulating ground state at half-filling due to nesting
for any $V > U \geq 0$). In the
simultaneous presence of interactions and random hopping, 
we have a stable metallic phase, and have found an interaction-stabilized, disorder-driven MIT
in $d = 2 + \epsilon >2$.
These features mutate if we break 
SLS, e.g.\ either by allowing ``diagonal'' (on-site) disorder or by tuning the 
filling fraction away from $1/2$. In the case of broken SLS
(keeping broken TRI, for simplicity), 
we recover the usual unitary 
metal symmetry class
(for spinless electrons),\cite{LeeRamakrishnan} 
described by an appropriate Finkel'stein NL$\sigma$M characterized by three 
coupling strengths, analogous to $\lambda$, $\gamma_{s}$, and $h$ appearing in Eqs.~(\ref{SD}) 
and (\ref{SI}). 
[Analogs of the ``dimerization'' disorder\cite{footnote-q}
parameter $\lambda_{A}$, 
as well as the CDW interaction strength $\gamma_{c}$ do \emph{not} 
appear in the unitary class
Finkel'stein sigma model;\cite{BK} see the discussion in Sec.~\ref{PandFR} 
following Eq.~(\ref{DiffLog}) for details.] In the case of short-ranged interparticle 
interactions, the FNL$\sigma$M predicts\cite{BK} that 
the unitary metal class is an 
insulator in $d = 2$ regardless of the presence of interactions, while 
the MIT in $d = 2 + \epsilon$ dimensions is actually 
in the same universality class as the 
\emph{non-interacting}, continuous (pure Anderson) transition, i.e.\ the 
interparticle interactions are irrelevant at the transition.
(Recall that we consider the spinless case.)


\subsection{Conclusion}\label{Conclusion}

As we have already summarized our main results in the Introduction 
(Sec.~\ref{Outline}), we conclude with a brief discussion of applications 
and extensions of our work, and we comment on some very recent results\cite{DellAnna}
for spin-1/2 fermion systems subject to the effects of both disorder and 
interactions.

As discussed in the last paragraph of Sec.~\ref{Results>2D}, the addition 
of a SLS-breaking perturbation to the Hubbard-like model given by 
Eqs.~(\ref{Hclean}) and (\ref{Hdis}) 
is expected to
push the system into 
(an interacting version of) the conventional, unitary metal class.\cite{LeeRamakrishnan} 
The (spinless) unitary metal with short-ranged interparticle interactions is 
known\cite{BK} to exhibit a MIT in $d = 2 + \epsilon$ dimensions, 
in the same universality class as
the associated \emph{non-interacting} (pure Anderson) 
transition in
this symmetry class.
It would be interesting to consider the crossover 
between the ``Anderson-Mott'' instability in the presence of SLS, 
identified in Sec.~\ref{Results>2D} and interpreted there as a disorder-driven, 
interaction-stabilized first order MIT, and the non-interacting, continuous 
unitary class Anderson transition expected in the absence of SLS. 

Let us next revisit possible connections to spin-$1/2$ fermion models.
As discussed in Sec.~\ref{Spin1/2} and elaborated in Appendices \ref{Spin1/2RHopping} 
and \ref{SC}, the Finkel'stein NL$\sigma$M calculation presented in this paper also 
applies to (i) a spin-$1/2$ Hubbard lattice model with SLS, broken TRI, and strong 
spin-orbit coupling, and to (ii) a continuum model of fermionic quasiparticles, native 
to a superconducting phase, possessing TRI and a remnant U(1) of spin SU(2) rotational symmetry 
(i.e.\ invariance under rotations about one axis in spin space),
preserved in every realization of the disorder 
(polar p-wave superconductor -- see Appendix D).
In the absence of interactions, both 
systems are realizations of the (quantum disorder) symmetry class AIII, within the
classification scheme of Ref.~\onlinecite{Zirnbauer}.

Consider first the interpretation in terms of the
{\it spin-1/2}
random (sublattice) hopping model 
mentioned above. 
Disorder may be realized, e.g.,\ 
via the application of a random orbital magnetic field to the otherwise 
clean model. (As discussed in Sec.~\ref{Spin1/2} and Appendix \ref{Spin1/2RHopping}, 
a random Zeeman field, on the other hand, realizes the \emph{completely different} 
disorder class C, while the combination of random orbital and Zeeman fields gives 
the ordinary unitary metal
symmetry class A.) The calculation presented in this paper 
also
applies to such a spin-1/2 random hopping model, subject to generic short-ranged
interparticle interactions, and to strong (homogenous or random) spin-orbit coupling. 
Spin-orbit coupling suppresses the hydrodynamic spin diffusion channel 
in the Finkel'stein NL$\sigma$M low energy field theory,
without modifying the symmetry class (AIII) of the disordered system
in the absence of interactions. 
[Without spin-orbit coupling, an additional interaction term
between the spin-densities would appear, which does not exist in the spinless
case, and which is therefore not included in our treatment here.]

On the other hand, the
superconductor quasiparticle interpretation of the AIII NL$\sigma$M,
mentioned above,
provides a 
very different
realization
of
our results. In particular, the 
superconductor interpretation does not require the disorder to take any special form 
(i.e.\ pure potential scattering will work), but requires only TRI and a remnant U(1) 
of spin SU(2) rotational symmetry
to be
preserved in every disorder realization.
As discussed in Appendix D, such
a situation
``naturally'' arises in the description of a spin-triplet, p-wave superconductor 
(lacking intrinsic spin-orbit coupling) in its 
polar, TRI phase.\cite{VollhardtWoelfle,Ho} This
tantalizing, but 
so far speculative connection could prove a more readily attainable connection to experiment.

As already mentioned, very recently 
Dell'Anna\cite{DellAnna} has independently studied several universality 
classes of Finkel'stein NL$\sigma$Ms, including realizations of both the particle-hole 
symmetric class C and the sublattice symmetric class AIII for spin-$1/2$ fermions.
In the AIII case, Dell'Anna includes the spin diffusion channel (i.e.\ assumes no 
spin-orbit coupling). We have compared our one-loop RG results Eqs.~(\ref{FlowEqs2}) 
and (\ref{DOSScaling2}) to his, and we find agreement if we suppress the interactions
associated with the spin degrees of freedom by hand. Dell'Anna has also computed
flow equations for a spin-$1/2$ realization of the BDI class, appropriate to the
half-filled spinful Hubbard model subject to real random hopping disorder [preserving TRI,
SLS, and spin SU(2) rotational symmetry in every realization]. This case may potentially
allow a detailed comparison with Monte Carlo simulations, because the lattice Hubbard model with
real random hopping manages to evade the infamous ``sign problem,'' and numerical results in 
$2\textrm{D}$ are already available.\cite{ScalettarOD1--BDI,ScalettarOD2--BDI+C}
At the time of this writing, however, Dell'Anna\cite{DellAnna} has not provided a
detailed analysis of his flow equations for this universality class (BDI).

\begin{acknowledgments}
	This work was supported in part by the NSF under Grant No.~DMR-00-75064. 
	One of us (MSF) acknowledges addition support by the UCSB Graduate Division
	(while at UCSB),
	by the Nanoscale Science and Engineering Initiative of the National Science 
	Foundation under NSF Award Number CHE-06-41523, and by the New 
	York State Office of Science, Technology, and Academic Research (NYSTAR).
	All Feynman diagrams in this paper 
	were created using the JaxoDraw\cite{JaxoDraw} program.
\end{acknowledgments}


\appendix


\section{Schwinger-Keldysh symmetry structure}\label{KeldyshSym}

In this appendix, we examine the symmetry content of the lattice fermion 
Keldysh action in Eqs.~(\ref{S1}) and (\ref{S2}). We establish a link between 
the \emph{non}-interacting sector [Eq.~(\ref{S1}); see also (\ref{S10}) and 
(\ref{DeltaS1})] and the transformation introduced in Eq.~(\ref{FermMatBilXfm}), 
justifying the promotion of the latter to a spatially-varying ``Goldstone 
fluctuation'' in Eq.~(\ref{QDef}).

For a given realization of the complex random hopping amplitudes $\{\delta t_{i, j}\}$, 
the only ``physical'' invariance possessed by Eqs.~(\ref{S1}) and (\ref{S2}) is 
the discrete SLS, defined by Eq.~(\ref{SLS}). Nevertheless, the 
\emph{non}-interacting sector of the Keldysh action, Eq.~(\ref{S1}), is 
actually invariant under a large group of \emph{continuous} transformations, 
to be defined below. An analysis of the symmetry content of Eq.~(\ref{S1}) 
clarifies our derivation of the Keldysh sigma model provided in Sec.~\ref{NLSM}, 
in which the diffusive fluctuations of heat and charge in the non-interacting 
random hopping model emerge as Goldstone modes of this (``spontaneously broken'') 
continuous symmetry. The full, interacting Finkel'stein NL$\sigma$M, defined by 
Eqs.~(\ref{Z})--(\ref{SI}), further incorporates the interparticle interactions 
of Eq.~(\ref{S2}), which formally obliterate this symmetry.

Adopting the matrix notations introduced in the paragraph preceding 
Eq.~(\ref{S1final}), we recast the non-interacting Keldysh action in 
Eqs.~(\ref{S10}) and (\ref{DeltaS1}):
\begin{equation}\label{AAS1part}
	S_{1} \equiv S_{1}^{\, h} + S_{1}^{\, \eta} + S_{1}^{\, \omega},
\end{equation}
where
\begin{subequations}\label{AAS1partdet}
\begin{align}
	S_{1}^{\, h} & = \sum_{\langle i j \rangle}
	\left\lgroup \bar{c}_{A i} \, t_{i, j} \, c_{B j} 
	+ \bar{c}_{B j} \, t_{i, j}^{*} \, c_{A i} \right\rgroup \label{AAS1h}\\
	S_{1}^{\, \eta} & = 
	i \eta 
	\sum_{i \in A} \bar{c}_{A i} \hat{\Sigma}_{3} \hat{\xi}_{3} c_{A i}
	+ i \eta 
	\sum_{j \in B} \bar{c}_{B j} \hat{\Sigma}_{3} \hat{\xi}_{3} c_{B j}, 
	\label{AAS1eta}\\
\intertext{and}
	S_{1}^{\, \omega} & = 
	\sum_{i \in A} \bar{c}_{A i} \hat{\Sigma}_{3} |\hat{\omega}| c_{A i}
	+ 
	\sum_{j \in B} \bar{c}_{B j} \hat{\Sigma}_{3} |\hat{\omega}| c_{B j}, \label{AAS1omega}
\end{align}
\end{subequations}
with $t^{\phantom{*}}_{i,j} \equiv t + \delta t^{\phantom{*}}_{i,j}$, as in Eq.~(\ref{S1}).

Consider a general, spatially uniform unitary transformation in 
$|\omega|\otimes\Sigma\otimes\xi$ (frequency$\otimes$Keldysh) space:
\begin{equation}\label{AAKeldyshXfm1}
\begin{array}{ll}
	c_{A i} \rightarrow \hat{U}_{A}^{\phantom{\dagger}} c_{A i},\qquad &  
	\bar{c}_{A i} \rightarrow \bar{c}_{A i} \hat{V}_{A}^{\phantom{\dagger}}, \\
	c_{B j} \rightarrow \hat{U}_{B}^{\phantom{\dagger}} c_{B j},\qquad & 
	\bar{c}_{B j} \rightarrow \bar{c}_{B j} \hat{V}_{B}^{\phantom{\dagger}}, 
\end{array}
\end{equation}
where
\begin{equation}\label{AAKeldyshXfm2}
	\hat{U}_{A}^{\dagger} \hat{U}_{A}^{\phantom{\dagger}} = 
	\hat{V}_{A}^{\dagger} \hat{V}_{A}^{\phantom{\dagger}} =
	\hat{U}_{B}^{\dagger} \hat{U}_{B}^{\phantom{\dagger}} =
	\hat{V}_{B}^{\dagger} \hat{V}_{B}^{\phantom{\dagger}} =
	\hat{1}_{|\omega|\otimes\Sigma\otimes\xi}.
\end{equation}
We now examine the symmetry content of the non-interacting Keldysh action 
$S_{1}$, Eq.~(\ref{AAS1part}), assembled in stages from the three pieces 
defined in Eqs.~(\ref{AAS1h})--(\ref{AAS1omega}).

Invariance of the ``Hamiltonian'' piece $S_{1}^{\, h}$, Eq.~(\ref{AAS1h}), 
under the transformation in Eq.~(\ref{AAKeldyshXfm1}) requires that
\begin{equation}\label{AAS1hSym}
	\hat{V}_{A}^{\phantom{\dagger}} = \hat{U}_{B}^{\dagger},
	\qquad 
	\hat{V}_{B}^{\phantom{\dagger}} = \hat{U}_{A}^{\dagger}.
\end{equation}
If we imagine discretizing and truncating the continuum of allowed absolute 
energies to a set of $n$ discrete values, 
$|\omega| \rightarrow |\omega|_{\alpha}$, with $\alpha \in \{1, \ldots, n\}$, 
then we can say that the conditions in Eqs.~(\ref{AAKeldyshXfm2}) and 
(\ref{AAS1hSym}) imply the action $S_{1}^{\, h}$ is invariant under transformations 
belonging to the space U(4n)$\times$U(4n). 

Next, we consider the ``Hamiltonian'' plus ``pole prescription'' pieces, 
Eqs.~(\ref{AAS1h}) and (\ref{AAS1eta}), respectively. Invariance of 
$S_{1}^{\, h} + S_{1}^{\, \eta}$ requires the restriction given by 
Eq.~(\ref{AAS1hSym}), as well as
\begin{equation}\label{AAS1hPetaSym}
	\hat{\Sigma}_{3} \, \hat{\xi}_{3} \, \hat{U}_{A}^{\phantom{\dagger}} \, \hat{\Sigma}_{3} \, \hat{\xi}_{3} = \hat{U}_{B}^{\phantom{\dagger}}
\end{equation}
Taken together, Eqs.~(\ref{AAS1hSym}) and (\ref{AAS1hPetaSym}) imply that 
the U(4n)$\times$U(4n) symmetry of $S_{1}^{\, h}$ is broken down to the 
diagonal subgroup U(4n) by the pole prescription term, $S_{1}^{\, \eta}$.

Finally, the full non-interacting action is assembled by adding the ``energy'' 
piece, defined by Eq.~(\ref{AAS1omega}). Under the transformation given by 
Eq.~(\ref{AAKeldyshXfm1}), invariance of 
$S_{1} \equiv S_{1}^{\, h} + S_{1}^{\, \eta} + S_{1}^{\, \omega}$ requires the 
conditions provided in Eqs.~(\ref{AAS1hSym}), (\ref{AAS1hPetaSym}), as well as
\begin{align}
	[\hat{\xi}_{3}, \hat{U}_{A}^{\phantom{\dagger}}] & = 0, \label{AAS1Sym1}\\
\intertext{and}
	[|\hat{\omega}|, \hat{U}_{A}^{\phantom{\dagger}}] & = 0, \label{AAS1Sym2}
\end{align}
with $[\hat{A},\hat{B}]\equiv \hat{A} \hat{B} - \hat{B} \hat{A}$. The 
non-interacting Keldysh action admits symmetry transformations off-diagonal 
in $\Sigma$ [$\sgn(\omega)$] space, but necessarily diagonal in both 
$|\omega|$ and $\xi$ (Keldysh) spaces, implying the (much smaller) symmetry 
group $[\textrm{U(2)} \times \textrm{U(2)}]^{n}$.

\begin{table}
\caption{\label{KeldyshSymTable}Symmetry content of the \emph{non}-interacting 
Keldysh action, Eqs.~(\ref{AAS1part}) and (\ref{AAS1partdet}).
The target space of the NL$\sigma$M is set by the quotient of the first and 
second entries, below.}
\begin{ruledtabular}
\begin{tabular}{lc}
Imposed condition & Implied invariance \\
\hline 
	1. Invc. of $S_{1}^{\, h}$: & U(4n)$\times$U(4n) \\
	2. Invc. of $S_{1}^{\, h} + S_{1}^{\, \eta}$: & U(4n) \\
	3. Invc. of $S_{1} = S_{1}^{\, h} + S_{1}^{\, \eta} + S_{1}^{\, \omega}$:\; & $[\textrm{U(2)} \times \textrm{U(2)}]^{n}$
\end{tabular}
\end{ruledtabular}
\end{table}

The pattern of symmetry breaking that we have found is summarized in Table \ref{KeldyshSymTable}.
The symmetry content of the Hamiltonian piece, $S_{1}^{\, h}$, of the 
Keldysh action is broken down into successively smaller remnants via the 
addition of the pole prescription and energy terms. We interpret the 
``pole prescription'' piece $S_{1}^{\, \eta}$, Eq.~(\ref{AAS1eta}) with 
$\eta \rightarrow 0^{+}$, as an infinitesimal source, analogous to an 
external field in an O(N) ferromagnet; in the diffusive metallic phase, 
this source selects a particular ground state from the U(4n)$\times$U(4n) 
saddle point manifold of the system. [The magnitude of the saddle point 
configuration is given by the elastic scattering rate $1/\tau$ due to the 
disorder, evaluated in the self-consistent Born approximation.\cite{LeeRamakrishnan}
See, e.g., Eqs.~(\ref{QCompSP}) and (\ref{SCBA}).] The (non-interacting)
NL$\sigma$M retains only the quantum Goldstone fluctuations about this 
symmetry-broken minimum. On the other hand, the energy piece defined in 
Eq.~(\ref{AAS1omega}) plays a decreasingly effective role in suppressing 
fluctuations in the low energy limit, $|\omega| \ll 1/{\tau}$; we therefore 
retain modes (weakly) off-diagonal in both the $|\omega|$ and  Keldysh ($\xi$) 
spaces. As a result, we expect that the non-interacting Keldysh NL$\sigma$M, 
appropriate to the complex random hopping model defined in Eqs.~(\ref{Hclean}) and 
(\ref{Hdis}), with $U = V = 0$, will possess the target manifold 
$\textrm{U(4n)}\times\textrm{U(4n)}/\textrm{U(4n)} \sim \textrm{U(4n)}$. 
The NL$\sigma$M with target manifold the Lie group $\textrm{U(4n)}$, for 
\emph{finite}, integer $n$, is termed the unitary principal chiral model 
in the field theory literature. An analogous result for the non-interacting, 
class AIII random hopping model, formulated in terms of bosonic replicas, 
was originally obtained by Gade and Wegner.\cite{Gade}
	
Let us make several remarks. First, consider tuning the fermion density away 
from half filling in the non-interacting random hopping model. This 
necessitates the addition of a finite chemical potential term to the 
Hamiltonian piece of the Keldysh action, Eq.~(\ref{AAS1h}). Going through 
the above analysis, one finds a smaller NL$\sigma$M target manifold 
$\textrm{U(4n)}/\textrm{U(2n)}\times\textrm{U(2n)}$, characteristic of the 
TRI-broken, unitary normal metal class, as expected. We also note that the 
symmetry operation given by Eqs.~(\ref{AAKeldyshXfm1}) and (\ref{AAKeldyshXfm2}) 
is \emph{not} the most general unitary transformation that one may implement 
upon the independent Grassmann fields $\bar{c}_{i A}^{\, a}$, 
${c}_{i A}^{\, a}$, $\bar{c}_{j B}^{\, a}$, and ${c}_{j B}^{\, a}$ appearing 
in the Keldysh action [Eqs.~(\ref{AAS1part}) and (\ref{AAS1partdet})]. 
In general, one should consider transformations that mix fields simultaneously 
in sublattice flavor ($c_{A} \leftrightarrow c_{B}$), particle-hole 
($c_{i} \leftrightarrow \bar{c}_{i}$), $\sgn(\omega)$ ($\Sigma$), Keldysh 
($\xi$), and $|\omega|$ spaces. This is most easily done by concatenating 
all field degrees of freedom at a particular location in position space 
into a single, many component Majorana fermion, upon which the largest 
unitary transformation possible is executed, and subsequently restricted 
in order to determine the symmetry content of the Keldysh action.\cite{Zirnbauer,AltlandZirnbauer} 
This procedure is always required in the presence of TRI, in order to obtain 
the Cooperon modes. For the sublattice-symmetric, broken TRI (class AIII) 
model studied here, only transformations diagonal in sublattice and 
particle-hole space preserve the Hamiltonian piece of the Keldysh action, 
$S_{1}^{\, h}$ in Eq.~(\ref{AAS1h}).


\section{Gradient expansion}\label{ExpTrick}

We derive here the gradient expansion expressed in Eqs.~(\ref{SDETexp}), 
(\ref{SDETdim}), and (\ref{SIrough1}) for the fermionic functional determinant 
in Eq.~(\ref{SDET}). Using the explicit construction [Eqs.~(\ref{QCompDef}) 
and (\ref{QDef})] for the continuum NL$\sigma$M matrix field 
$\hat{\mathcal{Q}}(\bm{r})$ in terms of its homogeneous saddle point 
value [Eq.~(\ref{QCompSP})], we have
\begin{equation}\label{ABQCompDef}
	\hat{\mathcal{Q}}(\bm{\mathrm{r}}) \equiv 
	\hat{\mathcal{U}}(\bm{\mathrm{r}}) \, \hat{\mathcal{Q}}_{\mathsf{SP}} \, \hat{\mathcal{U}}^{\dagger}(\bm{\mathrm{r}}), 
\end{equation}
where $\hat{\mathcal{Q}}_{\mathsf{SP}}$ and 
$\hat{\mathcal{U}}(\bm{\mathrm{r}})$ possess the sublattice space decompositions
\begin{equation}\label{ABQSPComp}
	\hat{\mathcal{Q}}_{\mathsf{SP}} = \frac{1}{2 \tau}	
	\begin{bmatrix}
		0 & \hat{\Sigma}_{3} \hat{\xi}_{3} \\
		\hat{\Sigma}_{3} \hat{\xi}_{3} & 0 
	\end{bmatrix}
\end{equation}
and 
\begin{equation}\label{ABUComp}
	\hat{\mathcal{U}}(\bm{\mathrm{r}}) =	
	\begin{bmatrix}
		\hat{U}_{A}^{\phantom{\dagger}}(\bm{\mathrm{r}}) & 0 \\
		0 & \hat{U}_{B}^{\phantom{\dagger}}(\bm{\mathrm{r}})
	\end{bmatrix},
\end{equation}
respectively.
Note that Eqs.~(\ref{ABQCompDef}) and (\ref{ABQSPComp}) imply that
\begin{equation}\label{QCompModConstraint} 
	\hat{\mathcal{Q}}^{2}(\bm{\mathrm{r}})=\hat{1}(1/2 \tau)^{2}.
\end{equation}

Now consider the determinant in Eq.~(\ref{SDET}). Abbreviating 
$G_{0}^{-1} +\rho \equiv G^{-1}$, we write
\begin{align}\label{ABSDETManip}
S_{\mathsf{DET}} = & 
	- \mathrm{Tr}
	\left\lgroup
		\ln\left[
		G^{-1} 
		+ i \hat{\sigma}_{1} \hat{\mathcal{U}} \, \hat{\mathcal{Q}}_{\mathsf{SP}} \, \hat{\mathcal{U}}^{\dagger}		
		\right]
	\right\rgroup \nonumber\\
	= &
	- \mathrm{Tr}
	\left\lgroup
		\ln\left[
		\hat{\sigma}_{1} \hat{\mathcal{U}} 
		\left(
		\hat{\mathcal{U}}^{\dagger}\hat{\sigma}_{1} 
		G^{-1}
		\hat{\mathcal{U}}
		+ i \hat{\mathcal{Q}}_{\mathsf{SP}}		
		\right)
		\hat{\mathcal{U}}^{\dagger}
		\right]
	\right\rgroup \nonumber\\
	= &
	- \mathrm{Tr}
	\left\lgroup
		\ln\left[
		\hat{1} 
		- i (2 \tau)^{2} \,
		\hat{\mathcal{U}}^{\dagger}\hat{\sigma}_{1} 
		G^{-1}
		\hat{\mathcal{U}}
		\hat{\mathcal{Q}}_{\mathsf{SP}}
		\right]
	\right\rgroup \nonumber\\
	& +
	\mathrm{Tr}
	\left\lgroup
	\ln\left[
	-i \hat{\mathcal{Q}}_{\mathsf{SP}} (2 \tau)^{2}
	\right]
	\right\rgroup.
\end{align}
Expanding the logarithm on the third line of Eq.~(\ref{ABSDETManip}), 
and dropping the constant term appearing on the fourth, we obtain
\begin{equation}\label{ABSDETexp}
	S_{\mathsf{DET}} \approx  
	S_{\mathsf{DET}}^{(1)} + \, S_{\mathsf{DET}}^{(2)}
	+ \ldots,
\end{equation}
where
\begin{equation}\label{ABSDETexpn}
	S_{\mathsf{DET}}^{(n)} \equiv 
	\frac{i^{n} (2 \tau)^{2 n}}{n} \, 
	\mathrm{Tr}
	\left[(G_{0}^{-1} + \rho) \hat{\mathcal{Q}} \hat{\sigma}_{1} \right]^{n}.
\end{equation}

Using the explicit form of the inverse Green's function $G_{0}^{-1}$ in 
Eq.~(\ref{CleanGF}), as well as the sublattice decomposition of $\hat{\mathcal{Q}}$ 
from Eq.~(\ref{QCompDef}), the first term on the right-hand side of 
Eq.~(\ref{ABSDETexp}) is
\begin{align}\label{ABSDETexp1}
	S_{\mathsf{DET}}^{(1)} \sim & 
	i 2 \tau
	\int d^{d} \bm{\mathrm{r}} \,
	\mathrm{Tr}
	\left\lgroup
	(\hat{\Sigma}_{3} |\hat{\omega}| + i \eta \hat{\Sigma}_{3} \hat{\xi}_{3})
	\,[\hat{Q}^{\dagger}({\bm{\mathrm{r}}})+\hat{Q}({\bm{\mathrm{r}}})]
	\right\rgroup
	\nonumber\\
	&+ 
	i 2 \tau
	\int d^{d}{\bm{\mathrm{r}}} \,
	\mathrm{Tr}
	\left\lgroup
	\rho_{A}({\bm{\mathrm{r}}}) \, \hat{Q}({\bm{\mathrm{r}}})
	+\rho_{B}({\bm{\mathrm{r}}}) \, \hat{Q}^{\dagger}({\bm{\mathrm{r}}})
	\right\rgroup,
\end{align}
where we have completed a partial trace over sublattice flavor $(\sigma)$ space.
The first line of Eq.~(\ref{ABSDETexp1}) gives the term on the first line of the 
expansion in Eq.~(\ref{SDETexp}). Writing out the trace on the second line of 
Eq.~(\ref{ABSDETexp1}) as a sum over the Keldysh species index $\{a\}$ 
and an integral over time $\{t\}$, and using the locality of the auxiliary fields 
$\rho_{A/B} \rightarrow \rho_{A/B}^{a}(t)$ in these indices [Eq.~(\ref{RhoVector})],
one obtains Eq.~(\ref{SIrough1}).

The evaluation of the second term $S_{\mathsf{DET}}^{(2)}$ in the expansion given by 
Eq.~(\ref{ABSDETexp}) involves the terms in Eq.~(\ref{CleanGF}) that are functions of 
the momentum operator $\hat{\bm{\mathrm{k}}}$, and are thus non-local in position space.
Consider the components of Eq.~(\ref{CleanGF}) arising from a non-zero mean bond 
dimerization ($\delta t_{\mathsf{dim}} \neq 0$); Eq.~(\ref{DimerFuncDef}) implies the 
microscopic definition
\begin{equation}\label{phiIDef}
	i\, \phi_{I}(\bm{\mathrm{k}}) = - i\, \delta t_{\mathsf{dim}} \sin(\bm{\mathrm{k}}\cdot\bm{\mathrm{n}})
\end{equation}
($\bm{\mathrm{n}}$ is a unit vector specifying the orientation of the average dimerization),
so that
\begin{subequations}\label{phiISym}
\begin{align}
	i\, \phi_{I}(-\bm{\mathrm{k}}) &= - i\, \phi_{I}(\bm{\mathrm{k}}), \\
	(i\, \phi_{I})^{*}(\bm{\mathrm{k}}) &= i\, \phi_{I}(-\bm{\mathrm{k}}).
\end{align}
\end{subequations}
We define $\phi_{\mathsf{dim}}(\bm{\mathrm{r}})$ as the following (coarse-grained, 
long wavelength) position space Fourier transform:
\begin{equation}\label{phidimDef}
	\phi_{\mathsf{dim}}(\bm{\mathrm{r}}) \equiv 
	\int \frac{d^{d}\bm{\mathrm{k}}}{(2 \pi)^{d}} \,   
	e^{i \bm{\mathrm{k}} \cdot \bm{\mathrm{r}}} \,i\, \phi_{I}(\bm{\mathrm{k}}),
\end{equation}
where the momentum integration should be taken up to some spherical cutoff 
of order the inverse mean free path. Eqs.~(\ref{phiISym}) and (\ref{phidimDef}) imply that 
$\phi_{\mathsf{dim}}(\bm{\mathrm{r}})$ obeys the conditions
\begin{subequations}\label{phidimSym}
\begin{align}
	\phi_{\mathsf{dim}}(-\bm{\mathrm{r}}) &= 
	- \phi_{\mathsf{dim}}(\bm{\mathrm{r}}), \label{phidimAntiSym}\\
	\phi_{\mathsf{dim}}^{*}(\bm{\mathrm{r}}) &= 
	\phi_{\mathsf{dim}}(\bm{\mathrm{r}}).
\end{align}
\end{subequations}
Thus $\phi_{\mathsf{dim}}(\bm{\mathrm{r}})$  is a real, antisymmetric function. Moreover,
we expect $\phi_{\mathsf{dim}}(\bm{\mathrm{r}})$ to possess a ``p-wave'' ($l = 1$ angular
momentum) structure as a function of $\bm{\mathrm{r}}/r$, given its momentum 
space representation, Eq.~(\ref{phiIDef}), which also implies that 
\begin{align}\label{phiPointer}
	\bm{\nabla} \phi_{\mathsf{dim}}(\bm{\mathrm{r}})|_{\bm{\mathrm{r}} = 0}
	=& - \int \frac{d^{d}\bm{\mathrm{k}}}{(2 \pi)^{d}} \,   
	\bm{\mathrm{k}} \, \phi_{I}(\bm{\mathrm{k}})
	\nonumber\\
	\sim&
	\mathfrak{c} \, \delta t_{\mathsf{dim}} \, \bm{\mathrm{n}},
\end{align}
where $\mathfrak{c}$ is some constant.

At second order in the expansion given by Eq.~(\ref{ABSDETexp}), Eq.~(\ref{ABSDETexpn}) 
gives 
\begin{align}\label{ABSDETexp2}
	S_{\mathsf{DET}}^{(2)} &\sim 
	\frac{-(2 \tau)^{4}}{2} \, 
	\mathrm{Tr}
	\left\lgroup
	\left[\varepsilon(\hat{\bm{\mathrm{k}}}) + \phi_{R}(\hat{\bm{\mathrm{k}}})
	+ \hat{\sigma}_{3} \, i \, \phi_{I}(\hat{\bm{\mathrm{k}}})\right]
	\hat{\mathcal{Q}}
	\right\rgroup^{2}
	\nonumber\\
	&
	\begin{aligned}[b]
		\sim
		\frac{-(2 \tau)^{4}}{2} \, 
		\big\lgroup &
		\mathrm{Tr}\left[
		\varepsilon(\hat{\bm{\mathrm{k}}}) \, \hat{\mathcal{Q}} \,
		\varepsilon(\hat{\bm{\mathrm{k}}}) \, \hat{\mathcal{Q}}
		\right]
		\\
		&+ 2\mathrm{Tr}\left[
		\varepsilon(\hat{\bm{\mathrm{k}}}) \, \hat{\mathcal{Q}} \,
		\hat{\sigma}_{3} \, i \, \phi_{I}(\hat{\bm{\mathrm{k}}}) \, \hat{\mathcal{Q}}
		\right]
		\big\rgroup.
	\end{aligned}
\end{align}
In Eq.~(\ref{ABSDETexp2}), we have freely discarded terms of no relevance. In particular, 
we ignore $\phi_{R}(\hat{\bm{\mathrm{k}}})$, since this function, although arising through 
the dimerization perturbation [Eqs.~(\ref{DeltaSdim}) and (\ref{DimerFuncDef})], is symmetric 
under reflection, and does not generate anything not already present in the FNL$\sigma$M action
(up to ``trivial'' anisotropy that may be removed via rescaling).
Writing the first term in Eq.~(\ref{ABSDETexp2}) in position space $\{\bm{\mathrm{r}}\}$, 
expanding the bandstructure 
$\varepsilon(\bm{\mathrm{k}}) \sim \overline{v}_{F} \bm{\mathrm{k}} 
\rightarrow - i \, \overline{v}_{F} \bm{\nabla}$, with $\overline{v}_{F}$ the average Fermi 
velocity (at half filling), and using Eq.~(\ref{QCompDef}) gives the usual ``stiffness'' term 
appearing on the second line of Eq.~(\ref{SDETexp}).

The second term in Eq.~(\ref{ABSDETexp2}) may be evaluated as follows:
\begin{subequations}\label{phiPointerExp}
\begin{widetext}
\begin{align}
	2\mathrm{Tr}\left[
	\varepsilon(\hat{\bm{\mathrm{k}}})\,\hat{\mathcal{Q}}
	\right.&\left. \!
	\hat{\sigma}_{3} \, i \,
	\phi_{I}(\hat{\bm{\mathrm{k}}}) 
	\hat{\mathcal{Q}}
	\right]
	\label{DimerAnsA}\\
	\sim & 
	t \int d^{d}{\bm{\mathrm{r}}} \, 
	\mathrm{Tr}\left\lgroup
	(-\bm{\nabla}^{2} \delta_{\bm{\mathrm{r}},\bm{\mathrm{r'}}})\,
	\hat{\mathcal{Q}}(\bm{\mathrm{r'}}) \,
	\hat{\sigma}_{3} \,
	\phi_{\mathsf{dim}}(\bm{\mathrm{r'}}-\bm{\mathrm{r}})\,
	\hat{\mathcal{Q}}(\bm{\mathrm{r}})
	\right\rgroup\label{DimerAnsB}\\
	\sim & 
	t \int d^{d}{\bm{\mathrm{r}}} \, 
	\mathrm{Tr}\left\lgroup
	(\bm{\nabla} \delta_{\bm{\mathrm{r}},\bm{\mathrm{r'}}})\cdot
	\left\{
	\hat{\mathcal{Q}}(\bm{\mathrm{r'}}) \,
	\hat{\sigma}_{3} \,
	\left[\bm{\nabla} \phi_{\mathsf{dim}}(\bm{\mathrm{r'}}-\bm{\mathrm{r}})\right]\,
	\hat{\mathcal{Q}}(\bm{\mathrm{r}})
	+
	\hat{\mathcal{Q}}(\bm{\mathrm{r'}}) \,
	\hat{\sigma}_{3} \,
	\phi_{\mathsf{dim}}(\bm{\mathrm{r'}}-\bm{\mathrm{r}})\,
	\left[\bm{\nabla} \hat{\mathcal{Q}}(\bm{\mathrm{r}})\right]
	\right\}
	\right\rgroup\label{DimerAnsC}\\
	\sim & 
	t \int d^{d}{\bm{\mathrm{r}}} \, 
	\mathrm{Tr}\left\lgroup
	(- \delta_{\bm{\mathrm{r}},\bm{\mathrm{r'}}})
	\left\{
	\hat{\mathcal{Q}}(\bm{\mathrm{r'}}) \,
	\hat{\sigma}_{3} \,
	\left[\bm{\nabla}^{2} \phi_{\mathsf{dim}}(\bm{\mathrm{r'}}-\bm{\mathrm{r}})\right]\,
	\hat{\mathcal{Q}}(\bm{\mathrm{r}})
	+
	\hat{\mathcal{Q}}(\bm{\mathrm{r'}}) \,
	\hat{\sigma}_{3} \,
	\left[\bm{\nabla} \phi_{\mathsf{dim}}(\bm{\mathrm{r'}}-\bm{\mathrm{r}})\right]\cdot
	\left[\bm{\nabla} \hat{\mathcal{Q}}(\bm{\mathrm{r}})\right]
	\right.\right.
	\nonumber\\
	& \left.\left.\qquad\quad
	+
	\hat{\mathcal{Q}}(\bm{\mathrm{r'}}) \,
	\hat{\sigma}_{3} \,
	\left[\bm{\nabla} \phi_{\mathsf{dim}}(\bm{\mathrm{r'}}-\bm{\mathrm{r}})\right]\cdot
	\left[\bm{\nabla} \hat{\mathcal{Q}}(\bm{\mathrm{r}})\right]
	+
	\hat{\mathcal{Q}}(\bm{\mathrm{r'}}) \,
	\hat{\sigma}_{3} \,
	\phi_{\mathsf{dim}}(\bm{\mathrm{r'}}-\bm{\mathrm{r}})\,
	\left[\bm{\nabla}^{2} \hat{\mathcal{Q}}(\bm{\mathrm{r}})\right]
	\right\}
	\right\rgroup\label{DimerAnsD}\\
	\sim & 
	2 t \int d^{d}{\bm{\mathrm{r}}} \,
	\mathrm{Tr}\left\lgroup
	\hat{\mathcal{Q}}(\bm{\mathrm{r}}) \,
	\hat{\sigma}_{3} \,
	\bm{\nabla} \hat{\mathcal{Q}}(\bm{\mathrm{r}})
	\right\rgroup
	\cdot \bm{\nabla'} \phi_{\mathsf{dim}}(\bm{\mathrm{r'}})|_{\bm{\mathrm{r'}} = 0}
	\label{DimerAnsE}\\
	\sim & 
	\frac{\mathfrak{c} \, t \, \delta t_{\mathsf{dim}}}{\tau^{2}} 
	\int d^{d}{\bm{\mathrm{r}}} \,
	\mathrm{Tr}\left\lgroup
	\hat{Q}^{\dagger}(\bm{\mathrm{r}})
	\bm{\nabla} \hat{Q}(\bm{\mathrm{r}})
	\right\rgroup
	\cdot \bm{\mathrm{n}},
	\label{DimerAnsF}
\end{align}
\end{widetext}
\end{subequations}
where we have used an expansion for the bandstructure,
$\varepsilon({\bm{\mathrm{k}}}) \propto t {\bm{\mathrm{k}}}^{2} +  \bm{\mathit{O}}({\bm{\mathrm{k}}}^{4})$,
valid for small $k$ (associated with the \emph{bottom} of the band), $t$ is the homogeneous 
hopping strength [Eq.~(\ref{Hclean})], and where we have employed the compact notation
$\delta^{(d)}(\bm{\mathrm{r}}-\bm{\mathrm{r'}})\equiv\delta_{\bm{\mathrm{r}},\bm{\mathrm{r'}}}$.  
In going between subequations (\ref{DimerAnsD}) and (\ref{DimerAnsE}), we have 
used Eqs.~(\ref{QCompModConstraint}) and (\ref{phidimAntiSym}), while between 
Eqs.~(\ref{DimerAnsE}) and (\ref{DimerAnsF}) we have employed the sublattice 
decomposition in Eq.~(\ref{QCompDef}), completed a partial trace over sublattice 
flavor space ($\sigma$), and used Eq.~(\ref{phiPointer}). Combining 
Eq.~(\ref{DimerAnsF}) with the prefactor from (\ref{ABSDETexp2}), 
we obtain Eq.~(\ref{SDETdim}) (with a suitable redefinition of the constant 
$\mathfrak{c}$).


\section{Disordered, bipartite lattice models for spin-1/2 electrons}\label{Spin1/2RHopping}

In this appendix, we identify the quantum disorder symmetry classes appropriate to 
models of non-interacting spin-1/2 fermions, with nearest-neighbor hopping on a 
bipartite lattice at half-filling, subject to random orbital or random Zeeman magnetic 
fields. Our starting point is the general bipartite lattice Hamiltonian
\begin{align}\label{ACLatticeHam}
	H = 
	\sum_{\alpha, \beta}
	& \Bigg[
	\sum_{i, j}
	c_{A i}^{\dagger \, \alpha} \,
	(\hat{t}_{AB}^{\phantom{\dagger}})_{i, j}^{\alpha, \beta} \,
	c_{B j}^{\beta} + \textrm{H.c.} 
	\nonumber\\
	\phantom{\Bigg[}
	& 
	+ \sum_{i, i'}
	c_{A i}^{\dagger \, \alpha} \,
	(\hat{t}_{AA}^{\phantom{\dagger}})_{i, i'}^{\alpha, \beta} \,
	c_{A i'}^{\beta} 
	\nonumber\\
	&
	+ \sum_{j, j'}
	c_{B j}^{\dagger \, \alpha} \,
	(\hat{t}_{BB}^{\phantom{\dagger}})_{j, j'}^{\alpha, \beta} \,
	c_{B j'}^{\beta}
	\Bigg],
\end{align}
where $c_{A i}^{\dagger \, \alpha}$ and $c_{B j}^{\beta}$ are creation 
and annihilation operators for fermions on the `$A$' and `$B$' 
sublattices, respectively, with $\alpha,\beta \in \{\uparrow,\downarrow\}$ spin-$1/2$ 
component indices. In Eq.~(\ref{ACLatticeHam}), indices $\{i,i'\}$ and $\{j,j'\}$ 
respectively denote sites on the $A$ and $B$ sublattices, so that the sum on ($i, j$) 
runs over \emph{all} (not just nearest neighbor) pairs of $A$ and $B$ sites, while 
the sums on ($i, i'$) and ($j, j'$) run over all same sublattice pairs of sites. 
Particle hopping is facilitated by the spin-dependent intersublattice 
$\{\hat{t}_{AB}^{\phantom{\dagger}}$, $\hat{t}_{AB}^{\dagger}\}$
and same sublattice
$\{\hat{t}_{AA}^{\phantom{\dagger}}$, $\hat{t}_{BB}^{\phantom{\dagger}}\}$
complex connectivity matrices. The same sublattice matrices are constrained 
by Hermiticity:
\begin{equation}\label{ACSameSubLatHerm}
	\hat{t}_{AA}^{\dagger} = \hat{t}_{AA}^{\phantom{\dagger}}, \qquad
	\hat{t}_{BB}^{\dagger} = \hat{t}_{BB}^{\phantom{\dagger}}. \quad
\end{equation}
Let us adopt a compact matrix notation reminiscent of that implemented in 
Eq.~(\ref{S10}), Sec.~\ref{SKPI}. We re-write Eq.~(\ref{ACLatticeHam}) as
\begin{equation}\label{ACLatHamCompact}
	H \equiv c^{\dagger} \hat{h} c,
\end{equation}
where $c$ ($c^{\dagger}$) is a column (row) vector,
with indices in position, sublattice flavor, and spin-$1/2$ spaces, i.e.\
$c \rightarrow c_{A i/B j}^{\alpha}$ with all indices displayed.
The single particle Hamiltonian $\hat{h}$ in Eq.~(\ref{ACLatHamCompact}) has 
the sublattice flavor space decomposition
\begin{equation}\label{ACSingPartHam}
	\hat{h} = 
	\begin{bmatrix}
	\hat{t}_{AA}^{\phantom{\dagger}} & \hat{t}_{AB}^{\phantom{\dagger}} \\
	\hat{t}_{AB}^{\dagger} & \hat{t}_{BB}^{\phantom{\dagger}} 
	\end{bmatrix}.
\end{equation}
If we assume that the model in Eq.~(\ref{ACLatticeHam}) resides upon a 
$d$-dimensional bipartite lattice of $2N$ sites, then $\hat{h}$ 
is a $4N\times4N$ Hermitian matrix.

In the presence of disorder, the elements of $\hat{h}$ will be random variables.
The structure of $\hat{h}$ can then be classified\cite{Zirnbauer,AltlandZirnbauer} 
using random matrix theory (RMT), according to the constraints imposed by 
any symmetries preserved in every realization of the static disorder. Consider 
the \emph{clean} case of real, homogeneous, spin-independent nearest-neighbor 
hopping on a bipartite lattice at half-filling. We will refer to this situation 
as pure ``sublattice'' hopping. Pure sublattice hopping corresponds to 
Eq.~(\ref{ACLatticeHam}) with
\begin{equation}\label{ACSublatticeHopping1}
	\hat{t}_{AA}^{\phantom{\dagger}} = \hat{t}_{BB}^{\phantom{\dagger}} = 0,
\end{equation}
and 
\begin{equation}\label{ACSublatticeHopping2}
	(\hat{t}_{AB}^{\phantom{\dagger}})_{i, j}^{\alpha, \beta}
	= t \, \delta_{\langle i, j \rangle} \delta^{\alpha, \beta},
\end{equation}
where $t = t^{*}$, and $\delta_{\langle i, j \rangle}$ gives one for $i$ and $j$ 
nearest-neighbors, vanishing otherwise. We introduce two commuting sets of Pauli 
matrices: the matrix $\hat{\sigma}_{m}$ acts in sublattice flavor space, while the 
matrix $\hat{J}_{n}$ acts in the spin-$1/2$ space, with $m,n \in \{1,2,3\}$. 
We use the conventional basis for all Pauli matrices. In addition to symmetries implied 
by the lattice space group, the single particle Hamiltonian with pure sublattice hopping 
[Eqs.~(\ref{ACSingPartHam}), (\ref{ACSublatticeHopping1}), and 
(\ref{ACSublatticeHopping2})] possesses spin SU(2) rotational invariance
\begin{equation}\label{ACSpinSU(2)}
	\hat{J}_{m} \, \hat{h} \, \hat{J}_{m} = \hat{h},	
\end{equation}
for $m = \{1,2,3\}$, as well as three discrete symmetries: time-reversal 
invariance (TRI), sublattice symmetry (SLS), and particle-hole symmetry (PH), 
defined as invariance under the following operations:
\begin{subequations}
\begin{align}
	c &\rightarrow -i \hat{J}_{2} c,	
		& c^{\dagger} & \rightarrow c^{\dagger} i \hat{J}_{2}  
			&& \textrm{(TRI)}, \label{ACTRIDef}\\
	c &\rightarrow \hat{\sigma}_{3} (c^{\dagger})^{\mathsf{T}},
		& c^{\dagger} & \rightarrow c^{\mathsf{T}} \hat{\sigma}_{3} 
			&& \textrm{(SLS)}, \label{ACSLSDef}\\
	c &\rightarrow -i \hat{J}_{2} \hat{\sigma}_{3} (c^{\dagger})^{\mathsf{T}}, 
		& c^{\dagger} & \rightarrow c^{\mathsf{T}} i \hat{J}_{2} \hat{\sigma}_{3}
			&& \textrm{(PH)}.\label{ACPHDef} 
\end{align}
\end{subequations}
In these equations, ``$\mathsf{T}$'' denotes the matrix transpose operation.
The TRI and SLS operations are antiunitary; the PH operation as defined in Eq.~(\ref{ACPHDef})
is a \emph{product} of TRI and SLS, and is therefore unitary. The PH operation so-defined
is unconventional, in that it involves a spin flip.
Eqs.~(\ref{ACTRIDef})--(\ref{ACPHDef}) equivalently imply the following conditions on $\hat{h}$:
\begin{subequations}
\begin{align}
	\hat{J}_{2} \, \hat{h}^{\mathsf{T}} \, \hat{J}_{2} &= \hat{h}
			&& \textrm{(TRI)}, \label{ACTRIhDef}\\
	-\hat{\sigma}_{3} \, \hat{h} \, \hat{\sigma}_{3} &= \hat{h}
			&& \textrm{(SLS)}, \label{ACSLShDef}\\
	- \hat{J}_{2} \sigma_{3} \, \hat{h}^{\mathsf{T}} \, \hat{J}_{2} \sigma_{3} &= \hat{h}
			&& \textrm{(PH)}.\label{ACPHhDef} 
\end{align}
\end{subequations}
Eq.~(\ref{ACSLShDef}) demonstrates that SLS translates into a ``chiral'' condition on 
$\hat{h}$: invariance under SLS forces $\hat{h}$ to be purely off-diagonal in sublattice 
flavor space (this condition provides an equivalent definition of SLS in the absence of 
interactions).   

For the case of disorder that breaks all of the ``internal'' symmetries detailed in 
Eqs.~(\ref{ACSpinSU(2)}) and (\ref{ACTRIhDef})--(\ref{ACPHhDef}), the only constraint 
placed upon $\hat{h}$ is Hermiticity.\cite{footnote-u}
The most precise statement that we can make, independent of the details of a 
particular disorder realization, is that $\hat{h}$ belongs to a matrix 
representation\cite{footnote-v}
of the Lie algebra u(4N)
of the unitary group:
we say that $\hat{h} \in \textrm{u(4N)}$. 
This places the model in Eq.~(\ref{ACLatticeHam}) into the ordinary ``unitary'' (Wigner-Dyson) 
class A.\cite{Zirnbauer} Adding one or more of the symmetry constraints in Eqs.~(\ref{ACSpinSU(2)}) 
and (\ref{ACTRIhDef})--(\ref{ACPHhDef}) will allow us to refine this statement, potentially 
altering the random matrix class.

Consider first the addition of a random orbital field to the pure sublattice hopping model defined
by Eqs.~(\ref{ACSingPartHam}), (\ref{ACSublatticeHopping1}), and (\ref{ACSublatticeHopping2}).
The presence of a random orbital field modifies only Eq.~(\ref{ACSublatticeHopping2}),
which now reads
\begin{equation}\label{ACSublatticeHopping2ROF}
	(\hat{t}_{AB}^{\phantom{\dagger}})_{i, j}^{\alpha, \beta}
	\rightarrow t e^{i \theta_{i,j}} \, \delta_{\langle i, j \rangle} \delta^{\alpha, \beta},
\end{equation}
with $\theta_{i,j}$ a bond-dependent real phase. Using Eqs.~(\ref{ACSublatticeHopping1})
and (\ref{ACSublatticeHopping2ROF}) in Eq.~(\ref{ACSingPartHam}), we see that $\hat{h}$ 
preserves spin SU(2) rotational symmetry [Eq.~(\ref{ACSpinSU(2)})], breaks TRI and PH 
[Eqs.~(\ref{ACTRIhDef}) and (\ref{ACPHhDef})], and preserves SLS [Eq.~(\ref{ACSLShDef})]. 
Imposing Eq.~(\ref{ACSpinSU(2)}) alone implies that $\hat{h} \in \textrm{u(2N)}$; 
subsequently enforcing the ``anti-SLS'' constraint
\begin{equation}\label{ACAntiSLSDef} 
	\hat{\sigma}_{3} \, \hat{h} \, \hat{\sigma}_{3} = \hat{h}
\end{equation}
further reduces the space to which $\hat{h}$ belongs to $\textrm{u(N)}\times\textrm{u(N)}$.
Since the conditions in Eqs.~(\ref{ACSLShDef}) and (\ref{ACAntiSLSDef}) are complementary, enforcing 
instead spin SU(2) and SLS leads to the identification $\hat{h} \in \textrm{u(2N)}/\textrm{u(N)}\times\textrm{u(N)}$,
associated with the ``chiral'' random matrix class AIII.\cite{Zirnbauer} Thus adding a random 
orbital magnetic field [Eq.~(\ref{ACSublatticeHopping2ROF})] to the spin-$1/2$ pure sublattice 
hopping model yields a system in the same class as the (non-interacting) spinless random hopping model 
[Eqs.~(\ref{Hclean}) and (\ref{Hdis}) with $U = V = 0$], discussed in Secs.~\ref{RandomHopping}
and \ref{NonIntDiscuss} of this paper. 

We may also consider imposing only SLS, i.e.\ removing the spin SU(2) constraint 
given by Eq.~(\ref{ACSpinSU(2)}). This is consistent with the addition of homogeneous spin-orbit coupling
to the orbital magnetic field case analyzed above. Using Eq.~(\ref{ACSLShDef}), we find 
$\hat{h} \in \textrm{u(4N)}/\textrm{u(2N)}\times\textrm{u(2N)}$, so that $\hat{h}$ belongs
to the same 
space of matrices
(and therefore the same random matrix class AIII), 
regardless of whether spin SU(2) rotational symmetry is preserved or destroyed (completely).
As discussed in Secs.~\ref{Spin1/2} and \ref{Conclusion}, the Finkel'stein NL$\sigma$M formulated in 
Sec.~\ref{NLSM}, defined by Eqs.~(\ref{Z})--(\ref{SI}), also applies to a spin-$1/2$ Hubbard model 
with sublattice hopping, subject to a random orbital magnetic field, and possessing spin-orbit 
coupling.

Next we turn to the case of a random Zeeman field. Given the assumption of
pure sublattice hopping [Eqs.~(\ref{ACSingPartHam}), (\ref{ACSublatticeHopping1}), and 
(\ref{ACSublatticeHopping2})], the introduction of a Zeeman field modifies 
Eq.~(\ref{ACSublatticeHopping1}) as follows:
\begin{equation}\label{ACSublatticeHopping1RZF}
\begin{aligned}
	(\hat{t}_{AA}^{\phantom{\dagger}})_{i, i'}^{\alpha, \beta} 
	& \rightarrow \bm{\mathrm{b}}_{A \,i}\cdot\bm{\mathrm{J}}^{\alpha,\beta} \, \delta_{i,i'}, \\
	(\hat{t}_{BB}^{\phantom{\dagger}})_{j, j'}^{\alpha, \beta} 
	& \rightarrow \bm{\mathrm{b}}_{B \,j}\cdot\bm{\mathrm{J}}^{\alpha,\beta} \, \delta_{j,j'}, \\
\end{aligned}
\end{equation}
where $\hat{\bm{\mathrm{J}}} = \{\hat{J}_{m}\} \rightarrow \{J_{m}^{\alpha,\beta}\}$, with 
$m \in \{1,2,3\}$, is a vector of spin-$1/2$ space Pauli matrices, and where
$\bm{\mathrm{b}}_{A \,i}$ and $\bm{\mathrm{b}}_{B \,j}$ denote real three-component vectors
with sublattice site-dependent orientations and magnitudes. Using Eqs.~(\ref{ACSublatticeHopping1RZF})
and (\ref{ACSublatticeHopping2}) in (\ref{ACSingPartHam}), we see that a random Zeeman field
breaks spin SU(2) rotational symmetry [Eq.~(\ref{ACSpinSU(2)})] (completely), TRI [Eq.~(\ref{ACTRIhDef})],
and SLS [Eq.~(\ref{ACSLShDef})], while preserving PH\cite{footnote-w}
[Eq.~(\ref{ACPHhDef})] in every static realization
of disorder. Eq.~(\ref{ACPHhDef}) provides a symplectic condition on $\hat{h}$; in isolation, PH
therefore implies that $\hat{h} \in \textrm{sp(4N)}$.
The 
Lie algebra $\textrm{sp(4N)}$
of the symplectic group
is associated with
the random matrix class C.\cite{Zirnbauer,AltlandZirnbauer} Class C is also realized by a spin-$1/2$ 
fermionic quasiparticle system native to a superconductor with singlet pairing, broken TRI, and spin 
SU(2) rotational symmetry preserved in every realization of 
disorder.\cite{Zirnbauer,AltlandZirnbauer,ClassCsc,VishveshwaraSenthilFisher,JengLudwigSenthilChamon,FabrizioDellAnnaCastellani,SpinQH} 

In summary, given the assumption of a clean system of spin-$1/2$ fermions with pure sublattice hopping 
[Eqs.~(\ref{ACSingPartHam}), (\ref{ACSublatticeHopping1}), and (\ref{ACSublatticeHopping2})], the 
introduction of a random ``orbital'' magnetic field, as in Eq.~(\ref{ACSublatticeHopping2ROF}), gives 
a non-interacting quantum disordered system with SLS, lacking TRI and PH, belonging to the random matrix
class AIII; the introduction of a random Zeeman field, as in Eq.~(\ref{ACSublatticeHopping1RZF}),
gives a system with PH, lacking spin SU(2) rotational symmetry, TRI, and SLS, belonging to class C.
The application of both random orbital and random Zeeman fields to a spin-$1/2$ system with pure sublattice
hopping breaks TRI, SLS, PH, and spin SU(2) rotational symmetry (completely), in which case
the ordinary unitary (Wigner-Dyson) class A is recovered.


\section{Polar superconductor interpretation}\label{SC}

Using the same type of random matrix theory (RMT) symmetry analysis\cite{AltlandZirnbauer} 
applied in Appendix \ref{Spin1/2RHopping} to classify systems of spin-$1/2$ lattice 
fermions, we demonstrate here a completely different ``microscopic'' interpretation of the 
Finkel'stein NL$\sigma$M studied in this paper, which provides a view of our results 
alternative to the spinless Hubbard model physics espoused in the main text.
The microscopic view adopted in this appendix concerns the fermionic quasiparticles 
associated with a particular type of superconductor, subject to quenched disorder.
(These quasiparticles  are initially taken to be non-interacting, within a mean field theory
treatment of pairing; subsequently, quasiparticle interactions
are added.)

Fermionic quasiparticles in a superconductor do not carry well-defined 
quantities of physical electric charge.\cite{Schrieffer} In the presence of 
disorder, this means that electric charge is not 
a ``slow,'' hydrodynamic 
variable; 
a low-energy non-linear sigma model description of such a
disordered non-interacting quasiparticle system
can 
therefore not describe any kind of conventional ``metal-insulator'' transition,
as might be observed experimentally through electric transport measurements. 
Disorder may yet dramatically influence other properties of the system, 
however, including thermal and spin transport coefficients [the 
latter only if 
at least a U(1) subgroup of spin SU(2) rotational symmetry is preserved],
as well as the behavior of the low energy, single particle (tunneling) 
density of states.\cite{footnote-x}
In particular, it is possible in principle to observe a ``thermal metal'' to
``thermal insulator'' transition,\cite{ClassCsc,VishveshwaraSenthilFisher,SpinQH,SenthilFisher}
indicative of Anderson localization of the single quasiparticle states due to the 
disorder.
This transition is described by a random Bogoliubov-De-Gennes equation,
and may possibly
be accessible\cite{VishveshwaraSenthilFisher}
by tuning,
e.g., the impurity concentration in a superconducting 
sample. To build a more realistic model of such a system, one should also include 
the effects of 
interactions\cite{JengLudwigSenthilChamon,FabrizioDellAnnaCastellani,DellAnna} 
between the quasiparticles of the superconductor; 
near $d = 2$, both disorder and interaction 
effects may be reliably described
within the Finkel'stein NL$\sigma$M framework.

In this appendix, we first show that a non-interacting, spin-$1/2$ 
quasiparticle system, native e.g.\ to 
a certain spin-triplet p-wave superconductor
and
subject to quenched disorder satisfying certain additional symmetry constraints, 
falls into the same random matrix class
(the ``chiral'' class AIII in the scheme of Ref.~\onlinecite{Zirnbauer}) 
as the non-interacting spinless random hopping model [Eqs.~(\ref{Hclean}) and 
(\ref{Hdis}) with $U = V = 0$] discussed in Secs.~\ref{RandomHopping} and 
\ref{NonIntDiscuss}. To obtain a
superconductor quasiparticle system in the ``chiral'' class 
AIII, the disorder 
must preserve,
in every static realization,
time-reversal 
invariance (TRI), as well as a remnant U(1) of the spin SU(2) rotational 
symmetry. Such a situation 
is expected to occur ``naturally'' for pure potential scattering 
(i.e.\ non-magnetic impurities,
and no spin-orbit scattering),
in a spin-triplet p-wave 
superconducting host residing in its 
TRI, ``polar'' phase.\cite{VollhardtWoelfle,Ho} 
We
emphasize that,
in contrast to the Hubbard model, 
this quasiparticle system may be defined directly in 
the continuum, without reference to a lattice or an additional sublattice 
symmetry. After demonstrating the equivalence of the non-interacting 
superconductor
quasiparticle and spinless random hopping
(normal) particle, or electron
 systems at the level of RMT, we will
briefly discuss the interpretation of the disorder and interaction parameters 
of the corresponding FNL$\sigma$M [Eqs.~(\ref{Z})--(\ref{SI})], studied in this paper,
in the superconductor quasiparticle context. 

In order to motivate the basic underlying idea, consider
first the
system of quasiparticles in a {\it two-dimensional spinless 
(or: spin-polarized)  $p_x$-}superconductor,\cite{VollhardtWoelfle,CommentMFTofPairing}
which is subject to static short-ranged disorder\cite{NersesyanTsvelikWengerNPB1994}
(in the potential and pair field). 
Weak disorder, acting on the pair of Dirac Fermions
at the two nodal points and preserving the TRI of the $p_x$ superconducting state,\cite{footnote-z}
is known\cite{GLL} to place this system in the (orthogonal) chiral\cite{Zirnbauer}
symmetry class BDI.\cite{FendleyKonik,FendleyKonikComment}
In the low-energy Dirac theory, the disorder occurs in two varieties: 
intra- 
and
internode scattering. 
The intranode randomness\cite{GLL} takes the form of a random U(1)
vector potential.\cite{LudwigFisherShankarGrinstein1994,NersesyanTsvelikWengerNPB1994}
In the $p_x$ superconductor realization,
quenched random U(1) 
vector potential fluctuations correspond 
to small random shifts of the positions of the nodal points,
and thus, in particular, to small random fluctuations of the \emph{orientation} 
of the Cooper pair wavefunction away from the $x$-axis.
Internode scattering appears as a pair of random masses for
the Dirac quasiparticles.\cite{GLL,NersesyanTsvelikWengerNPB1994}
In
the disorder-averaged Dirac theory, it is necessary to specify two parameters $g_{A}$ and 
$g$ 
in order to 
quantify the strength of intra- and internode scattering,
respectively.
As shown in Ref.~\onlinecite{GLL}, $g_{A}$ and $g$ each play a role in the class BDI 
disordered Dirac theory described above that
is 
analogous
to the parameters 
$\lambda_{A}$ and $\lambda$, respectively, in the NL$\sigma$M formulation of the (different) chiral 
disorder class AIII, as defined by Eqs.~(\ref{Z}), (\ref{SD}) and (\ref{QmatrixUnitary2})
in Sec.~\ref{NLSM}.  
We demonstrate below that this 
analogy
reflects a concrete realization of class AIII in terms of quasiparticles 
of a certain p-wave superconductor,
with spin 1/2, rather than
spin 0.
The
strength $g_{A}$ of the 
vector potential randomness in the spinless case is
analogous
to $\lambda_A$ in the sigma model description of the spin-1/2 quasiparticle system. 
Indeed, as we show below,
$\lambda_A$ 
quantifies
 the strength of random fluctuations
of the orientation of the Cooper pair wavefunction in the spin-1/2 p-wave superconductor
discussed below.

We turn now to the random matrix symmetry analysis. Consider a \emph{clean} p-wave, 
spin-triplet Bogoliubov-deGennes (pairing) Hamiltonian of the form
\begin{align}\label{ADPwaveHam}
	H =& \int d^{d} \bm{\mathrm{r}} \,
	c^{\dagger}(\bm{\mathrm{r}})
	\left[
		-\nabla^{2} - \mu
	\right]
	c(\bm{\mathrm{r}})
	\nonumber\\
	& + \frac{1}{2}\int d^{d} \bm{\mathrm{r}} \, d^{d} \bm{\mathrm{r'}} \,
	\left[
	\bm{\mathrm{b}}^{\dagger}(\bm{\mathrm{r}},\bm{\mathrm{r'}}) 
	\cdot \bm{\mathrm{\Delta}}(\bm{\mathrm{r}} - \bm{\mathrm{r'}}) + \textrm{H.c.} 	
	\right],
\end{align}
where we define the spin-triplet 
pair field
operator
\begin{equation}\label{ADPairon}
	\bm{\mathrm{b}}(\bm{\mathrm{r}},\bm{\mathrm{r'}}) \equiv 
	c^{\mathsf{T}}(\bm{\mathrm{r}})
	\, \hat{J}_{2} \, \hat{\bm{\mathrm{J}}}\, 
	c(\bm{\mathrm{r'}}).
\end{equation}
In Eq.~(\ref{ADPwaveHam}), $c^{\dagger}$ and $c$ denote two-component fermion creation and 
annihilation operators, i.e.\ $c \rightarrow c^{\alpha}$, where $\alpha \in \{\uparrow,\downarrow\}$
is a spin-$1/2$ component index; $\mu$ is the chemical potential. In the absence 
of disorder and quasiparticle interactions, the gap $\bm{\mathrm{\Delta}}(\bm{\mathrm{r}})$ 
in Eq.~(\ref{ADPwaveHam}) is a static, spatially antisymmetric vector-valued function. In 
Eq.~(\ref{ADPairon}), the superscript ``$\mathsf{T}$'' denotes the matrix transpose operation, 
and 
we denote the vector of Pauli matrices acting in spin-$1/2$ space by
$\hat{\bm{\mathrm{J}}} = \{\hat{J}_{m}\} \rightarrow \{J_{m}^{\alpha,\beta}\}$,
with $m \in \{1,2,3\}$. 
The 
pair field
operator defined by Eq.~(\ref{ADPairon}) carries electric charge $2 e$, 
and transforms like an O(3) vector 
under SU(2)-spin rotations. 
In the ``polar'' phase of a spin-triplet, p-wave superfluid, the gap function takes 
the form\cite{VollhardtWoelfle,Ho} (ignoring dynamical fluctuations):
\begin{equation}\label{ADGapFunc}
	\bm{\mathrm{\Delta}}(\bm{\mathrm{r}}) = \bm{\mathrm{\zeta}} \, \phi(\bm{\mathrm{r}}),
\end{equation}
where the fixed unit vector $\bm{\mathrm{\zeta}}$ and 
the spatially antisymmetric scalar function 
$\phi(\bm{\mathrm{r}})$ are both purely real. We take $\bm{\mathrm{\zeta}} = (0, 0, 1)$.
The Hamiltonian in Eq.~(\ref{ADPwaveHam}) then possesses time-reversal invariance (TRI) 
and a remnant U(1) of spin rotational symmetry 
(about the $m=3$ axis),
defined as invariance under the 
transformations
\begin{subequations}
\begin{align}
	c &\rightarrow -i \hat{J}_{2} c,	
		& c^{\dagger} & \rightarrow c^{\dagger} i \hat{J}_{2}  
			&& \textrm{[TRI]}, \label{ADTRIDef}\\
	c &\rightarrow e^{i  \hat{J}_{3} \theta } c,
		& c^{\dagger} & \rightarrow c^{\dagger} e^{- i  \hat{J}_{3} \theta} 
			&& \textrm{[spin U(1)]}. \label{ADSpinU(1)Def}
\end{align}
\end{subequations}
The time-reversal operation [Eq.~(\ref{ADTRIDef})] is antiunitary.

We implement the following compact notation:
first, we construct the Majorana spinor
\begin{equation}\label{ADMajDef}
	\chi \equiv 
	\begin{bmatrix}
		c \\
		\hat{J}_{2}\big(c^{\dagger}\big)^{\mathsf{T}}
	\end{bmatrix},
\end{equation}
which carries indices in particle-hole $\{\mathfrak{p}=1,2\}$, spin-$1/2$ 
$\{\alpha = \uparrow, \downarrow\}$, and position space $\bm{\mathrm{r}}$,
i.e.\ $\chi^{\alpha}_{\mathfrak{p} = 1}(\bm{\mathrm{r}}) = c^{\alpha}(\bm{\mathrm{r}})$
and 
$\chi^{\alpha}_{\mathfrak{p} = 2}(\bm{\mathrm{r}}) = 
	[\hat{J}_{2}\big(c^{\dagger}\big)^{\mathsf{T}}]^{\alpha}(\bm{\mathrm{r}})$
with all indices displayed. Note that $\chi$ transforms like $c$ under spin space
transformations [see e.g.\ Eq.~(\ref{ADSpinU(1)Def})]. Next, we introduce a second 
set standard of Pauli matrices $\{\hat{\Pi}_{m}\}$, $m \in \{1,2,3\}$, acting in the 
particle-hole space of $\chi$. The definition Eq.~(\ref{ADMajDef}) implies that
\begin{equation}\label{ADMajDefAdj}
	\chi^{\dagger} = \chi^{\mathsf{T}} i \, \hat{\Pi}_{2} \, \hat{J}_{2}. 
\end{equation}
Using Eqs.~(\ref{ADMajDef}) and (\ref{ADMajDefAdj}), we re-write Eq.~(\ref{ADPwaveHam}) as
\begin{equation}\label{ADHamCompact}
	H = \frac{1}{2} \chi^{\dagger} \hat{h} \chi,
\end{equation}
where we have introduced the single particle Hamiltonian
\begin{equation}\label{ADSingPartHam}
\hat{h} = 
	\begin{bmatrix}
	\hat{t} 
	& \hat{\bm{\mathrm{\Delta}}}\cdot\hat{\bm{\mathrm{J}}} \\	
	\hat{\bm{\mathrm{J}}}\cdot\hat{\bm{\mathrm{\Delta}}}^{\dagger} 
	& - \hat{J}_{2} \, \hat{t}^{\mathsf{T}} \, \hat{J}_{2}
	\end{bmatrix},
\end{equation}
with
\begin{align}
	&&\hat{t} & \rightarrow & 
	& t^{\alpha, \beta}(\bm{\mathrm{r}},\bm{\mathrm{r'}}) 
	= [(- \nabla^{2} - \mu)\delta^{d}(\bm{\mathrm{r}}-\bm{\mathrm{r'}})] 
	\delta^{\alpha,\beta}, 
	\nonumber\\
	&&\hat{\bm{\mathrm{\Delta}}}\cdot\hat{\bm{\mathrm{J}}} & \rightarrow & 
	& \hat{\bm{\mathrm{\Delta}}}(\bm{\mathrm{r}} - \bm{\mathrm{r'}})
	\cdot\hat{\bm{\mathrm{J}}}^{\alpha, \beta}.
\end{align}
($\alpha$ and $\beta$ are spin-$1/2$ component indices.) 

Combining Eqs.~(\ref{ADMajDefAdj}) and (\ref{ADHamCompact}), and noting that
the components of $\chi$ mutually anticommute up to irrelevant additive constants, 
we see that the single particle Hamiltonian $\hat{h}$ may be taken without loss
of generality to satisfy the ``Majorana'' condition
\begin{equation}\label{ADMajCond}
	- \hat{\Pi}_{2} \, \hat{J}_{2} \, \hat{h}^{\mathsf{T}}  \, \hat{J}_{2} \, \hat{\Pi}_{2}
	= \hat{h}.
\end{equation}
The transformations defined by Eqs.~(\ref{ADTRIDef}) and (\ref{ADSpinU(1)Def}) 
may be re-expressed in the $\chi$ language using Eq.~(\ref{ADMajDef}); doing so 
shows that TRI and spin U(1) rotational symmetry translate into the
following conditions upon $\hat{h}$ 
[via Eq.~(\ref{ADHamCompact}) and using Eq.~(\ref{ADMajCond})]:
\begin{subequations}
\begin{align}
	- \hat{\Pi}_{1} \, \hat{h} \, \hat{\Pi}_{1} =& \hat{h} & &\textrm{[TRI]}, \label{ADTRIhDef} \\
	\hat{J}_{3} \, \hat{h} \, \hat{J}_{3} =& \hat{h} & &\textrm{[spin U(1)]}. \label{ADSpinU(1)hDef}
\end{align}
\end{subequations}
Eq.~(\ref{ADTRIhDef}) shows that time-reversal invariance appears as
a ``chiral''\cite{Zirnbauer}
condition  [see Eq.(\ref{ACSLShDef}]
on the Bogoliubov-deGennes single particle Hamiltonian $\hat{h}$.

As discussed in Appendix \ref{Spin1/2RHopping}, in the presence of disorder, 
we may classify the non-interacting quasiparticle system introduced in Eq.~(\ref{ADPwaveHam}) 
by examining the algebraic structure of
the Hamiltonian
 $\hat{h}$ [Eqs.~(\ref{ADHamCompact}) and (\ref{ADSingPartHam})],
arising in the presence of an
arbitrary, static realization of impurities. If we imagine discretizing position space
into a lattice of N sites, then $\hat{h}$ is a $4N\times4N$ Hermitian square matrix. 
In the absence of both TRI and spin U(1) rotational invariance, the only constraint upon
this matrix is the ``Majorana'' condition, Eq.~(\ref{ADMajCond}), which implies
that $\hat{h}$ belongs to a matrix representation of the Lie algebra so(4N)
[in the defining (vector) representation].
We write this condition as $\hat{h} \in \textrm{so(4N)}$. This
space of matrices $\hat{h}$ 
is associated 
with the random matrix class D.\cite{Zirnbauer,AltlandZirnbauer,SenthilFisher}

Let us turn to the situation of interest, a quasiparticle system, subject to (e.g.\ pure
potential) disorder, that preserves the symmetry conditions native to the ``polar''
phase of a p-wave, spin-triplet superconducting 
host\cite{VollhardtWoelfle,Ho}: 
TRI and spin U(1) rotational
invariance. The ``Majorana'' [Eq.~(\ref{ADMajCond})] and spin U(1) [Eq.~(\ref{ADSpinU(1)hDef})]
conditions imply the following decomposition in spin-$1/2$ space:
\begin{equation}\label{ADAIII1}
	\hat{h} = \begin{bmatrix}
			\hat{h}_{2N}^{\phantom{\mathsf{T}}} & 0 \\
			0 & -\hat{\Pi}_{2} \, \hat{h}_{2N}^{\mathsf{T}} \, \hat{\Pi}_{2}
		\end{bmatrix},
\end{equation}
where $\hat{h}_{2N}^{\phantom{\mathsf{T}}} \in \textrm{u(2N)}$. Further imposing the 
``anti-TRI'' condition
\begin{equation}\label{ADAIII2}
	\hat{\Pi}_{1} \, \hat{h} \, \hat{\Pi}_{1} = \hat{h}
\end{equation}
would identify  $\hat{h}_{2N}^{\phantom{\mathsf{T}}} \in \textrm{u(N)}\times\textrm{u(N)}$. 
Eqs.~(\ref{ADTRIhDef}) and (\ref{ADAIII2}) are exact complements, so that 
instead, imposing
the ``Majorana'' condition, spin U(1) rotational invariance, and 
TRI leads to $\hat{h} \in \textrm{u(2N)}/\textrm{u(N)}\times\textrm{u(N)}$. This 
space of matrices
is associated with the ``chiral'' random matrix 
class AIII.\cite{Zirnbauer} 
The same space of matrices
also applies to the non-interacting, spinless random hopping model discussed 
in Secs.~\ref{RandomHopping} and \ref{NonIntDiscuss} of this paper.\cite{Gade,FC}
[See also the discussion below Eq.~(\ref{ACAntiSLSDef}) in Appendix \ref{Spin1/2RHopping}.]

At the level of random matrix theory, we have succeeded in identifying
(i) the system of non-interacting
spin-1/2 quasiparticles (treated within the mean field theory of pairing)
native to a superconducting host and subject to quenched disorder 
preserving TRI and a remnant U(1) of the SU(2) spin rotational symmetry, with
(ii)
the non-interacting limit of the 
spinless Hubbard-like model with random hopping, defined by Eqs.~(\ref{Hclean}) and (\ref{Hdis}) 
with $U = V = 0$. We have shown how such a system might ``naturally'' arise in the context 
of a p-wave, spin-triplet superconductor in its TRI, ``polar'' phase,\cite{VollhardtWoelfle,Ho}
subject to
pure potential scattering due to non-magnetic impurities. 

We can extend the analogy between these two systems to the detailed NL$\sigma$M formulation,
incorporating residual quasiparticle interactions
\`a la Finkel'stein.\cite{Finkelstein}
In contrast to the 2D case of Dirac quasiparticles in the spinless 
$p_x$
superconductor
mentioned
in the paragraph above Eq.~(\ref{ADPwaveHam}), we now restrict our discussion to the case 
$d > 2$, for which our RG calculation
of Sec.~\ref{Results>2D}
predicts a metal-insulator transition
due to the interplay of both disorder and interactions, 
as we will now explain.

In Sec.~\ref{NLSM}, we derived the class AIII Finkel'stein NL$\sigma$M (FNL$\sigma$M),
using the Hubbard-like model [Eqs.~(\ref{Hclean}) and (\ref{Hdis})] as our ``microscopic''
starting point. The action for the FNL$\sigma$M was defined in Sec.~\ref{NLSMSummary}
by Eqs.~(\ref{SD}) and (\ref{SI}). 
In order to understand the relevance of Sec.~\ref{Results>2D}
in the superconductor context,
we need to explain
the
re-interpretation
of the
FNL$\sigma$M parameters appropriate to the p-wave
superconductor quasiparticle
view, which we will now do.\cite{footnote-aa}
The parameter $\lambda$ still plays the role of the ``dimensionless resistance'', proportional
now to the U(1) spin-conductivity, associating positive and 
negative spin-U(1) charges
to particles with `up' and `down' $m=3$-component of spin, 
respectively.\cite{ClassCsc,VishveshwaraSenthilFisher,JengLudwigSenthilChamon} 
The second disorder strength $\lambda_{A}$ was attributed in Sec.~\ref{NLSMSummary} to quenched 
bond dimerization fluctuations in the random (sublattice) hopping model. 
As discussed in the paragraph above Eq.~(\ref{ADPwaveHam}), in the superconductor
$\lambda_{A}$ measures the strength of quenched random orientational 
fluctuations of the p-wave Cooper pairing wavefunction induced
by the disorder 
(the argument is outlined in footnote \onlinecite{footnote-ab}).

Turning to the interaction sector of the FNL$\sigma$M action, Eq.~(\ref{SI}),
we now re-interpret
 the interaction strengths $\Gamma_{s}$ and $\Gamma_{c}$
[Eq.~(\ref{SingletCDWGamma})] 
within the context of the quasiparticles of the superconductor.
The parameter
$\Gamma_{s}$
arises from
the
$S_{z}$-$S_{z}$
component of
spin-triplet\cite{footnote-ac}
interactions (in the particle-hole channel) inherited from the normal Fermi
liquid phase adjacent
to the BCS superconductor, and modifies the effective spin diffusion constant
in the
presence of the interactions.\cite{JengLudwigSenthilChamon} 
The parameter $\Gamma_{c}$, on the other
hand, characterizes the residual interaction in the particle-particle Cooper
channel,
and may be interpreted for $\Gamma_{c} < 0$ as an attractive BCS interaction
in a
different, spin-singlet (e.g.\ s-wave) pairing channel.\cite{footnote-ad}
Summarizing, the interaction sector of the Finkel'stein NL$\sigma$M defined
by Eq.~(\ref{SI})
can be re-expressed as follows:
\begin{align}
    S_{I} =&
    i \sum_{a=1,2} \xi^{a}
    \int dt \, d^{d}{\bm{\mathrm{r}}}
    \begin{aligned}[t]
    \bigg\lgroup&\Gamma_{s} \,
    \left[
    {Q}^{a, a}_{t, t}({\bm{\mathrm{r}}})+
    {Q}^{\dagger \, a, a}_{t, t}({\bm{\mathrm{r}}})
    \right]^{2}
    \\
    &+ \Gamma_{c} \,
    \left[
    {Q}^{a, a}_{t, t}({\bm{\mathrm{r}}})-
    {Q}^{\dagger \, a, a}_{t, t}({\bm{\mathrm{r}}})
    \right]^{2}
    \bigg\rgroup
    \end{aligned}
    \nonumber\\
    =&
    i \sum_{a=1,2} \xi^{a}
    \int dt \, d^{d}{\bm{\mathrm{r}}}
    \begin{aligned}[t]
    \bigg\lgroup&\Gamma_{s} \,
    \Big[S_{Z}^{a}(t,\vex{r})\Big]^2   
    \\
    &+ \Gamma_{c} \,
    \bar{\Delta}_{S}^{a} (t,\vex{r}) \, \Delta_{S}^{a} (t,\vex{r})   
    \bigg\rgroup,
    \end{aligned}
    \label{AD--SI}
\end{align}
where the fields $S_{Z}^{a}(t,\vex{r})$, $\Delta_{S}^{a}(t,\vex{r})$, and
$\bar{\Delta}_{S}^{a}(t,\vex{r})$
may be written in terms of $Q_{t,t}^{a,a}(\vex{r})$ and $Q^{\dagger \, a,
a}_{t, t}(\vex{r})$,
and represent the electron bilinear operators
\begin{align}
    S_{Z}^{a} (t,\vex{r}) &\sim \xi^{a}
    \left[
    \bar{c}_{\uparrow}^{a}(t,\vex{r}) \, c_{\uparrow}^{a}(t,\vex{r})
    -
    \bar{c}_{\downarrow}^{a}(t,\vex{r}) \, c_{\downarrow}^{a}(t,\vex{r})
    \right],\label{ADConsSpin}
    \\
    \Delta_{S}^{a} (t,\vex{r}) &\sim \xi^{a}
    c_{\downarrow}^{a}(t,\vex{r}) \,
    c_{\uparrow}^{a}(t,\vex{r}),\label{ADGap}
    \\
    \bar{\Delta}_{S}^{a} (t,\vex{r}) &\sim \xi^{a}
    \bar{c}_{\uparrow}^{a}(t,\vex{r}) \,
\bar{c}_{\downarrow}^{a}(t,\vex{r}).\label{ADbarGap}
\end{align}
As written, $\Delta_{S}^{a}(t,\vex{r})$ represents the simplest singlet
channel pairing amplitude,
$a\in\{1,2\}$ is the Keldysh index, and $\xi^{a}$ was defined by
Eq.~(\ref{KeldyshFactor}).


\section{Frequency-momentum shell integrals}\label{Integrals}

The frequency-momentum loop integrations needed in the RG calculation described in Sec.~\ref{oneloop} are cataloged in this appendix.
All integrations listed below are taken over a portion of the frequency-momentum shell, [Eq.~(\ref{FMShell}) and Fig.~\ref{FMShellFig}]
\begin{align}\label{FMShellInt}
	\int  \frac{d\omega \, d^{\textrm{2}}\bm{\mathrm{k}}}{(2 \pi)^{\textrm{3}}} 
	&= \frac{1}{2 D (2 \pi)^{2}}\int d\omega \, dx \nonumber\\
	&\equiv \frac{1}{2 D (2 \pi)^{2}}\left[\int_{0}^{\Lambda} d\omega \! \int_{\widetilde{\Lambda}}^{\Lambda} dx
	+\int_{0}^{\Lambda} dx \! \int_{\widetilde{\Lambda}}^{\Lambda} d\omega\right], 
\end{align}
unless stated otherwise. In this equation, we have made the change of integration variables $D \bm{\mathrm{k}}^{2} \equiv x$.
[$D$ is the (heat) diffusion constant defined by Eq.~(\ref{DDef}).]
The ratio of the cutoffs in Eq.~(\ref{FMShellInt}) is given by the expression
\begin{equation}\label{ShellRatioAppendix}
	\frac{\Lambda}{\widetilde{\Lambda}} \approx 1 + 2 dl,
\end{equation} 
with $0 < dl \ll 1$. 

The integrals are as follows:
\begin{align}\label{J0}
	J_{0}(z) 
	\equiv& \frac{1}{2 D (2 \pi)^{2}} 
	\int d\omega \, dx \,      
	\frac{1}{(x - i z \omega)^{2}}\nonumber\\
	=& \frac{1}{2 D (2 \pi)^{2}} \frac{i 2 dl}{z}
\end{align}
\begin{align}\label{J1}
	J_{1}(z,z') 
	\equiv& \frac{1}{2 D (2 \pi)^{2}} 
	\int d\omega \, dx \,      
	\frac{1}{x - i z \omega} \frac{1}{x - i z' \omega}\nonumber\\
	=& \frac{1}{2 D (2 \pi)^{2}} \frac{i 2 dl}{z-z'}\ln\left(\frac{z}{z'}\right)
\end{align}
\begin{align}\label{J2}
	J_{2}(z;z') 
	\equiv& -i \partial_{z} J_{1}(z,z')\nonumber\\
	=&\frac{1}{2 D (2 \pi)^{2}} 
	\int d\omega \, dx \,      
	\omega \frac{1}{(x - i z \omega)^{2}} \frac{1}{x - i z' \omega}\nonumber\\
	=& \frac{1}{2 D (2 \pi)^{2}} \frac{2 dl}{z-z'}\left[\frac{1}{z}-\frac{1}{z-z'}\ln\left(\frac{z}{z'}\right)\right]
\end{align}
\begin{align}\label{J3}
	J_{3}(z;z') 
	\equiv& \frac{1}{D} \partial_{z}[z J_{1}(z,z')]\nonumber\\
	=&\frac{1}{2 (2 \pi D)^{2}} 
	\int d\omega \, dx \,      
	x \frac{1}{(x - i z \omega)^{2}} \frac{1}{x - i z' \omega}\nonumber\\
	=& \frac{1}{2 (2 \pi D)^{2}} \frac{i 2 dl}{z-z'}\left[1 - \frac{z'}{z-z'}\ln\left(\frac{z}{z'}\right)\right]
\end{align}
\begin{align}\label{J4}
	J_{4}(z) 
	\equiv& J_{3}(z;z)\nonumber\\
	=& \frac{1}{2 (2 \pi D)^{2}} 
	\int d\omega \, dx \,       
	x \frac{1}{(x - i z \omega)^{3}}\nonumber\\
	=& \frac{1}{2 (2 \pi D)^{2}} \frac{i 2 dl}{2 z}
\end{align}
\begin{align}\label{J5}
	J_{5}(z;z') 
	\equiv& - \partial_{z} \partial_{z'} J_{1}(z,z')\nonumber\\
	=&\frac{1}{2 D (2 \pi)^{2}} 
	\int d\omega \, dx \,       
	\omega^{2} \frac{1}{(x - i z \omega)^{2}} \frac{1}{(x - i z' \omega)^{2}}\nonumber\\
	=& \frac{1}{2 D (2 \pi)^{2}} \frac{i 2 dl}{(z-z')^{3}}\left[2\ln\left(\frac{z}{z'}\right)+ \frac{z'^{2}-z^{2}}{z z'}\right]
\end{align}
\begin{align}\label{J6}
	J_{6}(z) 
	\equiv& J_{5}(z;z)\nonumber\\
	=&\frac{1}{2 D (2 \pi)^{2}} 
	\int d\omega \, dx \,       
	\omega^{2} \frac{1}{(x - i z \omega)^{4}} \nonumber\\
	=& \frac{-1}{2 D (2 \pi)^{2}} \frac{i 2 dl}{3 z^{3}}
\end{align}
\begin{align}\label{J7}
	J_{7}(z;z') 
	\equiv& -\frac{1}{2} \partial_{z}^{2} J_{1}(z,z')\nonumber\\
	=&\frac{1}{2 D (2 \pi)^{2}} 
	\int d\omega \, dx \,       
	\omega^{2} \frac{1}{(x - i z \omega)^{3}} \frac{1}{(x - i z' \omega)}\nonumber\\
	=& \frac{-1}{2 D (2 \pi)^{2}} \frac{i 2 dl}{2(z-z')^{3}}\nonumber\\
	 & \times\left[2\ln\left(\frac{z}{z'}\right)-\frac{z-z'}{z}(2+\frac{z-z'}{z})\right]
\end{align}
\begin{align}\label{J8}
	J_{8}(z;z') 
	\equiv& J_{2}(z;z)\nonumber\\
	=&\frac{1}{2 D (2 \pi)^{2}} 
	\int d\omega \, dx \,    
	\omega \frac{1}{(x - i z \omega)^{3}} \nonumber\\
	=& \frac{-1}{2 D (2 \pi)^{2}} \frac{2 dl}{2 z^{2}}
\end{align}

\begin{align}\label{J9}
	J_{9}(z)=&\frac{1}{2 D^{2} (2 \pi)^{3}}
	\nonumber\\
	&\times
	\begin{aligned}[t]
	\big[
	&\int_{0}^{\Lambda} d\omega \! \int_{0}^{\Lambda} d\omega' \! \int_{\widetilde{\Lambda}}^{\Lambda} dx
	\\
	&+\int_{0}^{\Lambda} dx \! \int_{0}^{\Lambda} d\omega \! \int_{\widetilde{\Lambda}}^{\Lambda} d\omega'
	\\
	&
	+\int_{0}^{\Lambda} dx \! \int_{0}^{\Lambda} d\omega' \! \int_{\widetilde{\Lambda}}^{\Lambda} d\omega
	\big]
		\begin{aligned}[t]
			&x \frac{1}{(x - i z \omega)^{2}}
			\nonumber\\
			&\times\frac{1}{(x - i z \omega')^{2}}
		\end{aligned}
	\end{aligned}
	\nonumber\\
	=& \frac{-1}{2 D^{2} (2 \pi)^{3}} \frac{2 dl}{z^{2}}
\end{align}

\begin{align}\label{J10}
	J_{10}(z;z') 
	\equiv& \frac{1}{2 i} \partial_{z} J_{3}(z;z')\nonumber\\
	=&\frac{1}{2 (2 \pi D)^{2}} 
	\int d\omega \, dx \,      
	x \, \omega \frac{1}{(x - i z \omega)^{3}} \frac{1}{x - i z' \omega}\nonumber\\
	=& \frac{1}{2 (2 \pi D)^{2}} \frac{2 dl}{2(z-z')^{2}}\nonumber\\
	 & \times\left[-1 - \frac{z'}{z} + \frac{2 z'}{z-z'}\ln\left(\frac{z}{z'}\right)\right]
\end{align}
\begin{align}\label{J11}
	J_{11}(z) 
	\equiv& \frac{1}{3 i} \partial_{z} J_{4}(z)\nonumber\\
	=& \frac{1}{2 (2 \pi D)^{2}} 
	\int d\omega \, dx \,       
	x \, \omega \frac{1}{(x - i z \omega)^{4}}\nonumber\\
	=& \frac{-1}{2 (2 \pi D)^{2}} \frac{2 dl}{6 z^{2}}
\end{align}
\begin{align}\label{J12}
	J_{12}(z;z') 
	\equiv& \frac{1}{3 i} \partial_{z} J_{10}(z;z')\nonumber\\
	=&\frac{1}{2 (2 \pi D)^{2}} 
	\int d\omega \, dx \,      
	x \, \omega^{2} \frac{1}{(x - i z \omega)^{4}} \frac{1}{x - i z' \omega}\nonumber\\
	=& \frac{-1}{2 (2 \pi D)^{2}} \frac{i 2 dl}{6 (z-z')^{3}}\nonumber\\
	 & \times\left[2 + \frac{z'}{z^{2}}(5z-z') - \frac{6 z'}{z-z'}\ln\left(\frac{z}{z'}\right)\right]
\end{align}



\begin{thebibliography}{99}
\bibitem{footnote-b}
	The presence of a Fermi surface with a finite extent, i.e.\ of codimension 1, 
	is a model-dependent detail. If a particular lattice model possesses an 
	extended Fermi surface, then the advent of SLS guarantees that surface is nested 
	[Eq.~(\ref{SLSandNesting})]. See e.g.\ Refs.~\onlinecite{FRADKIN,HWK,BDIpaper} for
	$2\textrm{D}$ counterexamples involving massless Dirac-like Fermi points.
\bibitem{footnote-c}
	The approximate transition temperature $T_{c}$ given by Eq.~(\ref{CDWTemp}) vanishes 
	more slowly with decreasing (absolute) interaction strength than does the usual 
	BCS superconducting temperature.\cite{Schrieffer} This occurs in $d=2$ because the 
	projection of the microscopic $U$ and $V$ density-density interactions 
	[Eq.~(\ref{Hclean})] upon the CDW interaction channel at the Fermi surface is 
	sensitive to the presence of logarithmic van Hove singularities, which occur 
	in the density of states at half filling. These van Hove singularities are 
	associated with the sharp corners of the Fermi line as it touches the zone 
	boundary. [See Fig.~\ref{BZandNesting}.]
\bibitem{footnote-d} 
	Note that for $U<0$ in Eqs.~(\ref{Hclean}) and (\ref{CDWcoupling}), we will also 
	have an additional, competing superconducting instability; in this case, although 
	the critical Fermi liquid phase is expected to be unstable, we cannot determine 
	the actual ground state without more detailed arguments.
\bibitem{footnote-e} 
	Invariance of the action in Eqs.~(\ref{S1}) and (\ref{S2}) under the 
	path integral version of the sublattice symmetry transformation prevents 
	the appearance of terms linear in $n_{A i / B j}^{\,a}$ in these equations.
	We take the SLS transformation of the Grassmann fields to be identical in 
	form to the operator-level deformation given by Eq.~(\ref{SLS}), with the 
	replacement $c_{A i / B j}^{\dagger} \rightarrow \bar{c}_{A i / B j}^{\,a}(t)$, 
	etc. Normal-ordering issues are most simply resolved with this convention.
\bibitem{footnote-f}
	In Eq.~(\ref{S10}), we have assumed, for simplicity of presentation, sublattice 
	inversion symmetry: $\varepsilon(-\bm{\mathrm{k}}) = \varepsilon(\bm{\mathrm{k}})$.
\bibitem{footnote-g}
	Moreover, the ``dimer covering'' trick is employed in the text only to yield 
	the simplest possible Hubbard-Stratonovich transformation, Eq.~(\ref{HSdis}), 
	involving the matrix field $\hat{\mathcal{Q}}_{i}$. A more complicated method 
	avoids this trick at the cost of introducing $z_{c}$ such matrix fields, one 
	for each nearest neighbor bond of a given sublattice site. Regardless, the 
	structure of the low-energy, continuum NL$\sigma$M produced is the same, because 
	the saddle point solution [Eq.~(\ref{QCompSP})] is independent of the chosen 
	``dimerization.''
\bibitem{footnote-h}
	Up to trivial anisotropy in the ``stiffness'' term that may be eliminated
	by rescaling; see Appendix \ref{ExpTrick} for details.
\bibitem{footnote-i}
	Recall that, even at the level of the effective field theory, 
	Eqs.~(\ref{SQ})--(\ref{SDET}), the Keldysh construction assures us that the 
	generating functional $Z$ [Eq.~(\ref{EffTheory})] is normalized to one.
\bibitem{footnote-j}
	It can be easily shown that the one loop, Wilsonian background field calculation 
	requires the expansion of Eqs.~(\ref{SDQmatrixSPLeft}) and (\ref{SIQmatrixSPLeft}) 
	only to second order in the $\hat{Y}$s.
\bibitem{footnote-k}
	We could also write the slow mode fields
	$\delta \hat{Q}_{\mathsf{S}}^{\protect\phantom{\dagger}}(\bm{\mathrm{r}}) 
	\equiv i \hat{Y}^{\protect\phantom{\dagger}}_{\mathsf{S}}(\bm{\mathrm{r}})
	+ \bm{\mathit{O}}(\hat{Y}_{\mathsf{S}}^{2})$ and
	$\delta \hat{Q}_{\mathsf{S}}^{\dagger}(\bm{\mathrm{r}}) 
	\equiv - i \hat{Y}^{\protect\phantom{\dagger}}_{\mathsf{S}}(\bm{\mathrm{r}})
	+ \bm{\mathit{O}}(\hat{Y}_{\mathsf{S}}^{2})$,
	where, in analogy to the fast mode parameterization in Eq.~(\ref{YFastParam}), 
	$\hat{Y}^{\protect\phantom{\dagger}}_{\mathsf{S}} = \hat{Y}^{\dagger}_{\mathsf{S}}$ 
	belongs to the Lie Algebra that generates the unitary slow mode field 
	$\hat{1} + \delta \hat{Q}^{\protect\phantom{\dagger}}_{\mathsf{S}}$. 
	Strictly speaking, the parameters $\lambda$, $\lambda_{A}$, $h$, 
	$\Gamma_{U}$, and $\Gamma_{V}$ couple to operators in the slow mode 
	theory [Eq.~(\ref{SSlow})] that are, to lowest non-vanishing order, 
	bilinear in $\hat{Y}^{\protect\phantom{\dagger}}_{\mathsf{S}}$; 
	Eq.~(\ref{FastSlowUnitary}) implies that 
	$\delta \hat{Q}_{\mathsf{S}}^{\protect\phantom{\dagger}} + \delta \hat{Q}_{\mathsf{S}}^{\dagger} 
	= - \delta \hat{Q}_{\mathsf{S}}^{\protect\phantom{\dagger}} \delta \hat{Q}_{\mathsf{S}}^{\dagger}$, 
	so that one must be careful classifying the order of terms polynomial 
	in $\delta \hat{Q}_{\mathsf{S}}^{\protect\phantom{\dagger}}$ and 
	$\delta \hat{Q}_{\mathsf{S}}^{\dagger}$. It is nevertheless useful to 
	work with Feynman vertices expressed in terms of these latter fields, 
	as in Table \ref{VertexTable}, because it allows us to employ diagrammatic 
	rules that distinguish $\delta \hat{Q}_{\mathsf{S}}^{\protect\phantom{\dagger}}$ 
	from its adjoint (as shown in Fig.~\ref{FeynmanRulesVertices}). Our 
	diagrammatic conventions make it easier to understand which diagram 
	corrects which coupling constant in the theory (Figs.~\ref{D1Fig}--\ref{D13Fig}).
\bibitem{footnote-m}
	Similar cancelations must occur at fourth order in 
	$\delta \hat{Q}_{\mathsf{S}}^{\protect\phantom{\dagger}}$ and $\delta \hat{Q}_{\mathsf{S}}^{\dagger}$, 
	although we have not checked this in detail.
	See also footnote \onlinecite{footnote-k}.
\bibitem{footnote-n}
	The interpretation of the operator in Eq.~(\ref{DimerDisorder}) as well as the coupling
	strength $\lambda_{A}$ was discussed in Secs.~\ref{SPGradient} and \ref{NLSMSummary}.
\bibitem{footnote-o} 
	At a critical point, $z$ may be uniquely determined 
	through the condition $d \ln h/dl \equiv 0$.\cite{BK,BDIpaper}
\bibitem{footnoteRGEquationsIn2D}
	Exact expressions for $f_{1}(\lambda)$ and $f_{2}(\lambda)$ 
	to all orders in $\lambda$ were obtained in Ref.~\onlinecite{GLL}, 
	in a particular RG scheme.
\bibitem{footnote-p}
	Note that in Eq.~(\ref{lambdaAFlow2}), the engineering dimension term $-\epsilon \lambda_{A}$ 
	arises through the sum of the first and third terms on the right-hand side of this equation.
\bibitem{footnoteDimensionalRegularization}
	In the Dimensional Regularization with Minimal
	Subtraction RG scheme, the RG equation for $\lambda$ in $d=2+\epsilon$ can be
	obtained in this way from that in $d=2$ dimensions.\cite{ZinnJustin}
	Furthermore, the vanishing of the RG function for $\lambda$ in one scheme
	implies the same in another scheme. (Moreover, $\lambda_A$ cannot feed into the
	RG equation for $\lambda$).
\bibitem{footnote-q}
	The physical interpretation of the coupling strength $\lambda_{A}$ was discussed in 
	Secs.~\ref{SPGradient} and \ref{NLSMSummary}.
\bibitem{footnote-r}
	In the asymptotic low energy regime of the closely related, non-interacting class BDI 
	NL$\sigma$M\cite{Gade,FC} in $d = 2$, appropriate to $2\textrm{D}$ random hopping models 
	for spinless fermions possessing both SLS and TRI,\cite{GLL} a ``freezing'' 
	transition\cite{MDH2D,MRF} occurs in the dynamic scaling relationship between energy 
	and length, yielding a low energy density of states as in Eq.~(\ref{DOSNI}), but with 
	the ``typical'' exponent $\alpha = 2/3$.\cite{DellAnnaMFC} 
	A similar phenomenon will occur in the broken TRI, sublattice symmetric class AIII 
	model studied in the present paper.
	In the present work 
	we do not discuss such a ``frozen'' regime, because we show in 
	Sec.~\ref{Results2D} that the interactions very rapidly flow to strong coupling
	in $2\textrm{D}$ as we renormalize to lower energy scales. We note, however, that 
	if the ``frozen'' regime could be reached before interactions destabilize the 
	non-interacting description, a previous result\cite{BDIpaper} for a disordered 
	Dirac\cite{FRADKIN,HWK,MDH2D,MRF} realization of class BDI suggests that this regime 
	would also be unstable to interaction effects.
\bibitem{footnote-s}
	Sharp band structure features (such as van Hove singularities) appearing in the clean 
	limit are typically smeared out in the presence of arbitrarily weak disorder, 
	as can be seen in the self-consistent Born approximation, Eq.~(\ref{SCBA}).
\bibitem{footnote-t}
	We do not consider here the regime
	$\gamma_{s} < 0$, which corresponds to the condition 
	$U + V < 0$, i.e.\ 
	where at least one of the microscopic interactions is attractive
	in the Hubbard-like model defined by Eqs.~(\ref{Hclean}) and (\ref{Hdis}).
\bibitem{footnote-u}
	An additional constraint implicit in the elements of $\hat{h}$ [Eq.~(\ref{ACSingPartHam})]
	is some notion of locality, i.e.\ that the connectivity matrices 
	$(\hat{t}_{AB}^{\protect\phantom{\dagger}})_{i, j}^{\alpha, \beta}$, etc.\ describe short-ranged
	hopping, so that the tight-binding Hamiltonian in Eq.~(\ref{ACLatticeHam}) has a well-defined
	``embedding'' dimensionality. At the level of the RMT classification, however, this additional 
	constraint has no consequences.
\bibitem{CommentMFTofPairing}
	The 2D superconductor is treated within the Mean Field Approximation of pairing, 
	and formed from a Fermi liquid with circular Fermi surface.
\bibitem{footnote-z}
	TRI dictates that the Dirac nodes always appear in pairs.
\bibitem{footnote-aa}
	In the superconductor context, the FNL$\sigma$M may be derived by
	embedding the clean Hamiltonian in Eqs.~(\ref{ADPwaveHam}) and
	(\ref{ADHamCompact}), augmented with a source of quenched disorder 
	(e.g.\ pure potential scattering) consistent with TRI and spin U(1) 
	symmetries [Eqs.~(\ref{ADTRIhDef}) and (\ref{ADSpinU(1)hDef})],
	in a Keldysh path integration analogous to that defined by
	Eqs.~(\ref{Zmicro}) and (\ref{S1}). Residual quasiparticle interactions 
	are encoded as in Eq.~(\ref{S2}). 
	(See also the discussion at the end of Appendix~\ref{SC}.)
	The structure of the FNL$\sigma$M is 
	almost entirely determined by Eq.~(\ref{QmatrixUnitary2}), which in turn 
	follows from the symmetry structure of the Keldysh action, as detailed in
	Appendix~\ref{KeldyshSym}. A corresponding analysis in terms of
	transformations upon the Majorana spinor in Eq.~(\ref{ADMajDef}) gives the 
	same constraint.
\bibitem{footnote-ab}
	For the p-wave superconductor, a gradient expansion very similar to that 
	presented in Appendix \ref{ExpTrick} shows that an antisymmetric gap function 
	$\phi(\bm{\mathrm{r}})$, as in Eq.~(\ref{ADGapFunc}), with a fixed, homogeneous 
	orbital orientation, specified by a unit vector $\bm{\mathrm{n}}$, leads to a 
	perturbation of the type in Eq.~(\ref{SDETdim}). [C.f.\ Eqs.~(\ref{phidimSym}), 
	(\ref{phiPointer}), and (\ref{phiPointerExp}) in Appendix \ref{ExpTrick}.] Following 
	the same logic used in Sec.~\ref{SPGradient}, we arrive at the stated interpretation 
	of $\lambda_{A}$.
\bibitem{footnote-ac}
	In the ``polar'' p-wave state\cite{VollhardtWoelfle,Ho},
	spin SU(2) rotational symmetry is broken down to a U(1) subgroup;
	as a result, spin triplet interactions will typically contain
	anisotropy in spin space, being either easy-axis (z) or easy-plane
	(xy)-dominated;
	in other words, there will be two interaction coupling constants. In the
	diffusive FNL$\sigma$M description of the disordered system, however,
	only
	the easy-axis interactions enter (up to irrelevant perturbations),
	because easy-axis spin is conserved, in analogy to the situation with 
	uniaxial spin symmetry discussed in Ref.~\onlinecite{JengLudwigSenthilChamon}.
\bibitem{footnote-ad}
	The spin singlet nature of the residual pairing channel characterized
	by $\Gamma_c$ is encoded at the coarse level of the FNL$\sigma$M;
	other details, esp.\ the orbital pairing symmetry, are not.
\end{thebibliography}
\end{document}